\documentclass[epj]{svjour}

\usepackage{longtable}
\usepackage{graphicx}% Include figure files
\usepackage{dcolumn}% Align table columns on decimal point
\usepackage{bm}% bold math
\usepackage{xcolor}
\usepackage{feynmp-auto}% Diagrams
\usepackage{adjustbox}
\usepackage{footnote}
\usepackage[breaklinks,colorlinks=true,urlcolor=black]{hyperref}
\usepackage[hyphenbreaks]{breakurl}%linebreak for the url in the bibliography
\usepackage{float} 
\usepackage{epsfig}
\usepackage{rotating}
\usepackage{braket}
\usepackage{amssymb}
\usepackage{amsmath}
\usepackage{afterpage}
\usepackage{xspace}

\usepackage{subfigure}
\usepackage{cuted}
\usepackage{cite}
\usepackage{array}
\newcolumntype{M}[1]{>{\centering\arraybackslash}m{#1}} %use m instead of p to get vertical centering as well
\makeatletter
\newsavebox{\@brx}
\newcommand{\llangle}[1][]{\savebox{\@brx}{\(\m@th{#1\langle}\)}%
  \mathopen{\copy\@brx\mkern2mu\kern-0.9\wd\@brx\usebox{\@brx}}}
\newcommand{\rrangle}[1][]{\savebox{\@brx}{\(\m@th{#1\rangle}\)}%
  \mathclose{\copy\@brx\mkern2mu\kern-0.9\wd\@brx\usebox{\@brx}}}
\makeatother
\usepackage{empheq}
\usepackage{cancel}
\usepackage{pgfplots}
\usetikzlibrary{plotmarks}
\usetikzlibrary{shapes.geometric}
\pgfdeclareplotmark{mystar}{
    \node[star,star point ratio=2.25,minimum size=4pt,
          inner sep=0pt,draw=violet,solid,fill=violet] {};
}
\pgfdeclareplotmark{mystaro}{
    \node[star,star point ratio=2.25,minimum size=4pt,
          inner sep=0pt,draw=violet,solid,fill=white] {};
}

\newenvironment{customlegend}[1][]{%
    \begingroup
    \csname pgfplots@init@cleared@structures\endcsname
    \pgfplotsset{#1}%
}{%
    \csname pgfplots@createlegend\endcsname
    \endgroup
}%
\def\addlegendimage{\csname pgfplots@addlegendimage\endcsname}

\newcommand{\addlegendimageintext}[1]{%
    \tikz {
        \begin{customlegend}[
            legend entries={\empty},
            legend style={
                draw=none,
                inner sep=0pt,
                column sep=0pt,
                nodes={inner sep=0pt}}]
        \addlegendimage{#1}
        \end{customlegend}
    }%
}

\newcommand{\padeborel}{\!
  \addlegendimageintext{color = gray, mark = *, mark options = {scale=1, solid}, only marks, line width = 0.5pt}\!
}
\newcommand{\pbviolet}{\!
  \addlegendimageintext{dotted, color = violet, mark = *, mark options = {scale=1, solid}, line width = 1.0pt}\!
}
\newcommand{\pbgreen}{\!
  \addlegendimageintext{dashed,color = green!40!gray, mark = *, mark options = {scale=1, solid},  line width = 1.0pt}\!
}
\newcommand{\pborange}{\!
  \addlegendimageintext{dashdotted, color = orange, mark = *, mark options = {scale=1, solid, fill=white}, line width = 1.0pt}\!
}
\newcommand{\cmviolet}{\!
  \addlegendimageintext{dotted, color = violet, mark = mystar, mark options = {scale=1, solid}, line width = 1.0pt}\!
}
\newcommand{\cmvioleto}{\!
  \addlegendimageintext{dotted, color = violet, mark = mystaro, mark options = {scale=1, solid, fill=white}, line width = 1.0pt}\!
}

\newcommand{\mgviolet}{\!
  \addlegendimageintext{dotted, color = violet, mark = diamond*, mark options = {scale=1, solid, fill=violet}, line width = 1.0pt}\!
}
\newcommand{\mgvioleto}{\!
  \addlegendimageintext{dotted, color = violet, mark = diamond*, mark options = {scale=1, solid, fill=white}, line width = 1.0pt}\!
}

\title{Addressing energy density functionals in the language of path-integrals I:\\Comparative study of diagrammatic techniques applied to the (0+0)\nobreakdash-D $O(N)$-symmetric $\varphi^{4}$-theory}

\titlerunning{Addressing energy density functionals in the language of path-integrals I}
\authorrunning{K. Fraboulet and J.-P. Ebran}

\author{
	Kilian Fraboulet\inst{1,2,3}
	\and
	Jean-Paul Ebran\inst{1,2}
}

\institute{
  CEA, DAM, DIF, 91297 Arpajon, France
  \and
  Universit\'e Paris-Saclay, CEA, Laboratoire Mati\`ere en Conditions Extr\^emes, 91680, Bruy\`eres-le-Ch\^atel, France
  \and
  Institut f\"ur Theoretische Physik and Center for Quantum Science, Universit\"at T\"ubingen, Auf der Morgenstelle 14, 72076 T\"ubingen, Germany
}

\date{\today}

\abstract{
The energy density functional (EDF) method is currently the only microscopic theoretical approach able to tackle the entire nuclear chart. Nevertheless, it suffers from limitations resulting from its empirical character and deteriorating its reliability. This paper is part of a larger program that aims at formulating the EDF approach as an effective field theory (EFT) in order to overcome these limitations. A relevant framework to achieve this is the path-integral (PI) formulation of quantum field theory (QFT). The latter indeed provides a wide variety of treatments of the many-body problem well suited to deal with non-perturbative interactions and to exploit a Lagrangian resulting from an EFT as a starting point. While developing the formalism in a general setting, we present below a comparative study of such techniques applied to a toy model, i.e. the (0+0)\nobreakdash-D $O(N)$-symmetric $\varphi^{4}$-theory. More specifically, our focus will be on the following diagrammatic techniques: loop expansion (LE), optimized perturbation theory (OPT) and self-consistent perturbation theory (SCPT). With these methods, we notably address the spontaneous breakdown of the $O(N)$ symmetry with care especially since spontaneous symmetry breakings (SSBs) play a paramount role in current implementations of the EDF approach.
}

\begin{document}
\maketitle

\sloppy %more flexible spacing between words to better break lines and ovoid overlapping between columns
\unitlength=1mm %determine the unit size of the diagrams

%%%%%%%%%%%%%%%%%%%%%%%%%%%%%%%
%%%%%%%%%%%%%%%%%%%%%%%%%%%%%%%
\section{\label{sec:Intro}Introduction}

The current energy density functional (EDF) approach relies on density functionals whose analytical forms are postulated consistently with the symmetries of the two-body nucleon interaction \cite{ben03,sch19}. These functionals depend on the nucleonic density as well as on free parameters fitted on nuclear data. Spontaneous symmetry breaking (SSB) is also at the heart of the EDF approach. It is known to be very efficient for capturing collective correlations in nuclear systems \cite{ben03,sch19}, in particular those related to superfluidity and deformation. Hence, the EDF method can be presented as a two-step procedure: i) A first step corresponding to the single-reference level (i.e. mean-field level) in which SSB is implemented; ii) A second step referred to as multi-reference level (i.e. beyond mean-field level) restoring the symmetries broken by the reference state chosen at the previous step. Using only the single-reference scheme, nuclear physicists have managed to calculate nuclear ground state (gs) observables across the whole nuclear chart since several decades \cite{ben03}, despite the complexity and the diversity of the nuclear phenomenology. Furthermore, multi-reference calculations were successfully exploited to access excited state properties.

It is also important to distinguish between the EDF method and density functional theory (DFT) governed by the Hohenberg-Kohn (HK) theorems \cite{hoh64} and the Kohn-Sham scheme \cite{koh65,koh65bis}. Both rely on density functionals but it should be stressed that EDF and DFT are however two different approaches, mainly for the following two reasons: i)~SSB is widely and efficiently used in the EDF framework but the HK theorems and the Kohn-Sham scheme are only formulated in a symmetry-conserving framework \cite{goe93}; ii)~The existence proofs for the DFT functionals, i.e. the HK theorems, introduced in DFT do not straightforwardly apply in the EDF framework (more specifically, one can actually show that the density functionals resulting from the EDF ansatz do satisfy the HK theorems for a specific choice of coordinates involving so-called internal densities \cite{eng07,bar07,gir08,gir08bis,gir08bis2,mes09,dug10,cha10,mes13}). This is in particular due to the fact that the HK theorems were initially formulated for many-body systems bound under the influence of an external potential. It is therefore not so clear how to formulate DFT for self-bound systems such as those encountered in nuclear physics. This highlights the difficulties of formulating solid theoretical foundations for a nuclear EDF framework.

As a matter of fact, the current EDF framework suffers from severe limitations: i)~No explicit connection with quantum chromodynamics (QCD); ii)~No possibility to systematically improve the obtained results in a reliable manner; iii)~Lack of robust evaluation of theoretical uncertainties. All of these limitations could be overcome by turning the EDF approach into an effective field theory (EFT) \cite{geo93,fur97,fur97bis,van99,mei05,bur07,fur08,epe09,hol16,li17,ham19,ham20,mac20}. To clarify this, we recall that an EFT is a recipe to construct a Lagrangian as a systematic expansion with chosen degrees of freedom (dofs) and symmetry constraints. Imposing the symmetry pattern of QCD to such a Lagrangian would thus enable us to construct a nuclear interaction in direct connection with the dynamics of quarks and gluons. Moreover, uncertainty quantification with Bayesian methods is a very active research area in the framework of EFTs, in nuclear physics in particular \cite{sch09,sch09bis,fur15,wes16}.

Technically, turning the EDF approach into an EFT translates into deriving density functionals from Lagrangians resulting from an EFT of QCD, or, in other words, developing many-body methods that could extract observables from such Lagrangians. A theoretical framework of choice for this task is the path-integral (PI) formulation (or functional integral formulation) of quantum field theory (QFT) \cite{par88,Leb91,neg98,zin02,fri20}. We will therefore follow this path. As the technical reasons for this choice will become more apparent throughout the paper, we can point out at this stage that the variety of the techniques presented below illustrates its relevance.

Reformulating the EDF method on more solid theoretical grounds has been an active area of research for years. Inspiration from DFT has fruitfully been investigated in alternative formulations of nuclear EDFs. A technique called density-matrix expansion \cite{car08,sto10}, inspired from the derivative expansion of DFT, managed to yield density functionals derived from two- and three-nucleon chiral interactions \cite{kai10,hol11}. We can also mention other works aiming at connecting the EDF framework with QCD via EFTs \cite{pug03,bha05,fur07,dru10,fur12,yan17,yan17bis,bou19,fur20}. Among these, we address in more detail the contribution of Furnstahl and collaborators \cite{pug03,bha05,fur07,dru10,fur12,fur20}, whose direction is actually similar to ours as they are also exploiting the PI formalism to turn the EDF approach into an EFT, using notably the 2-particle-point-irreducible (2PPI) effective action (EA) framework \cite{he90,ver92} which possesses all the ingredients of a DFT \cite{fuk94,val97}.

More specifically, our plan for this paper consists in applying different PI techniques to the (0+0)\nobreakdash-D $O(N)$-symmetric $\varphi^{4}$-theory (or (0+0)\nobreakdash-D $O(N)$ model for short), so as to benchmark them according to their ability of resumming correlations within the single channel of this toy model (i.e. the plasmonic channel). We can already justify at this stage the choice of such a toy model for the present study. In the past few years, various many-body techniques (not based on the PI formalism), coined as \textit{ab initio} approaches~\cite{her20}, have been designed to describe atomic nuclei using bare nuclear interactions resulting from EFTs of QCD. These techniques treat both two- and three-nucleon interactions, which is necessary for an accurate quantitative description of the nuclear many-body problem. They are thus based on a Hamiltonian of the form:
\begin{equation}
\begin{split}
H = & \ T_{\text{kin}} + V_{\text{2N}} + V_{\text{3N}} \\
= & \ T_{\text{kin}} + \frac{1}{(2!)^2} \sum_{pqrs} v_{pqrs} c_{p}^{\dagger} c_{q}^{\dagger} c_{s} c_{r} \\
& + \frac{1}{(3!)^2} \sum_{pqrstu} w_{pqrstu} c_{p}^{\dagger} c_{q}^{\dagger} c_{r}^{\dagger} c_{u} c_{t} c_{s} \;,
\end{split}
\label{eq:Hnucl}
\end{equation}
with summation over the relevant (single-particle) indices. $T_{\text{kin}}$ denotes the kinetic part whereas $c^{\dagger}$ and $c$ are fermionic creation and annihilation operators, respectively. Within current applications of these \textit{ab initio} approaches, the matrix elements $v_{pqrs}$ and $w_{pqrstu}$ are typically constructed from a chiral EFT (see e.g. Refs.~\cite{hue20,ent17} for more details on this construction and Refs.~\cite{fro22,fro22bis} for recent applications) and possess fairly rich symmetry properties. For the present study, we consider a model with only a very simple two-body interaction (only a $\varphi^4$-interaction and no three-body interaction), with a much less involved symmetry pattern (only an $O(N)$ symmetry). Furthermore, the chosen model is formulated within a zero-dimensional spacetime (i.e. it is a (0+0)\nobreakdash-D model). There are various reasons why we exploit such a simple model:
\begin{itemize}
\item It should actually be stressed that very few exhaustive comparative studies of PI techniques are available in the literature \cite{hay91}. The simplicity of the chosen (0+0)\nobreakdash-D toy model allows for testing the performances of various methods at low (numerical) costs and therefore makes it a relevant playground to achieve such an exhaustive comparison.
\item It is an exactly solvable model, which means that we have access to an exact solution to benchmark the tested methods.
\item We will also see that the presence of the $O(N)$ symmetry enables us to draw a parallel with the study of mesoscopic systems, and therefore of atomic nuclei. Furthermore, $O(N)$-symmetric models being exploited by several communities in theoretical physics (QCD, statistical physics, ...), the results obtained within this study can be of interest for various research works beyond nuclear theory.
\item Last but not least, the simplicity of the model, and in particular its (0+0)\nobreakdash-D character, enables us to understand and show more readily and more efficiently how the various tested PI techniques overcome the beforementioned limitations of the EDF method (like the impossibility to systematically improve the obtained results in a reliable manner) and how they exploit collective dofs to capture correlations. The role of collective dofs is a cornerstone for our study and our aim to reformulate the EDF approach. Within the latter, densities are exploited via a variational treatment of a density functional representing the energy of the system. Our study illustrates that the PI formalism allows for introducing various collective dofs and exploiting them in numerous ways: this involves notably Hubbard-Stratonovich transformations (HSTs) \cite{str57,hub59} and/or higher-order EAs \cite{lee60,lut60,ded62,ded64,ded64bis,bay62,cor74}.
\end{itemize}
However, after discussing the merits of the chosen model, we note that the performances of the tested PI approaches might differ as compared to those obtained from realistic nuclear interactions, as specified e.g. by the Hamiltonian~\eqref{eq:Hnucl}. There might be various reasons for this (inclusion of new symmetries, competing channels, ...). That said, we can still highlight throughout the present toy model study the predispositions of several families of PI approaches to grasp correlations via the treatment of collective dofs (and possibly the use of resummation theory). The work presented here remains in any case an intermediary step towards more realistic applications.

In section~\ref{sec:0DONmodel}, the studied toy model will be presented in detail, thus clarifying the latter remarks. Throughout this study, we will develop the formalisms of the tested methods in arbitrary dimensions and subsequently take the (0+0)\nobreakdash-D limit for our numerical applications. Hence, section~\ref{sec:Representations} will show regardless of the dimension that the partition function underlying the $O(N)$-symmetric $\varphi^4$-theory can be rewritten exactly in different manners thanks to a HST. From this, we will define the so-called original, mixed and collective representations of the studied model, the last two involving a collective dof. Section~\ref{sec:ComparativeStudy} contains our comparative study where the loop expansion (LE), optimized perturbation theory (OPT) and self-consistent perturbation theory (SCPT) are applied to the original representation of our $O(N)$ model and, when relevant, to its mixed and collective representations as well. It should be stressed that these PI techniques are all diagrammatic in the sense that Wick's theorem \cite{wic50} plays a central role in their formulation. This is not the case of functional renormalization group (FRG) approaches\footnote{We stress that our definition of diagrammatic techniques does not include FRG approaches even though their flow equations are often represented diagrammatically as well.} that will be treated in a subsequent study \cite{fra22}. Furthermore, since some of the tested PI techniques rely on asymptotic series (i.e. the LE and SCPT), resummation methods also play a fundamental role in our study, as will be thoroughly illustrated with our results in section~\ref{sec:ComparativeStudy}. Finally, our concluding remarks are outlined in section~\ref{sec:Conclusion}. Note also that further details are given afterwards in appendices, i.e.:
\begin{itemize}
\item Appendix~\ref{app:GeneratingFuncPIformalism}: an introduction to the PI formalism.
\item Appendix~\ref{app:1overNexpansion0DON}: an application of the $1/N$-expansion to the (0+0)\nobreakdash-D $O(N)$-symmetric $\varphi^{4}$-theory, already investigated in previous works \cite{sch94,kei12,ros16}, that we include in our comparative study.
\item Appendix~\ref{app:Resum}: a presentation of the formalism underlying the different resummation techniques (Pad\'e-Borel \cite{pad1892,bor28,ell96,ben99,kle01}, conformal mapping \cite{leg80} and Borel-hypergeometric \cite{mer15,mer16,ped16a,ped16b,san17,mer18,ant19} resummations) exploited in this study.
\item Appendix~\ref{app:Transseries}: a derivation of the transseries representation of the partition function of the (0+0)\nobreakdash-D $\varphi^{4}$-theory from perturbation theory based on a Lefschetz thimbles decomposition.
\item Appendix~\ref{app:SCPT1PIEA}: an application of SCPT based on 1PI EAs to the (0+0)\nobreakdash-D $O(N)$-symmetric $\varphi^{4}$-theory.
\item Appendix~\ref{app:Diagrams}: a discussion on the determination of the diagrams generated through Wick's theorem.
\end{itemize}
We stress as well that the presented study is part of a PhD project and we refer to the corresponding PhD thesis~\cite{fra21} for further details on the performed calculations.

%%%%%%%%%%%%%%%%%%%%%%%%%%%%%%%
%%%%%%%%%%%%%%%%%%%%%%%%%%%%%%%
\section{\label{sec:0DONmodel}(0+0)-D $O(N)$-symmetric $\varphi^{4}$-theory}

QFTs formulated in zero dimension feature a base manifold $\mathcal{M}$ reducing exactly to one point, $\mathcal{M} = \lbrace\bullet\rbrace$. All fields living on $\mathcal{M} = \lbrace\bullet\rbrace$ are completely specified by assigning a number (e.g. a real one) at this one point, such that the PI measure $\mathcal{D}\widetilde{\varphi}$ reduces to the standard Lebesgue measure $d\widetilde{\varphi}$. The tremendous simplifications brought by the latter feature explain why (0+0)\nobreakdash-D QFTs serve as safe, more controllable and hence useful didactic playgrounds for exploring various aspects of more complicated QFTs, as they allow for explicit solutions that can not be obtained in higher dimensions \cite{cai74,dos75,zin81,ben92,sch94,ban97,hoo99,ben00,mal01,kei12,kem13,bro15,ros16,lia18}. The dofs of the (0+0)\nobreakdash-D $O(N)$ model are represented by real fluctuating fields $\widetilde{\varphi}_a: \lbrace\bullet\rbrace\rightarrow\mathbb{R}$ living on the base manifold $\mathcal{M} = \lbrace\bullet\rbrace$, i.e. real random variables, with $O(N)$-symmetric quartic self-interaction. We store them in the $O(N)$ scalar multiplet:
\begin{equation}
\vec{\widetilde{\varphi}} \equiv \begin{pmatrix}
\widetilde{\varphi}_{1} \\
\vdots \\
\widetilde{\varphi}_{N}
\end{pmatrix} \;,
\end{equation}
and consider the Lie group action $O(N) \curvearrowright \mathbb{R}^N$ defined by left multiplication, with the infinitesimal transformation given by:
\begin{equation}
\delta_\epsilon \widetilde{\varphi}_{a} = \epsilon_{a b} \widetilde{\varphi}^{b} \;,
\end{equation}
characterized by real antisymmetric matrices (i.e. $\epsilon_{a b} = -\epsilon_{b a} \in\mathbb{R}$) and the indices $a$, $b$, ... (which label the $N$ orthogonal directions in the color space defined on $\mathbb{R}^N$) are referred to as color indices. Note that summation (from $1$ to $N$) over repeated color indices is always implied in our notations. The dynamics of the system is governed by the classical action $S:\mathbb{R}^N\rightarrow\mathbb{R}$, given by the expression:
\begin{equation}
S\Big(\vec{\widetilde{\varphi}}\Big)=\frac{m^{2}}{2}\vec{\widetilde{\varphi}}^{2}+\frac{\lambda}{4!}\left(\vec{\widetilde{\varphi}}^{2}\right)^{2} \;,
\label{eq:S0D}
\end{equation}
which is invariant under transformations of the $O(N)$ group, and where the 
real parameters $m^2$ and $\lambda$ stand for the bare squared mass and bare coupling constant, respectively. No kinetic terms contribute to this action owing to the (0+0)\nobreakdash-D nature of spacetime. In the present work, we notably consider two sources with $O(N)$ group structure, namely the local source $\vec{J}$ whose components $J_a$ are coupled to the fields $\widetilde{\varphi}_a$ and the bilocal source $\boldsymbol{K}$ whose elements $\boldsymbol{K}_{a b}$ are coupled to the composite fields $\widetilde{\varphi}^{a}\widetilde{\varphi}^{b}$. In this context, the Schwinger functional of the theory is given by:
\begin{equation}
Z\Big(\vec{J},\boldsymbol{K}\Big) = e^{\frac{1}{\hbar}W\big(\vec{J},\boldsymbol{K}\big)} = \int_{\mathbb{R}^N} d^N\vec{\widetilde{\varphi}} \ e^{-\frac{1}{\hbar}S_{JK}\big(\vec{\widetilde{\varphi}}\big)} \;,
\label{eq:ZJK}
\end{equation}
with
\begin{equation}
S_{JK}\Big(\vec{\widetilde{\varphi}}\Big)\equiv S\Big(\vec{\widetilde{\varphi}}\Big)-\vec{J}\cdot\vec{\widetilde{\varphi}}-\frac{1}{2} \widetilde{\varphi}^{a} \boldsymbol{K}_{a b} \widetilde{\varphi}^{b} \;.
\label{eq:SJK0D}
\end{equation}
where the symbol ``$\cdot$'' refers to the scalar product in color space defined as:
\begin{equation}
X \cdot Y \equiv X_{a} Y^{a} \;.
\label{eq:ScalarProduct0DON}
\end{equation}

\begin{figure}[!t]
  \begin{center}
      \includegraphics[width=0.95\linewidth]{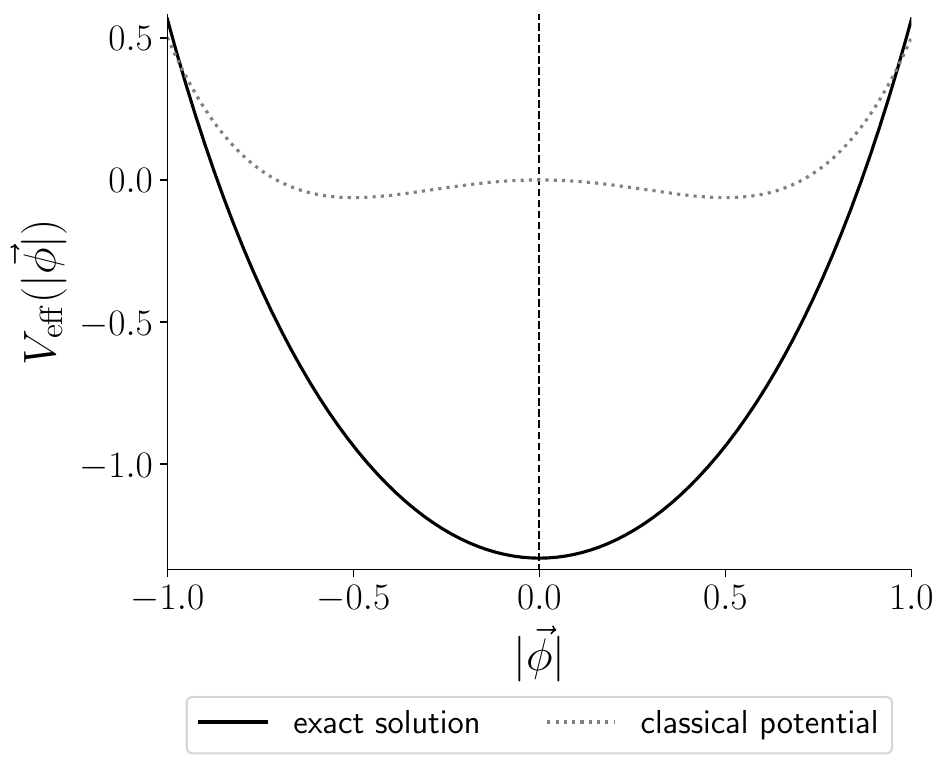}
      \includegraphics[width=0.95\linewidth]{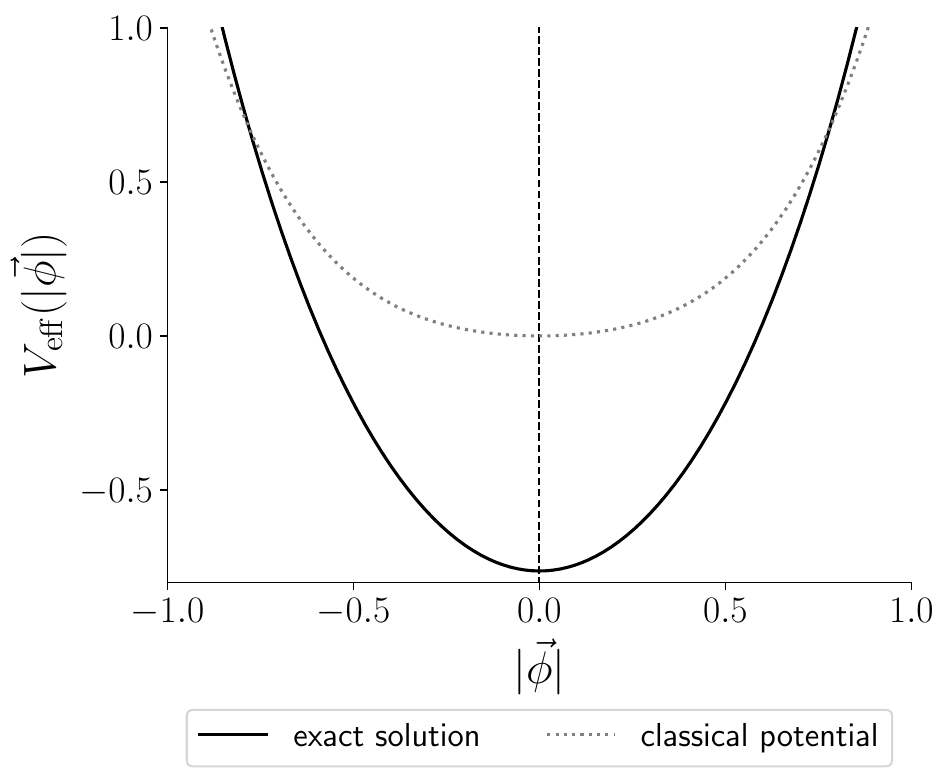}
    \caption{Classical and exact effective potentials as functions of the background constant field's modulus $\left|\vec{\phi}\right|$ at $N=2$, $\lambda/4! = 1$ and $m^2=-1$ (upper panel) or $m^2=+1$ (lower panel).}
    \label{fig:Exactpot}
  \end{center}
\end{figure}

The benefit of working in a (0+0)\nobreakdash-D spacetime is already manifest from the fact that expression~\eqref{eq:ZJK} admits an analytical representation \cite{kei12,ros16} in terms of the Kummer confluent hypergeometric function ${}_{1}F_{1}(a;b;z)$ \cite{abr65}. After rewriting the integral of Eq.~\eqref{eq:ZJK} in hyperspherical coordinates, the exact partition function reads:
\begin{equation}
\begin{split}
Z^\text{exact}\Big(\vec{J}=\vec{0},\boldsymbol{K}=\boldsymbol{0}\Big) = & \ e^{\frac{1}{\hbar}W^\text{exact}\big(\vec{J}=\vec{0},\boldsymbol{K}=\boldsymbol{0}\big)} \\
= & \ \Omega_{N}\mathcal{R}_{N-1} \;,
\end{split}
\label{eq:Z0DONexactsolution}
\end{equation}
with
\begin{equation}
\begin{split}
& \mathcal{R}_N\big(\hbar ;m^2;\lambda \big) \equiv \int_{0}^{\infty} d\widetilde{u} \ \widetilde{u}^{N} \ e^{-\frac{1}{\hbar}\left( \frac{m^{2}}{2}\widetilde{u}^{2}+\frac{\lambda}{4!}\widetilde{u}^{4}\right)} \\
& = \left\{
\begin{array}{lll}
        \displaystyle{2^\frac{N-1}{2} \left(\frac{\hbar}{m^2}\right)^\frac{N+1}{2}\Gamma\left(\frac{N+1}{2}\right) \hspace{0.1cm} \forall m^2 > 0 ~ \mathrm{and} ~ \lambda=0 \;,} \\
        \\
        \displaystyle{\infty \quad \forall m^2 \leq 0 ~ \mathrm{and} ~ \lambda=0 \;,} \\
        \\
        \displaystyle{2^{\frac{3N-5}{4}} 3^{\frac{N+1}{4}} \left(\frac{\lambda}{\hbar}\right)^{-\frac{N+3}{4}}} \\
      \displaystyle{\times\Bigg[\sqrt{\frac{\lambda}{\hbar}}\Gamma\bigg(\frac{N+1}{4}\bigg) {}_{1}F_{1}\bigg(\frac{N+1}{4};\frac{1}{2};\frac{3 m^{4}}{2\lambda\hbar}\bigg)}  \\
      \displaystyle{- \frac{m^{2}\sqrt{6}}{\hbar}\Gamma\bigg(\frac{N+3}{4}\bigg){}_{1}F_{1}\bigg(\frac{N+3}{4};\frac{3}{2};\frac{3 m^{4}}{2\lambda\hbar}\bigg)\Bigg] \hspace{0.1cm} \forall \lambda > 0 \;,}
    \end{array}
\right.
\end{split}
\label{eq:RN}
\end{equation}
for $N\in \mathbb{N}^*$, with $\Gamma(z)$ being Euler gamma function \cite{abr65}. Note also that $\Omega_{N}$ denotes the surface area of the $N$-dimensional unit sphere:
\begin{equation}
\Omega_{N}=\frac{2\pi^{\frac{N}{2}}}{\Gamma\big(\frac{N}{2}\big)} \;.
\label{eq:SN}
\end{equation}
The gs energy and density\footnote{Note that the denomination ``density'' is abusive in the case where $N=1$. Indeed, the $O(N)$ model does not exhibit any continuous symmetry in this situation, hence no conserved Noether current.} can be obtained from:
\begin{equation}
\begin{split}
E_{\mathrm{gs}} = & -\ln\Big( Z\Big(\vec{J}=\vec{0},\boldsymbol{K}=\boldsymbol{0}\Big)\Big) \\
= & - \frac{1}{\hbar} W\Big(\vec{J}=\vec{0},\boldsymbol{K}=\boldsymbol{0}\Big) \;,
\end{split}
\label{eq:DefEgsExactZexact0DON}
\end{equation}
\begin{equation}
\rho_{\mathrm{gs}} = \frac{1}{N}\left\langle\vec{\widetilde{\varphi}}^{2}\right\rangle = -\frac{2}{N} \left.\frac{\partial W\big(\vec{J},\boldsymbol{K}\big)}{\partial m^{2}}\right|_{\vec{J}=\vec{0}\atop\boldsymbol{K}=\boldsymbol{0}} \;,
\label{eq:DefrhogsExactwithExpectationValue0DON}
\end{equation}
with the expectation value defined as:
\begin{equation}
\big\langle \cdots \big\rangle \equiv \frac{1}{Z\big(\vec{J}=\vec{0},\boldsymbol{K}=\boldsymbol{0}\big)} \int d^{N}\vec{\widetilde{\varphi}} \ \cdots \ e^{- \frac{1}{\hbar} S\big(\vec{\widetilde{\varphi}}\big)}\;.
\label{eq:vacuumExpectationValue0DON}
\end{equation}
From these definitions, one can infer the exact solutions:
\begin{equation}
E^{\mathrm{exact}}_{\mathrm{gs}} = -\ln\big(\Omega_{N}\mathcal{R}_{N-1}\big) \;,
\label{eq:ExactSolutionEgs0DON}
\end{equation}
\begin{equation}
\rho^{\mathrm{exact}}_{\mathrm{gs}} = \frac{\mathcal{R}_{N+1}}{N\mathcal{R}_{N-1}} \;.
\label{eq:ExactSolutionRhogs0DON}
\end{equation}
On the other hand, the computation of the 1-point correlation function:
\begin{equation}
\vec{\overline{\phi}} \equiv \left\langle\vec{\widetilde{\varphi}}\right\rangle = \left.\frac{\partial W\big(\vec{J},\boldsymbol{K}\big)}{\partial\vec{J}}\right|_{\vec{J}=\vec{0}\atop\boldsymbol{K}=\boldsymbol{0}} \;,
\label{eq:DefPhiExactwithExpectationValue0DON}
\end{equation}
or of the effective potential $V_{\mathrm{eff}}\big(\vec{\phi}\big)$ provides information on the occurrence of SSB \cite{col73}. While the exact solution for the former reduces to $\vec{\overline{\phi}} \hspace{0.01cm} \rule{0cm}{0.35cm}^{\mathrm{exact}} = 0$ for all values of the coupling constant $\lambda$ and of the squared mass $m^2$, the latter derives from the 1-particle-irreducible (1PI) EA according to:
\begin{equation}
V_{\mathrm{eff}}\Big(\vec{\phi}\Big) = \Gamma^{(\mathrm{1PI})}\Big(\vec{\phi}\Big) = -W\Big(\vec{J},\boldsymbol{K}=\boldsymbol{0}\Big) + \vec{J}\cdot\vec{\phi} \;,
\label{eq:DefVeff0DON}
\end{equation}
with
\begin{equation}
\vec{\phi} = \left.\frac{\partial W\big(\vec{J},\boldsymbol{K}\big)}{\partial\vec{J}}\right|_{\boldsymbol{K}=\boldsymbol{0}} \;.
\label{eq:Defvecphi0DON}
\end{equation}
The exact effective potential $V_\text{eff}^\text{exact}\big(\vec{\phi}\big)$ is evaluated numerically from Eqs.~\eqref{eq:DefVeff0DON} and~\eqref{eq:Defvecphi0DON}. It is then plotted in Fig.~\ref{fig:Exactpot} for $N=2$ together with the classical potential:
\begin{equation}
U\Big(\vec{\phi}\Big) = \frac{m^2}{2} \vec{\phi}^2 + \frac{\lambda}{4!}\left(\vec{\phi}^{2}\right)^{2} \;,
\label{eq:classpot}
\end{equation}
which coincides with the classical action~\eqref{eq:S0D} due to the absence of kinetic terms in the present (0+0)\nobreakdash-D framework.

%%%%%%%%%%%%%%%%%%%
%%%%%%%%%%%%%%%%%%%

The lowest energy states of the system at the classical (i.e. at the tree) level are given by the minima of the classical potential~\eqref{eq:classpot}. In the situation where the coupling constant $\lambda$ is real and zero or positive (which is a restriction followed to obtain all numerical results presented in this study), we can show by minimizing the classical potential~\eqref{eq:classpot} that these are:
\begin{itemize}
\item For $m^2>0$, a unique vacuum $\vec{\phi}=\vec{0}$ where the $O(N)$ symmetry is conserved (spontaneously as well as explicitly).
\item For $m^2<0$ and $\lambda\neq 0$, a manifold of degenerate vacua (satisfying $\vec{\phi}^2=-3! m^2/\lambda > 0$) where the original $O(N)$ symmetry is spontaneously (but not explicitly) broken down to $O(N-1)$.
\end{itemize}
In the full theory, the lowest energy states are found by minimizing the exact effective potential $V_\text{eff}^\text{exact}\big(\vec{\phi}\big)$, which yields a unique gs conserving the $O(N)$ symmetry (spontaneously and explicitly), irrespectively of the sign of $m^2$ as shown by Fig.~\ref{fig:Exactpot}. This absence of broken symmetry in the exact solution of the toy model under consideration is consistent with the Mermin-Wagner theorem\footnote{See appendix~B of Ref.~\cite{koe22} for more details on that point.}~\cite{mer66,hoh67,col73bis} and enables us to make an analogy with the study of mesoscopic systems (and nuclei notably) which do not exhibit any SSB. We will refer to the phase with $m^{2}<0$ ($m^{2}>0$) as broken-symmetry (unbroken-symmetry) regime or phase, even though one must keep in mind that the $O(N)$ symmetry is broken down only spontaneously and only at the classical level. In what follows, the gs energy and density of the chosen toy model will be calculated using the tested PI techniques within each of these two phases (when possible) and subsequently compared with the corresponding exact solution for benchmarking.

%%%%%%%%%%%%%%%%%%%%%%%%%%%%%%%
%%%%%%%%%%%%%%%%%%%%%%%%%%%%%%%
\section{\label{sec:Representations}Different representations of the $O(N)$-symmetric $\varphi^4$-theory}
%==============================
\subsection{\label{sec:OrigRepr}Original representation}

As stressed earlier, most of our derivations are done in arbitrary dimensions even though all numerical applications presented below are performed in (0+0)\nobreakdash-D. We thus consider the finite-dimensional counterpart of the classical action~\eqref{eq:S0D}:
\begin{equation}
\begin{split}
S\Big[\vec{\widetilde{\varphi}}\Big] = & \ \int_x \bigg[ \frac{1}{2}\left(\nabla_x\vec{\widetilde{\varphi}}(x)\right)\cdot\left(\nabla_x\vec{\widetilde{\varphi}}(x)\right) +  \frac{m^2}{2} \vec{\widetilde{\varphi}}^{2}(x)\\
& + \frac{\lambda}{4!}\left(\vec{\widetilde{\varphi}}^{2}(x)\right)^2 \bigg] \;,
\end{split}
\label{eq:SfiniteD}
\end{equation}
which reduces to Eq.~\eqref{eq:S0D} in (0+0)\nobreakdash-D. Note also the shorthand notation for the integration:
\begin{equation}
\int_x \equiv \int_0^{\hbar/T} d\tau \int_{\mathbb{R}^{D-1}}d^{D-1}r \;,
\end{equation}
where $x$ collects both time and space coordinates in $D$-dimensional Euclidean spacetime and $T$ denotes the temperature. Introducing now a local source $\vec{J}(x)$ and a bilocal one $\boldsymbol{K}(x,y)$, a generating functional of the theory is given by:
\begin{equation}
Z\Big[\vec{J},\boldsymbol{K}\Big] = e^{\frac{1}{\hbar}W\big[\vec{J},\boldsymbol{K}\big]} = \int \mathcal{D}\vec{\widetilde{\varphi}} \ e^{-\frac{1}{\hbar}S_{JK}\big[\vec{\widetilde{\varphi}}\big]} \;,
\label{eq:ZJKfiniteD}
\end{equation}
with
\begin{equation}
\int \mathcal{D}\vec{\widetilde{\varphi}} = \int \mathcal{D}\widetilde{\varphi}_{1} \cdots \int \mathcal{D}\widetilde{\varphi}_{N} \;,
\end{equation}
and
\begin{equation}
\begin{split}
S_{JK}\Big[\vec{\widetilde{\varphi}}\Big] \equiv & \ S\Big[\vec{\widetilde{\varphi}}\Big]-\int_x J_a(x)\widetilde{\varphi}^a(x) \\
& -\frac{1}{2}\int_{x,y} \widetilde{\varphi}^a(x)\boldsymbol{K}_{ab}(x,y) \widetilde{\varphi}^b(y) \;.
\end{split}
\label{eq:SJKfiniteD}
\end{equation}
Except for the zero-dimensional situation, the integral in Eq.~\eqref{eq:ZJKfiniteD} can not be computed exactly, and therefore needs to be treated approximatively, e.g. within the expansion schemes discussed subsequently. Action~\eqref{eq:SfiniteD} defines the original representation of our $O(N)$ model. One can treat the latter via other classical actions by means of HSTs, thus defining other representations of this model.

%==============================
\subsection{\label{sec:MixRepr}Mixed representation}

In particular, the following HST is convenient in the studied case:
\begin{equation}
e^{-\frac{\lambda}{\hbar 4!}\int_x{\left(\vec{\widetilde{\varphi}}^{2}(x)\right)^2}} = \sqrt{\frac{6\hbar}{\pi\lambda}} \int \mathcal{D}\widetilde{\sigma} \ e^{-\int_x\left[\frac{6\hbar}{\lambda}\widetilde{\sigma}^2(x)+i\widetilde{\sigma}(x)\vec{\widetilde{\varphi}}^{2}(x)\right]} \;,
\label{eq:HSTONmodelFiniteDim}
\end{equation}
where $\widetilde{\sigma}(x)$ is a collective quantum field and a scalar in color space. From this, the original theory based on $\vec{\widetilde{\varphi}}(x)$ is then transformed into an equivalent one involving e.g. the following generating functional:
\begin{equation}
Z_{\mathrm{mix}}\big[\mathcal{J},\mathcal{K}\big] = e^{\frac{1}{\hbar}W_{\mathrm{mix}}[\mathcal{J},\mathcal{K}]} = \int \mathcal{D}\vec{\widetilde{\varphi}} \mathcal{D}\widetilde{\sigma} \ e^{-\frac{1}{\hbar}S_{\mathrm{mix},\mathcal{J}\mathcal{K}}\big[\vec{\widetilde{\varphi}},\widetilde{\sigma}\big]}\;,
\label{eq:Zmix}
\end{equation}
with
\begin{equation}
\begin{split}
S_{\text{mix},\mathcal{J}\mathcal{K}}\Big[\widetilde{\Psi}\Big] \equiv & \ S_\text{mix}\Big[\widetilde{\Psi}\Big] - \int_{x} \mathcal{J}_{\alpha}(x) \widetilde{\Psi}^{\alpha}(x) \\
& - \frac{1}{2} \int_{x,y} \widetilde{\Psi}^{\alpha}(x) \mathcal{K}_{\alpha\beta}(x,y) \widetilde{\Psi}^{\beta}(y)\;,
\end{split}
\label{eq:SmixJK}
\end{equation}
and
\begin{equation}
\begin{split}
S_\text{mix}\Big[\widetilde{\Psi}\Big] = & \ \frac{1}{2} \int_x \bigg[ \left(\nabla_x\vec{\widetilde{\varphi}}(x)\right)\cdot\left(\nabla_x\vec{\widetilde{\varphi}}(x)\right) \\
& +\left(m^2+i\sqrt{\frac{\lambda}{3}}\widetilde{\sigma}(x)\right)\vec{\widetilde{\varphi}}^{2}(x)+\widetilde{\sigma}^2(x)\bigg] \;,
\end{split}
\label{eq:Smix}
\end{equation}
where summation over repeated Greek indices is implied from $1$ to $N+1$ (which contrasts with color indices labeled by Latin letters). We have just introduced superfields living in a ($N+1$)\nobreakdash-dimensional extended color space aggregating the $N$ components of the original dofs and the collective one, i.e.:
\begin{equation}
\widetilde{\Psi}(x) \equiv \begin{pmatrix}
\vec{\widetilde{\varphi}}(x) \\
\widetilde{\sigma}(x)
\end{pmatrix} \;,
\label{eq:supernotationPsi}
\end{equation}
\begin{equation}
\mathcal{J}(x) \equiv \begin{pmatrix}
\vec{J}(x) \\
 j(x)
\end{pmatrix} \;,
\label{eq:supernotationJ}
\end{equation}
\begin{equation}
\mathcal{K}(x,y) \equiv \begin{pmatrix}
\boldsymbol{K}(x,y) & \vec{0} \\
\vec{0}^{\mathrm{T}} & k(x,y)
\end{pmatrix} \;,
\label{eq:supernotationK}
\end{equation}
where sources $\vec{J}(x)$ and $\boldsymbol{K}(x,y)$ ($j(x)$ and $k(x,y)$) are associated with the original (collective) field sector. Therefore, the mixed system thus obtained involves both $\vec{\widetilde{\varphi}}$ and $\widetilde{\sigma}$ as dofs. In this case, the original quartic interaction between the original dofs in Eq.~\eqref{eq:SfiniteD} has been replaced by a Yukawa interaction between $\vec{\widetilde{\varphi}}$ and the collective field $\widetilde{\sigma}$ in Eq.~\eqref{eq:Smix}, which defines the mixed representation of the studied model.

%==============================
\subsection{\label{sec:CollRepr}Collective representation}

Finally, we will also exploit the fact that the mixed action~\eqref{eq:Smix} is now quadratic in the field $\vec{\widetilde{\varphi}}$, which can therefore be integrated out in Eq.~\eqref{eq:Zmix}, thus leading to:
\begin{equation}
Z_{\mathrm{col}}[\mathcal{J}] = e^{\frac{1}{\hbar}W_{\mathrm{col}}[\mathcal{J}]} = \int \mathcal{D}\widetilde{\sigma} \ e^{-\frac{1}{\hbar}S_{\mathrm{col},\mathcal{J}}[\widetilde{\sigma}]}\;,
\label{eq:Zcoll}
\end{equation}
with
\begin{equation}
\begin{split}
S_{\mathrm{col},\mathcal{J}}[\widetilde{\sigma}] = & \ S_{\mathrm{col}}[\widetilde{\sigma}] - \int_{x} j(x) \widetilde{\sigma}(x) \\
& - \frac{1}{2} \int_{x,y} J^a(x) \boldsymbol{G}_{\widetilde{\sigma};ab}(x,y) J^b(y) \;,
\end{split}
\label{eq:ScollJ}
\end{equation}
and
\begin{equation}
S_{\text{col}}[\widetilde{\sigma}]= \frac{1}{2}\int_x\widetilde{\sigma}^2(x) -\frac{1}{2} \mathrm{STr}\left[\ln(\boldsymbol{G}_{\widetilde{\sigma}})\right] \;,
\label{eq:Scoll}
\end{equation}
where the propagator $\boldsymbol{G}_{\widetilde{\sigma}}$ is given by:
\begin{equation}
\boldsymbol{G}^{-1}_{\widetilde{\sigma};ab}(x,y) = \left(-\nabla_x^2 + m^2 + i\sqrt{\frac{\lambda}{3}}\widetilde{\sigma}(x)\right)\delta_{ab} \delta(x-y) \;,
\label{eq:Gcoll}
\end{equation}
and the supertrace $\mathrm{STr}$ is taken with respect to both color and spacetime indices, i.e. $\mathrm{STr}=\mathrm{Tr}_{a}\mathrm{Tr}_{x}$. Action~\eqref{eq:Scoll} defines what we refer to as the collective representation of our $O(N)$ model, since it only depends on the collective field $\widetilde{\sigma}$. Owing to the exactness of the HST, the generating functionals $Z_{\mathrm{mix}}$ and $Z_{\mathrm{col}}$ both reduce to $Z$ at vanishing sources, ignoring a physically irrelevant numerical factor. The local or bilocal sources involved in Eqs.~\eqref{eq:SJKfiniteD},~\eqref{eq:SmixJK} and~\eqref{eq:ScollJ} are just introduced in accordance with the generating functionals exploited in the comparative study presented in section~\ref{sec:ComparativeStudy}.

%%%%%%%%%%%%%%%%%%%%%%%%%%%%%%%
%%%%%%%%%%%%%%%%%%%%%%%%%%%%%%%
\section{\label{sec:ComparativeStudy}Comparative study}
%==============================
\subsection{\label{sec:LE}Loop expansion}
%~~~
\subsubsection{\label{sec:OrigLE}Original loop expansion}

Many approaches aiming at describing many-body systems rely on a partitioning step, i.e. a splitting of the classical action (or related functionals such as the Hamiltonian) between an unperturbed and a residual part, the unperturbed part corresponding to an exactly solvable system taken usually as leading order of the underlying expansion. This procedure is at the heart of all diagrammatic PI techniques discussed in this study. In the framework of the LE, the partitioning is applied to the classical action in presence of the source(s) by expanding the latter around (one of) its saddle point(s). For the original theory based on Eq.~\eqref{eq:SJKfiniteD}, the saddle point under consideration, denoted as $\vec{\varphi}_\text{cl}(x)$, satisfies:
\begin{equation}
\left.\frac{\delta S_{JK}\big[\vec{\widetilde{\varphi}}\big]}{\delta \vec{\widetilde{\varphi}}(x)}\right|_{\vec{\widetilde{\varphi}}=\vec{\varphi}_\text{cl}} = \vec{0} \quad \forall x \;.
\label{eq:extrSJK}
\end{equation}
We then introduce the reduced Planck's constant $\hbar$ by setting $\vec{\widetilde{\varphi}} = \vec{\varphi}_\text{cl} + \sqrt{\hbar} \ \vec{\widetilde{\chi}}$. From this, the Taylor expansion of $S_{JK}\big[\vec{\widetilde{\varphi}}\big]$ around $\vec{\widetilde{\varphi}}=\vec{\varphi}_\text{cl}$ leads to the following splitting:
\begin{equation}
S_{JK}\Big[\vec{\widetilde{\varphi}}=\vec{\varphi}_\text{cl} + \sqrt{\hbar} \ \vec{\widetilde{\chi}}\Big] = S_{\varphi_\text{cl};JK}^0\Big[\vec{\widetilde{\chi}}\Big] + S_{\varphi_\text{cl};JK}^1\Big[\vec{\widetilde{\chi}}\Big] \;,
\label{eq:PartitioningOriginalLE}
\end{equation}
where
\begin{equation}
\begin{split}
S_{\varphi_\text{cl};JK}^0\Big[\vec{\widetilde{\chi}}\Big] = & \ S_{JK}[\vec{\varphi}_\text{cl}] \\
& + \frac{\hbar}{2}\int_{x,y}\widetilde{\chi}^a(x) \boldsymbol{G}^{-1}_{\varphi_\text{cl};JK;ab}(x,y) \widetilde{\chi}^b(y) \;,
\end{split}
\label{eq:LES1}
\end{equation}
\begin{equation}
\begin{split}
S_{\varphi_\text{cl};JK}^1\Big[\vec{\widetilde{\chi}}\Big] = & \ \frac{\hbar^{\frac{3}{2}}\lambda}{3!}\int_x \vec{\widetilde{\chi}}^{2}(x) \vec{\widetilde{\chi}}(x)\cdot\vec{\varphi}_{\text{cl}}(x) \\
& + \frac{\hbar^{2}\lambda}{4!}\int_x \left(\vec{\widetilde{\chi}}^{2}(x)\right)^{2} \;,
\end{split}
\label{eq:LES2}
\end{equation}
with $\boldsymbol{G}_{\varphi_\text{cl};JK}^{-1}$ being the unperturbed inverse propagator in presence of the sources:
\begin{equation}
\begin{split}
\boldsymbol{G}^{-1}_{\varphi_\text{cl};JK;ab}(x,y) \equiv & \ \left.\frac{\delta^{2} S_{JK}\big[\vec{\widetilde{\varphi}}\big]}{\delta \widetilde{\varphi}^{a}(x)\delta \widetilde{\varphi}^{b}(y)}\right|_{\vec{\widetilde{\varphi}}=\vec{\varphi}_\text{cl}} \\
= & \left(-\nabla_x^2 + m^2 + \frac{\lambda}{6} \vec{\varphi}_\text{cl}^{2}(x)\right)\delta_{ab} \delta(x-y) \\
& + \frac{\lambda}{3}\varphi_{\text{cl};a}(x)\varphi_{\text{cl};b}(x)\delta(x-y) \\
& - \boldsymbol{K}_{ab}(x,y) \;.
\end{split}
\label{eq:LEG}
\end{equation}
Eqs.~\eqref{eq:PartitioningOriginalLE} to~\eqref{eq:LES2} thus set the partitioning for the original LE, i.e. for the LE in the original representation of the studied $O(N)$ model. The role of the unperturbed and residual parts is played respectively by $S_{\varphi_\text{cl};JK}^0$ and $S_{\varphi_\text{cl};JK}^1$ in this case. The LE then consists in incorporating systematically the effects of the residual part $S_{\varphi_\text{cl};JK}^1$ on top of the exactly solvable system defined by the quadratic measure $S_{\varphi_\text{cl};JK}^0$. This translates into a Taylor expansion of the exponential of $S_{\varphi_\text{cl};JK}^1$ in the generating functional~\eqref{eq:ZJKfiniteD} combined with Eqs.~\eqref{eq:PartitioningOriginalLE} to~\eqref{eq:LES2}, thus yielding:
\begin{equation}
\begin{split}
& Z^\text{LE;orig}\Big[\vec{J},\boldsymbol{K}\Big] \\
& = e^{-\frac{1}{\hbar}S_{JK}[\vec{\varphi}_\text{cl}]} \left(\int \mathcal{D}\vec{\widetilde{\chi}} \ e^{-\frac{1}{2}\int_{x,y} \widetilde{\chi}^{a}(x) \boldsymbol{G}^{-1}_{\varphi_\text{cl};JK;ab}(x,y)\widetilde{\chi}^{b}(y)}\right) \\
& \hspace{0.3cm} \times\Bigg[ 1 + \sum^{\infty}_{n=1} \frac{\left(-1\right)^n}{\left(3!\right)^n n!} \sum_{q=0}^n \begin{pmatrix}
n \\
q
\end{pmatrix} \frac{\hbar^\frac{n+q}{2}}{4^q} \\
& \hspace{0.3cm} \times \Bigg\langle \left(\lambda\int_x \vec{\widetilde{\chi}}^2(x)\vec{\widetilde{\chi}}(x)\cdot\vec{\varphi}_\text{cl}(x) \right)^{n-q} \\
& \hspace{0.3cm} \times \left(\lambda \int_x \left(\vec{\widetilde{\chi}}^2(x) \right)^2 \right)^{q}\Bigg\rangle_{0,JK}\Bigg] \;,
\end{split}
\label{eq:ZJKPT3}
\end{equation}
with the source-dependent expectation value defined as:
\begin{equation}
\big\langle\cdots\big\rangle_{0,JK} = \frac{1}{Z_{0}\big[\vec{J},\boldsymbol{K}\big]} \int \mathcal{D}\vec{\widetilde{\chi}} \ \cdots \ e^{-\frac{1}{\hbar}S_{\varphi_\text{cl};JK}^0\big[\vec{\widetilde{\chi}}\big]} \;,
\label{eq:SourceDepExpValueOriginalLE}
\end{equation}
and
\begin{equation}
Z_{0}\Big[\vec{J},\boldsymbol{K}\Big]=\int \mathcal{D}\vec{\widetilde{\chi}} \ e^{-\frac{1}{\hbar}S_{\varphi_\text{cl};JK}^0\big[\vec{\widetilde{\chi}}\big]} \;.
\label{eq:Z0JKoriginalLE}
\end{equation}
An expansion of the Schwinger functional $W^\text{LE;orig}\big[\vec{J},\boldsymbol{K}\big] = \hbar \ln\Big( Z^\text{LE;orig}\big[\vec{J},\boldsymbol{K}\big] \Big)$, organized with respect to $\hbar$ as well, derives from Eq.~\eqref{eq:ZJKPT3} together with the linked-cluster theorem \cite{neg98}, by virtue of which one can substitute the correlation functions in Eq.~\eqref{eq:ZJKPT3} by their connected counterparts. This leads to:
\begin{equation}
\begin{split}
W^\text{LE;orig}\Big[\vec{J},\boldsymbol{K}\Big] = & -S_{JK}\big[\vec{\varphi}_\text{cl}\big] +\frac{\hbar}{2} \mathrm{STr}\left[\ln\big(\boldsymbol{G}_{\varphi_\text{cl};JK}\big) \right] \\
& + \sum^{\infty}_{n=1} \frac{\left(-1\right)^n}{\left(3!\right)^n n!}\sum_{q=0}^n 
\begin{pmatrix}
n \\
q
\end{pmatrix} \frac{\hbar^\frac{n+q+2}{2}}{4^q} \\
& \times \Bigg\langle \left(\lambda\int_x \vec{\widetilde{\chi}}^2(x)\vec{\widetilde{\chi}}(x)\cdot\vec{\varphi}_\text{cl}(x) \right)^{n-q} \\
& \times\left(\lambda \int_x \left( \vec{\widetilde{\chi}}^2(x) \right)^2 \right)^{q}\Bigg\rangle^\text{c}_{0,JK} \;.
\end{split}
\label{eq:WJKPT}
\end{equation}
Denoting the modulus of the classical solution $\vec{\varphi}_\text{cl}$ as $\varrho(x)\equiv\left|\vec{\varphi}_\text{cl}(x)\right|$, we can choose $a=N$ as the direction along which the SSB occurs in the broken-symmetry phase without any loss of generality, i.e.:
\begin{equation}
\vec{\varphi}_\text{cl}(x) = \varrho(x)\begin{pmatrix}
0 \\
\vdots\\
0 \\
1
\end{pmatrix} \;.
\label{eq:DefPhiclModulusRho}
\end{equation}
The propagator~\eqref{eq:LEG} can then be separated into the one associated to the $O(N-1)$ subspace (the Goldstone manifold in the broken-symmetry phase when $N\geq 2$), namely:
\begin{equation}
\begin{split}
G^{-1}_{\varphi_\text{cl};JK;\mathfrak{g};ab}(x,y) = & \left(-\nabla_x^2 + m^2 + \frac{\lambda}{6} \varrho^2(x)\right)\delta_{ab}\delta(x-y) \\
& - \boldsymbol{K}_{ab}(x,y) \quad \forall a,b\in [1,N-1] \;,
\end{split}
\label{eq:LEGG}
\end{equation}
and the one of the remaining massive (or Higgs) mode for $a=N$, i.e.:
\begin{equation}
\begin{split}
\boldsymbol{G}^{-1}_{\varphi_\text{cl};JK;NN}(x,y) = & \left(-\nabla_x^2 + m^2 + \frac{\lambda}{2} \varrho^2(x)\right)\delta(x-y) \\
& - \boldsymbol{K}_{NN}(x,y) \;.
\end{split}
\label{eq:LEGN}
\end{equation}
By means of Wick's theorem, we then represent the connected correlation functions in Eq.~\eqref{eq:WJKPT} by a set of Feynman diagrams with the rules:
\begin{subequations}
\begin{align}
%\begin{equation}
\begin{gathered}
\begin{fmffile}{Diagrams/LoopExpansion1_FeynRuleGbis}
\begin{fmfgraph*}(20,20)
\fmfleft{i0,i1,i2,i3}
\fmfright{o0,o1,o2,o3}
\fmflabel{$x, a$}{v1}
\fmflabel{$y, b$}{v2}
\fmf{phantom}{i1,v1}
\fmf{phantom}{i2,v1}
\fmf{plain,tension=0.6}{v1,v2}
\fmf{phantom}{v2,o1}
\fmf{phantom}{v2,o2}
\end{fmfgraph*}
\end{fmffile}
\end{gathered} \quad &\rightarrow \boldsymbol{G}_{\varphi_\text{cl};JK;ab}(x,y)\;,
\label{eq:FeynRulesLoopExpansionPropagator} \\
%\end{equation} 
%\begin{equation}
\begin{gathered}
\begin{fmffile}{Diagrams/LoopExpansion1_FeynRuleV3bis}
\begin{fmfgraph*}(20,20)
\fmfleft{i0,i1,i2,i3}
\fmfright{o0,o1,o2,o3}
\fmfv{decor.shape=cross,decor.angle=45,decor.size=3.5thick,foreground=(0,,0,,1)}{o2}
\fmf{phantom,tension=2.0}{i1,i1bis}
\fmf{plain,tension=2.0}{i1bis,v1}
\fmf{phantom,tension=2.0}{i2,i2bis}
\fmf{plain,tension=2.0}{i2bis,v1}
\fmf{dots,label=$x$,tension=0.6,foreground=(0,,0,,1)}{v1,v2}
\fmf{phantom,tension=2.0}{o1bis,o1}
\fmf{plain,tension=2.0}{v2,o1bis}
\fmf{phantom,tension=2.0}{o2bis,o2}
\fmf{phantom,tension=2.0}{v2,o2bis}
\fmf{dashes,tension=0.0,foreground=(0,,0,,1)}{v2,o2}
\fmflabel{$a$}{i1bis}
\fmflabel{$b$}{i2bis}
\fmflabel{$c$}{o1bis}
\fmflabel{$N$}{o2bis}
\end{fmfgraph*}
\end{fmffile}
\end{gathered} \quad &\rightarrow \lambda\varrho(x)\delta_{a b}\delta_{c N}\;,
\label{eq:FeynRulesLoopExpansion3legVertex} \\
%\end{equation}
%\begin{equation}
\begin{gathered}
\begin{fmffile}{Diagrams/LoopExpansion1_FeynRuleV4bis}
\begin{fmfgraph*}(20,20)
\fmfleft{i0,i1,i2,i3}
\fmfright{o0,o1,o2,o3}
\fmf{phantom,tension=2.0}{i1,i1bis}
\fmf{plain,tension=2.0}{i1bis,v1}
\fmf{phantom,tension=2.0}{i2,i2bis}
\fmf{plain,tension=2.0}{i2bis,v1}
\fmf{zigzag,label=$x$,tension=0.6,foreground=(0,,0,,1)}{v1,v2}
\fmf{phantom,tension=2.0}{o1bis,o1}
\fmf{plain,tension=2.0}{v2,o1bis}
\fmf{phantom,tension=2.0}{o2bis,o2}
\fmf{plain,tension=2.0}{v2,o2bis}
\fmflabel{$a$}{i1bis}
\fmflabel{$b$}{i2bis}
\fmflabel{$c$}{o1bis}
\fmflabel{$d$}{o2bis}
\end{fmfgraph*}
\end{fmffile}
\end{gathered} \quad &\rightarrow \lambda\delta_{a b}\delta_{c d}\;.
\label{eq:FeynRulesLoopExpansion4legVertex}
%\end{equation}
\end{align}
\end{subequations}
Up to second order in $\hbar$, the diagrammatic series thus obtained for the Schwinger functional of Eq.~\eqref{eq:WJKPT} reads:
\begin{equation}
\begin{split}
& W^\text{LE;orig}\Big[\vec{J},\boldsymbol{K}\Big] \\
& = -S_{JK}\big[\vec{\varphi}_{\mathrm{cl}}\big] + \frac{\hbar}{2} \mathrm{STr}\left[\ln\big(\boldsymbol{G}_{\varphi_\text{cl};JK}\big)\right] \\
& + \hbar^{2} \left(\rule{0cm}{1.2cm}\right. -\frac{1}{24} \hspace{0.08cm} \begin{gathered}
\begin{fmffile}{Diagrams/LoopExpansion1_Hartree}
\begin{fmfgraph}(30,20)
\fmfleft{i}
\fmfright{o}
\fmf{phantom,tension=10}{i,i1}
\fmf{phantom,tension=10}{o,o1}
\fmf{plain,left,tension=0.5}{i1,v1,i1}
\fmf{plain,right,tension=0.5}{o1,v2,o1}
\fmf{zigzag,foreground=(0,,0,,1)}{v1,v2}
\end{fmfgraph}
\end{fmffile}
\end{gathered} \hspace{0.05cm} -\frac{1}{12}\begin{gathered}
\begin{fmffile}{Diagrams/LoopExpansion1_Fock}
\begin{fmfgraph}(15,15)
\fmfleft{i}
\fmfright{o}
\fmf{phantom,tension=11}{i,v1}
\fmf{phantom,tension=11}{v2,o}
\fmf{plain,left,tension=0.4}{v1,v2,v1}
\fmf{zigzag,foreground=(0,,0,,1)}{v1,v2}
\end{fmfgraph}
\end{fmffile}
\end{gathered} \\
\hspace{0.3cm} & + \frac{1}{18} \ \ \begin{gathered}
\begin{fmffile}{Diagrams/LoopExpansion1_Diag1}
\begin{fmfgraph}(34,20)
\fmfleft{i}
\fmfright{o}
\fmfv{decor.shape=cross,decor.size=3.5thick,foreground=(0,,0,,1)}{i}
\fmfv{decor.shape=cross,decor.size=3.5thick,foreground=(0,,0,,1)}{o}
\fmf{dashes,tension=2.0,foreground=(0,,0,,1)}{i,v3}
\fmf{dashes,tension=2.0,foreground=(0,,0,,1)}{o,v4}
\fmf{plain,right,tension=0.7}{v2,v4}
\fmf{dots,left,tension=0.7,foreground=(0,,0,,1)}{v2,v4}
\fmf{plain,left,tension=0.7}{v1,v3}
\fmf{dots,right,tension=0.7,foreground=(0,,0,,1)}{v1,v3}
\fmf{plain,tension=1.5}{v1,v2}
\end{fmfgraph}
\end{fmffile}
\end{gathered} \hspace{0.2cm} + \frac{1}{18} \begin{gathered}
\begin{fmffile}{Diagrams/LoopExpansion1_Diag2}
\begin{fmfgraph}(27,15)
\fmfleft{i}
\fmfright{o}
\fmftop{vUp}
\fmfbottom{vDown}
\fmfv{decor.shape=cross,decor.size=3.5thick,foreground=(0,,0,,1)}{v1}
\fmfv{decor.shape=cross,decor.size=3.5thick,foreground=(0,,0,,1)}{v2}
\fmf{phantom,tension=10}{i,i1}
\fmf{phantom,tension=10}{o,o1}
\fmf{phantom,tension=2.2}{vUp,v5}
\fmf{phantom,tension=2.2}{vDown,v6}
\fmf{phantom,tension=0.5}{v3,v4}
\fmf{phantom,tension=10.0}{i1,v1}
\fmf{phantom,tension=10.0}{o1,v2}
\fmf{dashes,tension=2.0,foreground=(0,,0,,1)}{v1,v3}
\fmf{dots,left=0.4,tension=0.5,foreground=(0,,0,,1)}{v3,v5}
\fmf{plain,left=0.4,tension=0.5}{v5,v4}
\fmf{plain,right=0.4,tension=0.5}{v3,v6}
\fmf{dots,right=0.4,tension=0.5,foreground=(0,,0,,1)}{v6,v4}
\fmf{dashes,tension=2.0,foreground=(0,,0,,1)}{v4,v2}
\fmf{plain,tension=0}{v5,v6}
\end{fmfgraph}
\end{fmffile}
\end{gathered} \\
\hspace{0.3cm} & +\frac{1}{18} \ \begin{gathered}
\begin{fmffile}{Diagrams/LoopExpansion1_Diag4}
\begin{fmfgraph}(35,18)
\fmfleft{i}
\fmfright{o}
\fmftop{vUp}
\fmfbottom{vDown}
\fmfv{decor.shape=cross,decor.size=3.5thick,foreground=(0,,0,,1)}{v3bis}
\fmfv{decor.shape=cross,decor.size=3.5thick,foreground=(0,,0,,1)}{o}
\fmf{phantom,tension=10}{i,i1}
\fmf{dashes,tension=1.2,foreground=(0,,0,,1)}{o,v4}
\fmf{phantom,tension=0.5}{v3bis,i}
\fmf{phantom,tension=2.7}{v3bis,vUp}
\fmf{dashes,tension=0.9,foreground=(0,,0,,1)}{v3,v3bis}
\fmf{phantom,tension=0.5}{v4bis,i}
\fmf{phantom,tension=2.7}{v4bis,vDown}
\fmf{phantom,tension=0.9}{v3,v4bis}
\fmf{plain,left,tension=0.5}{i1,v1,i1}
\fmf{plain,right,tension=0.5}{v2,v4}
\fmf{dots,left,tension=0.5,foreground=(0,,0,,1)}{v2,v4}
\fmf{dots,foreground=(0,,0,,1)}{v1,v3}
\fmf{plain}{v3,v2}
\end{fmfgraph}
\end{fmffile}
\end{gathered} \hspace{0.2cm} + \frac{1}{36} \hspace{-0.15cm} \begin{gathered}
\begin{fmffile}{Diagrams/LoopExpansion1_Diag3}
\begin{fmfgraph}(25,20)
\fmfleft{i}
\fmfright{o}
\fmftop{vUp}
\fmfbottom{vDown}
\fmfv{decor.shape=cross,decor.angle=45,decor.size=3.5thick,foreground=(0,,0,,1)}{vUpbis}
\fmfv{decor.shape=cross,decor.angle=45,decor.size=3.5thick,foreground=(0,,0,,1)}{vDownbis}
\fmf{phantom,tension=0.8}{vUp,vUpbis}
\fmf{phantom,tension=0.8}{vDown,vDownbis}
\fmf{dashes,tension=0.5,foreground=(0,,0,,1)}{v3,vUpbis}
\fmf{phantom,tension=0.5}{v4,vUpbis}
\fmf{phantom,tension=0.5}{v3,vDownbis}
\fmf{dashes,tension=0.5,foreground=(0,,0,,1)}{v4,vDownbis}
\fmf{phantom,tension=11}{i,v1}
\fmf{phantom,tension=11}{v2,o}
\fmf{plain,left,tension=0.5}{v1,v2,v1}
\fmf{dots,tension=1.7,foreground=(0,,0,,1)}{v1,v3}
\fmf{plain}{v3,v4}
\fmf{dots,tension=1.7,foreground=(0,,0,,1)}{v4,v2}
\end{fmfgraph}
\end{fmffile}
\end{gathered} \\
\hspace{0.3cm} & +\frac{1}{72} \ \begin{gathered}
\begin{fmffile}{Diagrams/LoopExpansion1_Diag5}
\begin{fmfgraph}(40,18)
\fmfleft{i}
\fmfright{o}
\fmftop{vUp}
\fmfbottom{vDown}
\fmf{phantom,tension=1.0}{vUp,vUpbis}
\fmf{phantom,tension=1.0}{vDown,vDownbis}
\fmf{dashes,tension=0.5,foreground=(0,,0,,1)}{v3,vUpbis}
\fmf{phantom,tension=0.5}{v4,vUpbis}
\fmf{phantom,tension=0.5}{v3,vDownbis}
\fmf{dashes,tension=0.5,foreground=(0,,0,,1)}{v4,vDownbis}
\fmfv{decor.shape=cross,decor.angle=45,decor.size=3.5thick,foreground=(0,,0,,1)}{vUpbis}
\fmfv{decor.shape=cross,decor.angle=45,decor.size=3.5thick,foreground=(0,,0,,1)}{vDownbis}
\fmf{phantom,tension=10}{i,i1}
\fmf{phantom,tension=10}{o,o1}
\fmf{plain,left,tension=0.5}{i1,v1,i1}
\fmf{plain,right,tension=0.5}{o1,v2,o1}
\fmf{dots,tension=1.2,foreground=(0,,0,,1)}{v1,v3}
\fmf{plain,tension=0.6}{v3,v4}
\fmf{dots,tension=1.2,foreground=(0,,0,,1)}{v4,v2}
\end{fmfgraph}
\end{fmffile}
\end{gathered} \left.\rule{0cm}{1.2cm}\right) \\
& + \mathcal{O}\Big(\hbar^3\Big)\;.
\end{split}
\label{eq:WKjLoopExpansionStep3}
\end{equation}

In the (0+0)\nobreakdash-D situation, the original LE result for the Schwinger functional given by Eq.~\eqref{eq:WKjLoopExpansionStep3} reduces to:
\begin{equation}
\begin{split}
W^\text{LE;orig}\Big(\vec{J},\boldsymbol{K}\Big) = & -S_{JK}\big(\vec{\varphi}_{\mathrm{cl}}\big) \\
& + \frac{\hbar}{2}\Big[(N-1)\ln\big(2\pi \mathfrak{G}_{\varphi_\text{cl};JK;\mathfrak{g}}\big) \\
& +\ln\big(2\pi \boldsymbol{G}_{\varphi_\text{cl};JK;NN}\big)\Big] \\
& + \frac{\hbar^{2}\lambda}{72} \Big[ -3 \mathfrak{G}_{\varphi_\text{cl};JK;\mathfrak{g}}^2 \left(-1 + N^2\right) \\
& + 15 \boldsymbol{G}_{\varphi_\text{cl};JK;NN}^3 \lambda \varrho^2 + \boldsymbol{G}_{\varphi_\text{cl};JK;NN}^2 (-9 \\
& + 6 \mathfrak{G}_{\varphi_\text{cl};JK;\mathfrak{g}} \left(-1 + N\right) \lambda \varrho^2) \\
& + \boldsymbol{G}_{\varphi_\text{cl};JK;NN} \mathfrak{G}_{\varphi_\text{cl};JK;\mathfrak{g}} \left(-1 + N\right) (-6 \\
& + \mathfrak{G}_{\varphi_\text{cl};JK;\mathfrak{g}} \left(1 + N\right) \lambda \varrho^2) \Big] \\
&+ \mathcal{O}\big(\hbar^3\big)\;,
\end{split}
\label{eq:Worig}
\end{equation}
where we have assumed that $\boldsymbol{K}_{ab}=K\delta_{ab}$ so that Eq.~\eqref{eq:LEGG} becomes:
\begin{equation}
G^{-1}_{\varphi_\text{cl};JK;\mathfrak{g};ab} = \mathfrak{G}^{-1}_{\varphi_\text{cl};JK;\mathfrak{g}} \delta_{ab} \quad \forall a,b\in [1,N-1] \;,
\label{eq:origLEGoldstoneProp}
\end{equation}
with
\begin{equation}
\mathfrak{G}^{-1}_{\varphi_\text{cl};JK;\mathfrak{g}} = m^2 + \frac{\lambda}{6} \varrho^2 - K \;.
\label{eq:origLEGoldstonePropBis}
\end{equation}
When $\vec{J}=\vec{0}$ and $\boldsymbol{K}=\boldsymbol{0}$, the modulus $\varrho$ satisfies:
\begin{equation}
\varrho^2\big(\vec{J}=\vec{0},\boldsymbol{K}=\boldsymbol{0}\big) = \begin{cases} \displaystyle{0} \quad \forall m^2\geq 0 \;, \\
\\
\displaystyle{-\frac{6m^2}{\lambda}} \quad \forall m^2< 0 ~ \text{and} ~ \lambda\neq 0 \;,  \end{cases}
\label{eq:SolutionsmodulusrhoLE}
\end{equation}
which yields the following expressions for $\mathfrak{G}_{\varphi_\text{cl};JK;\mathfrak{g}}$ and $\boldsymbol{G}_{\varphi_\text{cl};JK;NN}$ at vanishing sources:
\begin{equation}
\mathfrak{G}_{\varphi_\text{cl};\mathfrak{g}} = \begin{cases} \displaystyle{\frac{1}{m^2} \quad \forall m^2> 0 \;,} \\
\\
\displaystyle{\infty \quad \forall m^2\leq 0 ~ \text{and} ~ \lambda\neq 0 \;,}  \end{cases} \forall N\geq 2 \;,
\end{equation}
\begin{equation}
\boldsymbol{G}_{\varphi_\text{cl},NN} = \begin{cases} \displaystyle{\frac{1}{m^2} \quad \forall m^2\geq 0 \;,} \\
\\
\displaystyle{-\frac{1}{2m^2} \quad \forall m^2< 0 ~ \text{and} ~ \lambda\neq 0 \;,} \end{cases} \forall N\geq 1 \;.
\label{eq:origLEmassiveProp}
\end{equation}
From Eqs.~\eqref{eq:DefEgsExactZexact0DON} and~\eqref{eq:DefrhogsExactwithExpectationValue0DON}, one can then obtain the corresponding series for the gs energy and density. Setting $g\equiv \hbar\lambda/m^4$ (showing in this way that our loop-wise expansion amounts to an expansion in powers of the coupling constant $\lambda$ and the original LE is thus simply an implementation of perturbation theory in the PI formalism), we get in the regime with $m^{2}>0$ and for $N\in\mathbb{N}^{*}$:
\begin{equation}
\begin{split}
E_\text{gs}^\text{LE;orig} = & -\frac{N}{2} \ln\bigg(\frac{2\pi}{m^{2}}\bigg) +  \frac{N\left(2+N\right)}{24}g \\
& - \frac{N \left(6 + 5 N + N^{2}\right)}{144}g^2 \\
& + \frac{N \left(120 + 128 N + 44 N^{2} + 5 N^{3}\right)}{2592}g^3 \\
& + \mathcal{O}\big(g^4\big) \;,
\end{split}
\label{eq:EgsPTpos}
\end{equation}
and
\begin{equation}
\begin{split}
\rho_\mathrm{gs}^\text{LE;orig} = & \ \frac{\hbar}{m^{2}}\bigg( 1 - \frac{2+N}{6}g+\frac{6 + 5 N + N^{2}}{18}g^2 \\
& - \frac{120 + 128 N + 44 N^{2} + 5 N^{3}}{216}g^3 \\
& + \mathcal{O}\big(g^4\big) \bigg)\;.
\end{split}
\end{equation}
In the regime with $m^{2}<0$, we have for $N=1$:
\begin{equation}
E^\text{LE;orig}_\mathrm{gs} = - \frac{3}{2 g} - \frac{1}{2} \ln\bigg(\frac{\pi}{-m^{2}}\bigg)-\frac{g}{8} - \frac{g^{2}}{12} - \frac{11 g^{3}}{96} + \mathcal{O}\big(g^{4}\big)\;,
\label{eq:ResultPTEgsm2neg}
\end{equation}
and
\begin{equation}
\rho^\text{LE;orig}_\mathrm{gs}= \frac{\hbar}{m^{2}}\left( -\frac{6}{g}+ 1  + \frac{g}{2} + \frac{2 g^{2}}{3} + \frac{11 g^{3}}{8 }+\mathcal{O}\big(g^{4}\big)\right)\;.
\label{eq:ResultPTrhogsm2neg}
\end{equation}
For $N>1$, no finite results can be obtained in the phase with $m^2<0$ for the gs energy and density, as long as we stick with the original dofs $\vec{\widetilde{\varphi}}$, because in this case the Goldstone propagator $\mathfrak{G}_{\varphi_\text{cl};\mathfrak{g}}$ exhibits (infrared) divergences \cite{col74} which preclude the spontaneous breakdown of the continuous $O(N)$ symmetry. Furthermore, the 1-point correlation function of the studied model can also be determined within the original LE from Eq.~\eqref{eq:WKjLoopExpansionStep3} or~\eqref{eq:Worig} according to definition~\eqref{eq:DefPhiExactwithExpectationValue0DON}. In the broken-symmetry phase at $m^{2}=-1$ and $N=1$, we have:
\begin{equation}
\begin{split}
\overline{\phi} \equiv \left|\vec{\overline{\phi}}\right| = & \frac{1}{49152\sqrt{6\lambda}}\Big(294912\hbar-36684\hbar^2\lambda-7872\hbar^3\lambda^2 \\
& -30816\hbar^4\lambda^3-64573\hbar^5\lambda^4 + \mathcal{O}\big(\hbar^{6}\big)\Big)\;.
\end{split}
\label{eq:ResultphiOriginalLE}
\end{equation}

%%%%%%%%%%%%%%%%%%%
\begin{figure}[!t]
  \begin{center}
      \includegraphics[width=0.95\linewidth]{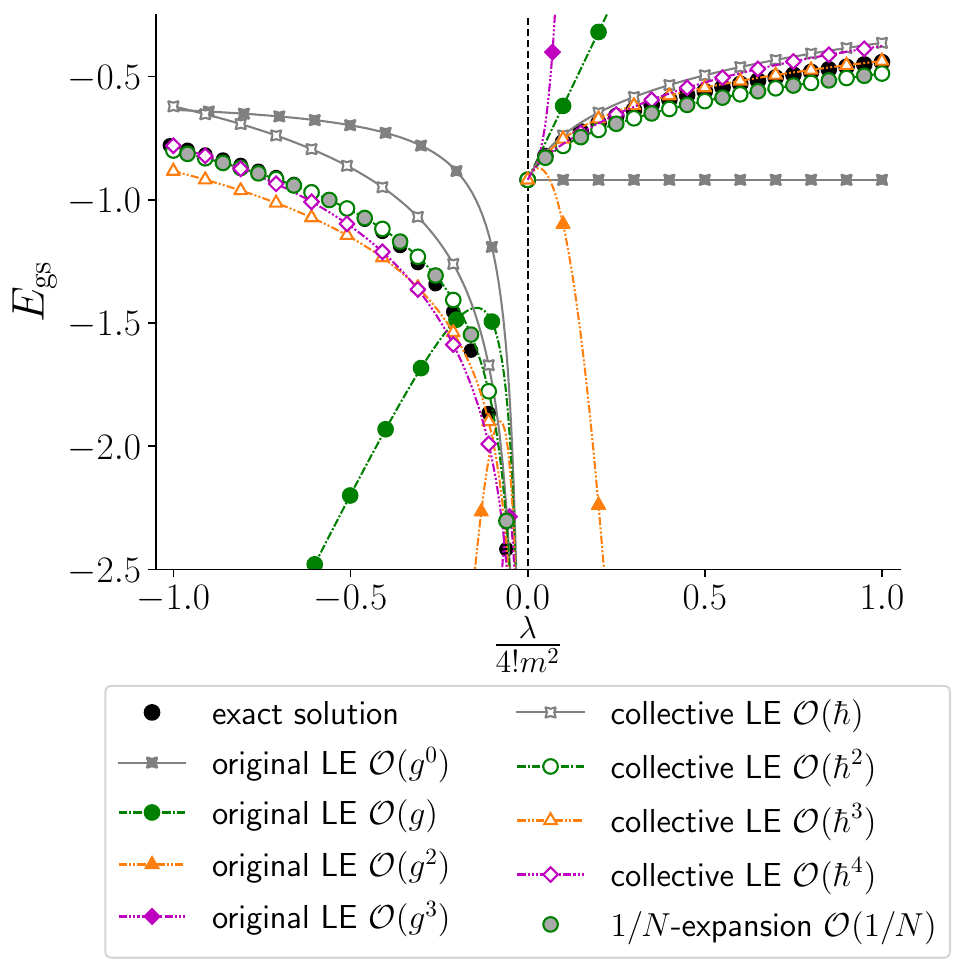}
      \includegraphics[width=0.95\linewidth]{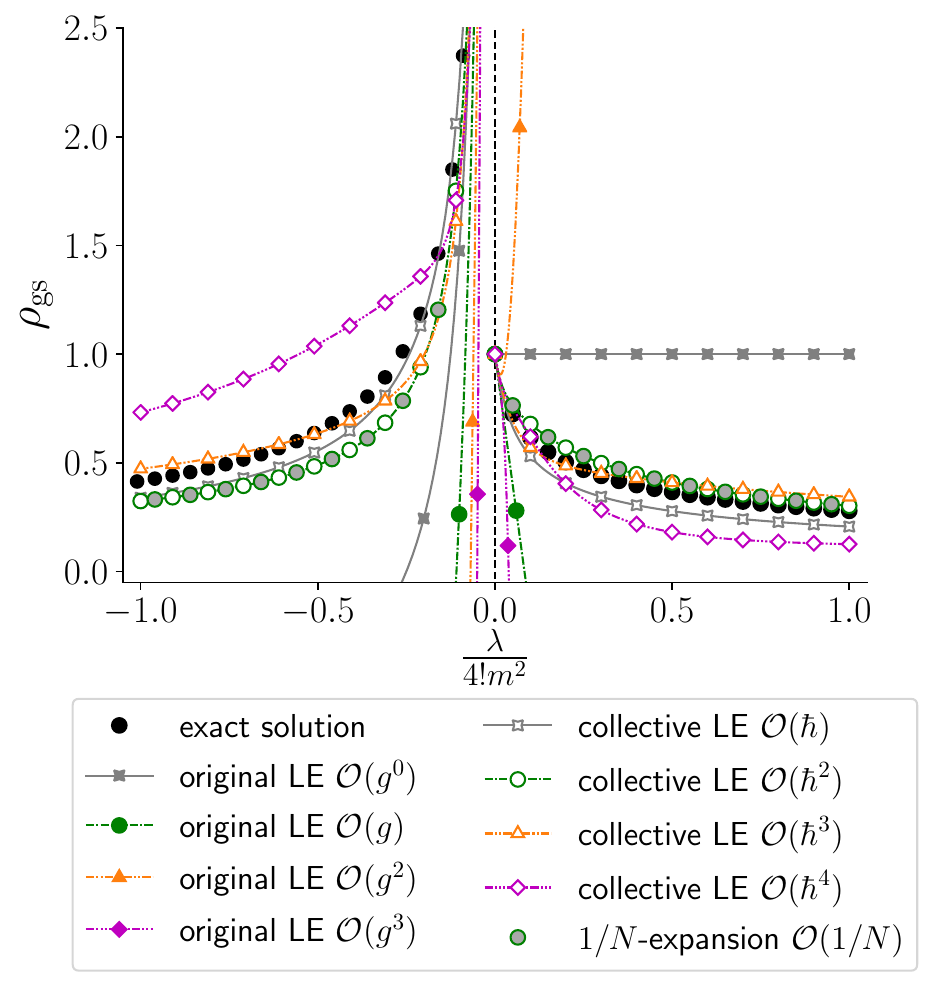}
    \caption{Gs energy $E_{\mathrm{gs}}$ (upper panel) or density $\rho_{\mathrm{gs}}$ (lower panel) calculated at $\hbar=1$, $m^{2}=\pm 1$ and $N=1$, and compared with the corresponding exact solution (black dots). The indication ``$\mathcal{O}\big(\hbar^{n}\big)$'' for the collective LE results specifies that the series representing $W^\text{LE;col}$ has been exploited up to order $\mathcal{O}\big(\hbar^{n}\big)$ (which implies notably that the corresponding series for $E_{\mathrm{gs}}^\text{LE;col}$ is calculated up to order $\mathcal{O}(\hbar^{n-1})$ since $E^\text{LE;col}_\text{gs} =  -\frac{1}{\hbar}W^\text{LE;col}(\mathcal{J}=0)$). Recall also that the expansion parameter $g$ is defined as $g\equiv\hbar\lambda/m^4$.}
    \label{fig:O1PTcoll}
  \end{center}
\end{figure}
%%%%%%%%%%%%%%%%%%%
%%%%%%%%%%%%%%%%%%%
\begin{figure}[!t]
  \begin{center}
      \includegraphics[width=0.95\linewidth]{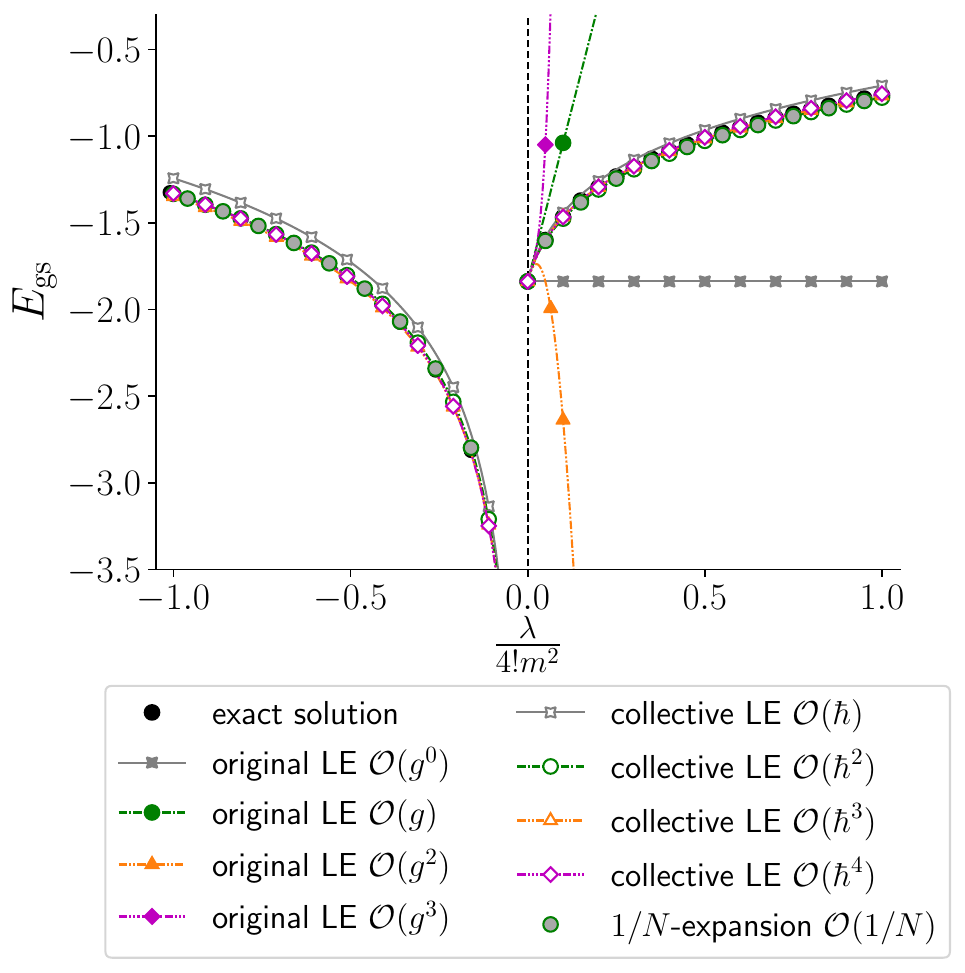}
      \includegraphics[width=0.95\linewidth]{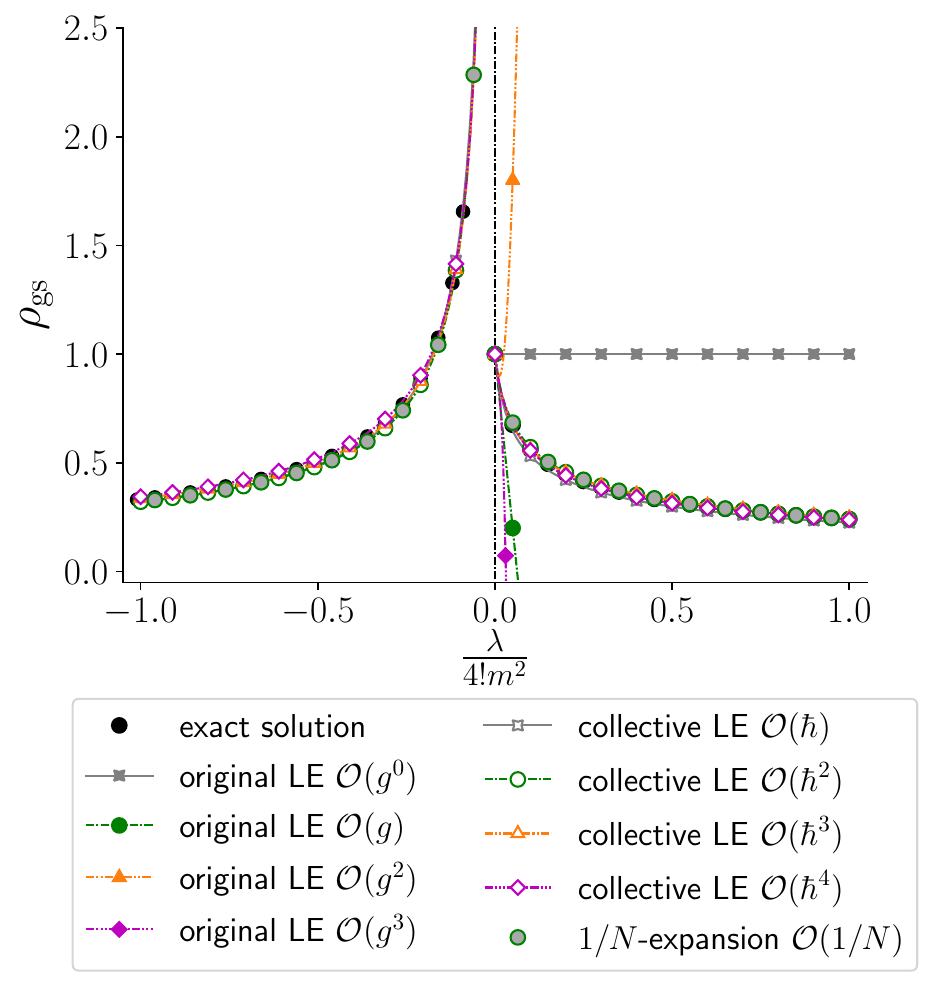}
    \caption{Same as Fig.~\ref{fig:O1PTcoll} with $N=2$ instead. Note that no finite results can be obtained in the broken-symmetry phase (which corresponds to the left panel of both plots) from the LE in the original (and mixed) representation(s).}
    \label{fig:O2PTcoll}
  \end{center}
\end{figure}
%%%%%%%%%%%%%%%%%%%

\begin{figure}[!t]
\begin{center}
      \includegraphics[width=0.95\linewidth]{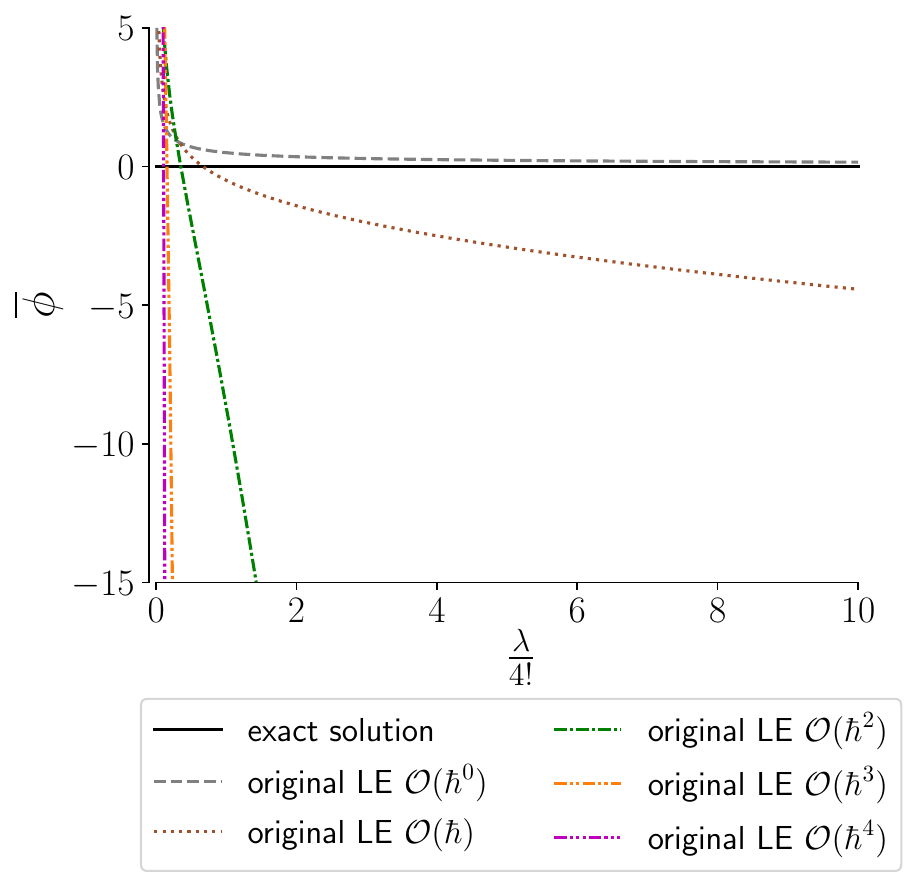}
      \end{center}
      \caption{1-point correlation function $\vec{\overline{\phi}}$ (defined from Eq.~\eqref{eq:DefPhiExactwithExpectationValue0DON}) calculated at $\hbar=1$, $m^2=-1$ and $N=1$ from the first orders of the original (and mixed) LE(s) as a function of the coupling constant $\lambda/4!$. At $N=1$, $\vec{\overline{\phi}}$ coincides with $\overline{\phi}\equiv\left\lvert\vec{\overline{\phi}}\right\rvert$.}
      \label{fig:vev}
\end{figure}

Results~\eqref{eq:EgsPTpos} to~\eqref{eq:ResultphiOriginalLE} are represented in Figs.~\ref{fig:O1PTcoll} to~\ref{fig:vev}. We can see from all of those plots that the LE series derived so far show no signs of convergence: the results obtained for $E_{\mathrm{gs}}$ and $\rho_{\mathrm{gs}}$ worsen as the truncation order (with respect to $g$ or $\hbar$) of these series increases, except for very small values of $\lambda$. The original LE only yields a reasonable description of the gs energy and density for $\lambda\lesssim 0.2$ (hence $\lambda/4!\lesssim 8.10^{-3}$) and quickly deteriorates for larger $\lambda$. Such a behavior signals the illegitimate application of perturbation theory to a system where the fundamental phenomena are non-perturbative in nature. Indeed, in quantum mechanics and QFT, perturbation theory typically produces asymptotic series with a zero radius of convergence, whose origin lies in instanton-like effects, i.e. an instability of the theory at some phase of the coupling (here for $\lambda<0$, where the potential becomes unbounded), which translates into a factorial growth of the number of Feynman diagrams with the order of the expansion \cite{hur52,ben76}. Furthermore, Fig.~\ref{fig:vev} illustrates that finite truncation orders within the original LE never yield a vanishing 1-point correlation function. This is problematic according to our discussion of section~\ref{sec:0DONmodel} showing that the exact solution of the studied model never exhibits spontaneous breakdowns of its $O(N)$ symmetry.

%%%%%%%%%%%%%%%%%%%
%%%%%%%%%%%%%%%%%%%

Furthermore, one can show that the LE in the mixed representation coincides with the original LE \cite{fra21}. The series underlying the mixed LE are therefore also equivalently organized with respect to $\lambda$ and $\hbar$, and yield divergent results for $E_{\mathrm{gs}}$ and $\rho_{\mathrm{gs}}$ in the broken-symmetry phase when $N>1$ as well. We will thus exploit the collective representation as a last attempt to tackle the broken-symmetry phase for all values of $N$ within the LE.

%~~~
\subsubsection{\label{sec:CollLE}Collective loop expansion}

In the nuclear EDF formalism, the use of a collective dof in the form of the nucleonic density turned out to be very effective in the description of nuclear systems, as mentioned earlier. We start here our investigation of PI approaches based on collective dof(s) with the LE in the collective representation, i.e. the collective LE, also referred to as the $\epsilon$-expansion \cite{ben77,daw12}. With definitions~\eqref{eq:Zcoll} to~\eqref{eq:Gcoll} in mind, we first discuss the partitioning underlying the collective LE. As for the original LE, one must first introduce a saddle point to that end:
\begin{equation}
\textcolor{white}{\quad \forall x} \left.\frac{\delta S_{\mathrm{col},\mathcal{J}}[\widetilde{\sigma}]}{\delta \widetilde{\sigma}(x)}\right|_{\widetilde{\sigma}=\sigma_{\mathrm{cl}}} = 0 \quad \forall x \;.
\label{eq:minimizationSbosonicK}
\end{equation}
Setting $\widetilde{\sigma} = \sigma_\text{cl}+\sqrt{\hbar} \ \widetilde{\zeta}$, a Taylor expansion of $S_{\mathrm{col},\mathcal{J}}$ around $\widetilde{\sigma}=\sigma_\text{cl}$ leads to the splitting:
\begin{equation}
S_{\mathrm{col},\mathcal{J}}\Big[\widetilde{\sigma} = \sigma_\text{cl}+\sqrt{\hbar} \ \widetilde{\zeta}\Big] = S^0_{\mathrm{col},\sigma_\text{cl};\mathcal{J}}\Big[\widetilde{\zeta}\Big]+ S^1_{\mathrm{col},\sigma_\text{cl};\mathcal{J}}\Big[\widetilde{\zeta}\Big] \;,
\end{equation}
\begin{equation}
S^0_{\mathrm{col},\sigma_\text{cl};\mathcal{J}}\Big[\widetilde{\zeta}\Big] = S_{\mathrm{col},\mathcal{J}}[\sigma_{\mathrm{cl}}] + \frac{\hbar}{2} \int_{x,y} \widetilde{\zeta}(x) D^{-1}_{\sigma_\text{cl};\mathcal{J}}(x,y) \widetilde{\zeta}(y) \;,
\end{equation}
\begin{equation}
\begin{split}
S^1_{\mathrm{col},\sigma_\text{cl};\mathcal{J}}\Big[\widetilde{\zeta}\Big] = & \ \sum_{n=3}^{\infty} \frac{\hbar^{\frac{n}{2}}}{n!} \int_{x_{1},\cdots, x_{n}} S_{\mathrm{col}, \mathcal{J}}^{(n)}(x_{1}, \cdots ,x_{n}) \\
& \times \widetilde{\zeta}(x_{1}) \cdots \widetilde{\zeta}(x_{n}) \;,
\end{split}
\label{eq:ResildualPartCollLE}
\end{equation}
with
\begin{equation}
S_{\mathrm{col}, \mathcal{J}}^{(n)}(x_{1},\cdots,x_{n}) \equiv \left.\frac{\delta^n S_{\mathrm{col},\mathcal{J}}[\widetilde{\sigma}]}{\delta \widetilde{\sigma}(x_1)\cdots\delta \widetilde{\sigma}(x_n)}\right|_{\widetilde{\sigma}=\sigma_{\mathrm{cl}}} \;,
\end{equation}
and $D_{\sigma_\text{cl};\mathcal{J}}(x,y)$ being the propagator of the collective field in presence of the sources, i.e.:
\begin{equation}
D^{-1}_{\sigma_\text{cl};\mathcal{J}}(x,y) = S_{\mathrm{col}, \mathcal{J}}^{(2)}(x,y) \;.
\label{eq:DefDpropagCollectiveLE}
\end{equation}
The unperturbed and residual channels both involve the propagator of the original field $\vec{\widetilde{\varphi}}(x)$:
\begin{equation}
\boldsymbol{G}^{-1}_{\sigma_\text{cl};\mathcal{J};ab}(x,y) = \left(-\nabla_x^2 + m^2 + i\sqrt{\frac{\lambda}{3}}\sigma_\text{cl}(x)\right)\delta_{ab}\delta(x-y) \;.
\label{eq:DefGpropagCollectiveLE}
\end{equation} 
The residual part~\eqref{eq:ResildualPartCollLE} now contains an infinite number of terms due to the logarithm structure of the collective classical action $S_{\mathrm{col}}$. This induces that the resulting diagrammatic construction is more demanding to determine as compared to that of the original (or mixed) LE. We will then examine if this additional effort is worth.

As for the original LE, we will outline how to construct the collective LE series up to their first non-trivial order for our $O(N)$ model at arbitrary dimensions but we will perform applications of this method up to its third non-trivial order and combine it with resummation procedures in (0+0)\nobreakdash-D. The collective LE is usually not exploited beyond its first non-trivial order. To our knowledge, this approach has never been pushed up to its third non-trivial order so far, regardless of the model under consideration. Following the same steps as for the original LE, the partition function of the theory based on Eq.~\eqref{eq:ScollJ} reads up to the first non-trivial order (i.e. up to order $\mathcal{O}(\hbar)$):
\begin{equation}
\begin{split}
& Z^\text{LE;col}\big[\mathcal{J}\big] \\
& = e^{-\frac{1}{\hbar}S_{\mathrm{col},\mathcal{J}}[\sigma_{\mathrm{cl}}]} \left( \int\mathcal{D}\widetilde{\zeta} \ e^{-\frac{1}{2}\int_{x,y}\widetilde{\zeta}(x) D^{-1}_{\sigma_\text{cl};\mathcal{J}}(x,y) \widetilde{\zeta}(y)} \right) \Bigg[1 \\
& \hspace{0.3cm} -\frac{\hbar}{24} \int_{x,y,z,u} S_{\mathrm{col}, \mathcal{J}}^{(4)}(x,y,z,u) \ \left\langle \widetilde{\zeta}(x) \widetilde{\zeta}(y) \widetilde{\zeta}(z) \widetilde{\zeta}(u) \right\rangle_{0,\mathcal{J}} \\
& \hspace{0.3cm} + \frac{\hbar}{72} \int_{{x_{\scalebox{0.4}{1}},y_{\scalebox{0.4}{1}},z_{\scalebox{0.4}{1}}}\atop{x_{\scalebox{0.4}{2}},y_{\scalebox{0.4}{2}},z_{\scalebox{0.4}{2}}}} \hspace{-0.2cm} S_{\mathrm{col}, \mathcal{J}}^{(3)}(x_{1},y_{1},z_{1}) S_{\mathrm{col}, \mathcal{J}}^{(3)}(x_{2},y_{2},z_{2}) \\
& \hspace{0.3cm} \times \left\langle \widetilde{\zeta}(x_{1}) \widetilde{\zeta}(y_{1}) \widetilde{\zeta}(z_{1}) \widetilde{\zeta}(x_{2}) \widetilde{\zeta}(y_{2}) \widetilde{\zeta}(z_{2}) \right\rangle_{0,\mathcal{J}} \\
& \hspace{0.3cm} + \mathcal{O}\big(\hbar^{2}\big)\Bigg]\;,
\end{split}
\label{eq:Zlecol}
\end{equation}
where
\begin{equation}
\big\langle\cdots\big\rangle_{0,\mathcal{J}} = \frac{1}{Z_{\mathrm{col},0}\big[\mathcal{J}\big]} \int \mathcal{D}\widetilde{\zeta} \ \cdots \ e^{-\frac{1}{\hbar} S^0_{\mathrm{col},\sigma_\text{cl};\mathcal{J}}\big[\widetilde{\zeta}\big]}\;,
\end{equation}
and
\begin{equation}
Z_{\mathrm{col},0}\big[\mathcal{J}\big] = \int \mathcal{D}\widetilde{\zeta} \  e^{-\frac{1}{\hbar} S^0_{\mathrm{col},\sigma_\text{cl};\mathcal{J}}\big[\widetilde{\zeta}\big]}\;.
\end{equation}
Introducing the Feynman rules:
\begin{subequations}
\begin{align}
\begin{gathered}
\begin{fmffile}{Diagrams/LEcol-G}
\begin{fmfgraph*}(20,12)
\fmfleft{i0,i1,i2,i3}
\fmfright{o0,o1,o2,o3}
\fmflabel{$x, a$}{v1}
\fmflabel{$y, b$}{v2}
\fmf{phantom}{i1,v1}
\fmf{phantom}{i2,v1}
\fmf{plain,tension=0.6}{v1,v2}
\fmf{phantom}{v2,o1}
\fmf{phantom}{v2,o2}
\end{fmfgraph*}
\end{fmffile}
\end{gathered} \quad &\rightarrow \boldsymbol{G}_{\sigma_\text{cl};\mathcal{J};ab}(x,y)\;,
\label{eq:FeynRulesLoopExpansionBosonicHSG}\\
\begin{gathered}
\begin{fmffile}{Diagrams/LEcol-D}
\begin{fmfgraph*}(20,16)
\fmfleft{i0,i1,i2,i3}
\fmfright{o0,o1,o2,o3}
\fmfv{label=$x$}{v1}
\fmfv{label=$y$}{v2}
\fmf{phantom}{i1,v1}
\fmf{phantom}{i2,v1}
\fmf{wiggly,tension=0.6}{v1,v2}
\fmf{phantom}{v2,o1}
\fmf{phantom}{v2,o2}
\end{fmfgraph*}
\end{fmffile}
\end{gathered} \quad &\rightarrow D_{\sigma_\text{cl};\mathcal{J}}(x,y)\;,
\label{eq:FeynRulesLoopExpansionBosonicHSH}\\
\begin{gathered}
\begin{fmffile}{Diagrams/LEcol-V}
\begin{fmfgraph*}(5,5)
\fmfleft{i1}
\fmfright{o1}
\fmfv{label=$x$,label.angle=-90,label.dist=4,foreground=(0,,0,,1)}{v1}
\fmf{plain}{i1,v1}
\fmf{plain}{v1,o1}
\fmflabel{$a$}{i1}
\fmflabel{$b$}{o1}
\fmfdot{v1}
\end{fmfgraph*}
\end{fmffile}
\end{gathered}\qquad &\rightarrow i\sqrt{\frac{\lambda}{3}}\delta_{ab}\;,
\label{eq:FeynRulesLoopExpansionBosonicHSvertexDot} \\
\nonumber\\
\begin{gathered}
\begin{fmffile}{Diagrams/LEcol-J}
\begin{fmfgraph*}(10,5)
\fmfleft{i1}
\fmfright{o1}
\fmfv{decor.shape=circle,decor.filled=empty,decor.size=.26w,l=$\times$,label.dist=0}{v1}
\fmfv{label.angle=-90,label.dist=6}{v2}
\fmf{plain,tension=2.5}{i1,v1}
\fmf{phantom}{v1,o1}
\fmf{phantom,tension=2.5}{i1,v2}
\fmf{phantom}{v2,o1}
\fmflabel{$x, a$}{v2}
\end{fmfgraph*}
\end{fmffile}
\end{gathered} \hspace{-0.2cm} &\rightarrow J_{a}(x)\;,
\label{eq:FeynRulesLoopExpansionBosonicHSK} \\
\nonumber\\
\nonumber\\
\begin{gathered}
\begin{fmffile}{Diagrams/LoopExpansionBosonicHS_FeynRuleVertexS3}
\begin{fmfgraph*}(4,4)
\fmfleft{i0,i1,i2,i3}
\fmfright{o0,o1,o2,o3}
\fmfv{label=$x$,label.angle=-135,label.dist=11,decor.shape=triangle,decor.filled=shaded,decor.size=1.6w,foreground=(0,,0,,1)}{v1}
\fmfv{label=$y$,label.angle=-45,label.dist=11}{v2}
\fmfv{label=$z$,label.angle=90,label.dist=11}{v3}
\fmfbottom{v1bis}
\fmf{phantom}{i1,v1}
\fmf{phantom}{i2,v1}
\fmf{phantom}{v1,o1}
\fmf{phantom}{v1,o2}
\fmf{phantom}{i1,v2}
\fmf{phantom}{i2,v2}
\fmf{phantom}{v2,o1}
\fmf{phantom}{v2,o2}
\fmf{phantom}{i1,v3}
\fmf{phantom}{i2,v3}
\fmf{phantom}{v3,o1}
\fmf{phantom}{v3,o2}
\fmf{phantom,tension=0.6}{v1,v1bis}
\end{fmfgraph*}
\end{fmffile}
\end{gathered} \quad &\rightarrow S_{\mathrm{col},\mathcal{J}}^{(3)}(x,y,z)\;,
\label{eq:FeynRulesLoopExpansionBosonicHSvertexS3} \\
\nonumber \\
\nonumber\\
\begin{gathered}
\begin{fmffile}{Diagrams/LoopExpansionBosonicHS_FeynRuleVertexS4}
\begin{fmfgraph*}(4,4)
\fmfleft{i0,i1,i2,i3}
\fmfright{o0,o1,o2,o3}
\fmfv{label=$x$,label.angle=-135,label.dist=11,decor.shape=square,decor.filled=shaded,decor.size=1.2w,foreground=(0,,0,,1)}{v1}
\fmfv{label=$y$,label.angle=-45,label.dist=11}{v2}
\fmfv{label=$z$,label.angle=45,label.dist=11}{v3}
\fmfv{label=$u$,label.angle=135,label.dist=11}{v4}
\fmfbottom{v1bis}
\fmf{phantom}{i1,v1}
\fmf{phantom}{i2,v1}
\fmf{phantom}{v1,o1}
\fmf{phantom}{v1,o2}
\fmf{phantom}{i1,v2}
\fmf{phantom}{i2,v2}
\fmf{phantom}{v2,o1}
\fmf{phantom}{v2,o2}
\fmf{phantom}{i1,v3}
\fmf{phantom}{i2,v3}
\fmf{phantom}{v3,o1}
\fmf{phantom}{v3,o2}
\fmf{phantom}{i1,v4}
\fmf{phantom}{i2,v4}
\fmf{phantom}{v4,o1}
\fmf{phantom}{v4,o2}
\fmf{phantom,tension=0.6}{v1,v1bis}
\end{fmfgraph*}
\end{fmffile}
\end{gathered} \quad &\rightarrow S_{\mathrm{col},\mathcal{J}}^{(4)}(x,y,z,u)\;, \\
\nonumber
\label{eq:FeynRulesLoopExpansionBosonicHSvertexS4}
\end{align}
\end{subequations}
the terms involved in the brackets of the right-hand side of Eq.~\eqref{eq:Zlecol} read:
\begin{equation}
\begin{split}
& \int_{x,y,z,u} S_{\mathrm{col}, \mathcal{J}}^{(4)}(x,y,z,u) \left\langle \widetilde{\zeta}(x) \widetilde{\zeta}(y) \widetilde{\zeta}(z) \widetilde{\zeta}(u) \right\rangle_{0,\mathcal{J}} \\
& = 3 \ \ \begin{gathered}
\begin{fmffile}{Diagrams/LoopExpansionBosonicHS_ApplicationWickTheoremS4_Diag4}
\begin{fmfgraph}(20,20)
\fmfleft{i}
\fmfright{o}
\fmfv{decor.shape=square,decor.filled=shaded,decor.size=0.18w,foreground=(0,,0,,1)}{v1}
\fmf{phantom,tension=10}{i,i1}
\fmf{phantom,tension=10}{o,o1}
\fmf{wiggly,left,tension=0.4}{i1,v1,i1}
\fmf{wiggly,right,tension=0.4}{o1,v1,o1}
\end{fmfgraph}
\end{fmffile}
\end{gathered} \;,
\end{split}
\label{eq:ApplicationWickTheoremBosonicActionS4}
\end{equation}
and
\begin{equation}
\begin{split}
& \int_{{x_{\scalebox{0.4}{1}},y_{\scalebox{0.4}{1}},z_{\scalebox{0.4}{1}}}\atop{x_{\scalebox{0.4}{2}},y_{\scalebox{0.4}{2}},z_{\scalebox{0.4}{2}}}} S_{\mathrm{col},\mathcal{J}}^{(3)}(x_{1},y_{1},z_{1}) S_{\mathrm{col}, \mathcal{J}}^{(3)}(x_{2},y_{2},z_{2}) \\
& \times \left\langle \widetilde{\zeta}(x_{1}) \widetilde{\zeta}(y_{1}) \widetilde{\zeta}(z_{1}) \widetilde{\zeta}(x_{2}) \widetilde{\zeta}(y_{2}) \widetilde{\zeta}(z_{2}) \right\rangle_{0,\mathcal{J}} \\
& = 9 \hspace{0.2cm} \begin{gathered}
\begin{fmffile}{Diagrams/LoopExpansionBosonicHS_ApplicationWickTheoremS4_Diag5}
\begin{fmfgraph}(28,20)
\fmfleft{i}
\fmfright{o}
\fmfv{decor.shape=triangle,decor.filled=shaded,decor.size=0.16w,foreground=(0,,0,,1)}{v1}
\fmfv{decor.shape=triangle,decor.filled=shaded,decor.size=0.16w,foreground=(0,,0,,1)}{v2}
\fmf{phantom,tension=25}{i,i1}
\fmf{phantom,tension=25}{o,o1}
\fmf{wiggly,left,tension=0.7}{i1,v1,i1}
\fmf{wiggly,right,tension=0.7}{o1,v2,o1}
\fmf{wiggly}{v1,v2}
\end{fmfgraph}
\end{fmffile}
\end{gathered} \hspace{0.05cm} + 6 \hspace{-0.4cm} \begin{gathered}
\begin{fmffile}{Diagrams/LoopExpansionBosonicHS_ApplicationWickTheoremS4_Diag6}
\begin{fmfgraph}(28,20)
\fmfleft{i}
\fmfright{o}
\fmfv{decor.shape=triangle,decor.filled=shaded,decor.size=0.16w,foreground=(0,,0,,1)}{v1}
\fmfv{decor.shape=triangle,decor.filled=shaded,decor.size=0.16w,foreground=(0,,0,,1)}{v2}
\fmf{phantom,tension=5}{i,v1}
\fmf{phantom,tension=5}{v2,o}
\fmf{wiggly,left,tension=0.8}{v1,v2,v1}
\fmf{wiggly}{v1,v2}
\end{fmfgraph}
\end{fmffile}
\end{gathered} \hspace{-0.5cm} \;.
\end{split}
\label{eq:ApplicationWickTheoremBosonicActionS3}
\end{equation}
In these expressions, the propagator of the collective field as well as the vertex functions can be evaluated after exploiting the following expression for the derivative of the 
original field propagator $\boldsymbol{G}_{\widetilde{\sigma}}$ defined by Eq.~\eqref{eq:Gcoll}:
\begin{equation}
\begin{split}
& \frac{\delta \boldsymbol{G}_{\widetilde{\sigma};ab}(x,y)}{\delta \widetilde{\sigma}(z)} \\
& = \frac{\delta \left(\boldsymbol{G}^{-1}_{\widetilde{\sigma}}\right)_{ab}^{-1}(x,y)}{\delta \widetilde{\sigma}(z)} \\
& = - \int_{u,v} \left.\boldsymbol{G}_{\widetilde{\sigma};a}\right.^{c}(x,u) \frac{\delta \boldsymbol{G}^{-1}_{\widetilde{\sigma};cd}(u,v)}{\delta \widetilde{\sigma}(z)} \left.\left.\boldsymbol{G}_{\widetilde{\sigma};}\right.^{d}\right._{b}(v,y) \\
& = -i\sqrt{\frac{\lambda}{3}} \left.\boldsymbol{G}_{\widetilde{\sigma};a}\right.^c(x,z)\boldsymbol{G}_{\widetilde{\sigma};cb}(z,y) \;.
\end{split}
\end{equation}
From this, we show:
\begin{equation}
\begin{split}
D^{-1}_{\sigma_\text{cl};\mathcal{J}}(x,y) = & - \hspace{0.2cm} \begin{gathered}
\begin{fmffile}{Diagrams/LoopExpansionBosonicHS_Vertex_Diag1}
\begin{fmfgraph*}(15,20)
\fmfleft{i1,i2}
\fmfright{o1,o2}
\fmfv{label=$x$,label.dist=4,foreground=(0,,0,,1)}{v1}
\fmfv{label=$y$,label.dist=4,foreground=(0,,0,,1)}{v2}
\fmfv{decor.shape=circle,decor.filled=empty,decor.size=.17w,l=$\times$,label.dist=0}{v3}
\fmfv{decor.shape=circle,decor.filled=empty,decor.size=.17w,l=$\times$,label.dist=0}{v4}
\fmf{phantom}{i1,v1}
\fmf{phantom}{i2,v3}
\fmf{phantom}{o1,v2}
\fmf{phantom}{o2,v4}
\fmf{plain,tension=0.2}{v1,v2}
\fmf{phantom,tension=0.2}{v3,v4}
\fmf{plain}{v1,v3}
\fmf{plain}{v2,v4}
\fmfdot{v1,v2}
\end{fmfgraph*}
\end{fmffile}
\end{gathered} \hspace{0.15cm} - \frac{1}{2} \hspace{0.15cm} \begin{gathered}
\begin{fmffile}{Diagrams/LoopExpansionBosonicHS_Vertex_Diag2}
\begin{fmfgraph*}(20,20)
\fmfleft{i0,i1,i2,i3}
\fmfright{o0,o1,o2,o3}
\fmfv{label=$x$,label.dist=4,foreground=(0,,0,,1)}{v1}
\fmfv{label=$y$,label.dist=4,foreground=(0,,0,,1)}{v2}
\fmf{phantom}{i1,v1}
\fmf{phantom}{i2,v1}
\fmf{plain,left,tension=0.4}{v1,v2,v1}
\fmf{phantom}{v2,o1}
\fmf{phantom}{v2,o2}
\fmfdot{v1,v2}
\end{fmfgraph*}
\end{fmffile}
\end{gathered} \\
& + \delta(x-y)\;,
\end{split}
\label{eq:SbosonicKLoopExpansionH}
\end{equation}
\begin{equation}
\begin{split}
\begin{gathered}
\begin{fmffile}{Diagrams/LoopExpansionBosonicHS_FeynRuleVertexS3}
\begin{fmfgraph*}(4,4)
\fmfleft{i0,i1,i2,i3}
\fmfright{o0,o1,o2,o3}
\fmfv{label=$x$,label.angle=-135,label.dist=11,decor.shape=triangle,decor.filled=shaded,decor.size=1.6w,foreground=(0,,0,,1)}{v1}
\fmfv{label=$y$,label.angle=-45,label.dist=11}{v2}
\fmfv{label=$z$,label.angle=90,label.dist=11}{v3}
\fmfbottom{v1bis}
\fmf{phantom}{i1,v1}
\fmf{phantom}{i2,v1}
\fmf{phantom}{v1,o1}
\fmf{phantom}{v1,o2}
\fmf{phantom}{i1,v2}
\fmf{phantom}{i2,v2}
\fmf{phantom}{v2,o1}
\fmf{phantom}{v2,o2}
\fmf{phantom}{i1,v3}
\fmf{phantom}{i2,v3}
\fmf{phantom}{v3,o1}
\fmf{phantom}{v3,o2}
\fmf{phantom,tension=0.6}{v1,v1bis}
\end{fmfgraph*}
\end{fmffile}
\end{gathered} \hspace{0.5cm} = & \hspace{0.2cm} \begin{gathered}
\begin{fmffile}{Diagrams/LoopExpansionBosonicHS_Vertex_Diag3}
\begin{fmfgraph*}(22,20)
\fmfleft{i1,i2}
\fmfright{o1,o2}
\fmfbottom{i0,o0}
\fmfv{label=$x$,label.dist=4,label.angle=-90,foreground=(0,,0,,1)}{v1}
\fmfv{label=$y$,label.dist=4,foreground=(0,,0,,1)}{v2}
\fmfv{label=$z$,label.dist=4,label.angle=180,foreground=(0,,0,,1)}{v3}
\fmfv{decor.shape=circle,decor.filled=empty,decor.size=.12w,l=$\times$,label.dist=0}{v1b}
\fmfv{decor.shape=circle,decor.filled=empty,decor.size=.12w,l=$\times$,label.dist=0}{v3b}
\fmf{phantom}{i1,v1}
\fmf{phantom}{i2,v3b}
\fmf{phantom}{o1,v2}
\fmf{phantom}{o2,v3b}
\fmf{phantom}{i0,v1b}
\fmf{phantom}{o0,v2b}
\fmf{plain,tension=0.5}{v1,v2}
\fmf{phantom}{v1,v3}
\fmf{plain}{v2,v3}
\fmf{plain,tension=0.2}{v1,v1b}
\fmf{phantom,tension=0.2}{v2,v2b}
\fmf{plain,tension=1.5}{v3,v3b}
\fmfdot{v1,v2,v3}
\end{fmfgraph*}  
\end{fmffile}
\end{gathered} \hspace{-0.4cm} + \hspace{-0.4cm} \begin{gathered}
\begin{fmffile}{Diagrams/LoopExpansionBosonicHS_Vertex_Diag4}
\begin{fmfgraph*}(22,20)
\fmfleft{i1,i2}
\fmfright{o1,o2}
\fmfbottom{i0,o0}
\fmfv{label=$x$,label.dist=4,foreground=(0,,0,,1)}{v1}
\fmfv{label=$y$,label.dist=4,label.angle=-90,foreground=(0,,0,,1)}{v2}
\fmfv{label=$z$,label.dist=4,label.angle=180,foreground=(0,,0,,1)}{v3}
\fmfv{decor.shape=circle,decor.filled=empty,decor.size=.12w,l=$\times$,label.dist=0}{v2b}
\fmfv{decor.shape=circle,decor.filled=empty,decor.size=.12w,l=$\times$,label.dist=0}{v3b}
\fmf{phantom}{i1,v1}
\fmf{phantom}{i2,v3b}
\fmf{phantom}{o1,v2}
\fmf{phantom}{o2,v3b}
\fmf{phantom}{i0,v1b}
\fmf{phantom}{o0,v2b}
\fmf{plain,tension=0.5}{v1,v2}
\fmf{plain}{v1,v3}
\fmf{phantom}{v2,v3}
\fmf{phantom,tension=0.2}{v1,v1b}
\fmf{plain,tension=0.2}{v2,v2b}
\fmf{plain,tension=1.5}{v3,v3b}
\fmfdot{v1,v2,v3}
\end{fmfgraph*}
\end{fmffile}
\end{gathered} \hspace{0.1cm} + \hspace{0.1cm} \begin{gathered}
\begin{fmffile}{Diagrams/LoopExpansionBosonicHS_Vertex_Diag5}
\begin{fmfgraph*}(22,20)
\fmfleft{i1,i2}
\fmfright{o1,o2}
\fmfbottom{i0,o0}
\fmfv{label=$x$,label.dist=4,label.angle=-90,foreground=(0,,0,,1)}{v1}
\fmfv{label=$y$,label.dist=4,label.angle=-90,foreground=(0,,0,,1)}{v2}
\fmfv{label=$z$,label.dist=4,label.angle=90,foreground=(0,,0,,1)}{v3}
\fmfv{decor.shape=circle,decor.filled=empty,decor.size=.12w,l=$\times$,label.dist=0}{v1b}
\fmfv{decor.shape=circle,decor.filled=empty,decor.size=.12w,l=$\times$,label.dist=0}{v2b}
\fmf{phantom}{i1,v1}
\fmf{phantom}{i2,v3b}
\fmf{phantom}{o1,v2}
\fmf{phantom}{o2,v3b}
\fmf{phantom}{i0,v1b}
\fmf{phantom}{o0,v2b}
\fmf{phantom,tension=0.5}{v1,v2}
\fmf{plain}{v1,v3}
\fmf{plain}{v2,v3}
\fmf{plain,tension=0.2}{v1,v1b}
\fmf{plain,tension=0.2}{v2,v2b}
\fmf{phantom,tension=1.5}{v3,v3b}
\fmfdot{v1,v2,v3}
\end{fmfgraph*}
\end{fmffile}
\end{gathered} \\
& + \hspace{-0.4cm} \begin{gathered}
\begin{fmffile}{Diagrams/LoopExpansionBosonicHS_Vertex_Diag6}
\begin{fmfgraph*}(22,20)
\fmfleft{i1,i2}
\fmfright{o1,o2}
\fmfbottom{i0,o0}
\fmfv{label=$x$,label.dist=4,foreground=(0,,0,,1)}{v1}
\fmfv{label=$y$,label.dist=4,foreground=(0,,0,,1)}{v2}
\fmfv{label=$z$,label.dist=4,label.angle=90,foreground=(0,,0,,1)}{v3}
\fmf{phantom}{i1,v1}
\fmf{phantom}{i2,v3b}
\fmf{phantom}{o1,v2}
\fmf{phantom}{o2,v3b}
\fmf{phantom}{i0,v1b}
\fmf{phantom}{o0,v2b}
\fmf{plain,tension=0.5}{v1,v2}
\fmf{plain}{v1,v3}
\fmf{plain}{v2,v3}
\fmf{phantom,tension=0.2}{v1,v1b}
\fmf{phantom,tension=0.2}{v2,v2b}
\fmf{phantom,tension=1.5}{v3,v3b}
\fmfdot{v1,v2,v3}
\end{fmfgraph*}
\end{fmffile}
\end{gathered} \hspace{-0.3cm} \;,
\end{split}
\label{eq:SbosonicKLoopExpansionS3}
\end{equation}
and
\pagebreak
\begin{strip}
\begin{equation}
\begin{split}
\begin{gathered}
\begin{fmffile}{Diagrams/LoopExpansionBosonicHS_FeynRuleVertexS4}
\begin{fmfgraph*}(4,4)
\fmfleft{i0,i1,i2,i3}
\fmfright{o0,o1,o2,o3}
\fmfv{label=$x$,label.angle=-135,label.dist=11,decor.shape=square,decor.filled=shaded,decor.size=1.2w,foreground=(0,,0,,1)}{v1}
\fmfv{label=$y$,label.angle=-45,label.dist=11}{v2}
\fmfv{label=$z$,label.angle=45,label.dist=11}{v3}
\fmfv{label=$u$,label.angle=135,label.dist=11}{v4}
\fmfbottom{v1bis}
\fmf{phantom}{i1,v1}
\fmf{phantom}{i2,v1}
\fmf{phantom}{v1,o1}
\fmf{phantom}{v1,o2}
\fmf{phantom}{i1,v2}
\fmf{phantom}{i2,v2}
\fmf{phantom}{v2,o1}
\fmf{phantom}{v2,o2}
\fmf{phantom}{i1,v3}
\fmf{phantom}{i2,v3}
\fmf{phantom}{v3,o1}
\fmf{phantom}{v3,o2}
\fmf{phantom}{i1,v4}
\fmf{phantom}{i2,v4}
\fmf{phantom}{v4,o1}
\fmf{phantom}{v4,o2}
\fmf{phantom,tension=0.6}{v1,v1bis}
\end{fmfgraph*}
\end{fmffile}
\end{gathered} \hspace{0.5cm} = & -\left(\rule{0cm}{1.1cm}\right. \hspace{-0.31cm} \begin{gathered}
\begin{fmffile}{Diagrams/LoopExpansionBosonicHS_Vertex_Diag7}
\begin{fmfgraph*}(25,25)
\fmfleft{i1,i2}
\fmfright{o1,o2}
\fmfbottom{i0,o0}
\fmftop{i3,o3}
\fmfv{label=$x$,label.dist=4,label.angle=180,foreground=(0,,0,,1)}{v1}
\fmfv{label=$y$,label.dist=4,foreground=(0,,0,,1)}{v2}
\fmfv{label=$z$,label.dist=4,foreground=(0,,0,,1)}{v3}
\fmfv{label=$u$,label.dist=4,label.angle=180,foreground=(0,,0,,1)}{v4}
\fmfv{decor.shape=circle,decor.filled=empty,decor.size=.108w,l=$\times$,label.dist=0}{v1b}
\fmfv{decor.shape=circle,decor.filled=empty,decor.size=.108w,l=$\times$,label.dist=0}{v4b}
\fmf{phantom}{i1,v1}
\fmf{phantom}{i2,v4}
\fmf{phantom}{o1,v2}
\fmf{phantom}{o2,v3}
\fmf{phantom}{i3,v4b}
\fmf{phantom}{o3,v3b}
\fmf{phantom}{i0,v1b}
\fmf{phantom}{o0,v2b}
\fmf{plain,tension=1.6}{v1,v2}
\fmf{plain,tension=1.6}{v3,v4}
\fmf{phantom,tension=2.0}{v1,v4}
\fmf{plain,tension=2.0}{v2,v3}
\fmf{phantom,tension=0}{v1,v3}
\fmf{phantom,tension=0}{v2,v4}
\fmf{plain}{v1,v1b}
\fmf{phantom}{v2,v2b}
\fmf{phantom}{v3,v3b}
\fmf{plain}{v4,v4b}
\fmfdot{v1,v2,v3,v4}
\end{fmfgraph*}
\end{fmffile}
\end{gathered} + \begin{gathered}
\begin{fmffile}{Diagrams/LoopExpansionBosonicHS_Vertex_Diag8}
\begin{fmfgraph*}(25,25)
\fmfleft{i1,i2}
\fmfright{o1,o2}
\fmfbottom{i0,o0}
\fmftop{i3,o3}
\fmfv{label=$x$,label.dist=4,foreground=(0,,0,,1)}{v1}
\fmfv{label=$y$,label.dist=4,label.angle=0,foreground=(0,,0,,1)}{v2}
\fmfv{label=$z$,label.dist=4,foreground=(0,,0,,1)}{v3}
\fmfv{label=$u$,label.dist=4,label.angle=180,foreground=(0,,0,,1)}{v4}
\fmfv{decor.shape=circle,decor.filled=empty,decor.size=.108w,l=$\times$,label.dist=0}{v2b}
\fmfv{decor.shape=circle,decor.filled=empty,decor.size=.108w,l=$\times$,label.dist=0}{v4b}
\fmf{phantom}{i1,v1}
\fmf{phantom}{i2,v4}
\fmf{phantom}{o1,v2}
\fmf{phantom}{o2,v3}
\fmf{phantom}{i3,v4b}
\fmf{phantom}{o3,v3b}
\fmf{phantom}{i0,v1b}
\fmf{phantom}{o0,v2b}
\fmf{plain,tension=1.6}{v1,v2}
\fmf{plain,tension=1.6}{v3,v4}
\fmf{phantom,tension=2.0}{v1,v4}
\fmf{phantom,tension=2.0}{v2,v3}
\fmf{plain,tension=0}{v1,v3}
\fmf{phantom,tension=0}{v2,v4}
\fmf{phantom}{v1,v1b}
\fmf{plain}{v2,v2b}
\fmf{phantom}{v3,v3b}
\fmf{plain}{v4,v4b}
\fmfdot{v1,v2,v3,v4}
\end{fmfgraph*}
\end{fmffile}
\end{gathered} + \begin{gathered}
\begin{fmffile}{Diagrams/LoopExpansionBosonicHS_Vertex_Diag9}
\begin{fmfgraph*}(25,25)
\fmfleft{i1,i2}
\fmfright{o1,o2}
\fmfbottom{i0,o0}
\fmftop{i3,o3}
\fmfv{label=$x$,label.dist=4,foreground=(0,,0,,1)}{v1}
\fmfv{label=$y$,label.dist=4,label.angle=0,foreground=(0,,0,,1)}{v2}
\fmfv{label=$z$,label.dist=4,foreground=(0,,0,,1)}{v3}
\fmfv{label=$u$,label.dist=4,label.angle=180,foreground=(0,,0,,1)}{v4}
\fmfv{decor.shape=circle,decor.filled=empty,decor.size=.108w,l=$\times$,label.dist=0}{v2b}
\fmfv{decor.shape=circle,decor.filled=empty,decor.size=.108w,l=$\times$,label.dist=0}{v4b}
\fmf{phantom}{i1,v1}
\fmf{phantom}{i2,v4}
\fmf{phantom}{o1,v2}
\fmf{phantom}{o2,v3}
\fmf{phantom}{i3,v4b}
\fmf{phantom}{o3,v3b}
\fmf{phantom}{i0,v1b}
\fmf{phantom}{o0,v2b}
\fmf{phantom,tension=1.6}{v1,v2}
\fmf{phantom,tension=1.6}{v3,v4}
\fmf{plain,tension=2.0}{v1,v4}
\fmf{plain,tension=2.0}{v2,v3}
\fmf{plain,tension=0}{v1,v3}
\fmf{phantom,tension=0}{v2,v4}
\fmf{phantom}{v1,v1b}
\fmf{plain}{v2,v2b}
\fmf{phantom}{v3,v3b}
\fmf{plain}{v4,v4b}
\fmfdot{v1,v2,v3,v4}
\end{fmfgraph*}
\end{fmffile}
\end{gathered} + \begin{gathered}
\begin{fmffile}{Diagrams/LoopExpansionBosonicHS_Vertex_Diag10}
\begin{fmfgraph*}(25,25)
\fmfleft{i1,i2}
\fmfright{o1,o2}
\fmfbottom{i0,o0}
\fmftop{i3,o3}
\fmfv{label=$x$,label.dist=4,label.angle=180,foreground=(0,,0,,1)}{v1}
\fmfv{label=$y$,label.dist=4,label.angle=0,foreground=(0,,0,,1)}{v2}
\fmfv{label=$z$,label.dist=4,foreground=(0,,0,,1)}{v3}
\fmfv{label=$u$,label.dist=4,foreground=(0,,0,,1)}{v4}
\fmfv{decor.shape=circle,decor.filled=empty,decor.size=.108w,l=$\times$,label.dist=0}{v1b}
\fmfv{decor.shape=circle,decor.filled=empty,decor.size=.108w,l=$\times$,label.dist=0}{v2b}
\fmf{phantom}{i1,v1}
\fmf{phantom}{i2,v4}
\fmf{phantom}{o1,v2}
\fmf{phantom}{o2,v3}
\fmf{phantom}{i3,v4b}
\fmf{phantom}{o3,v3b}
\fmf{phantom}{i0,v1b}
\fmf{phantom}{o0,v2b}
\fmf{phantom,tension=1.6}{v1,v2}
\fmf{plain,tension=1.6}{v3,v4}
\fmf{plain,tension=2.0}{v1,v4}
\fmf{plain,tension=2.0}{v2,v3}
\fmf{phantom,tension=0}{v1,v3}
\fmf{phantom,tension=0}{v2,v4}
\fmf{plain}{v1,v1b}
\fmf{plain}{v2,v2b}
\fmf{phantom}{v3,v3b}
\fmf{phantom}{v4,v4b}
\fmfdot{v1,v2,v3,v4}
\end{fmfgraph*}
\end{fmffile}
\end{gathered} + \begin{gathered}
\begin{fmffile}{Diagrams/LoopExpansionBosonicHS_Vertex_Diag11}
\begin{fmfgraph*}(25,25)
\fmfleft{i1,i2}
\fmfright{o1,o2}
\fmfbottom{i0,o0}
\fmftop{i3,o3}
\fmfv{label=$x$,label.dist=4,label.angle=180,foreground=(0,,0,,1)}{v1}
\fmfv{label=$y$,label.dist=4,foreground=(0,,0,,1)}{v2}
\fmfv{label=$z$,label.dist=4,foreground=(0,,0,,1)}{v3}
\fmfv{label=$u$,label.dist=4,label.angle=180,foreground=(0,,0,,1)}{v4}
\fmfv{decor.shape=circle,decor.filled=empty,decor.size=.108w,l=$\times$,label.dist=0}{v1b}
\fmfv{decor.shape=circle,decor.filled=empty,decor.size=.108w,l=$\times$,label.dist=0}{v4b}
\fmf{phantom}{i1,v1}
\fmf{phantom}{i2,v4}
\fmf{phantom}{o1,v2}
\fmf{phantom}{o2,v3}
\fmf{phantom}{i3,v4b}
\fmf{phantom}{o3,v3b}
\fmf{phantom}{i0,v1b}
\fmf{phantom}{o0,v2b}
\fmf{phantom,tension=1.6}{v1,v2}
\fmf{phantom,tension=1.6}{v3,v4}
\fmf{phantom,tension=2.0}{v1,v4}
\fmf{plain,tension=2.0}{v2,v3}
\fmf{plain,tension=0}{v1,v3}
\fmf{plain,tension=0}{v2,v4}
\fmf{plain}{v1,v1b}
\fmf{phantom}{v2,v2b}
\fmf{phantom}{v3,v3b}
\fmf{plain}{v4,v4b}
\fmfdot{v1,v2,v3,v4}
\end{fmfgraph*}
\end{fmffile}
\end{gathered} \\
& + \begin{gathered}
\begin{fmffile}{Diagrams/LoopExpansionBosonicHS_Vertex_Diag12}
\begin{fmfgraph*}(25,25)
\fmfleft{i1,i2}
\fmfright{o1,o2}
\fmfbottom{i0,o0}
\fmftop{i3,o3}
\fmfv{label=$x$,label.dist=4,foreground=(0,,0,,1)}{v1}
\fmfv{label=$y$,label.dist=4,foreground=(0,,0,,1)}{v2}
\fmfv{label=$z$,label.dist=4,label.angle=0,foreground=(0,,0,,1)}{v3}
\fmfv{label=$u$,label.dist=4,label.angle=180,foreground=(0,,0,,1)}{v4}
\fmfv{decor.shape=circle,decor.filled=empty,decor.size=.108w,l=$\times$,label.dist=0}{v3b}
\fmfv{decor.shape=circle,decor.filled=empty,decor.size=.108w,l=$\times$,label.dist=0}{v4b}
\fmf{phantom}{i1,v1}
\fmf{phantom}{i2,v4}
\fmf{phantom}{o1,v2}
\fmf{phantom}{o2,v3}
\fmf{phantom}{i3,v4b}
\fmf{phantom}{o3,v3b}
\fmf{phantom}{i0,v1b}
\fmf{phantom}{o0,v2b}
\fmf{plain,tension=1.6}{v1,v2}
\fmf{phantom,tension=1.6}{v3,v4}
\fmf{phantom,tension=2.0}{v1,v4}
\fmf{phantom,tension=2.0}{v2,v3}
\fmf{plain,tension=0}{v1,v3}
\fmf{plain,tension=0}{v2,v4}
\fmf{phantom}{v1,v1b}
\fmf{phantom}{v2,v2b}
\fmf{plain}{v3,v3b}
\fmf{plain}{v4,v4b}
\fmfdot{v1,v2,v3,v4}
\end{fmfgraph*}
\end{fmffile}
\end{gathered} + \begin{gathered}
\begin{fmffile}{Diagrams/LoopExpansionBosonicHS_Vertex_Diag13}
\begin{fmfgraph*}(25,25)
\fmfleft{i1,i2}
\fmfright{o1,o2}
\fmfbottom{i0,o0}
\fmftop{i3,o3}
\fmfv{label=$x$,label.dist=4,foreground=(0,,0,,1)}{v1}
\fmfv{label=$y$,label.dist=4,foreground=(0,,0,,1)}{v2}
\fmfv{label=$z$,label.dist=4,label.angle=0,foreground=(0,,0,,1)}{v3}
\fmfv{label=$u$,label.dist=4,label.angle=180,foreground=(0,,0,,1)}{v4}
\fmfv{decor.shape=circle,decor.filled=empty,decor.size=.108w,l=$\times$,label.dist=0}{v3b}
\fmfv{decor.shape=circle,decor.filled=empty,decor.size=.108w,l=$\times$,label.dist=0}{v4b}
\fmf{phantom}{i1,v1}
\fmf{phantom}{i2,v4}
\fmf{phantom}{o1,v2}
\fmf{phantom}{o2,v3}
\fmf{phantom}{i3,v4b}
\fmf{phantom}{o3,v3b}
\fmf{phantom}{i0,v1b}
\fmf{phantom}{o0,v2b}
\fmf{plain,tension=1.6}{v1,v2}
\fmf{phantom,tension=1.6}{v3,v4}
\fmf{plain,tension=2.0}{v1,v4}
\fmf{plain,tension=2.0}{v2,v3}
\fmf{phantom,tension=0}{v1,v3}
\fmf{phantom,tension=0}{v2,v4}
\fmf{phantom}{v1,v1b}
\fmf{phantom}{v2,v2b}
\fmf{plain}{v3,v3b}
\fmf{plain}{v4,v4b}
\fmfdot{v1,v2,v3,v4}
\end{fmfgraph*}
\end{fmffile}
\end{gathered} + \begin{gathered}
\begin{fmffile}{Diagrams/LoopExpansionBosonicHS_Vertex_Diag14}
\begin{fmfgraph*}(25,25)
\fmfleft{i1,i2}
\fmfright{o1,o2}
\fmfbottom{i0,o0}
\fmftop{i3,o3}
\fmfv{label=$x$,label.dist=4,label.angle=180,foreground=(0,,0,,1)}{v1}
\fmfv{label=$y$,label.dist=4,foreground=(0,,0,,1)}{v2}
\fmfv{label=$z$,label.dist=4,label.angle=0,foreground=(0,,0,,1)}{v3}
\fmfv{label=$u$,label.dist=4,foreground=(0,,0,,1)}{v4}
\fmfv{decor.shape=circle,decor.filled=empty,decor.size=.108w,l=$\times$,label.dist=0}{v1b}
\fmfv{decor.shape=circle,decor.filled=empty,decor.size=.108w,l=$\times$,label.dist=0}{v3b}
\fmf{phantom}{i1,v1}
\fmf{phantom}{i2,v4}
\fmf{phantom}{o1,v2}
\fmf{phantom}{o2,v3}
\fmf{phantom}{i3,v4b}
\fmf{phantom}{o3,v3b}
\fmf{phantom}{i0,v1b}
\fmf{phantom}{o0,v2b}
\fmf{phantom,tension=1.6}{v1,v2}
\fmf{phantom,tension=1.6}{v3,v4}
\fmf{plain,tension=2.0}{v1,v4}
\fmf{plain,tension=2.0}{v2,v3}
\fmf{phantom,tension=0}{v1,v3}
\fmf{plain,tension=0}{v2,v4}
\fmf{plain}{v1,v1b}
\fmf{phantom}{v2,v2b}
\fmf{plain}{v3,v3b}
\fmf{phantom}{v4,v4b}
\fmfdot{v1,v2,v3,v4}
\end{fmfgraph*}
\end{fmffile}
\end{gathered} + \begin{gathered}
\begin{fmffile}{Diagrams/LoopExpansionBosonicHS_Vertex_Diag15}
\begin{fmfgraph*}(25,25)
\fmfleft{i1,i2}
\fmfright{o1,o2}
\fmfbottom{i0,o0}
\fmftop{i3,o3}
\fmfv{label=$x$,label.dist=4,label.angle=180,foreground=(0,,0,,1)}{v1}
\fmfv{label=$y$,label.dist=4,label.angle=0,foreground=(0,,0,,1)}{v2}
\fmfv{label=$z$,label.dist=4,foreground=(0,,0,,1)}{v3}
\fmfv{label=$u$,label.dist=4,foreground=(0,,0,,1)}{v4}
\fmfv{decor.shape=circle,decor.filled=empty,decor.size=.108w,l=$\times$,label.dist=0}{v1b}
\fmfv{decor.shape=circle,decor.filled=empty,decor.size=.108w,l=$\times$,label.dist=0}{v2b}
\fmf{phantom}{i1,v1}
\fmf{phantom}{i2,v4}
\fmf{phantom}{o1,v2}
\fmf{phantom}{o2,v3}
\fmf{phantom}{i3,v4b}
\fmf{phantom}{o3,v3b}
\fmf{phantom}{i0,v1b}
\fmf{phantom}{o0,v2b}
\fmf{phantom,tension=1.6}{v1,v2}
\fmf{plain,tension=1.6}{v3,v4}
\fmf{phantom,tension=2.0}{v1,v4}
\fmf{phantom,tension=2.0}{v2,v3}
\fmf{plain,tension=0}{v1,v3}
\fmf{plain,tension=0}{v2,v4}
\fmf{plain}{v1,v1b}
\fmf{plain}{v2,v2b}
\fmf{phantom}{v3,v3b}
\fmf{phantom}{v4,v4b}
\fmfdot{v1,v2,v3,v4}
\end{fmfgraph*}
\end{fmffile}
\end{gathered} + \begin{gathered}
\begin{fmffile}{Diagrams/LoopExpansionBosonicHS_Vertex_Diag16}
\begin{fmfgraph*}(25,25)
\fmfleft{i1,i2}
\fmfright{o1,o2}
\fmfbottom{i0,o0}
\fmftop{i3,o3}
\fmfv{label=$x$,label.dist=4,foreground=(0,,0,,1)}{v1}
\fmfv{label=$y$,label.dist=4,label.angle=0,foreground=(0,,0,,1)}{v2}
\fmfv{label=$z$,label.dist=4,label.angle=0,foreground=(0,,0,,1)}{v3}
\fmfv{label=$u$,label.dist=4,foreground=(0,,0,,1)}{v4}
\fmfv{decor.shape=circle,decor.filled=empty,decor.size=.108w,l=$\times$,label.dist=0}{v2b}
\fmfv{decor.shape=circle,decor.filled=empty,decor.size=.108w,l=$\times$,label.dist=0}{v3b}
\fmf{phantom}{i1,v1}
\fmf{phantom}{i2,v4}
\fmf{phantom}{o1,v2}
\fmf{phantom}{o2,v3}
\fmf{phantom}{i3,v4b}
\fmf{phantom}{o3,v3b}
\fmf{phantom}{i0,v1b}
\fmf{phantom}{o0,v2b}
\fmf{phantom,tension=1.6}{v1,v2}
\fmf{phantom,tension=1.6}{v3,v4}
\fmf{plain,tension=2.0}{v1,v4}
\fmf{phantom,tension=2.0}{v2,v3}
\fmf{plain,tension=0}{v1,v3}
\fmf{plain,tension=0}{v2,v4}
\fmf{phantom}{v1,v1b}
\fmf{plain}{v2,v2b}
\fmf{plain}{v3,v3b}
\fmf{phantom}{v4,v4b}
\fmfdot{v1,v2,v3,v4}
\end{fmfgraph*}
\end{fmffile}
\end{gathered} \\
& + \begin{gathered}
\begin{fmffile}{Diagrams/LoopExpansionBosonicHS_Vertex_Diag17}
\begin{fmfgraph*}(25,25)
\fmfleft{i1,i2}
\fmfright{o1,o2}
\fmfbottom{i0,o0}
\fmftop{i3,o3}
\fmfv{label=$x$,label.dist=4,foreground=(0,,0,,1)}{v1}
\fmfv{label=$y$,label.dist=4,label.angle=0,foreground=(0,,0,,1)}{v2}
\fmfv{label=$z$,label.dist=4,label.angle=0,foreground=(0,,0,,1)}{v3}
\fmfv{label=$u$,label.dist=4,foreground=(0,,0,,1)}{v4}
\fmfv{decor.shape=circle,decor.filled=empty,decor.size=.108w,l=$\times$,label.dist=0}{v2b}
\fmfv{decor.shape=circle,decor.filled=empty,decor.size=.108w,l=$\times$,label.dist=0}{v3b}
\fmf{phantom}{i1,v1}
\fmf{phantom}{i2,v4}
\fmf{phantom}{o1,v2}
\fmf{phantom}{o2,v3}
\fmf{phantom}{i3,v4b}
\fmf{phantom}{o3,v3b}
\fmf{phantom}{i0,v1b}
\fmf{phantom}{o0,v2b}
\fmf{plain,tension=1.6}{v1,v2}
\fmf{plain,tension=1.6}{v3,v4}
\fmf{plain,tension=2.0}{v1,v4}
\fmf{phantom,tension=2.0}{v2,v3}
\fmf{phantom,tension=0}{v1,v3}
\fmf{phantom,tension=0}{v2,v4}
\fmf{phantom}{v1,v1b}
\fmf{plain}{v2,v2b}
\fmf{plain}{v3,v3b}
\fmf{phantom}{v4,v4b}
\fmfdot{v1,v2,v3,v4}
\end{fmfgraph*}
\end{fmffile}
\end{gathered}+ \begin{gathered}
\begin{fmffile}{Diagrams/LoopExpansionBosonicHS_Vertex_Diag18}
\begin{fmfgraph*}(25,25)
\fmfleft{i1,i2}
\fmfright{o1,o2}
\fmfbottom{i0,o0}
\fmftop{i3,o3}
\fmfv{label=$x$,label.dist=4,label.angle=180,foreground=(0,,0,,1)}{v1}
\fmfv{label=$y$,label.dist=4,foreground=(0,,0,,1)}{v2}
\fmfv{label=$z$,label.dist=4,label.angle=0,foreground=(0,,0,,1)}{v3}
\fmfv{label=$u$,label.dist=4,foreground=(0,,0,,1)}{v4}
\fmfv{decor.shape=circle,decor.filled=empty,decor.size=.108w,l=$\times$,label.dist=0}{v1b}
\fmfv{decor.shape=circle,decor.filled=empty,decor.size=.108w,l=$\times$,label.dist=0}{v3b}
\fmf{phantom}{i1,v1}
\fmf{phantom}{i2,v4}
\fmf{phantom}{o1,v2}
\fmf{phantom}{o2,v3}
\fmf{phantom}{i3,v4b}
\fmf{phantom}{o3,v3b}
\fmf{phantom}{i0,v1b}
\fmf{phantom}{o0,v2b}
\fmf{plain,tension=1.6}{v1,v2}
\fmf{plain,tension=1.6}{v3,v4}
\fmf{phantom,tension=2.0}{v1,v4}
\fmf{phantom,tension=2.0}{v2,v3}
\fmf{phantom,tension=0}{v1,v3}
\fmf{plain,tension=0}{v2,v4}
\fmf{plain}{v1,v1b}
\fmf{phantom}{v2,v2b}
\fmf{plain}{v3,v3b}
\fmf{phantom}{v4,v4b}
\fmfdot{v1,v2,v3,v4}
\end{fmfgraph*}
\end{fmffile}
\end{gathered} + \begin{gathered}
\begin{fmffile}{Diagrams/LoopExpansionBosonicHS_Vertex_Diag19}
\begin{fmfgraph*}(25,25)
\fmfleft{i1,i2}
\fmfright{o1,o2}
\fmfbottom{i0,o0}
\fmftop{i3,o3}
\fmfv{label=$x$,label.dist=4,foreground=(0,,0,,1)}{v1}
\fmfv{label=$y$,label.dist=4,foreground=(0,,0,,1)}{v2}
\fmfv{label=$z$,label.dist=4,foreground=(0,,0,,1)}{v3}
\fmfv{label=$u$,label.dist=4,foreground=(0,,0,,1)}{v4}
\fmf{phantom}{i1,v1}
\fmf{phantom}{i2,v4}
\fmf{phantom}{o1,v2}
\fmf{phantom}{o2,v3}
\fmf{phantom}{i3,v4b}
\fmf{phantom}{o3,v3b}
\fmf{phantom}{i0,v1b}
\fmf{phantom}{o0,v2b}
\fmf{plain,tension=1.6}{v1,v2}
\fmf{plain,tension=1.6}{v3,v4}
\fmf{phantom,tension=2.0}{v1,v4}
\fmf{phantom,tension=2.0}{v2,v3}
\fmf{plain,tension=0}{v1,v3}
\fmf{plain,tension=0}{v2,v4}
\fmf{phantom}{v1,v1b}
\fmf{phantom}{v2,v2b}
\fmf{phantom}{v3,v3b}
\fmf{phantom}{v4,v4b}
\fmfdot{v1,v2,v3,v4}
\end{fmfgraph*}
\end{fmffile}
\end{gathered} + \begin{gathered}
\begin{fmffile}{Diagrams/LoopExpansionBosonicHS_Vertex_Diag20}
\begin{fmfgraph*}(25,25)
\fmfleft{i1,i2}
\fmfright{o1,o2}
\fmfbottom{i0,o0}
\fmftop{i3,o3}
\fmfv{label=$x$,label.dist=4,foreground=(0,,0,,1)}{v1}
\fmfv{label=$y$,label.dist=4,foreground=(0,,0,,1)}{v2}
\fmfv{label=$z$,label.dist=4,foreground=(0,,0,,1)}{v3}
\fmfv{label=$u$,label.dist=4,foreground=(0,,0,,1)}{v4}
\fmf{phantom}{i1,v1}
\fmf{phantom}{i2,v4}
\fmf{phantom}{o1,v2}
\fmf{phantom}{o2,v3}
\fmf{phantom}{i3,v4b}
\fmf{phantom}{o3,v3b}
\fmf{phantom}{i0,v1b}
\fmf{phantom}{o0,v2b}
\fmf{phantom,tension=1.6}{v1,v2}
\fmf{phantom,tension=1.6}{v3,v4}
\fmf{plain,tension=2.0}{v1,v4}
\fmf{plain,tension=2.0}{v2,v3}
\fmf{plain,tension=0}{v1,v3}
\fmf{plain,tension=0}{v2,v4}
\fmf{phantom}{v1,v1b}
\fmf{phantom}{v2,v2b}
\fmf{phantom}{v3,v3b}
\fmf{phantom}{v4,v4b}
\fmfdot{v1,v2,v3,v4}
\end{fmfgraph*}
\end{fmffile}
\end{gathered}+ \begin{gathered}
\begin{fmffile}{Diagrams/LoopExpansionBosonicHS_Vertex_Diag21}
\begin{fmfgraph*}(25,25)
\fmfleft{i1,i2}
\fmfright{o1,o2}
\fmfbottom{i0,o0}
\fmftop{i3,o3}
\fmfv{label=$x$,label.dist=4,foreground=(0,,0,,1)}{v1}
\fmfv{label=$y$,label.dist=4,foreground=(0,,0,,1)}{v2}
\fmfv{label=$z$,label.dist=4,foreground=(0,,0,,1)}{v3}
\fmfv{label=$u$,label.dist=4,foreground=(0,,0,,1)}{v4}
\fmf{phantom}{i1,v1}
\fmf{phantom}{i2,v4}
\fmf{phantom}{o1,v2}
\fmf{phantom}{o2,v3}
\fmf{phantom}{i3,v4b}
\fmf{phantom}{o3,v3b}
\fmf{phantom}{i0,v1b}
\fmf{phantom}{o0,v2b}
\fmf{plain,tension=1.6}{v1,v2}
\fmf{plain,tension=1.6}{v3,v4}
\fmf{plain,tension=2.0}{v1,v4}
\fmf{plain,tension=2.0}{v2,v3}
\fmf{phantom,tension=0}{v1,v3}
\fmf{phantom,tension=0}{v2,v4}
\fmf{phantom}{v1,v1b}
\fmf{phantom}{v2,v2b}
\fmf{phantom}{v3,v3b}
\fmf{phantom}{v4,v4b}
\fmfdot{v1,v2,v3,v4}
\end{fmfgraph*}
\end{fmffile}
\end{gathered} \hspace{-0.31cm} \left.\rule{0cm}{1.1cm}\right)\;.
\end{split}
\label{eq:SbosonicKLoopExpansionS4}
\end{equation}
\pagebreak
After plugging the vertex functions~\eqref{eq:SbosonicKLoopExpansionS3} and~\eqref{eq:SbosonicKLoopExpansionS4} into Eqs.~\eqref{eq:ApplicationWickTheoremBosonicActionS4} and~\eqref{eq:ApplicationWickTheoremBosonicActionS3} combined with Eq.~\eqref{eq:Zlecol}, the Schwinger functional in the collective representation is expressed up to order $\mathcal{O}\big(\hbar^2\big)$ as follows:
\begin{equation}
\begin{split}
W^\text{LE;col}\big[\mathcal{J}\big] = & -S_{\mathrm{col},\mathcal{J}}[\sigma_\text{cl}]+\frac{\hbar}{2}\mathrm{Tr}\left[\ln\big(D_{\sigma_\text{cl};\mathcal{J}}\big)\right] \\
& + \hbar^{2} \left[\rule{0cm}{1.1cm}\right. \frac{1}{8} \left(\rule{0cm}{1.1cm}\right. 4 \hspace{-0.2cm} \begin{gathered}
\begin{fmffile}{Diagrams/LoopExpansionBosonicHS_W_Diag1}
\begin{fmfgraph*}(25,25)
\fmfleft{i1,i2}
\fmfright{o1,o2}
\fmfbottom{i0,o0}
\fmftop{i3,o3}
\fmfv{decor.shape=circle,decor.size=2.0thick,foreground=(0,,0,,1)}{v1}
\fmfv{decor.shape=circle,decor.size=2.0thick,foreground=(0,,0,,1)}{v2}
\fmfv{decor.shape=circle,decor.size=2.0thick,foreground=(0,,0,,1)}{v3}
\fmfv{decor.shape=circle,decor.size=2.0thick,foreground=(0,,0,,1)}{v4}
\fmfv{decor.shape=circle,decor.filled=empty,decor.size=.1w,l=$\times$,label.dist=0}{v3b}
\fmfv{decor.shape=circle,decor.filled=empty,decor.size=.1w,l=$\times$,label.dist=0}{v4b}
\fmf{phantom}{i1,v1}
\fmf{phantom}{i2,v4}
\fmf{phantom}{o1,v2}
\fmf{phantom}{o2,v3}
\fmf{phantom}{i3,v4b}
\fmf{phantom}{o3,v3b}
\fmf{phantom}{i0,v1b}
\fmf{phantom}{o0,v2b}
\fmf{plain,tension=1.6}{v1,v2}
\fmf{phantom,tension=1.6}{v3,v4}
\fmf{wiggly,tension=2.0}{v1,v4}
\fmf{wiggly,tension=2.0}{v2,v3}
\fmf{phantom,tension=0}{v1,v3}
\fmf{phantom,tension=0}{v2,v4}
\fmf{plain,right=0.8,tension=0}{v2,v3}
\fmf{plain,left=0.8,tension=0}{v1,v4}
\fmf{phantom}{v1,v1b}
\fmf{phantom}{v2,v2b}
\fmf{plain}{v3,v3b}
\fmf{plain}{v4,v4b}
\end{fmfgraph*}
\end{fmffile}
\end{gathered} \hspace{-0.2cm} + 4 \hspace{-0.2cm} \begin{gathered}
\begin{fmffile}{Diagrams/LoopExpansionBosonicHS_W_Diag2}
\begin{fmfgraph*}(25,25)
\fmfleft{i1,i2}
\fmfright{o1,o2}
\fmfbottom{i0,o0}
\fmftop{i3,o3}
\fmfv{decor.shape=circle,decor.size=2.0thick,foreground=(0,,0,,1)}{v1}
\fmfv{decor.shape=circle,decor.size=2.0thick,foreground=(0,,0,,1)}{v2}
\fmfv{decor.shape=circle,decor.size=2.0thick,foreground=(0,,0,,1)}{v3}
\fmfv{decor.shape=circle,decor.size=2.0thick,foreground=(0,,0,,1)}{v4}
\fmfv{decor.shape=circle,decor.filled=empty,decor.size=.1w,l=$\times$,label.dist=0}{v1b}
\fmfv{decor.shape=circle,decor.filled=empty,decor.size=.1w,l=$\times$,label.dist=0}{v4b}
\fmf{phantom}{i1,v1}
\fmf{phantom}{i2,v4}
\fmf{phantom}{o1,v2}
\fmf{phantom}{o2,v3}
\fmf{phantom}{i3,v4b}
\fmf{phantom}{o3,v3b}
\fmf{phantom}{i0,v1b}
\fmf{phantom}{o0,v2b}
\fmf{plain,tension=1.6}{v1,v2}
\fmf{plain,tension=1.6}{v3,v4}
\fmf{wiggly,tension=2.0}{v1,v4}
\fmf{wiggly,tension=2.0}{v2,v3}
\fmf{phantom,tension=0}{v1,v3}
\fmf{phantom,tension=0}{v2,v4}
\fmf{plain,right=0.8,tension=0}{v2,v3}
\fmf{phantom,left=0.8,tension=0}{v1,v4}
\fmf{plain}{v1,v1b}
\fmf{phantom}{v2,v2b}
\fmf{phantom}{v3,v3b}
\fmf{plain}{v4,v4b}
\end{fmfgraph*}
\end{fmffile}
\end{gathered} \hspace{-0.5cm} + 4 \hspace{-0.3cm} \begin{gathered}
\begin{fmffile}{Diagrams/LoopExpansionBosonicHS_W_Diag3}
\begin{fmfgraph*}(25,25)
\fmfleft{i1,i2}
\fmfright{o1,o2}
\fmfbottom{i0,o0}
\fmftop{i3,o3}
\fmfv{decor.shape=circle,decor.size=2.0thick,foreground=(0,,0,,1)}{v1}
\fmfv{decor.shape=circle,decor.size=2.0thick,foreground=(0,,0,,1)}{v2}
\fmfv{decor.shape=circle,decor.size=2.0thick,foreground=(0,,0,,1)}{v3}
\fmfv{decor.shape=circle,decor.size=2.0thick,foreground=(0,,0,,1)}{v4}
\fmfv{decor.shape=circle,decor.filled=empty,decor.size=.1w,l=$\times$,label.dist=0}{v3b}
\fmfv{decor.shape=circle,decor.filled=empty,decor.size=.1w,l=$\times$,label.dist=0}{v4b}
\fmf{phantom}{i1,v1}
\fmf{phantom}{i2,v4}
\fmf{phantom}{o1,v2}
\fmf{phantom}{o2,v3}
\fmf{phantom}{i3,v4b}
\fmf{phantom}{o3,v3b}
\fmf{phantom}{i0,v1b}
\fmf{phantom}{o0,v2b}
\fmf{plain,tension=1.6}{v1,v2}
\fmf{phantom,tension=1.6}{v3,v4}
\fmf{wiggly,tension=2.0}{v1,v4}
\fmf{wiggly,tension=2.0}{v2,v3}
\fmf{plain,tension=0}{v1,v3}
\fmf{plain,tension=0}{v2,v4}
\fmf{phantom,right=0.8,tension=0}{v2,v3}
\fmf{phantom,left=0.8,tension=0}{v1,v4}
\fmf{phantom}{v1,v1b}
\fmf{phantom}{v2,v2b}
\fmf{plain}{v3,v3b}
\fmf{plain}{v4,v4b}
\end{fmfgraph*}
\end{fmffile}
\end{gathered} \hspace{-0.4cm} + 2 \hspace{-0.4cm} \begin{gathered}
\begin{fmffile}{Diagrams/LoopExpansionBosonicHS_W_Diag5}
\begin{fmfgraph*}(25,25)
\fmfleft{i1,i2}
\fmfright{o1,o2}
\fmfbottom{i0,o0}
\fmftop{i3,o3}
\fmfv{decor.shape=circle,decor.size=2.0thick,foreground=(0,,0,,1)}{v1}
\fmfv{decor.shape=circle,decor.size=2.0thick,foreground=(0,,0,,1)}{v2}
\fmfv{decor.shape=circle,decor.size=2.0thick,foreground=(0,,0,,1)}{v3}
\fmfv{decor.shape=circle,decor.size=2.0thick,foreground=(0,,0,,1)}{v4}
\fmf{phantom}{i1,v1}
\fmf{phantom}{i2,v4}
\fmf{phantom}{o1,v2}
\fmf{phantom}{o2,v3}
\fmf{phantom}{i3,v4b}
\fmf{phantom}{o3,v3b}
\fmf{phantom}{i0,v1b}
\fmf{phantom}{o0,v2b}
\fmf{plain,tension=1.6}{v1,v2}
\fmf{plain,tension=1.6}{v3,v4}
\fmf{wiggly,tension=2.0}{v1,v4}
\fmf{wiggly,tension=2.0}{v2,v3}
\fmf{phantom,tension=0}{v1,v3}
\fmf{phantom,tension=0}{v2,v4}
\fmf{plain,right=0.8,tension=0}{v2,v3}
\fmf{plain,left=0.8,tension=0}{v1,v4}
\fmf{phantom}{v1,v1b}
\fmf{phantom}{v2,v2b}
\fmf{phantom}{v3,v3b}
\fmf{phantom}{v4,v4b}
\end{fmfgraph*}
\end{fmffile}
\end{gathered} \hspace{-0.5cm} + \hspace{-0.7cm} \begin{gathered}
\begin{fmffile}{Diagrams/LoopExpansionBosonicHS_W_Diag6}
\begin{fmfgraph*}(25,25)
\fmfleft{i1,i2}
\fmfright{o1,o2}
\fmfbottom{i0,o0}
\fmftop{i3,o3}
\fmfv{decor.shape=circle,decor.size=2.0thick,foreground=(0,,0,,1)}{v1}
\fmfv{decor.shape=circle,decor.size=2.0thick,foreground=(0,,0,,1)}{v2}
\fmfv{decor.shape=circle,decor.size=2.0thick,foreground=(0,,0,,1)}{v3}
\fmfv{decor.shape=circle,decor.size=2.0thick,foreground=(0,,0,,1)}{v4}
\fmf{phantom}{i1,v1}
\fmf{phantom}{i2,v4}
\fmf{phantom}{o1,v2}
\fmf{phantom}{o2,v3}
\fmf{phantom}{i3,v4b}
\fmf{phantom}{o3,v3b}
\fmf{phantom}{i0,v1b}
\fmf{phantom}{o0,v2b}
\fmf{plain,tension=1.6}{v1,v2}
\fmf{plain,tension=1.6}{v3,v4}
\fmf{wiggly,tension=2.0}{v1,v4}
\fmf{wiggly,tension=2.0}{v2,v3}
\fmf{plain,tension=0}{v1,v3}
\fmf{plain,tension=0}{v2,v4}
\fmf{phantom,right=0.8,tension=0}{v2,v3}
\fmf{phantom,left=0.8,tension=0}{v1,v4}
\fmf{phantom}{v1,v1b}
\fmf{phantom}{v2,v2b}
\fmf{phantom}{v3,v3b}
\fmf{phantom}{v4,v4b}
\end{fmfgraph*}
\end{fmffile}
\end{gathered} \hspace{-0.6cm} \left.\rule{0cm}{1.1cm}\right) \\
& \hspace{0.9cm} + \frac{1}{12} \left(\rule{0cm}{1.1cm}\right. 6 \begin{gathered}
\begin{fmffile}{Diagrams/LoopExpansionBosonicHS_W_Diag7}
\begin{fmfgraph*}(35,20)
\fmfleft{i1,i2}
\fmfright{o1,o2}
\fmfbottom{i0,o0}
\fmfbottom{b0}
\fmfbottom{b1}
\fmfbottom{b2}
\fmftop{i3,o3}
\fmfv{decor.shape=circle,decor.size=2.0thick,foreground=(0,,0,,1)}{v1}
\fmfv{decor.shape=circle,decor.size=2.0thick,foreground=(0,,0,,1)}{v2}
\fmfv{decor.shape=circle,decor.size=2.0thick,foreground=(0,,0,,1)}{v3}
\fmfv{decor.shape=circle,decor.size=2.0thick,foreground=(0,,0,,1)}{v4}
\fmfv{decor.shape=circle,decor.size=2.0thick,foreground=(0,,0,,1)}{v5}
\fmfv{decor.shape=circle,decor.size=2.0thick,foreground=(0,,0,,1)}{v6}
\fmfv{decor.shape=circle,decor.filled=empty,decor.size=.073w,l=$\times$,label.dist=0}{v1b}
\fmfv{decor.shape=circle,decor.filled=empty,decor.size=.073w,l=$\times$,label.dist=0}{v3b}
\fmfv{decor.shape=circle,decor.filled=empty,decor.size=.073w,l=$\times$,label.dist=0}{v4b}
\fmfv{decor.shape=circle,decor.filled=empty,decor.size=.073w,l=$\times$,label.dist=0}{v6b}
\fmf{phantom,tension=1.4}{i1,v1}
\fmf{phantom}{i2,v3b}
\fmf{phantom}{i0,v1b}
\fmf{phantom}{o2,v6b}
\fmf{phantom,tension=1.4}{o1,v5}
\fmf{phantom}{o0,v5b}
\fmf{phantom,tension=1.11}{v3b,v6b}
\fmf{phantom,tension=1.38}{i0,v2}
\fmf{phantom,tension=1.38}{o0,v2}
\fmf{phantom,tension=1.8}{i0,v2b}
\fmf{phantom,tension=1.2}{o0,v2b}
\fmf{phantom,tension=1.2}{b0,v2b}
\fmf{phantom,tension=1.2}{b1,v2b}
\fmf{phantom,tension=1.2}{b2,v2b}
\fmf{phantom,tension=1.38}{i0,v4}
\fmf{phantom,tension=1.38}{o0,v4}
\fmf{phantom,tension=1.2}{i0,v4b}
\fmf{phantom,tension=1.8}{o0,v4b}
\fmf{phantom,tension=1.2}{b0,v4b}
\fmf{phantom,tension=1.2}{b1,v4b}
\fmf{phantom,tension=1.2}{b2,v4b}
\fmf{phantom,tension=2}{i3,v3}
\fmf{phantom,tension=2}{o3,v6}
\fmf{phantom,tension=2}{i3,v3b}
\fmf{phantom,tension=0.8}{o3,v3b}
\fmf{phantom,tension=0.8}{i3,v6b}
\fmf{phantom,tension=2}{o3,v6b}
\fmf{plain,tension=1.4}{v1,v2}
\fmf{plain,tension=1.4}{v4,v5}
\fmf{phantom}{v1,v3}
\fmf{phantom,left=0.8,tension=0}{v1,v3}
\fmf{plain}{v5,v6}
\fmf{phantom,right=0.8,tension=0}{v5,v6}
\fmf{plain}{v2,v3}
\fmf{phantom}{v4,v6}
\fmf{wiggly,tension=0.5}{v2,v4}
\fmf{wiggly,tension=2}{v3,v6}
\fmf{wiggly,right=0.8,tension=0}{v1,v5}
\fmf{plain,tension=1}{v1,v1b}
\fmf{phantom,tension=0.2}{v2,v2b}
\fmf{plain,tension=1.5}{v3,v3b}
\fmf{plain,tension=0.2}{v4,v4b}
\fmf{phantom,tension=1}{v5,v5b}
\fmf{plain,tension=1.5}{v6,v6b}
\end{fmfgraph*}
\end{fmffile}
\end{gathered} \hspace{-0.6cm} + 3 \begin{gathered}
\begin{fmffile}{Diagrams/LoopExpansionBosonicHS_W_Diag8}
\begin{fmfgraph*}(35,20)
\fmfleft{i1,i2}
\fmfright{o1,o2}
\fmfbottom{i0,o0}
\fmfbottom{b0}
\fmfbottom{b1}
\fmfbottom{b2}
\fmftop{i3,o3}
\fmfv{decor.shape=circle,decor.size=2.0thick,foreground=(0,,0,,1)}{v1}
\fmfv{decor.shape=circle,decor.size=2.0thick,foreground=(0,,0,,1)}{v2}
\fmfv{decor.shape=circle,decor.size=2.0thick,foreground=(0,,0,,1)}{v3}
\fmfv{decor.shape=circle,decor.size=2.0thick,foreground=(0,,0,,1)}{v4}
\fmfv{decor.shape=circle,decor.size=2.0thick,foreground=(0,,0,,1)}{v5}
\fmfv{decor.shape=circle,decor.size=2.0thick,foreground=(0,,0,,1)}{v6}
\fmfv{decor.shape=circle,decor.filled=empty,decor.size=.073w,l=$\times$,label.dist=0}{v1b}
\fmfv{decor.shape=circle,decor.filled=empty,decor.size=.073w,l=$\times$,label.dist=0}{v3b}
\fmfv{decor.shape=circle,decor.filled=empty,decor.size=.073w,l=$\times$,label.dist=0}{v5b}
\fmfv{decor.shape=circle,decor.filled=empty,decor.size=.073w,l=$\times$,label.dist=0}{v6b}
\fmf{phantom,tension=1.4}{i1,v1}
\fmf{phantom}{i2,v3b}
\fmf{phantom}{i0,v1b}
\fmf{phantom}{o2,v6b}
\fmf{phantom,tension=1.4}{o1,v5}
\fmf{phantom}{o0,v5b}
\fmf{phantom,tension=1.11}{v3b,v6b}
\fmf{phantom,tension=1.38}{i0,v2}
\fmf{phantom,tension=1.38}{o0,v2}
\fmf{phantom,tension=1.8}{i0,v2b}
\fmf{phantom,tension=1.2}{o0,v2b}
\fmf{phantom,tension=1.2}{b0,v2b}
\fmf{phantom,tension=1.2}{b1,v2b}
\fmf{phantom,tension=1.2}{b2,v2b}
\fmf{phantom,tension=1.38}{i0,v4}
\fmf{phantom,tension=1.38}{o0,v4}
\fmf{phantom,tension=1.2}{i0,v4b}
\fmf{phantom,tension=1.8}{o0,v4b}
\fmf{phantom,tension=1.2}{b0,v4b}
\fmf{phantom,tension=1.2}{b1,v4b}
\fmf{phantom,tension=1.2}{b2,v4b}
\fmf{phantom,tension=2}{i3,v3}
\fmf{phantom,tension=2}{o3,v6}
\fmf{phantom,tension=2}{i3,v3b}
\fmf{phantom,tension=0.8}{o3,v3b}
\fmf{phantom,tension=0.8}{i3,v6b}
\fmf{phantom,tension=2}{o3,v6b}
\fmf{plain,tension=1.4}{v1,v2}
\fmf{plain,tension=1.4}{v4,v5}
\fmf{phantom}{v1,v3}
\fmf{phantom,left=0.8,tension=0}{v1,v3}
\fmf{phantom}{v5,v6}
\fmf{phantom,right=0.8,tension=0}{v5,v6}
\fmf{plain}{v2,v3}
\fmf{plain}{v4,v6}
\fmf{wiggly,tension=0.5}{v2,v4}
\fmf{wiggly,tension=2}{v3,v6}
\fmf{wiggly,right=0.8,tension=0}{v1,v5}
\fmf{plain,tension=1}{v1,v1b}
\fmf{phantom,tension=0.2}{v2,v2b}
\fmf{plain,tension=1.5}{v3,v3b}
\fmf{phantom,tension=0.2}{v4,v4b}
\fmf{plain,tension=1}{v5,v5b}
\fmf{plain,tension=1.5}{v6,v6b}
\end{fmfgraph*}
\end{fmffile}
\end{gathered} \hspace{-0.3cm} + 6 \begin{gathered}
\begin{fmffile}{Diagrams/LoopExpansionBosonicHS_W_Diag9}
\begin{fmfgraph*}(35,20)
\fmfleft{i1,i2}
\fmfright{o1,o2}
\fmfbottom{i0,o0}
\fmfbottom{b0}
\fmfbottom{b1}
\fmfbottom{b2}
\fmftop{i3,o3}
\fmfv{decor.shape=circle,decor.size=2.0thick,foreground=(0,,0,,1)}{v1}
\fmfv{decor.shape=circle,decor.size=2.0thick,foreground=(0,,0,,1)}{v2}
\fmfv{decor.shape=circle,decor.size=2.0thick,foreground=(0,,0,,1)}{v3}
\fmfv{decor.shape=circle,decor.size=2.0thick,foreground=(0,,0,,1)}{v4}
\fmfv{decor.shape=circle,decor.size=2.0thick,foreground=(0,,0,,1)}{v5}
\fmfv{decor.shape=circle,decor.size=2.0thick,foreground=(0,,0,,1)}{v6}
\fmfv{decor.shape=circle,decor.filled=empty,decor.size=.073w,l=$\times$,label.dist=0}{v1b}
\fmfv{decor.shape=circle,decor.filled=empty,decor.size=.073w,l=$\times$,label.dist=0}{v3b}
\fmf{phantom,tension=1.4}{i1,v1}
\fmf{phantom}{i2,v3b}
\fmf{phantom}{i0,v1b}
\fmf{phantom}{o2,v6b}
\fmf{phantom,tension=1.4}{o1,v5}
\fmf{phantom}{o0,v5b}
\fmf{phantom,tension=1.11}{v3b,v6b}
\fmf{phantom,tension=1.38}{i0,v2}
\fmf{phantom,tension=1.38}{o0,v2}
\fmf{phantom,tension=1.8}{i0,v2b}
\fmf{phantom,tension=1.2}{o0,v2b}
\fmf{phantom,tension=1.2}{b0,v2b}
\fmf{phantom,tension=1.2}{b1,v2b}
\fmf{phantom,tension=1.2}{b2,v2b}
\fmf{phantom,tension=1.38}{i0,v4}
\fmf{phantom,tension=1.38}{o0,v4}
\fmf{phantom,tension=1.2}{i0,v4b}
\fmf{phantom,tension=1.8}{o0,v4b}
\fmf{phantom,tension=1.2}{b0,v4b}
\fmf{phantom,tension=1.2}{b1,v4b}
\fmf{phantom,tension=1.2}{b2,v4b}
\fmf{phantom,tension=2}{i3,v3}
\fmf{phantom,tension=2}{o3,v6}
\fmf{phantom,tension=2}{i3,v3b}
\fmf{phantom,tension=0.8}{o3,v3b}
\fmf{phantom,tension=0.8}{i3,v6b}
\fmf{phantom,tension=2}{o3,v6b}
\fmf{plain,tension=1.4}{v1,v2}
\fmf{plain,tension=1.4}{v4,v5}
\fmf{phantom}{v1,v3}
\fmf{phantom,left=0.8,tension=0}{v1,v3}
\fmf{plain}{v5,v6}
\fmf{phantom,right=0.8,tension=0}{v5,v6}
\fmf{plain}{v2,v3}
\fmf{plain}{v4,v6}
\fmf{wiggly,tension=0.5}{v2,v4}
\fmf{wiggly,tension=2}{v3,v6}
\fmf{wiggly,right=0.8,tension=0}{v1,v5}
\fmf{plain,tension=1}{v1,v1b}
\fmf{phantom,tension=0.2}{v2,v2b}
\fmf{plain,tension=1.5}{v3,v3b}
\fmf{phantom,tension=0.2}{v4,v4b}
\fmf{phantom,tension=1}{v5,v5b}
\fmf{phantom,tension=1.5}{v6,v6b}
\end{fmfgraph*}
\end{fmffile}
\end{gathered} \hspace{-0.6cm} + \hspace{-0.7cm} \begin{gathered}
\begin{fmffile}{Diagrams/LoopExpansionBosonicHS_W_Diag10}
\begin{fmfgraph*}(35,20)
\fmfleft{i1,i2}
\fmfright{o1,o2}
\fmfbottom{i0,o0}
\fmfbottom{b0}
\fmfbottom{b1}
\fmfbottom{b2}
\fmftop{i3,o3}
\fmfv{decor.shape=circle,decor.size=2.0thick,foreground=(0,,0,,1)}{v1}
\fmfv{decor.shape=circle,decor.size=2.0thick,foreground=(0,,0,,1)}{v2}
\fmfv{decor.shape=circle,decor.size=2.0thick,foreground=(0,,0,,1)}{v3}
\fmfv{decor.shape=circle,decor.size=2.0thick,foreground=(0,,0,,1)}{v4}
\fmfv{decor.shape=circle,decor.size=2.0thick,foreground=(0,,0,,1)}{v5}
\fmfv{decor.shape=circle,decor.size=2.0thick,foreground=(0,,0,,1)}{v6}
\fmf{phantom,tension=1.4}{i1,v1}
\fmf{phantom}{i2,v3b}
\fmf{phantom}{i0,v1b}
\fmf{phantom}{o2,v6b}
\fmf{phantom,tension=1.4}{o1,v5}
\fmf{phantom}{o0,v5b}
\fmf{phantom,tension=1.11}{v3b,v6b}
\fmf{phantom,tension=1.38}{i0,v2}
\fmf{phantom,tension=1.38}{o0,v2}
\fmf{phantom,tension=1.8}{i0,v2b}
\fmf{phantom,tension=1.2}{o0,v2b}
\fmf{phantom,tension=1.2}{b0,v2b}
\fmf{phantom,tension=1.2}{b1,v2b}
\fmf{phantom,tension=1.2}{b2,v2b}
\fmf{phantom,tension=1.38}{i0,v4}
\fmf{phantom,tension=1.38}{o0,v4}
\fmf{phantom,tension=1.2}{i0,v4b}
\fmf{phantom,tension=1.8}{o0,v4b}
\fmf{phantom,tension=1.2}{b0,v4b}
\fmf{phantom,tension=1.2}{b1,v4b}
\fmf{phantom,tension=1.2}{b2,v4b}
\fmf{phantom,tension=2}{i3,v3}
\fmf{phantom,tension=2}{o3,v6}
\fmf{phantom,tension=2}{i3,v3b}
\fmf{phantom,tension=0.8}{o3,v3b}
\fmf{phantom,tension=0.8}{i3,v6b}
\fmf{phantom,tension=2}{o3,v6b}
\fmf{plain,tension=1.4}{v1,v2}
\fmf{plain,tension=1.4}{v4,v5}
\fmf{plain}{v1,v3}
\fmf{phantom,left=0.8,tension=0}{v1,v3}
\fmf{plain}{v5,v6}
\fmf{phantom,right=0.8,tension=0}{v5,v6}
\fmf{plain}{v2,v3}
\fmf{plain}{v4,v6}
\fmf{wiggly,tension=0.5}{v2,v4}
\fmf{wiggly,tension=2}{v3,v6}
\fmf{wiggly,right=0.8,tension=0}{v1,v5}
\fmf{phantom,tension=1}{v1,v1b}
\fmf{phantom,tension=0.2}{v2,v2b}
\fmf{phantom,tension=1.5}{v3,v3b}
\fmf{phantom,tension=0.2}{v4,v4b}
\fmf{phantom,tension=1}{v5,v5b}
\fmf{phantom,tension=1.5}{v6,v6b}
\end{fmfgraph*}
\end{fmffile}
\end{gathered} \hspace{-0.6cm} \left.\rule{0cm}{1.1cm}\right) \\
\\
& \hspace{0.9cm} + \frac{1}{8} \left(\rule{0cm}{1.1cm}\right. 4 \begin{gathered}
\begin{fmffile}{Diagrams/LoopExpansionBosonicHS_W_Diag11}
\begin{fmfgraph*}(35,20)
\fmfleft{i1,i2}
\fmfright{o1,o2}
\fmfbottom{i0,o0}
\fmfbottom{b0}
\fmfbottom{b1}
\fmfbottom{b2}
\fmftop{i3,o3}
\fmfv{decor.shape=circle,decor.size=2.0thick,foreground=(0,,0,,1)}{v1}
\fmfv{decor.shape=circle,decor.size=2.0thick,foreground=(0,,0,,1)}{v2}
\fmfv{decor.shape=circle,decor.size=2.0thick,foreground=(0,,0,,1)}{v3}
\fmfv{decor.shape=circle,decor.size=2.0thick,foreground=(0,,0,,1)}{v4}
\fmfv{decor.shape=circle,decor.size=2.0thick,foreground=(0,,0,,1)}{v5}
\fmfv{decor.shape=circle,decor.size=2.0thick,foreground=(0,,0,,1)}{v6}
\fmfv{decor.shape=circle,decor.filled=empty,decor.size=.073w,l=$\times$,label.dist=0}{v1b}
\fmfv{decor.shape=circle,decor.filled=empty,decor.size=.073w,l=$\times$,label.dist=0}{v3b}
\fmfv{decor.shape=circle,decor.filled=empty,decor.size=.073w,l=$\times$,label.dist=0}{v4b}
\fmfv{decor.shape=circle,decor.filled=empty,decor.size=.073w,l=$\times$,label.dist=0}{v6b}
\fmf{phantom,tension=1.4}{i1,v1}
\fmf{phantom}{i2,v3b}
\fmf{phantom}{i0,v1b}
\fmf{phantom}{o2,v6b}
\fmf{phantom,tension=1.4}{o1,v5}
\fmf{phantom}{o0,v5b}
\fmf{phantom,tension=1.11}{v3b,v6b}
\fmf{phantom,tension=1.38}{i0,v2}
\fmf{phantom,tension=1.38}{o0,v2}
\fmf{phantom,tension=1.8}{i0,v2b}
\fmf{phantom,tension=1.2}{o0,v2b}
\fmf{phantom,tension=1.2}{b0,v2b}
\fmf{phantom,tension=1.2}{b1,v2b}
\fmf{phantom,tension=1.2}{b2,v2b}
\fmf{phantom,tension=1.38}{i0,v4}
\fmf{phantom,tension=1.38}{o0,v4}
\fmf{phantom,tension=1.2}{i0,v4b}
\fmf{phantom,tension=1.8}{o0,v4b}
\fmf{phantom,tension=1.2}{b0,v4b}
\fmf{phantom,tension=1.2}{b1,v4b}
\fmf{phantom,tension=1.2}{b2,v4b}
\fmf{phantom,tension=2}{i3,v3}
\fmf{phantom,tension=2}{o3,v6}
\fmf{phantom,tension=2}{i3,v3b}
\fmf{phantom,tension=0.8}{o3,v3b}
\fmf{phantom,tension=0.8}{i3,v6b}
\fmf{phantom,tension=2}{o3,v6b}
\fmf{plain,tension=1.4}{v1,v2}
\fmf{plain,tension=1.4}{v4,v5}
\fmf{wiggly}{v1,v3}
\fmf{phantom,left=0.8,tension=0}{v1,v3}
\fmf{wiggly}{v5,v6}
\fmf{plain,right=0.8,tension=0}{v5,v6}
\fmf{plain}{v2,v3}
\fmf{phantom}{v4,v6}
\fmf{wiggly,tension=0.5}{v2,v4}
\fmf{phantom,tension=2}{v3,v6}
\fmf{phantom,right=0.8,tension=0}{v1,v5}
\fmf{plain,tension=1}{v1,v1b}
\fmf{phantom,tension=0.2}{v2,v2b}
\fmf{plain,tension=1.5}{v3,v3b}
\fmf{plain,tension=0.2}{v4,v4b}
\fmf{phantom,tension=1}{v5,v5b}
\fmf{plain,tension=1.5}{v6,v6b}
\end{fmfgraph*}
\end{fmffile}
\end{gathered} \hspace{-0.4cm} + 4 \begin{gathered}
\begin{fmffile}{Diagrams/LoopExpansionBosonicHS_W_Diag12}
\begin{fmfgraph*}(35,20)
\fmfleft{i1,i2}
\fmfright{o1,o2}
\fmfbottom{i0,o0}
\fmfbottom{b0}
\fmfbottom{b1}
\fmfbottom{b2}
\fmftop{i3,o3}
\fmfv{decor.shape=circle,decor.size=2.0thick,foreground=(0,,0,,1)}{v1}
\fmfv{decor.shape=circle,decor.size=2.0thick,foreground=(0,,0,,1)}{v2}
\fmfv{decor.shape=circle,decor.size=2.0thick,foreground=(0,,0,,1)}{v3}
\fmfv{decor.shape=circle,decor.size=2.0thick,foreground=(0,,0,,1)}{v4}
\fmfv{decor.shape=circle,decor.size=2.0thick,foreground=(0,,0,,1)}{v5}
\fmfv{decor.shape=circle,decor.size=2.0thick,foreground=(0,,0,,1)}{v6}
\fmfv{decor.shape=circle,decor.filled=empty,decor.size=.073w,l=$\times$,label.dist=0}{v2b}
\fmfv{decor.shape=circle,decor.filled=empty,decor.size=.073w,l=$\times$,label.dist=0}{v3b}
\fmfv{decor.shape=circle,decor.filled=empty,decor.size=.073w,l=$\times$,label.dist=0}{v4b}
\fmfv{decor.shape=circle,decor.filled=empty,decor.size=.073w,l=$\times$,label.dist=0}{v6b}
\fmf{phantom,tension=1.4}{i1,v1}
\fmf{phantom}{i2,v3b}
\fmf{phantom}{i0,v1b}
\fmf{phantom}{o2,v6b}
\fmf{phantom,tension=1.4}{o1,v5}
\fmf{phantom}{o0,v5b}
\fmf{phantom,tension=1.11}{v3b,v6b}
\fmf{phantom,tension=1.38}{i0,v2}
\fmf{phantom,tension=1.38}{o0,v2}
\fmf{phantom,tension=1.8}{i0,v2b}
\fmf{phantom,tension=1.2}{o0,v2b}
\fmf{phantom,tension=1.2}{b0,v2b}
\fmf{phantom,tension=1.2}{b1,v2b}
\fmf{phantom,tension=1.2}{b2,v2b}
\fmf{phantom,tension=1.38}{i0,v4}
\fmf{phantom,tension=1.38}{o0,v4}
\fmf{phantom,tension=1.2}{i0,v4b}
\fmf{phantom,tension=1.8}{o0,v4b}
\fmf{phantom,tension=1.2}{b0,v4b}
\fmf{phantom,tension=1.2}{b1,v4b}
\fmf{phantom,tension=1.2}{b2,v4b}
\fmf{phantom,tension=2}{i3,v3}
\fmf{phantom,tension=2}{o3,v6}
\fmf{phantom,tension=2}{i3,v3b}
\fmf{phantom,tension=0.8}{o3,v3b}
\fmf{phantom,tension=0.8}{i3,v6b}
\fmf{phantom,tension=2}{o3,v6b}
\fmf{plain,tension=1.4}{v1,v2}
\fmf{plain,tension=1.4}{v4,v5}
\fmf{wiggly}{v1,v3}
\fmf{plain,left=0.8,tension=0}{v1,v3}
\fmf{wiggly}{v5,v6}
\fmf{plain,right=0.8,tension=0}{v5,v6}
\fmf{phantom}{v2,v3}
\fmf{phantom}{v4,v6}
\fmf{wiggly,tension=0.5}{v2,v4}
\fmf{phantom,tension=2}{v3,v6}
\fmf{phantom,right=0.8,tension=0}{v1,v5}
\fmf{phantom,tension=1}{v1,v1b}
\fmf{plain,tension=0.2}{v2,v2b}
\fmf{plain,tension=1.5}{v3,v3b}
\fmf{plain,tension=0.2}{v4,v4b}
\fmf{phantom,tension=1}{v5,v5b}
\fmf{plain,tension=1.5}{v6,v6b}
\end{fmfgraph*}
\end{fmffile}
\end{gathered} \hspace{-0.4cm} + \hspace{-0.2cm} \begin{gathered}
\begin{fmffile}{Diagrams/LoopExpansionBosonicHS_W_Diag13}
\begin{fmfgraph*}(35,20)
\fmfleft{i1,i2}
\fmfright{o1,o2}
\fmfbottom{i0,o0}
\fmfbottom{b0}
\fmfbottom{b1}
\fmfbottom{b2}
\fmftop{i3,o3}
\fmfv{decor.shape=circle,decor.size=2.0thick,foreground=(0,,0,,1)}{v1}
\fmfv{decor.shape=circle,decor.size=2.0thick,foreground=(0,,0,,1)}{v2}
\fmfv{decor.shape=circle,decor.size=2.0thick,foreground=(0,,0,,1)}{v3}
\fmfv{decor.shape=circle,decor.size=2.0thick,foreground=(0,,0,,1)}{v4}
\fmfv{decor.shape=circle,decor.size=2.0thick,foreground=(0,,0,,1)}{v5}
\fmfv{decor.shape=circle,decor.size=2.0thick,foreground=(0,,0,,1)}{v6}
\fmfv{decor.shape=circle,decor.filled=empty,decor.size=.073w,l=$\times$,label.dist=0}{v1b}
\fmfv{decor.shape=circle,decor.filled=empty,decor.size=.073w,l=$\times$,label.dist=0}{v3b}
\fmfv{decor.shape=circle,decor.filled=empty,decor.size=.073w,l=$\times$,label.dist=0}{v5b}
\fmfv{decor.shape=circle,decor.filled=empty,decor.size=.073w,l=$\times$,label.dist=0}{v6b}
\fmf{phantom,tension=1.4}{i1,v1}
\fmf{phantom}{i2,v3b}
\fmf{phantom}{i0,v1b}
\fmf{phantom}{o2,v6b}
\fmf{phantom,tension=1.4}{o1,v5}
\fmf{phantom}{o0,v5b}
\fmf{phantom,tension=1.11}{v3b,v6b}
\fmf{phantom,tension=1.38}{i0,v2}
\fmf{phantom,tension=1.38}{o0,v2}
\fmf{phantom,tension=1.8}{i0,v2b}
\fmf{phantom,tension=1.2}{o0,v2b}
\fmf{phantom,tension=1.2}{b0,v2b}
\fmf{phantom,tension=1.2}{b1,v2b}
\fmf{phantom,tension=1.2}{b2,v2b}
\fmf{phantom,tension=1.38}{i0,v4}
\fmf{phantom,tension=1.38}{o0,v4}
\fmf{phantom,tension=1.2}{i0,v4b}
\fmf{phantom,tension=1.8}{o0,v4b}
\fmf{phantom,tension=1.2}{b0,v4b}
\fmf{phantom,tension=1.2}{b1,v4b}
\fmf{phantom,tension=1.2}{b2,v4b}
\fmf{phantom,tension=2}{i3,v3}
\fmf{phantom,tension=2}{o3,v6}
\fmf{phantom,tension=2}{i3,v3b}
\fmf{phantom,tension=0.8}{o3,v3b}
\fmf{phantom,tension=0.8}{i3,v6b}
\fmf{phantom,tension=2}{o3,v6b}
\fmf{plain,tension=1.4}{v1,v2}
\fmf{plain,tension=1.4}{v4,v5}
\fmf{wiggly}{v1,v3}
\fmf{phantom,left=0.8,tension=0}{v1,v3}
\fmf{wiggly}{v5,v6}
\fmf{phantom,right=0.8,tension=0}{v5,v6}
\fmf{plain}{v2,v3}
\fmf{plain}{v4,v6}
\fmf{wiggly,tension=0.5}{v2,v4}
\fmf{phantom,tension=2}{v3,v6}
\fmf{phantom,right=0.8,tension=0}{v1,v5}
\fmf{plain,tension=1}{v1,v1b}
\fmf{phantom,tension=0.2}{v2,v2b}
\fmf{plain,tension=1.5}{v3,v3b}
\fmf{phantom,tension=0.2}{v4,v4b}
\fmf{plain,tension=1}{v5,v5b}
\fmf{plain,tension=1.5}{v6,v6b}
\end{fmfgraph*}
\end{fmffile}
\end{gathered} \\
& \hspace{1.7cm} + 4 \hspace{-0.4cm} \begin{gathered}
\begin{fmffile}{Diagrams/LoopExpansionBosonicHS_W_Diag14}
\begin{fmfgraph*}(35,20)
\fmfleft{i1,i2}
\fmfright{o1,o2}
\fmfbottom{i0,o0}
\fmfbottom{b0}
\fmfbottom{b1}
\fmfbottom{b2}
\fmftop{i3,o3}
\fmfv{decor.shape=circle,decor.size=2.0thick,foreground=(0,,0,,1)}{v1}
\fmfv{decor.shape=circle,decor.size=2.0thick,foreground=(0,,0,,1)}{v2}
\fmfv{decor.shape=circle,decor.size=2.0thick,foreground=(0,,0,,1)}{v3}
\fmfv{decor.shape=circle,decor.size=2.0thick,foreground=(0,,0,,1)}{v4}
\fmfv{decor.shape=circle,decor.size=2.0thick,foreground=(0,,0,,1)}{v5}
\fmfv{decor.shape=circle,decor.size=2.0thick,foreground=(0,,0,,1)}{v6}
\fmfv{decor.shape=circle,decor.filled=empty,decor.size=.073w,l=$\times$,label.dist=0}{v2b}
\fmfv{decor.shape=circle,decor.filled=empty,decor.size=.073w,l=$\times$,label.dist=0}{v3b}
\fmf{phantom,tension=1.4}{i1,v1}
\fmf{phantom}{i2,v3b}
\fmf{phantom}{i0,v1b}
\fmf{phantom}{o2,v6b}
\fmf{phantom,tension=1.4}{o1,v5}
\fmf{phantom}{o0,v5b}
\fmf{phantom,tension=1.11}{v3b,v6b}
\fmf{phantom,tension=1.38}{i0,v2}
\fmf{phantom,tension=1.38}{o0,v2}
\fmf{phantom,tension=1.8}{i0,v2b}
\fmf{phantom,tension=1.2}{o0,v2b}
\fmf{phantom,tension=1.2}{b0,v2b}
\fmf{phantom,tension=1.2}{b1,v2b}
\fmf{phantom,tension=1.2}{b2,v2b}
\fmf{phantom,tension=1.38}{i0,v4}
\fmf{phantom,tension=1.38}{o0,v4}
\fmf{phantom,tension=1.2}{i0,v4b}
\fmf{phantom,tension=1.8}{o0,v4b}
\fmf{phantom,tension=1.2}{b0,v4b}
\fmf{phantom,tension=1.2}{b1,v4b}
\fmf{phantom,tension=1.2}{b2,v4b}
\fmf{phantom,tension=2}{i3,v3}
\fmf{phantom,tension=2}{o3,v6}
\fmf{phantom,tension=2}{i3,v3b}
\fmf{phantom,tension=0.8}{o3,v3b}
\fmf{phantom,tension=0.8}{i3,v6b}
\fmf{phantom,tension=2}{o3,v6b}
\fmf{plain,tension=1.4}{v1,v2}
\fmf{plain,tension=1.4}{v4,v5}
\fmf{wiggly}{v1,v3}
\fmf{plain,left=0.8,tension=0}{v1,v3}
\fmf{wiggly}{v5,v6}
\fmf{plain,right=0.8,tension=0}{v5,v6}
\fmf{phantom}{v2,v3}
\fmf{plain}{v4,v6}
\fmf{wiggly,tension=0.5}{v2,v4}
\fmf{phantom,tension=2}{v3,v6}
\fmf{phantom,right=0.8,tension=0}{v1,v5}
\fmf{phantom,tension=1}{v1,v1b}
\fmf{plain,tension=0.2}{v2,v2b}
\fmf{plain,tension=1.5}{v3,v3b}
\fmf{phantom,tension=0.2}{v4,v4b}
\fmf{phantom,tension=1}{v5,v5b}
\fmf{phantom,tension=1.5}{v6,v6b}
\end{fmfgraph*}
\end{fmffile}
\end{gathered} \hspace{-0.4cm} + 2 \begin{gathered}
\begin{fmffile}{Diagrams/LoopExpansionBosonicHS_W_Diag15}
\begin{fmfgraph*}(35,20)
\fmfleft{i1,i2}
\fmfright{o1,o2}
\fmfbottom{i0,o0}
\fmfbottom{b0}
\fmfbottom{b1}
\fmfbottom{b2}
\fmftop{i3,o3}
\fmfv{decor.shape=circle,decor.size=2.0thick,foreground=(0,,0,,1)}{v1}
\fmfv{decor.shape=circle,decor.size=2.0thick,foreground=(0,,0,,1)}{v2}
\fmfv{decor.shape=circle,decor.size=2.0thick,foreground=(0,,0,,1)}{v3}
\fmfv{decor.shape=circle,decor.size=2.0thick,foreground=(0,,0,,1)}{v4}
\fmfv{decor.shape=circle,decor.size=2.0thick,foreground=(0,,0,,1)}{v5}
\fmfv{decor.shape=circle,decor.size=2.0thick,foreground=(0,,0,,1)}{v6}
\fmfv{decor.shape=circle,decor.filled=empty,decor.size=.073w,l=$\times$,label.dist=0}{v1b}
\fmfv{decor.shape=circle,decor.filled=empty,decor.size=.073w,l=$\times$,label.dist=0}{v3b}
\fmf{phantom,tension=1.4}{i1,v1}
\fmf{phantom}{i2,v3b}
\fmf{phantom}{i0,v1b}
\fmf{phantom}{o2,v6b}
\fmf{phantom,tension=1.4}{o1,v5}
\fmf{phantom}{o0,v5b}
\fmf{phantom,tension=1.11}{v3b,v6b}
\fmf{phantom,tension=1.38}{i0,v2}
\fmf{phantom,tension=1.38}{o0,v2}
\fmf{phantom,tension=1.8}{i0,v2b}
\fmf{phantom,tension=1.2}{o0,v2b}
\fmf{phantom,tension=1.2}{b0,v2b}
\fmf{phantom,tension=1.2}{b1,v2b}
\fmf{phantom,tension=1.2}{b2,v2b}
\fmf{phantom,tension=1.38}{i0,v4}
\fmf{phantom,tension=1.38}{o0,v4}
\fmf{phantom,tension=1.2}{i0,v4b}
\fmf{phantom,tension=1.8}{o0,v4b}
\fmf{phantom,tension=1.2}{b0,v4b}
\fmf{phantom,tension=1.2}{b1,v4b}
\fmf{phantom,tension=1.2}{b2,v4b}
\fmf{phantom,tension=2}{i3,v3}
\fmf{phantom,tension=2}{o3,v6}
\fmf{phantom,tension=2}{i3,v3b}
\fmf{phantom,tension=0.8}{o3,v3b}
\fmf{phantom,tension=0.8}{i3,v6b}
\fmf{phantom,tension=2}{o3,v6b}
\fmf{plain,tension=1.4}{v1,v2}
\fmf{plain,tension=1.4}{v4,v5}
\fmf{wiggly}{v1,v3}
\fmf{phantom,left=0.8,tension=0}{v1,v3}
\fmf{wiggly}{v5,v6}
\fmf{plain,right=0.8,tension=0}{v5,v6}
\fmf{plain}{v2,v3}
\fmf{plain}{v4,v6}
\fmf{wiggly,tension=0.5}{v2,v4}
\fmf{phantom,tension=2}{v3,v6}
\fmf{phantom,right=0.8,tension=0}{v1,v5}
\fmf{plain,tension=1}{v1,v1b}
\fmf{phantom,tension=0.2}{v2,v2b}
\fmf{plain,tension=1.5}{v3,v3b}
\fmf{phantom,tension=0.2}{v4,v4b}
\fmf{phantom,tension=1}{v5,v5b}
\fmf{phantom,tension=1.5}{v6,v6b}
\end{fmfgraph*}
\end{fmffile}
\end{gathered} \hspace{-0.4cm} + \hspace{-0.4cm} \begin{gathered}
\begin{fmffile}{Diagrams/LoopExpansionBosonicHS_W_Diag16}
\begin{fmfgraph*}(35,20)
\fmfleft{i1,i2}
\fmfright{o1,o2}
\fmfbottom{i0,o0}
\fmfbottom{b0}
\fmfbottom{b1}
\fmfbottom{b2}
\fmftop{i3,o3}
\fmfv{decor.shape=circle,decor.size=2.0thick,foreground=(0,,0,,1)}{v1}
\fmfv{decor.shape=circle,decor.size=2.0thick,foreground=(0,,0,,1)}{v2}
\fmfv{decor.shape=circle,decor.size=2.0thick,foreground=(0,,0,,1)}{v3}
\fmfv{decor.shape=circle,decor.size=2.0thick,foreground=(0,,0,,1)}{v4}
\fmfv{decor.shape=circle,decor.size=2.0thick,foreground=(0,,0,,1)}{v5}
\fmfv{decor.shape=circle,decor.size=2.0thick,foreground=(0,,0,,1)}{v6}
\fmf{phantom,tension=1.4}{i1,v1}
\fmf{phantom}{i2,v3b}
\fmf{phantom}{i0,v1b}
\fmf{phantom}{o2,v6b}
\fmf{phantom,tension=1.4}{o1,v5}
\fmf{phantom}{o0,v5b}
\fmf{phantom,tension=1.11}{v3b,v6b}
\fmf{phantom,tension=1.38}{i0,v2}
\fmf{phantom,tension=1.38}{o0,v2}
\fmf{phantom,tension=1.8}{i0,v2b}
\fmf{phantom,tension=1.2}{o0,v2b}
\fmf{phantom,tension=1.2}{b0,v2b}
\fmf{phantom,tension=1.2}{b1,v2b}
\fmf{phantom,tension=1.2}{b2,v2b}
\fmf{phantom,tension=1.38}{i0,v4}
\fmf{phantom,tension=1.38}{o0,v4}
\fmf{phantom,tension=1.2}{i0,v4b}
\fmf{phantom,tension=1.8}{o0,v4b}
\fmf{phantom,tension=1.2}{b0,v4b}
\fmf{phantom,tension=1.2}{b1,v4b}
\fmf{phantom,tension=1.2}{b2,v4b}
\fmf{phantom,tension=2}{i3,v3}
\fmf{phantom,tension=2}{o3,v6}
\fmf{phantom,tension=2}{i3,v3b}
\fmf{phantom,tension=0.8}{o3,v3b}
\fmf{phantom,tension=0.8}{i3,v6b}
\fmf{phantom,tension=2}{o3,v6b}
\fmf{plain,tension=1.4}{v1,v2}
\fmf{plain,tension=1.4}{v4,v5}
\fmf{wiggly}{v1,v3}
\fmf{plain,left=0.8,tension=0}{v1,v3}
\fmf{wiggly}{v5,v6}
\fmf{plain,right=0.8,tension=0}{v5,v6}
\fmf{plain}{v2,v3}
\fmf{plain}{v4,v6}
\fmf{wiggly,tension=0.5}{v2,v4}
\fmf{phantom,tension=2}{v3,v6}
\fmf{phantom,right=0.8,tension=0}{v1,v5}
\fmf{phantom,tension=1}{v1,v1b}
\fmf{phantom,tension=0.2}{v2,v2b}
\fmf{phantom,tension=1.5}{v3,v3b}
\fmf{phantom,tension=0.2}{v4,v4b}
\fmf{phantom,tension=1}{v5,v5b}
\fmf{phantom,tension=1.5}{v6,v6b}
\end{fmfgraph*}
\end{fmffile}
\end{gathered} \hspace{-0.5cm} \left.\rule{0cm}{1.1cm}\right) \left.\rule{0cm}{1.2cm}\right] \\
& + \mathcal{O}\big(\hbar^{3}\big)\;.
\end{split}
\label{eq:SbosonicKLoopExpansionStep5}
\end{equation}
\end{strip}

In the (0+0)\nobreakdash-D limit, the collective Schwinger functional expanded up to second order in $\hbar$ satisfies:
\begin{equation}
\begin{split}
W^\text{LE;col}(\mathcal{J})= & - S_{\mathrm{col},\mathcal{J}}(\sigma_{\mathrm{cl}})+{\frac{\hbar}{2}}\ln(D_{\sigma_\text{cl};\mathcal{J}}) \\
&  -{\frac{\hbar^{2}}{648}} \ G_{\sigma_\text{cl};\mathcal{J}}^4 D_{\sigma_\text{cl};\mathcal{J}}^2 N \lambda^2 \Big[-27 \\
& - 108 G_{\sigma_\text{cl};\mathcal{J}} J^2 + 5 G_{\sigma_\text{cl};\mathcal{J}}^2 D_{\sigma_\text{cl};\mathcal{J}} N \lambda \\
& + 30 G_{\sigma_\text{cl};\mathcal{J}}^3 D_{\sigma_\text{cl};\mathcal{J}} J^2 N \lambda \\
& + 45 G_{\sigma_\text{cl};\mathcal{J}}^4 D_{\sigma_\text{cl};\mathcal{J}} J^4 N \lambda\Big] \\
& + \mathcal{O}\big(\hbar^{3}\big) \;,
\end{split}
\label{eq:ResultWSbosonicKLoopExpansion0DON}
\end{equation}
where it follows from Eq.~\eqref{eq:SbosonicKLoopExpansionH} that:
\begin{equation}
D^{-1}_{\sigma_\text{cl};\mathcal{J}} = \frac{\lambda N}{6} \ G_{\sigma_\text{cl};\mathcal{J}}^{2} \left(2 G_{\sigma_\text{cl};\mathcal{J}} J^{2} + 1\right) + 1\;,
\label{eq:DsigmaJCollRepre}
\end{equation}
where we have used the relations $J_{a}=J$ $\forall a$ and $\boldsymbol{G}_{\sigma_\text{cl};\mathcal{J};ab}=G_{\sigma_\text{cl};\mathcal{J}}\delta_{ab}$ with $G_{\sigma_\text{cl};\mathcal{J}}$ given by:
\begin{equation}
G_{\sigma_\text{cl};\mathcal{J}}^{-1} = m^{2} + i\sqrt{\frac{\lambda}{3}}\sigma_{\mathrm{cl}}\;,
\label{eq:GsigmaCollRepreDiag}
\end{equation}
as can be deduced from Eq.~\eqref{eq:DefGpropagCollectiveLE}. In contrast with the original (and mixed) representation(s), the LE in the collective representation is not organized with respect to the coupling constant $\lambda$, thus making it non-perturbative. The corresponding expressions for the gs energy and density are deduced once again after imposing that all sources vanish. The saddle points of the collective classical action $S_{\mathrm{col}}$ in (0+0)\nobreakdash-D read:
\begin{equation}
\overline{\sigma}_{\mathrm{cl}}\equiv\sigma_{\mathrm{cl}}[\mathcal{J}=0]=i\left(\frac{\sqrt{3}m^{2}\pm\sqrt{3m^{4}+2N\lambda}}{2\sqrt{\lambda}}\right)\;,
\label{eq:ClassicalSolutionBosonicAction}
\end{equation}
thus yielding a non-perturbative expression for $G_{\sigma_\text{cl}}$:
\begin{equation}
G^{-1}_{\sigma_{\mathrm{cl}}} \equiv G^{-1}_{\sigma_\text{cl};\mathcal{J} = 0} = \frac{1}{2}\left(m^{2}\mp\sqrt{m^{4}+\frac{2}{3}\lambda N}\right) \;,
\label{eq:GsigmaBosonicActionClassicalSolution0DON}
\end{equation}
and for $D^{-1}_{\sigma_\text{cl};\mathcal{J}}$:
\begin{equation}
D^{-1}_{\sigma_\text{cl}} \equiv D^{-1}_{\sigma_\text{cl};\mathcal{J} = 0} = \frac{2\lambda N}{3\left(m^2\mp\sqrt{m^4 + \frac{2}{3}\lambda N}\right)^{2}} + 1 \;,
\end{equation}
at vanishing $\mathcal{J}$. We then deduce the series expansions for the gs energy and density valid for all $N \geq 1$ and for both signs of $m^2$:
\pagebreak
\begin{strip}
\begin{equation}
\begin{split}
E^\text{LE;col}_{\mathrm{gs}} = & -\frac{3 m^4 + N \lambda - m^{2} \sqrt{9 m^{4} + 6 N \lambda} + 2 N \lambda \ln\bigg(\frac{12 \pi}{3 m^{2} + \sqrt{9 m^{4} + 6 N \lambda}}\bigg)}{4 \hbar \lambda} \\
& + \frac{1}{2} \ln\Bigg(1 + \frac{6 N \lambda}{\left(3 m^{2} + \sqrt{9 m^{4} + 6 N \lambda}\right)^2}\Bigg) \\
& -\hbar\frac{N \lambda^2}{6 \left(3 m^{4} + 2 N \lambda + m^{2} \sqrt{9 m^{4} + 6 N \lambda}\right)^{3}}\Big[27 m^{4} + 8 N \lambda + 9 m^{2} \sqrt{9 m^{4} + 6 N \lambda}\Big] \\
& +\hbar^{2}\frac{10 m^{2} N \lambda^3}{\left(3 m^{4} + 2 N \lambda + m^{2} \sqrt{9 m^{4} + 6 N \lambda}\right)^{6}} \\
& \hspace{0.4cm} \times \Big[108 m^{10}+ 90 m^{6} N \lambda + 15 m^{2} N^2 \lambda^2 + \left( N^2 \lambda^2 + 18 m^{4} N \lambda + 36 m^{8}\right) \sqrt{9 m^{4} + 6 N \lambda} \Big] \\
& -\hbar^{3}\frac{2 N \lambda^{4}}{45 \left(3 m^{4} + 2 N \lambda + m^{2} \sqrt{9 m^{4} + 6 N \lambda}\right)^{9}} \\
& \hspace{0.4cm} \times \Big[9185400 m^{20} + 11136204 m^{16} N \lambda + 3610818 m^{12} N^{2} \lambda^{2} + 52650 m^{8} N^3 \lambda^3 - 70605 m^{4} N^{4} \lambda^{4} \\
& \hspace{0.85cm} - 1792 N^{5} \lambda^{5} + \big( - 7119 m^{2} N^{4} \lambda^{4} - 48474 m^{6} N^{3} \lambda^{3} + 476550 m^{10} N^{2} \lambda^{2} + 2691468 m^{14} N \lambda \\
& \hspace{0.85cm} + 3061800 m^{18} \big) \sqrt{9 m^{4} + 6 N \lambda} \Big] \\
& + \mathcal{O}\big(\hbar^{4}\big) \;,
\end{split}
\label{eq:ResultEgsBosonicAction0DON}
\end{equation}
and
\begin{equation}
\begin{split}
\rho^\text{LE;col}_{\mathrm{gs}} = & \ \frac{6}{3 m^{2} + \sqrt{9 m^{4} + 6 N \lambda}} \\
& + \frac{\sqrt{3} m^{2} + \sqrt{3 m^{4} + 2 N \lambda}}{\sqrt{3 m^{4} + 2 N \lambda}} \\
& \hspace{0.4cm} \times \Bigg\lbrace -\hbar\frac{6 \lambda }{ \left(3 m^{2} + \sqrt{9 m^{4} + 6 N \lambda}\right) \left(3 m^{4} + 2 N \lambda + m^{2} \sqrt{9 m^{4} + 6 N \lambda}\right)} \\
& \hspace{1.105cm} +\hbar^{2}\frac{2 \lambda^{2} \left(3 m^{2} + \sqrt{9 m^{4} + 6 N \lambda}\right)\left(9 m^{4} + N \lambda + 3 m^{2} \sqrt{9 m^{4} + 6 N \lambda}\right)}{\left(3 m^{4} + 2 N \lambda + m^{2} \sqrt{9 m^{4} + 6 N \lambda}\right)^{4}} \\
& \hspace{1.105cm} -\hbar^{3}\frac{10 \lambda^{3}\left(3 m^{2} + \sqrt{9 m^{4} + 6 N \lambda}\right)^{3}}{\left(3 m^{4} + 2 N \lambda + m^{2} \sqrt{9 m^{4} + 6 N \lambda}\right)^{7}} \\
& \hspace{1.505cm} \times \bigg[54 m^{8} + 15 m^{4} N \lambda - 2 N^{2} \lambda^{2} + \left(- m^{2} N \lambda + 18 m^{6}\right) \sqrt{9 m^{4} + 6 N \lambda}\bigg] \\
& \hspace{1.105cm} + \hbar^{4}\frac{2 \lambda^{4}  \left(3 m^{2} + \sqrt{9 m^{4} + 6 N \lambda}\right)^{5}}{9\left(3 m^{4} + 2 N \lambda + m^{2} \sqrt{9 m^{4} + 6 N \lambda}\right)^{10}} \\
& \hspace{1.505cm} \times \bigg[11340 m^{12} + 2538 m^{8} N \lambda - 1722 m^{4} N^{2} \lambda^2 - 21 N^{3} \lambda^{3} + \big(- 226 m^{2} N^{2} \lambda^2 \\
& \hspace{1.95cm} - 414 m^{6} N \lambda + 3780 m^{10}\big) \sqrt{9 m^{4} + 6 N \lambda} \bigg] + \mathcal{O}\big(\hbar^{5}\big) \Bigg\rbrace\;,
\end{split}
\label{eq:ResultrhogsBosonicAction0DON}
\end{equation}
\end{strip}
where relations homologous to Eqs.~\eqref{eq:DefEgsExactZexact0DON} and~\eqref{eq:DefrhogsExactwithExpectationValue0DON} were exploited. Regarding the 1-point correlation function of the original field, we deduce from the derivative of the generating functional $Z_{\mathrm{col}}[\mathcal{J}]$ with respect to the source $\vec{J}$ that:
\begin{equation}
\begin{split}
\vec{\overline{\phi}}(x) \propto & \ \left.\frac{\delta Z_{\mathrm{col}}\big[\vec{J},j\big]}{\delta\vec{J}(x)}\right|_{\vec{J}=\vec{0} \atop j=0} \\
\propto & \ \left[\int \mathcal{D}\widetilde{\sigma} \left(\int_{y}\boldsymbol{G}_{\widetilde{\sigma}}(x,y)\vec{J}(y)\right) e^{-\frac{1}{\hbar}S_{\text{col},\mathcal{J}}[\widetilde{\sigma}]} \right]_{\vec{J}=\vec{0} \atop j=0} \\
= & \ \vec{0} \;,
\end{split}
\end{equation}
to all orders of the collective LE, regardless of the dimension. In other words, the $O(N)$ symmetry, although possibly (spontaneously) broken down at the classical level, always gets exactly restored from the first non-trivial order of the collective LE, whereas its original counterpart shows no sign of such a restoration (at $N=1$) according to Fig.~\ref{fig:vev}.

Results obtained from Eqs.~\eqref{eq:ResultEgsBosonicAction0DON} and~\eqref{eq:ResultrhogsBosonicAction0DON} are plotted in Figs.~\ref{fig:O1PTcoll} and~\ref{fig:O2PTcoll}, respectively at $N=1$ and $N=2$. The first orders of the collective LE lie fairly close to the exact solutions, even in the broken-symmetry phase where no description can be obtained for $N\geq 2$ within the original (and mixed) LE(s). Hence, the introduction of a collective dof in the framework of the LE can lead to excellent results already at the first non-trivial order (i.e. at order $\mathcal{O}(\hbar)$ for $E_{\mathrm{gs}}$ and $\mathcal{O}(\hbar^{2})$ for $\rho_{\mathrm{gs}}$) in the strongly-coupled regime of the studied model: a parallel can thus be drawn with the nuclear EDF approach and its excellent results at the Hartree-Fock(-Bogoliubov) level. However, the collective LE series are also asymptotic and divergent, even though this remains less apparent from Figs.~\ref{fig:O1PTcoll} and~\ref{fig:O2PTcoll} as compared with the original LE. Note furthermore that the $1/N$-expansion coincides with the collective LE at their first non-trivial orders according to Figs.~\ref{fig:O1PTcoll} and~\ref{fig:O2PTcoll} (recall that the derivation of the series underlying the $1/N$-expansion for the studied toy model is outlined in Appendix~\ref{app:1overNexpansion0DON}). Although this equivalence breaks down at higher truncation orders, it illustrates the non-perturbative character of the collective LE. From this connection, we also expect the collective LE to be more and more performing as $N$ increases, which is in accordance with Figs.~\ref{fig:O1PTcoll} and~\ref{fig:O2PTcoll}.

It is acknowledged that asymptotic series hide relevant information about the system, which needs to be deciphered through proper resummation techniques. Within this frame, perturbation theory is typically combined with a meticulously crafted analytic continuation function, yielding accurate results far beyond the weakly-interacting regime, and even allowing for the computation of genuinely non-perturbative features from low-order perturbation theory, as discussed in the next section.

%==============================
\subsection{\label{sec:Resummation}Resummation procedures}
%~~~
\subsubsection{Pad\'e-Borel resummation}

\begin{figure}[!t]
  \begin{center}
      \includegraphics[width=0.95\linewidth]{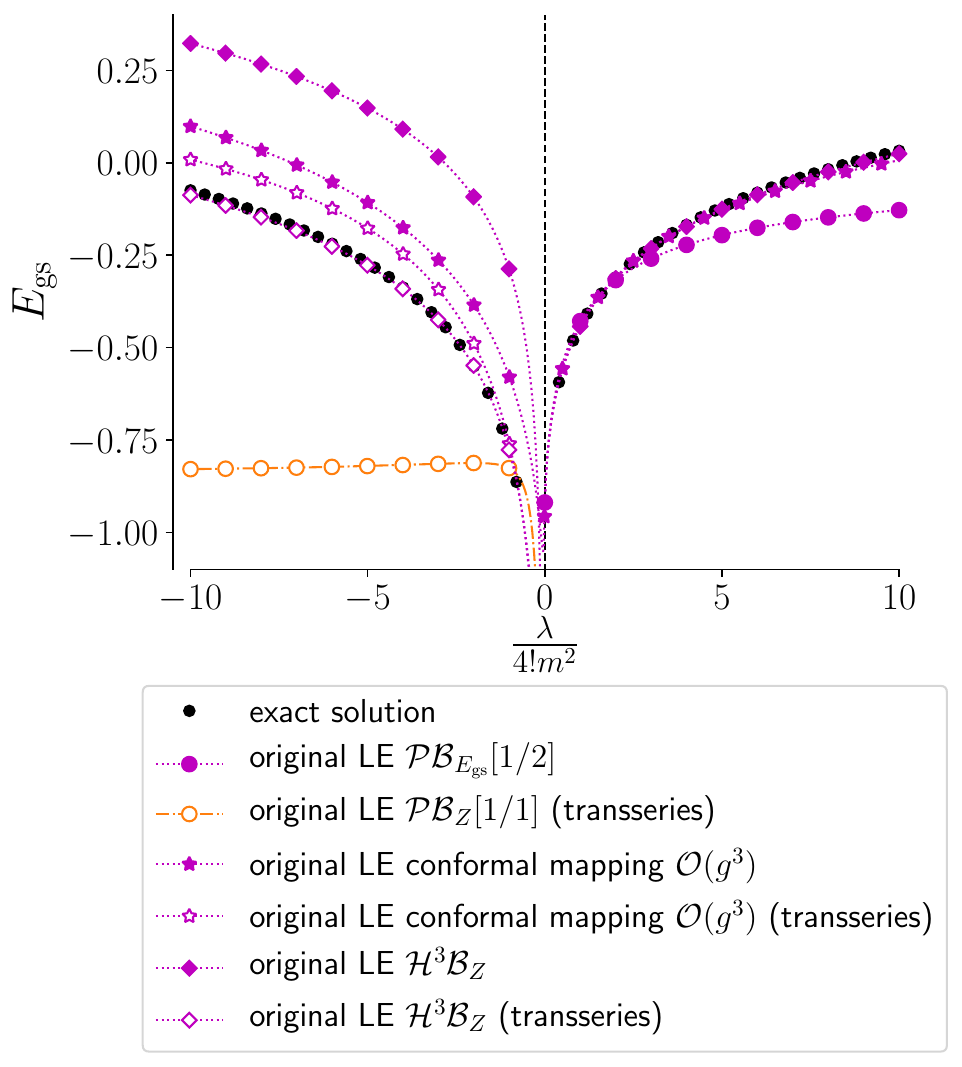}
      \includegraphics[width=0.95\linewidth]{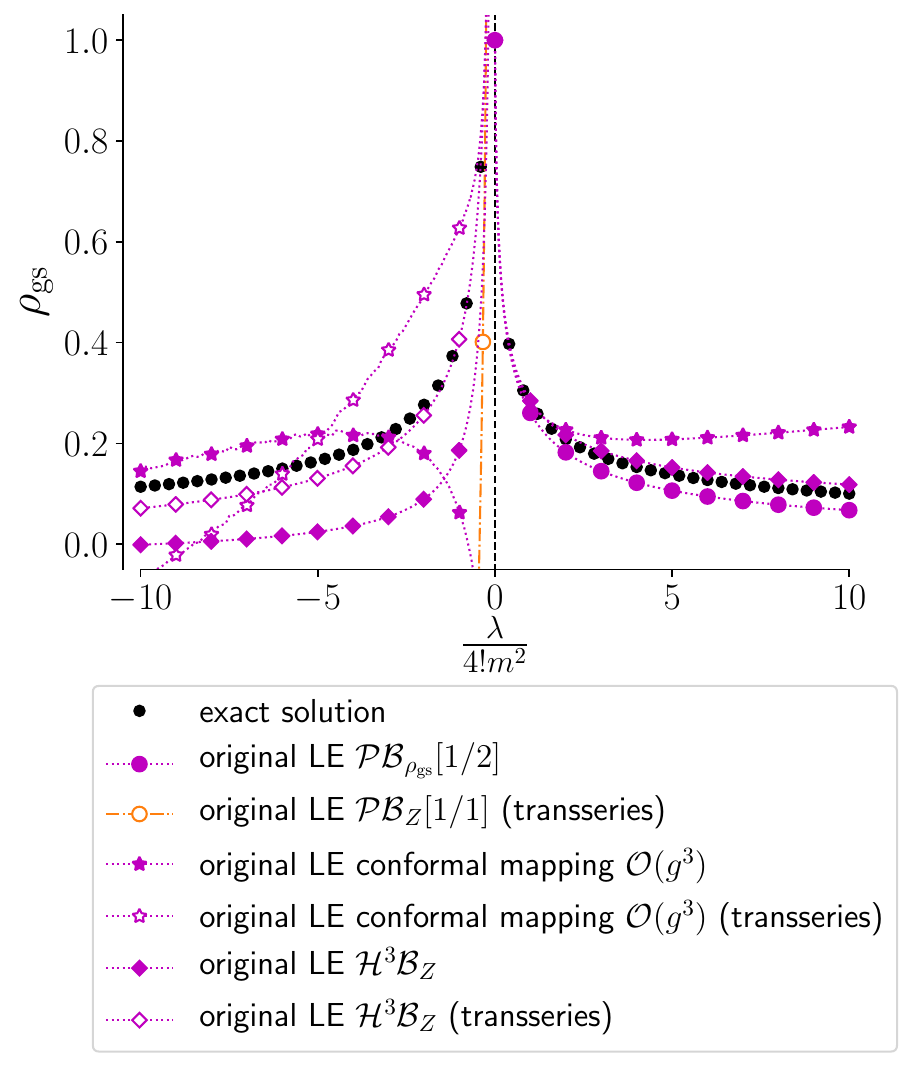}
    \caption{Gs energy $E_{\mathrm{gs}}$ (upper panel) or density $\rho_{\mathrm{gs}}$ (lower panel) calculated at $\hbar=1$, $m^{2}=\pm 1$ and $N=1$, and compared with the corresponding exact solution (black dots). All presented results are obtained from series determined via the LE in the original representation.}
    \label{fig:O1ResumPTori}
  \end{center}
\end{figure}

\begin{figure}[!t]
  \begin{center}
      \includegraphics[width=0.95\linewidth]{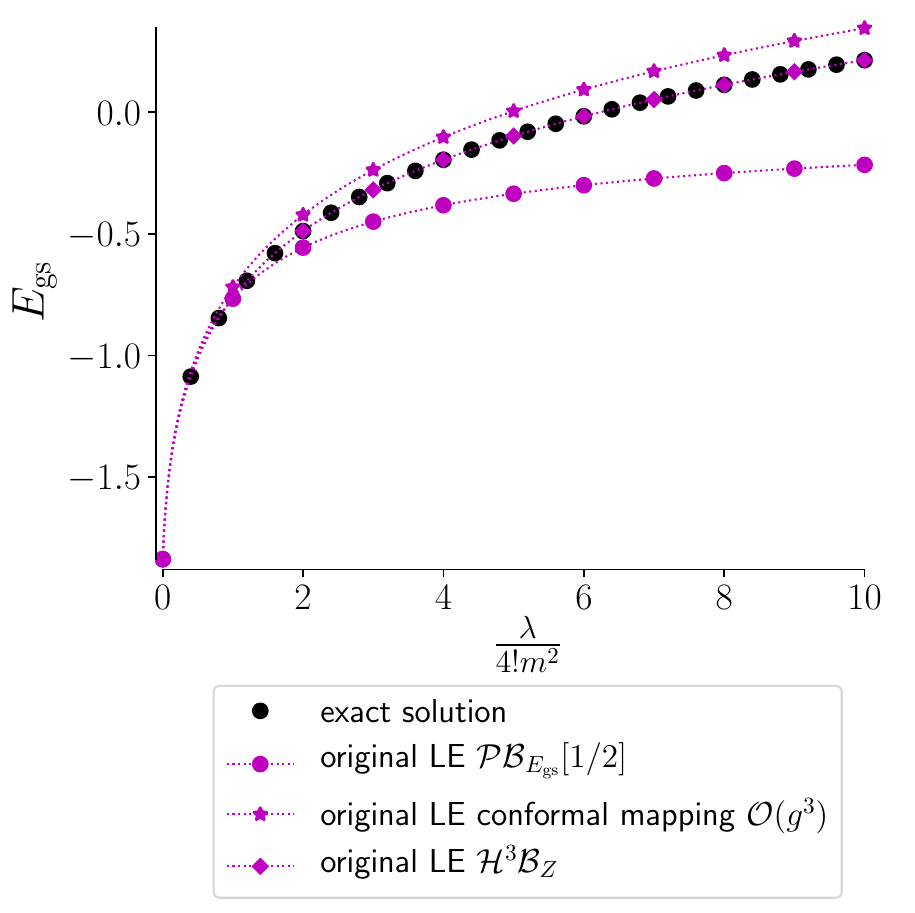}
      \includegraphics[width=0.95\linewidth]{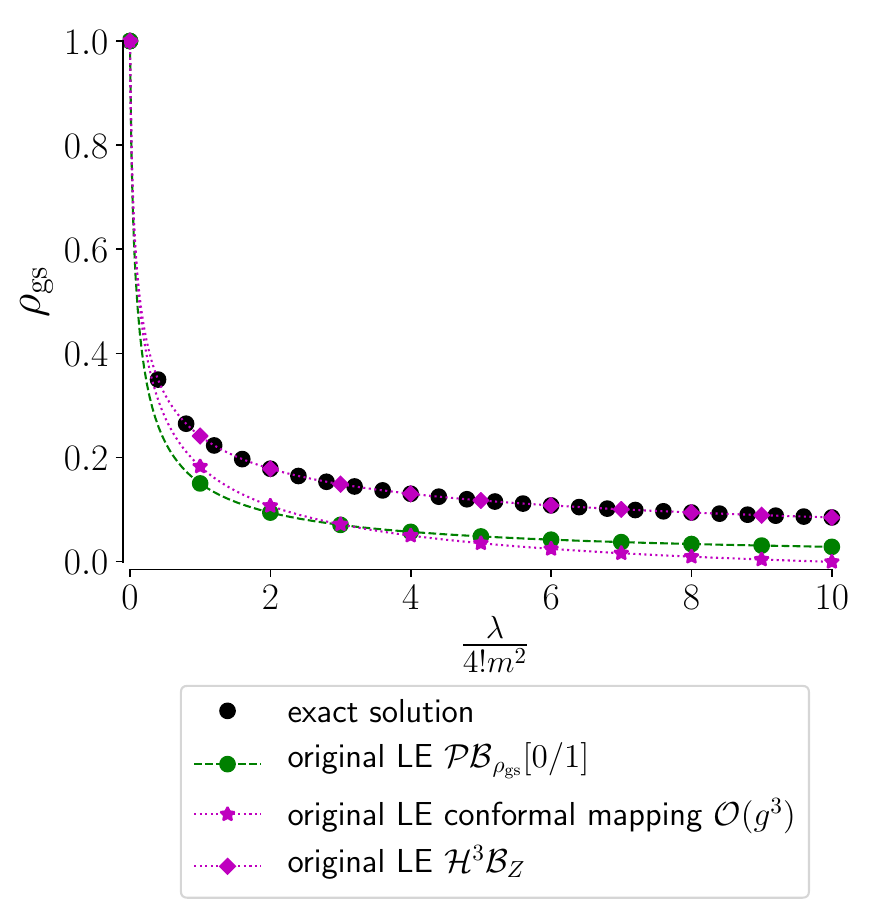}
    \caption{Same as Fig.~\ref{fig:O1ResumPTori} with $m^{2}=+1$ and $N=2$ instead. As in Fig.~\ref{fig:O2PTcoll}, no finite results can be obtained in the broken-symmetry phase from the LE in the original (and mixed) representation(s).}
    \label{fig:O2ResumPTori}
  \end{center}
\end{figure}

We first consider one of the most conventional and widespread resummation procedure, i.e. the Pad\'e-Borel approach \cite{pad1892,bor28,ell96,ben99,kle01}. The latter notably enriches the structure of the asymptotic series taken as input by constructing from it a rational function known as Pad\'{e} approximant (see appendix~\ref{app:PBResum} for further details). We compared the performances of the Pad\'e-Borel resummation procedure in reproducing the gs energy and density for the studied toy model under various settings, namely by following the step-by-step procedure: i) change the order $M$ of the original LE up to order $\mathcal{O}\big(g^3\big)$; ii) consider all the possible Pad\'e approximants\footnote{In the present study, all Pad\'e approximants are determined with the $\mathtt{PadeApproximant}$ function of $\mathtt{Mathematica~12.1}$.} at a given order $M$; iii) either resum the power series or transseries representation of the partition function $Z$ and, from there, compute the gs energy and density (indicated by $\mathcal{PB}_{Z}$ in Figs.~\ref{fig:O1ResumPTori} and~\ref{fig:O2ResumPTori}), or directly consider the perturbative expansion of the gs energy and density (given by Eqs.~\eqref{eq:EgsPTpos} to~\eqref{eq:ResultPTrhogsm2neg}) and then proceed with their resummation (indicated by $\mathcal{PB}_{E_{\mathrm{gs}}}$ and $\mathcal{PB}_{\rho_{\mathrm{gs}}}$ in Figs.~\ref{fig:O1ResumPTori} and~\ref{fig:O2ResumPTori}). The best results are displayed as lines with a \padeborel symbol (the color indicates the truncation order in the original LE while the filled/open aspect refers to the power series/transseries representation of the resummed quantity) in Figs.~\ref{fig:O1ResumPTori} and~\ref{fig:O2ResumPTori}, at $N=1$ and $2$ respectively. Note that, although the series representation of the partition function directly follows from Eq.~\eqref{eq:ZJKPT3}, the determination of its transseries representation requires some extra work involving e.g. Picard-Lefschetz theory \cite{wit10,tan15} (recall that this derivation is detailed in appendix~\ref{app:Transseries}). In the unbroken-symmetry regime, the best description of the gs energy on the one hand is obtained at $N=1$ and $2$ with the Pad\'e-Borel resummation of the perturbative series for $E_{\mathrm{gs}}$ pushed up to order $\mathcal{O}\big(g^3\big)$ (\pbviolet), with $[1/2]$ Pad\'e approximants. On the other hand, the best reproduction of the gs density in the unbroken-symmetry regime is achieved via Pad\'e-Borel resummation of the original LE series for $\rho_{\mathrm{gs}}$ pushed up to order $\mathcal{O}\big(g^3\big)$ (\pbviolet) (up to order $\mathcal{O}(g)$ (\pbgreen)) with $[1/2]$ ($[0/1]$) Pad\'e approximants at $N=1$ ($N=2$). In the broken-symmetry regime at $N=1$, the best description of the gs energy and density is given by the Pad\'{e}-Borel-\'{E}calle resummation\footnote{The Pad\'{e}-Borel-\'{E}calle resummation is obtained from a slight modification of the integration path underlying the inverse Borel transforms in the corresponding Pad\'{e}-Borel resummation procedure, thus avoiding singularities in the Borel plane.}~\cite{eca81} of the transseries representation of $Z$ at order $\mathcal{O}\big(g^2\big)$ in the original LE (\pborange), with $[1/1]$ Pad\'e approximants. In all these cases, a major improvement over the bare LE results can be noticed, i.e. (except for $m^2<0$) the global behavior of the gs energy and density with respect to the coupling strength $\lambda/4!$ is now consistent with the exact trend over the whole range of tested values (i.e. for $\lambda/4!\in[0,10]$, which is wider than $[0,1]$ considered in Figs.~\ref{fig:O1PTcoll} and~\ref{fig:O2PTcoll}), and even quantitatively reproduced up to $\lambda/4! \sim 2$.

%%%%%%%%%%%%%%%%%%%
%%%%%%%%%%%%%%%%%%%
%%%%%%%%%%%%%%%%%%%
%%%%%%%%%%%%%%%%%%%

%~~~
\subsubsection{Conformal mapping}

While the Pad\'e-Borel resummation procedure only involves the knowledge about the first terms of the initial asymptotic series, the method of Borel transform with conformal mapping also exploits the knowledge of the large-order behavior of this series by continuing the Borel transform beyond its circle of convergence with a mapping of the Borel plane to a circle of unit radius~\cite{leg80} (see appendix~\ref{app:ConfMappResum} for further details). The gs energy and density obtained for our toy model after a conformal mapping resummation of the partition function $Z$ are displayed as \cmviolet (\cmvioleto) when $Z$ is represented by a power series (transseries) in Figs.~\ref{fig:O1ResumPTori} and~\ref{fig:O2ResumPTori}, at $N=1$ and $2$ respectively. As far as the energy is concerned, the conformal mapping resummation applied to the partition function yields results in better agreement with the exact ones, especially in the strongly-interacting regime (for $\lambda/4!\gtrsim 2$ more specifically), in comparison with the estimates obtained via Pad\'e-Borel resummation of the energy perturbative series at the same order of the original LE. The results are however not as good for the gs density. In the phase with $m^2<0$ where the partition function is not Borel-summable, Fig.~\ref{fig:O1ResumPTori} shows the gs energy and density of the system obtained both from $Z$ represented by an ambiguous power series and by a resurgent transseries. In the former case, the global behavior of the gs energy with the coupling strength is fairly well reproduced, but a quantitative reproduction of the exact result is not achieved, even in the weakly-interacting limit. The gs density, when deduced from a power series representation of $Z$, misses the monotonic decreasing displayed by the exact result when going from weak to strong couplings. Representing $Z$ by a transseries slightly improves the description of the gs energy and density of the system, with now a correct description of the range set by $\lambda/4!\lesssim 1$.

%~~~
\subsubsection{Borel-hypergeometric resummation}

Borel-hypergeometric or Meijer-G resummation \cite{mer15,mer16,ped16a,ped16b,san17,mer18,ant19} extends the idea behind Pad\'e approximants while trying to overcome the known issues of the latter\footnote{As rational functions, Pad\'e approximants built-in singularities are poles. Hence, since many poles are needed to mimic a branch cut, the Pad\'e-Borel resummation procedure converges slowly when the Borel transform to be approximated displays branch cuts.} by working with more sophisticated continuation functions, i.e. hypergeometric functions, which can notably mimic branch cuts in the complex plane, and whose Borel integral is known and conveniently represented by Meijer G-functions \cite{mar83,bea13}. To our knowledge and unlike the other resummation methods discussed previously, the Borel-hypergeometric resummation has never been applied to any $O(N)$-symmetric theory, putting aside the (0+0)\nobreakdash-D case at $N=1$ \cite{mer18}. We will push our investigations in (0+0)\nobreakdash-D up to $N=4$ in the present study. The gs energy and density of the system obtained from the Borel-hypergeometric resummation of the partition function are reported in Figs.~\ref{fig:O1ResumPTori} and~\ref{fig:O2ResumPTori} at $N=1$ and $2$ respectively. Results corresponding to the third non-trivial order of the original LE are displayed as \mgviolet (\mgvioleto) when $Z$ is represented by a power series (transseries). The Borel-hypergeometric resummation of the partition function yields the best results among all the resummation schemes, as could have been expected from the fact that the exact solutions for $E_{\mathrm{gs}}$ and $\rho_{\mathrm{gs}}$ (given by Eqs.~\eqref{eq:ExactSolutionEgs0DON} and~\eqref{eq:ExactSolutionRhogs0DON}) are both related through Eq.~\eqref{eq:RN} to hypergeometric functions which exhibit branch cuts themselves. The Borel-hypergeometric resummation even leads to an exact description of the partition function of the $O(2)$- and $O(4)$-symmetric theories from the third non-trivial order of the original LE, as shown by Fig.~\ref{fig:MeijerON} (see appendix~\ref{app:BorelHyperGeoResum} for more details on this point). The present analysis thus illustrates that particularly high accuracies can be reached with the Borel-hypergeometric resummation at fairly low orders, when trying to describe functions exhibiting branch cuts.

%%%%%%%%%%%%%%%%%%%
\begin{figure}[!t]
  \begin{center}
      \includegraphics[width=0.95\linewidth]{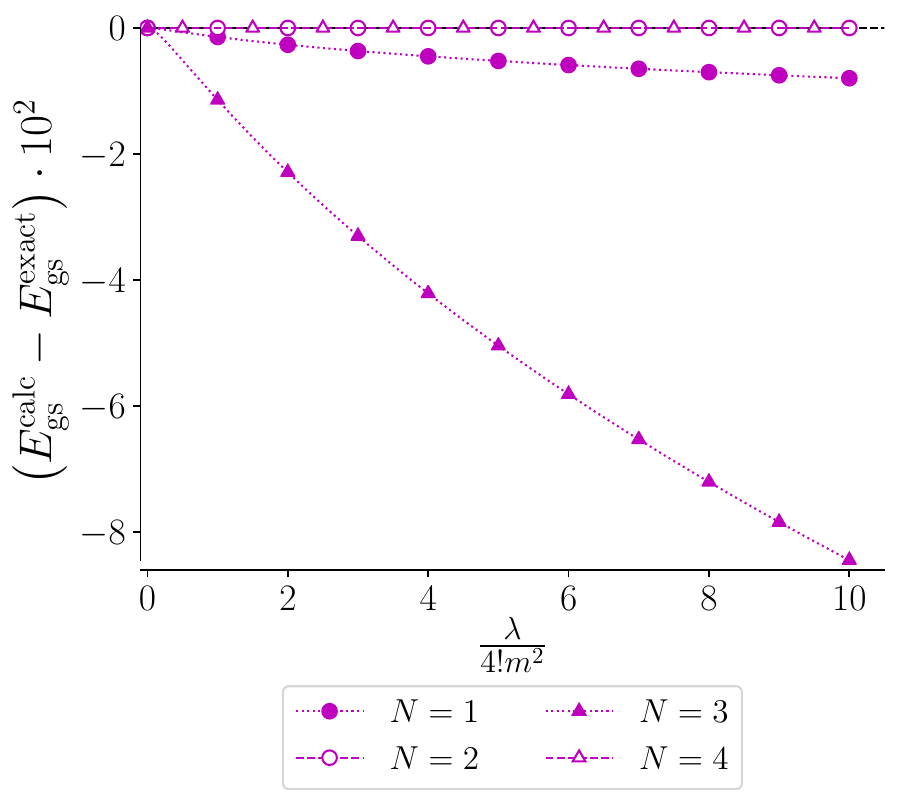}
      \includegraphics[width=0.95\linewidth]{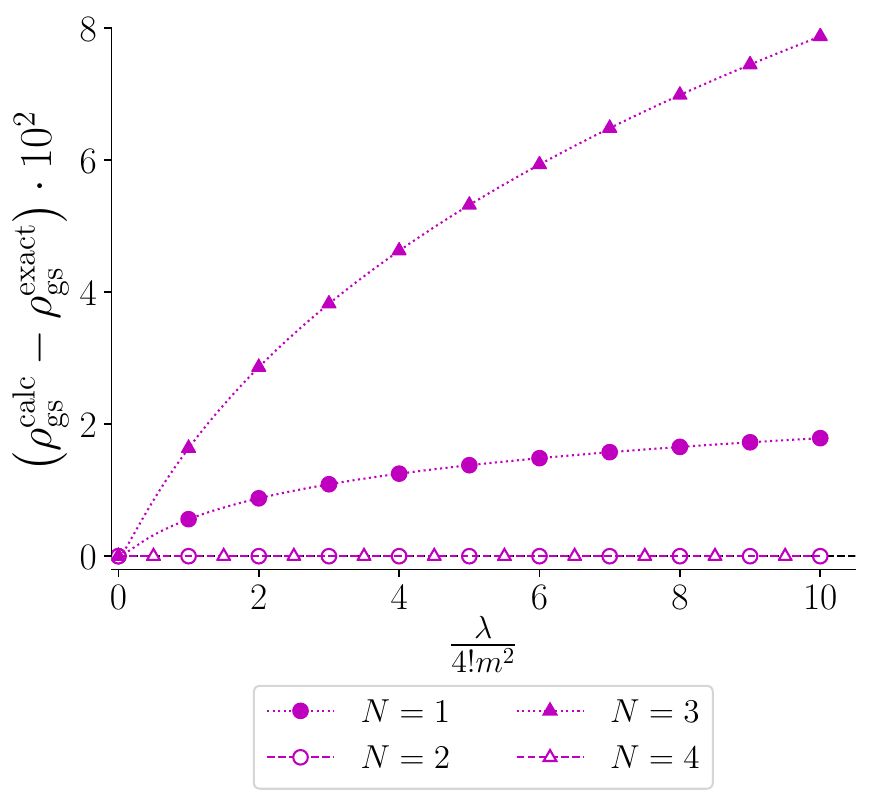}
    \caption{Gs energy $E_{\mathrm{gs}}^{\mathrm{calc}}$ (upper panel) or density $\rho_{\mathrm{gs}}^{\mathrm{calc}}$ (lower panel) calculated from Borel-hypergeometric resummation applied to the original (or mixed) LE series up to their third non-trivial order (notably labeled ``LE $\mathcal{H}^{3}\mathcal{B}_{Z}$'' in Figs.~\ref{fig:O1ResumPTori} and~\ref{fig:O2ResumPTori}, which correspond to the present results (in the unbroken-symmetry regime) labeled ``$N=1$'' and ``$N=2$'', respectively). More specifically, we show here the difference between these results and the corresponding exact solution $E_{\mathrm{gs}}^{\mathrm{exact}}$ or $\rho_{\mathrm{gs}}^{\mathrm{exact}}$ at $\hbar=1$, $m^{2}=+1$ and $N=1,2,3$ and $4$.}
    \label{fig:MeijerON}
  \end{center}
\end{figure}
%%%%%%%%%%%%%%%%%%%

Hence, resummation techniques offer an impressive way of extracting sensible results from the very simple ordinary perturbation theory (i.e. the original LE) over a wide range of values for the coupling constant $\lambda/4!$, including the strongly-coupled regime. The various resummation techniques at our disposal actually render the LE (and all other techniques based on asymptotic series) rather versatile. The description of the gs energy and density (as well as the 1-point correlation function $\vec{\overline{\phi}}$ at $N=1$ in the phase with $m^2<0$) are significantly improved at trivial cost. However, a very accurate reproduction of the system's features requires to reach at least the third non-trivial order of the original LE, which can be difficult to determine in realistic cases. Besides, the theoretical foundation of the nuclear EDF can not be found in standard perturbation theory (even when completed by a resummation procedure), as no self-consistent dressing of, e.g., the field propagator is achieved and the energy is not obtained as a functional of the density in the spirit of DFT. We now turn to an optimized version of perturbation theory which bears stronger resemblance with EDFs as it involves a collective dof (even in the original representation of our $O(N)$ model) and relies on self-consistent expansions.

%==============================
\subsection{\label{sec:OPT}Optimized perturbation theory}

Just like the LE, OPT relies on a partitioning step defining an unperturbed and a residual part of the classical action under consideration. However, OPT offers the possibility to optimize the chosen reference part. We actually name OPT all strategies involving an optimized splitting into unperturbed and residual parts via some conditions, which includes more approaches than those exploited in Refs.~\cite{ste81,kil81,oko87,oko87bis,oko88,dun88,oko89,dun89,jon89,jon90,coo91,kli91,buc92,buc92bis,sis91,sis92,sis92bis,kor93,kor93bis,sis93,sis94,sis94bis,kle98,kle11}, such as variational perturbation theory (VPT) \cite{kle01,kle06,kle11bis}, linear delta-expansion (LDE) \cite{jon90bis}, self-consistent expansion (SCE) \cite{sch98,sch08}, self-similar perturbation theory \cite{yuk99} or order-dependent mapping (ODM) \cite{sez79,zin10}. Hence, in the framework of OPT, there exists an infinite number of acceptable reference parts $S^0_\sigma$ depending on some parameter $\sigma(x)$ (which is a collective dof and a classical field, not to be confused with the collective quantum field $\widetilde{\sigma}(x)$ introduced via HST), thus yielding the following splitting:
\begin{equation}
\begin{split}
S\Big[\vec{\widetilde{\varphi}}\Big] = & \ S^0_\sigma\Big[\vec{\widetilde{\varphi}}\Big]+ S^1_\sigma\Big[\vec{\widetilde{\varphi}}\Big] \\
\equiv & \ \left(S^0\Big[\vec{\widetilde{\varphi}}\Big]-\frac{1}{2}\int_x \sigma(x) \vec{\widetilde{\varphi}}^2(x)\right) \\
& + \left(S^1\Big[\vec{\widetilde{\varphi}}\Big]+\frac{1}{2}\int_x \sigma(x) \vec{\widetilde{\varphi}}^2(x)\right) \;,
\end{split}
\label{eq:GeneralPartiotioningOPT}
\end{equation}
where the original splitting appears as the particular case where $\sigma(x) = 0$. Formally, one has done nothing but adding and subtracting an arbitrary quadratic term in the classical action. The idea behind OPT is then to exploit the introduction of such a Gaussian kernel to reorganize the partitioning into unperturbed and residual parts in a more flexible fashion, where non-perturbative correlations are shifted towards the easily solvable unperturbed channel. Indeed, when the perturbative expansion around $S^0_\sigma$ is truncated at some finite order, physical quantities exhibit an artificial dependence in the parameter $\sigma(x)$ that must be fixed. Relevant choices of $\sigma(x)$, such that $S^0_\sigma$ mimics as faithfully as possible the full action $S$, allow us to dress the propagator of the original field with non-trivial physics and turn the original divergent perturbative series into an exponentially-fast convergent one. The OPT expansion is thus self-consistent, hence the aforementioned connection with the nuclear EDF approach.

Several optimization criteria can be devised to this end:
\begin{itemize}
\item In the ODM approach, $\sigma(x)$ is determined according to mathematical convergence properties of the series~\cite{zin10}.

\item Other classes of strategies are based on the fact that a given physical quantity $\mathcal{O}^{(k)}$ computed at $k$th order of the OPT expansion exhibits an artificial dependence with respect to $\sigma(x)$, so that one should try to make $\mathcal{O}^{(k)}$ minimally sensitive to it. This can be achieved via the so-called \emph{principle of minimal sensitivity} (PMS)~\cite{ste81}, i.e. a variational principle imposing:
\begin{equation}
\left.\frac{\delta\mathcal{O}^{(k)}}{\delta\sigma(x)}\right|_{\sigma=\sigma^{(k)}_\text{PMS}} = 0 \;,
\end{equation}
or via the \emph{turning point} (TP) method \cite{kle05} where one rather looks for a plateau in the behavior of $\mathcal{O}^{(k)}$ with respect to $\sigma(x)$:
\begin{equation}
\left.\frac{\delta^2\mathcal{O}^{(k)}}{\delta\sigma(x)^2}\right|_{\sigma=\sigma^{(k)}_\text{TP}} = 0 \;,
\end{equation}
or via the \emph{fastest apparent convergence} (FAC) \cite{ste81} where $\sigma(x)$ is fixed so that $\mathcal{O}$ calculated at two subsequent OPT orders yields the same result, i.e.:
\begin{equation}
\left[\mathcal{O}^{(k)}-\mathcal{O}^{(k-1)}\right]_{\sigma=\sigma^{(k)}_\text{FAC}} = 0 \;,
\end{equation}
which amounts to imposing that the $k$th coefficient in the OPT expansion is zero.

\item Another kind of optimization procedure involves a self-consistent condition (SCC) for $\sigma(x)$ where some physical features of the system are asked to be faithfully reproduced from the zeroth order of the description. In the spirit of Kohn-Sham DFT, one can ask that the 2-point correlation function of the system calculated at $k$th order of the OPT expansion $\left\langle\vec{\widetilde{\varphi}}(x)\cdot\vec{\widetilde{\varphi}}(y)\right\rangle^{(k)}$ (which reduces to the density at $y=x$) coincides with the zeroth-order one $\left\langle\vec{\widetilde{\varphi}}(x)\cdot\vec{\widetilde{\varphi}}(y)\right\rangle^{(0)}$:
\begin{equation}
\left[\left\langle\vec{\widetilde{\varphi}}(x)\cdot\vec{\widetilde{\varphi}}(y)\right\rangle^{(k)} -\left\langle\vec{\widetilde{\varphi}}(x)\cdot\vec{\widetilde{\varphi}}(y)\right\rangle^{(0)}\right]_{\sigma=\sigma^{(k)}_\text{SCC}} = 0 \;.
\end{equation}
Such an optimization procedure, like the previous ones (PMS, TP, FAC), requires the calculation of a physical quantity at $k$th order of the OPT expansion, which is often difficult to achieve. We can use instead the following alternative implementation of the SCC:
\begin{equation}
\begin{split}
\Bigg[&\left\langle\left(\vec{\widetilde{\varphi}}(x)\cdot\vec{\widetilde{\varphi}}(y)\right)^{m(k)}\right\rangle^{(1)} \\
& -\left\langle\left(\vec{\widetilde{\varphi}}(x)\cdot\vec{\widetilde{\varphi}}(y)\right)^{m(k)}\right\rangle^{(0)}\Bigg]_{\sigma=\sigma^{(k)}_\text{SCC}} = 0 \;,
\end{split}
\end{equation}
which is only a first-order relation, therefore easy to compute. The dependence in the working expansion order $k$ appears via the exponent of the correlation function $m(k)$. In particular, a dependence of the form $m(k)=k$ was studied in Ref.~\cite{rem18} and shown to yield an exponentially-fast convergent series representation of physical quantities.
\end{itemize}

OPT has been widely used for decades via the above optimization criteria \cite{ste81,kil81,oko87,oko87bis,oko88,dun88,oko89,dun89,jon89,jon90,coo91,kli91,buc92,buc92bis,sis91,sis92,sis92bis,kor93,kor93bis,sis93,sis94,sis94bis,kle98,kle11,kle01,kle06,kle11bis,jon90bis,sch98,sch08,yuk99,sez79,zin10}. In particular, we can mention a previous study \cite{ros16} of the unbroken-symmetry phase of the (0+0)\nobreakdash-D $O(N)$ model considered in the present work using OPT based on the PMS, the TP method and the FAC. However, this work exploits the (0+0)\nobreakdash-D nature of the problem to directly expand the quantities of interest, thus bypassing notably the diagrammatic constructions underlying Wick's theorem that can hardly be avoided in finite dimensions. In what follows, we will not use such a shortcut and construct the diagrammatic series underlying OPT for our $O(N)$ model in arbitrary dimensions\footnote{Note that the construction of diagrammatic series in the framework of OPT has already been discussed in Ref.~\cite{oko87} for a $\varphi^{4}$-theory, but not for the $O(N)$-symmetric case.} as usual. We will investigate the SCC as well and extend results of Ref.~\cite{ros16} to the broken-symmetry regime of the studied (0+0)\nobreakdash-D $O(N)$ model.

We thus turn back to our $O(N)$ model with classical action~\eqref{eq:SfiniteD} for which the partitioning~\eqref{eq:GeneralPartiotioningOPT} reduces to:
\begin{equation}
S\Big[\vec{\widetilde{\varphi}}\Big] = S^0_{\sigma}\Big[\vec{\widetilde{\varphi}}\Big] + S^1_{\sigma}\Big[\vec{\widetilde{\varphi}}\Big] \;,
\end{equation}
with
\begin{equation}
S^0_{\sigma}\Big[\vec{\widetilde{\varphi}}\Big] = \frac{1}{2}\int_{x,y}\widetilde{\varphi}^a(x) \boldsymbol{G}^{-1}_{\sigma;ab}(x,y)\widetilde{\varphi}^b(y) \;,
\end{equation}
\begin{equation}
S^1_{\sigma}\Big[\vec{\widetilde{\varphi}}\Big] = \int_x \left[ \frac{\lambda}{4!} \left(\vec{\widetilde{\varphi}}^2(x)\right)^2+\frac{1}{2}\sigma(x)\vec{\widetilde{\varphi}}^2(x)\right] \;,
\end{equation}
and the OPT propagator defined as follows:
\begin{equation}
\boldsymbol{G}^{-1}_{\sigma;ab}(x,y) = \left(-\nabla^2_x + m^2 - \sigma(x)\right)\delta_{ab}\delta(x-y) \;.
\label{eq:OPTpropagatorGsigma}
\end{equation}
The propagator underlying the OPT expansion is thus dressed by $\sigma(x)$ instead of the classical solution $\vec{\varphi}_\text{cl}(x)$ in the original LE. We then proceed as for the LE with the expansion of the partition function:
\begin{equation}
Z=\int\mathcal{D}\vec{\widetilde{\varphi}} \ e^{-\left(S^0_{\sigma}\big[\vec{\widetilde{\varphi}}\big] + \delta S^1_{\sigma}\big[\vec{\widetilde{\varphi}}\big]\right)} \;,
\label{eq:PartitionFunctionOPT}
\end{equation}
where a fictitious factor $\delta$ has been introduced in order to keep track of the order for the OPT expansion ($\delta$ must therefore be set equal to 1 at the end of all calculations, which is the condition for Eq.~\eqref{eq:PartitionFunctionOPT} to reduce to the original partition function of the studied $O(N)$ model) and we set $\hbar=1$ in this entire section on OPT. Taylor expanding the exponential of the residual action $S^1_\sigma$ in Eq.~\eqref{eq:PartitionFunctionOPT} yields the following expressions:
\begin{equation}
\begin{split}
& Z^\text{OPT} \\
& = \left(\int\mathcal{D}\vec{\widetilde{\varphi}} \ e^{-\int_{x,y} \widetilde{\varphi}^a(x) \boldsymbol{G}_{\sigma;ab}^{-1}(x,y)\widetilde{\varphi}^b(y)}\right) \\
& \hspace{0.3cm} \times \Bigg[ 1 +\sum_{k=1}^{\infty}\frac{\left(-\delta\right)^k}{k!}\sum_{l=0}^k\begin{pmatrix}
k \\
l
\end{pmatrix} \\
& \hspace{0.3cm} \times \left\langle \left(\frac{1}{2}\int_x\sigma(x)\vec{\widetilde{\varphi}}^{2}(x)\right)^{k-l} \left( \frac{\lambda}{4!} \int_x\left(\vec{\widetilde{\varphi}}^{2}(x)\right)^2\right)^l \right\rangle_{0,\sigma} \Bigg] \;,
\end{split}
\end{equation}
and
\begin{equation}
\begin{split}
& W^\text{OPT} \\
& = \frac{1}{2}\mathrm{STr}\left[\ln (\boldsymbol{G}_\sigma)\right] \\
& \hspace{0.3cm} +\sum_{k=1}^{\infty}\frac{\left(-\delta\right)^k}{k!}\sum_{l=0}^k\begin{pmatrix}
k \\
l
\end{pmatrix} \\
& \hspace{0.3cm} \times \left\langle\left(\frac{1}{2}\int_x\sigma(x)\vec{\widetilde{\varphi}}^{2}(x)\right)^{k-l} \left( \frac{\lambda}{4!} \int_x\left(\vec{\widetilde{\varphi}}^{2}(x)\right)^2\right)^l\right\rangle_{0,\sigma}^\text{c} \;,
\end{split}
\label{eq:WOPTbeforeapplyingWick}
\end{equation}
where the $\sigma$-dependent expectation value is defined by:
\begin{equation}
\big\langle\cdots\big\rangle_{0,\sigma} \equiv \frac{1}{Z_{0,\sigma}}\int\mathcal{D}\vec{\widetilde{\varphi}} \ \cdots \ e^{-S^0_{\sigma}\big[\vec{\widetilde{\varphi}}\big]} \;,
\label{eq:OPTsigmaDepExpValueNumber1}
\end{equation}
with
\begin{equation}
Z_{0,\sigma} = \int\mathcal{D}\vec{\widetilde{\varphi}} \ e^{-S^0_{\sigma}\big[\vec{\widetilde{\varphi}}\big]} \;.
\label{eq:OPTsigmaDepExpValueNumber2}
\end{equation}
The connected correlation functions in Eq.~\eqref{eq:WOPTbeforeapplyingWick} are then rewritten with the help of Wick's theorem together with the Feynman rules:
\begin{subequations}
\begin{align}
%\begin{equation}
\begin{gathered}
\begin{fmffile}{Diagrams/LoopExpansion1_FeynRuleGbis2}
\begin{fmfgraph*}(20,5)
\fmfleft{i0,i1,i2,i3}
\fmfright{o0,o1,o2,o3}
\fmflabel{$x, a$}{v1}
\fmflabel{$y, b$}{v2}
\fmf{phantom}{i1,v1}
\fmf{phantom}{i2,v1}
\fmf{plain,tension=0.6}{v1,v2}
\fmf{phantom}{v2,o1}
\fmf{phantom}{v2,o2}
\end{fmfgraph*}
\end{fmffile}
\end{gathered} \quad &\rightarrow \boldsymbol{G}_{\sigma;ab}(x,y)\;,
\label{eq:FeynRulesOPTPropagator} \\
\begin{gathered}
\begin{fmffile}{Diagrams/LoopExpansion1_FeynRuleV4bis}
\begin{fmfgraph*}(20,20)
\fmfleft{i0,i1,i2,i3}
\fmfright{o0,o1,o2,o3}
\fmf{phantom,tension=2.0}{i1,i1bis}
\fmf{plain,tension=2.0}{i1bis,v1}
\fmf{phantom,tension=2.0}{i2,i2bis}
\fmf{plain,tension=2.0}{i2bis,v1}
\fmf{zigzag,label=$x$,tension=0.6,foreground=(0,,0,,1)}{v1,v2}
\fmf{phantom,tension=2.0}{o1bis,o1}
\fmf{plain,tension=2.0}{v2,o1bis}
\fmf{phantom,tension=2.0}{o2bis,o2}
\fmf{plain,tension=2.0}{v2,o2bis}
\fmflabel{$a$}{i1bis}
\fmflabel{$b$}{i2bis}
\fmflabel{$c$}{o1bis}
\fmflabel{$d$}{o2bis}
\end{fmfgraph*}
\end{fmffile}
\end{gathered} \quad &\rightarrow \lambda\delta_{a b}\delta_{c d}\;,
\label{eq:FeynRuleOPT4legVertex} \\
\begin{gathered}
\begin{fmffile}{Diagrams/OPT_FeynRuleSig2}
\begin{fmfgraph*}(4,4)
\fmfleft{i1}
\fmfright{o1}
\fmfv{decor.shape=square,decor.size=2.5thick,label=$x$,label.angle=-90,foreground=(0,,0,,1)}{v3}
\fmflabel{$a$}{v1}
\fmflabel{$b$}{v2}
\fmf{plain}{i1,v1}
\fmf{plain}{v1,v3}
\fmf{plain}{v3,v2}
\fmf{plain}{v2,o1}
\end{fmfgraph*}
\end{fmffile}
\end{gathered} \hspace{0.5cm} &\rightarrow \sigma(x) \delta_{a b}\;.
\label{eq:FeynRulesOPTvertexSquare}
\end{align}
\end{subequations}
In this way, the Schwinger functional reads up to the first order in $\delta$:
\begin{equation}
\begin{split}
W^\text{OPT} = & \ \frac{1}{2} \mathrm{STr}\left[\ln(\boldsymbol{G}_{\sigma})\right] \\
& - \delta \left(\rule{0cm}{0.8cm}\right. \frac{1}{2} \hspace{0.23cm} \begin{gathered}
\begin{fmffile}{Diagrams/OPT_Diag1}
\begin{fmfgraph}(10,10)
\fmfleft{i}
\fmfright{o}
\fmfv{decor.shape=square,decor.size=2.5thick,foreground=(0,,0,,1)}{v1}
\fmftop{v3}
\fmfbottom{v4}
\fmf{phantom,tension=30}{i,v1}
\fmf{phantom,tension=30}{v2,o}
\fmf{plain,left=0.43,tension=0.5}{v1,v3}
\fmf{plain,left=0.43,tension=0.5}{v3,v2}
\fmf{plain,left=0.43,tension=0.5}{v2,v4}
\fmf{plain,left=0.43,tension=0.5}{v4,v1}
\end{fmfgraph}
\end{fmffile}
\end{gathered} + \frac{1}{24} \hspace{0.08cm} \begin{gathered}
\begin{fmffile}{Diagrams/LoopExpansion1_Hartree}
\begin{fmfgraph}(30,20)
\fmfleft{i}
\fmfright{o}
\fmf{phantom,tension=10}{i,i1}
\fmf{phantom,tension=10}{o,o1}
\fmf{plain,left,tension=0.5}{i1,v1,i1}
\fmf{plain,right,tension=0.5}{o1,v2,o1}
\fmf{zigzag,foreground=(0,,0,,1)}{v1,v2}
\end{fmfgraph}
\end{fmffile}
\end{gathered} \\
& +\frac{1}{12}\begin{gathered}
\begin{fmffile}{Diagrams/LoopExpansion1_Fock}
\begin{fmfgraph}(15,15)
\fmfleft{i}
\fmfright{o}
\fmf{phantom,tension=11}{i,v1}
\fmf{phantom,tension=11}{v2,o}
\fmf{plain,left,tension=0.4}{v1,v2,v1}
\fmf{zigzag,foreground=(0,,0,,1)}{v1,v2}
\end{fmfgraph}
\end{fmffile}
\end{gathered} \left.\rule{0cm}{0.8cm}\right) \\
& + \mathcal{O}\Big(\delta^2\Big)\;.
\end{split}
\label{eq:WKjLoopExpansionStep3OPT}
\end{equation}

We deduce the expressions of the partition function and the Schwinger functional of our $O(N)$ model in the (0+0)\nobreakdash-D limit (pushed up to order $\mathcal{O}(\delta^3)$ in the OPT expansion and setting $\delta=1$):
\begin{equation}
\begin{split}
& Z^{\text{OPT};(3)} \\
& = (2\pi G_\sigma)^\frac{N}{2}\Bigg[1-\frac{N\sigma G_\sigma}{2} \\
& \hspace{0.3cm} + \frac{N(N+2)}{8} \left(\sigma^2-\frac{\lambda}{3}\right) G_\sigma^2 \\
& \hspace{0.3cm} +\frac{N(N+2)(N+4)}{48}\left(\lambda-\sigma^2\right)\sigma G_\sigma^3 \\
& \hspace{0.3cm} +\frac{N(N+2)(N+4)(N+6)}{192}\left(\frac{\lambda}{6}-\sigma^2\right)\lambda G_\sigma^4 \\
& \hspace{0.3cm} -\frac{N(N+2)(N+4)(N+6)(N+8)}{2304}\lambda^2\sigma G_\sigma^5 \\
& \hspace{0.3cm} -\frac{N(N+2)(N+4)(N+6)(N+8)(N+10)}{82944}\lambda^3G_\sigma^6\Bigg] \;,
\end{split}
\end{equation}
and
\begin{equation}
\begin{split}
& W^{\text{OPT};(3)} \\
& = \frac{N}{2}\ln(2\pi G_{\sigma}) - \frac{N\sigma G_{\sigma}}{2} - \frac{1}{24} N \left(\lambda\left(N+2\right)-6\sigma^2\right)G_{\sigma}^2 \\
& + \frac{1}{12} N \sigma \left(\lambda\left(N+2\right)-2\sigma^2\right)G_{\sigma}^3 \\
& +\frac{1}{144}\lambda N \left(N+2\right)\left(\lambda\left(N+3\right)-18\sigma^2\right)G_{\sigma}^4 \\
& -\frac{1}{36}\lambda^2 N\sigma \left(N^2 + 5N +6\right)G_{\sigma}^5 \\
&-\frac{1}{2592}\lambda^3 N\left(5 N^3 + 44 N^2 + 128 N + 120\right)G_{\sigma}^6 \;,
\end{split}
\end{equation}
with the dressed propagator:
\begin{equation}
G_\sigma=\frac{1}{m^2-\sigma} \;,
\end{equation}
and $\sigma$ to be determined via one of the optimization conditions discussed previously. The gs energy and density of our (0+0)\nobreakdash-D $O(N)$ model (for all $N\geq 1$ and for both the unbroken- and broken-symmetry phases) obtained from OPT thus read:
\begin{equation}
\begin{split}
E_\text{gs}^{\text{OPT};(3)} = & -\frac{N}{2}\ln\bigg(\frac{2\pi}{m^2-\sigma}\bigg) + \frac{N}{2592 (m^2 - \sigma)} \Bigg( 1296 \sigma \\
& + \frac{108 (\lambda (2 + N) - 6 \sigma^2)}{m^2 - \sigma} \\
& - \frac{216 \sigma (\lambda (2 + N) - 2 \sigma^2)}{(m^2 - \sigma)^2} \\
& - \frac{18 \lambda (2 + N) (\lambda (3 + N) - 18 \sigma^2)}{(m^2 - \sigma)^3} \\
& + \frac{72 \lambda^2 (6 + 5 N + N^2) \sigma}{(m^2 - \sigma)^4} \\
& + \frac{\lambda^3 (120 + 128 N + 44 N^2 + 5 N^3)}{(m^2 - \sigma)^5} \Bigg) \;,
\end{split}
\end{equation}
and
\begin{equation}
\begin{split}
\rho_\text{gs}^{\text{OPT};(3)} = & \ \frac{1}{m^2-\sigma} - \frac{1}{216 (m^2 - \sigma)^2} \Bigg( 216 \sigma \\
& + \frac{36 (\lambda (2 + N) - 6 \sigma^2)}{m^2 - \sigma} \\
& - \frac{108 (\lambda (2 + N) \sigma - 2 \sigma^3)}{(m^2 - \sigma)^2} \\
& - \frac{12 \lambda (2 + N) (\lambda (3 + N) - 18 \sigma^2)}{(m^2 - \sigma)^3} \\
& + \frac{60 \lambda^2 (6 + 5 N + N^2) \sigma}{(m^2 - \sigma)^4} \\
& + \frac{\lambda^3 (120 + 128 N + 44 N^2 + 5 N^3)}{(m^2 - \sigma)^5} \Bigg) \;.
\end{split}
\end{equation}
A definite estimate of the gs energy and density is only obtained after fixing a value for $\sigma$, which we now discuss.

\begin{figure}[!t]
  \begin{center}
      \includegraphics[width=0.95\linewidth]{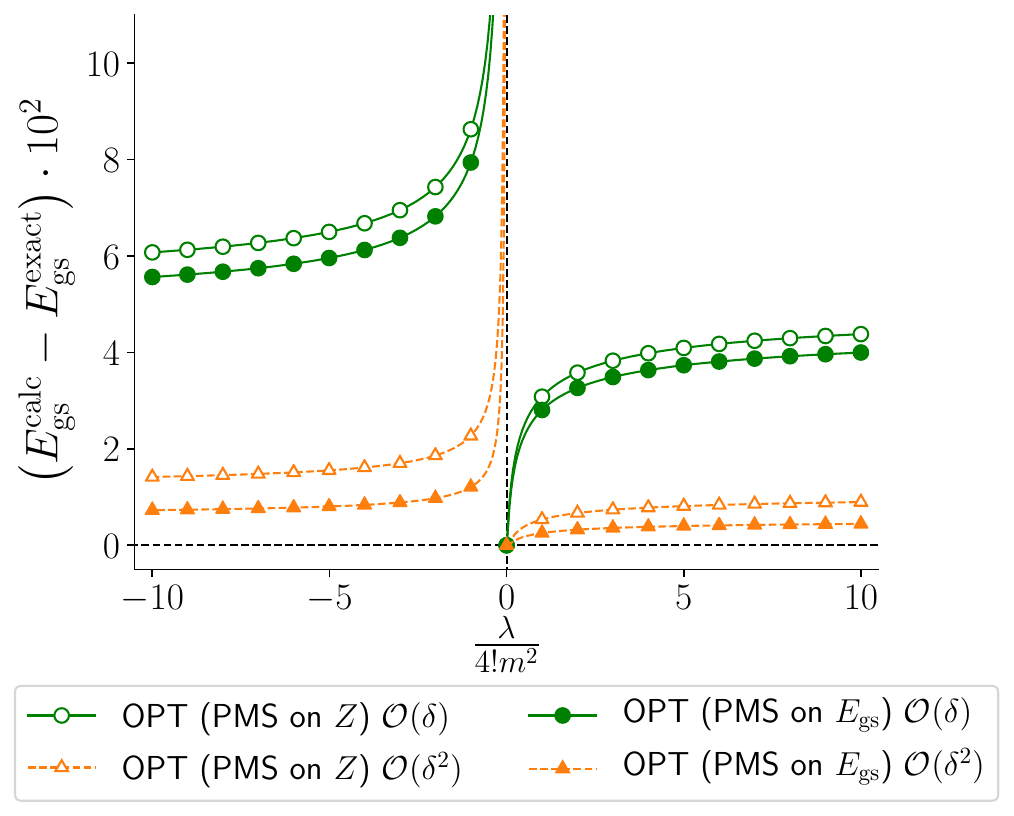}
      \includegraphics[width=0.95\linewidth]{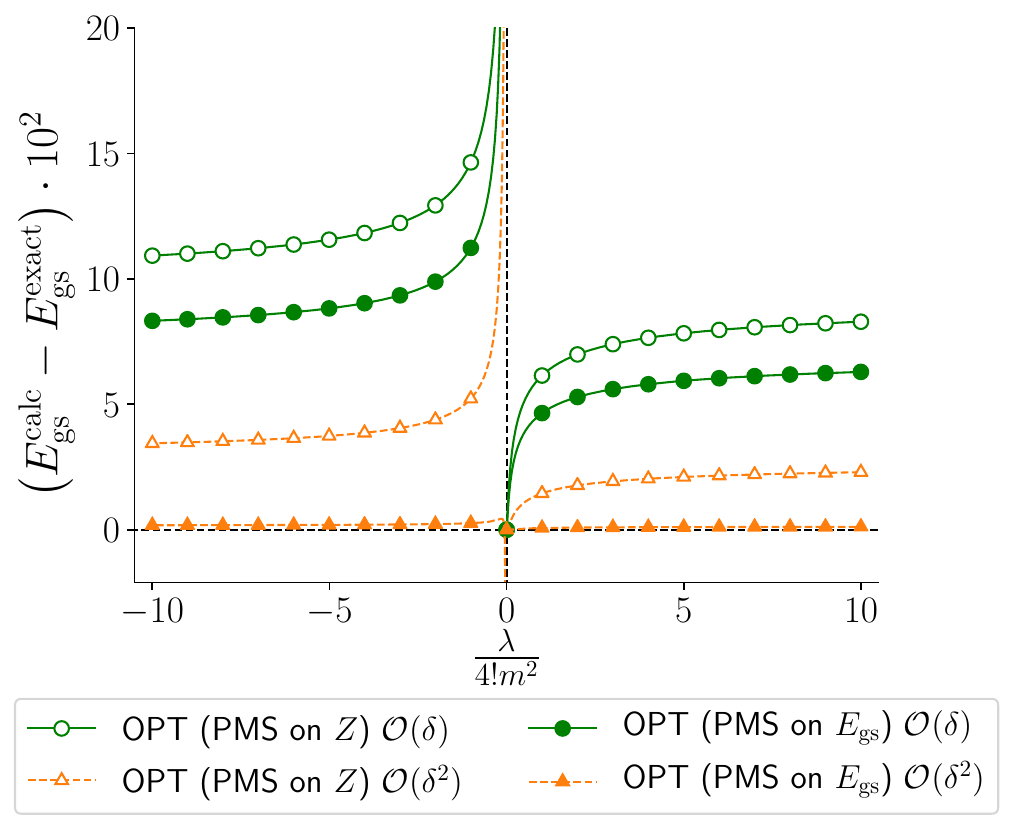}
    \caption{Difference between the calculated gs energy $E_{\mathrm{gs}}^{\mathrm{calc}}$ and the corresponding exact solution $E_{\mathrm{gs}}^{\mathrm{exact}}$ at $N=1$ (upper panel) and $N=2$ (lower panel), with $\hbar=1$ and $m^{2}=\pm 1$.}
    \label{fig:OPTPMSonZorE}
  \end{center}
\end{figure}

What distinguishes OPT from many other approaches is the fact that the optimization of the field $\sigma$ depends on the working order $k$. We focus on three classes of optimization criteria for $\sigma$, namely the PMS, the TP method and the SCC\footnote{The study of Ref.~\cite{ros16} has already shown that the FAC optimization procedure is less performing than the PMS approach when determining the gs energy, the self-energy and the fourth-order vertex function $\Gamma^{(\mathrm{1PI})(4)}\big(\vec{\phi}=\vec{0}\big)$ in the framework of the unbroken-symmetry phase of the toy model considered here.}. We first consider the PMS, where we either ask the partition function $Z^\text{OPT;(k)}$, the gs energy $E^\text{OPT;(k)}_\text{gs}$ or the gs density $\rho^\text{OPT;(k)}_\text{gs}$, computed at order $k$ (i.e. up to order $\mathcal{O}(\delta^k)$) of the OPT expansion, to be extremal with respect to $\sigma$, i.e.:
\begin{equation}
\left.\frac{\partial Z^\text{OPT;(k)}}{\partial\sigma}\right|_{\sigma = \sigma^{(k)}_{\mathrm{PMS};Z}} = 0 \;,
\label{eq:PMSeqZ0DON}
\end{equation}
\begin{equation}
\left.\frac{\partial E^\text{OPT;(k)}_\text{gs}}{\partial\sigma}\right|_{\sigma = \sigma^{(k)}_{\mathrm{PMS};E}} = 0 \;,
\label{eq:PMSeqEgs0DON}
\end{equation}
\begin{equation}
\left.\frac{\partial\rho^\text{OPT;(k)}_\text{gs}}{\partial\sigma}\right|_{\sigma = \sigma^{(k)}_{\mathrm{PMS};\rho}} = 0 \;.
\label{eq:PMSeqrhogs0DON}
\end{equation}
For instance, at the first non-trivial order (i.e. up to order $\mathcal{O}(\delta)$) of the OPT expansion, the three above equations are polynomial and second-order with respect to $\sigma$ and their solutions read:
\begin{equation}
\begin{split}
\sigma^{(1)}_{\mathrm{PMS};Z} = & \frac{1}{6\left(N+2\right)} \Big(3 m^2\left(N+4\right) \\
& \pm \sqrt{3N\left(\lambda\left(N^2 +4N +4\right) + 3 N m^4\right)}\Big) \;,
\end{split}
\end{equation}
\begin{equation}
\sigma^{(1)}_{\mathrm{PMS};E} = \frac{1}{2}\left(m^2 \pm \sqrt{m^{4} + \frac{2\lambda}{3}\left(N+2\right)}\right) \;,
\label{eq:solutionPMSEgs0DON}
\end{equation}
\begin{equation}
\sigma^{(1)}_{\mathrm{PMS};\rho} = \frac{1}{2}\left(m^2 \pm  \sqrt{m^4+\lambda\left(N+2\right)}\right) \;.
\label{eq:solutionPMSrhogs0DON}
\end{equation}
The complexity of the PMS equations~\eqref{eq:PMSeqZ0DON} to~\eqref{eq:PMSeqrhogs0DON} increases with the truncation order, i.e. with the working order $k$. We illustrate in Fig.~\ref{fig:OPTPMSonZorE} at $N=1$ and $2$ that, for the purpose of determining $E_{\mathrm{gs}}$ or $\rho_{\mathrm{gs}}$, it is in general more efficient to apply the PMS directly on the OPT series representing $E_{\mathrm{gs}}$ or $\rho_{\mathrm{gs}}$ respectively (i.e. to exploit Eqs.~\eqref{eq:PMSeqEgs0DON} and~\eqref{eq:PMSeqrhogs0DON}), rather than on $Z$ (i.e. rather than using Eq.~\eqref{eq:PMSeqZ0DON}). We can already appreciate in this figure the nice convergence properties of OPT at its first two non-trivial orders, which will be discussed further later in this section. Considering then the TP method, the optimization of $\sigma$ relies on the equations:
\begin{equation}
\left.\frac{\partial^2 E^\text{OPT;(k)}_\text{gs}}{\partial\sigma^2}\right|_{\sigma=\sigma^{(k)}_{\mathrm{TP};E}} = 0 \;,
\label{eq:solutionTPEgs0DON}
\end{equation}
\begin{equation}
\left.\frac{\partial^2\rho^\text{OPT;(k)}_\text{gs}}{\partial\sigma^2}\right|_{\sigma=\sigma^{(k)}_{\mathrm{TP};\rho}} = 0 \;,
\label{eq:solutionTPrhogs0DON}
\end{equation}
whose solutions are, at the first non-trivial order of the OPT expansion:
\begin{equation}
\sigma^{(1)}_{\mathrm{TP};E} = \pm \sqrt{m^4 + \frac{\lambda}{2}\left(N+2\right)} \;,
\end{equation}
\begin{equation}
\sigma^{(1)}_{\mathrm{TP};\rho} = \frac{1}{4}\left(m^2 \pm \sqrt{9 m^4 + 8 \lambda\left(N+2\right)}\right) \;.
\end{equation}
As for the PMS, the complexity of the TP optimization equations such as Eqs.~\eqref{eq:solutionTPEgs0DON} and~\eqref{eq:solutionTPrhogs0DON} grows with the working order $k$. Finally, we implement the optimization of $\sigma$ via the SCC by solving the equation:
\begin{equation}
\left\langle\left(\vec{\widetilde{\varphi}}^2\right)^k\right\rangle_{0,\sigma=\sigma^{(k)}_\text{SCC}}^{(1)} = \left\langle\left(\vec{\widetilde{\varphi}}^2\right)^k\right\rangle_{0,\sigma=\sigma^{(k)}_\text{SCC}}^{(0)} \;,
\label{eq:SCCequation0DON}
\end{equation}
where the $\sigma$-dependent expectation value is already defined by Eqs.~\eqref{eq:OPTsigmaDepExpValueNumber1} and~\eqref{eq:OPTsigmaDepExpValueNumber2} in arbitrary dimensions. Whatever the working order $k$ in the OPT expansion, the complexity of the SCC equation~\eqref{eq:SCCequation0DON} remains that of a second-order polynomial equation in $\sigma$. It is the order of the correlation functions involved in both the right-hand and left-hand sides of this SCC equation that changes with $k$. One can then write the solutions of Eq.~\eqref{eq:SCCequation0DON} for all $k$ as:
\begin{equation}
\sigma^{(k)}_\text{SCC} = \frac{1}{2}\left(m^2\pm\sqrt{m^4+\frac{2\lambda}{3}(N+k+1)}\right) \;.
\label{eq:SolutionSCC0DON}
\end{equation}
Hence, the solutions~\eqref{eq:solutionPMSEgs0DON} and~\eqref{eq:SolutionSCC0DON}, obtained respectively from the PMS and the SCC, coincide at $k=1$.

%%%%%%%%%%%%%%%%%%%
%%%%%%%%%%%%%%%%%%%

It is also instructive to analyze from the latter solutions how the propagator gets renormalized in the frameworks of both OPT and the collective LE. To that end, we write the dressed unperturbed propagators in the generic form:
\begin{equation}
\boldsymbol{G}^\star_{0;ab} = \frac{1}{m_\star^2} \delta_{ab} \;,
\end{equation}
with $m_\star$ being a renormalized mass. On the one hand, we have obtained for the collective LE:
\begin{equation}
\begin{split}
m_{\star;\text{LE};\text{col}}^2 = & \ m^2+i\sqrt{\frac{\lambda}{3}}\overline{\sigma}_\text{cl} \\
= & \ \frac{1}{2}\left(m^2 + \sqrt{m^4+\frac{2\lambda}{3}N}\right) \;,
\end{split}
\label{eq:renormalizedmassmstart}
\end{equation}
as follows from Eq.~\eqref{eq:ClassicalSolutionBosonicAction}\footnote{Only the minus sign solution (which is the physical solution) is taken in Eq.~\eqref{eq:ClassicalSolutionBosonicAction} to obtain the second line of Eq.~\eqref{eq:renormalizedmassmstart}.}. However, the LE propagators are the same whatever the working order $k$ of the LE whereas the propagator involved in OPT depends on the collective dof $\sigma$, which itself depends on the working order $k$ (i.e. it depends up to which order $\mathcal{O}(\delta^k)$ the OPT is carried out), namely $m_{\star;\text{OPT}}^2 = m_{\star;\text{OPT}}^2(\sigma(k))$. Focusing on the PMS and SCC optimization procedures only, we obtain\footnote{Similarly to Eq.~\eqref{eq:renormalizedmassmstart}, expressions~\eqref{eq:mstarOPTPMSE},~\eqref{eq:mstarOPTPMSrho} and~\eqref{eq:mstarOPTSCC} are all physical solutions. They are obtained by taking the minus sign solutions of~\eqref{eq:solutionPMSEgs0DON},~\eqref{eq:solutionPMSrhogs0DON} and~\eqref{eq:SolutionSCC0DON}, respectively. Their physical character can be seen from the fact that they all reduce to $m_{\star}^2=m^2$ at $\lambda=0$ and $m^2 \geq 0$.} at the first non-trivial order of the OPT expansion:
\begin{equation}
m_{\star;\text{OPT;PMS;$E$;(1)}}^2 = \frac{1}{2}\left(m^2 + \sqrt{m^{4} + \frac{2\lambda}{3}\left(N+2\right)}\right) \;,
\label{eq:mstarOPTPMSE}
\end{equation}
\begin{equation}
m_{\star;\text{OPT;PMS;$\rho$;(1)}}^2 = \frac{1}{2}\left(m^2 + \sqrt{m^4+\lambda\left(N+2\right)}\right) \;,
\label{eq:mstarOPTPMSrho}
\end{equation}
and, up to order $\mathcal{O}\big(\delta^k\big)$ of the OPT expansion,
\begin{equation}
m_{\star;\text{OPT;SCC;(k)}}^2 =\frac{1}{2}\left(m^2 + \sqrt{m^4+\frac{2\lambda}{3}(N+k+1)}\right) \;.
\label{eq:mstarOPTSCC}
\end{equation}
We recover of course the property $m_{\star;\text{OPT;SCC;(1)}}^2=m_{\star;\text{OPT;PMS;$E$;(1)}}^2$ discussed before but we can also see a significant resemblance between the renormalized masses~\eqref{eq:mstarOPTPMSE} and~\eqref{eq:mstarOPTSCC} obtained from OPT via PMS and SCC on the one hand and, on the other hand, that given by Eq.~\eqref{eq:renormalizedmassmstart} for the collective LE.

%%%%%%%%%%%%%%%%%%%
\begin{figure}[!t]
  \begin{center}
      \includegraphics[width=0.95\linewidth]{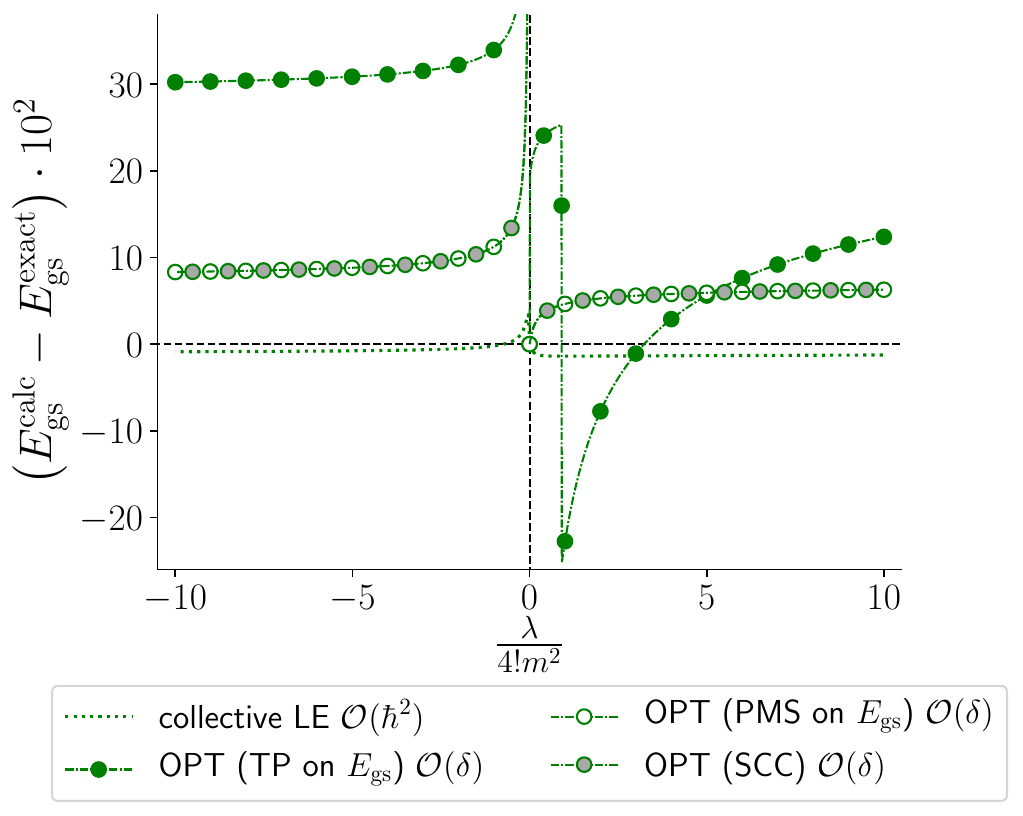}
      \includegraphics[width=0.95\linewidth]{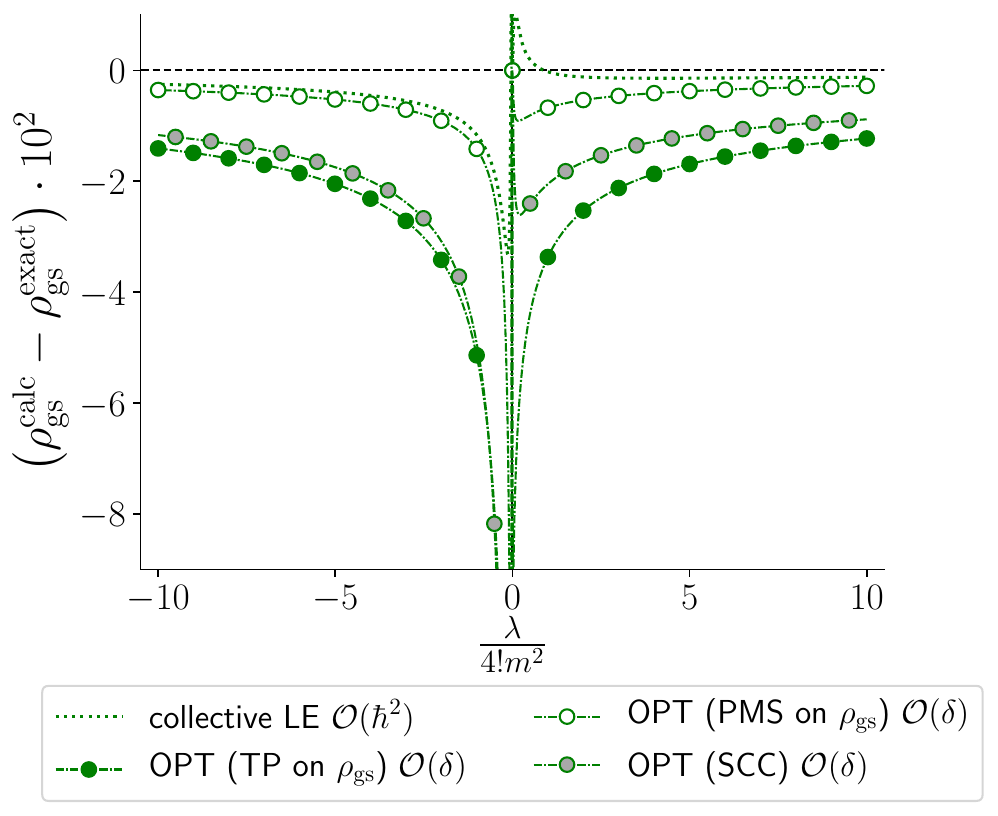}
    \caption{Difference between the calculated gs energy $E_{\mathrm{gs}}^{\mathrm{calc}}$ (upper panel) or density $\rho_{\mathrm{gs}}^{\mathrm{calc}}$ (lower panel) and the corresponding exact solution $E_{\mathrm{gs}}^{\mathrm{exact}}$ or $\rho_{\mathrm{gs}}^{\mathrm{exact}}$ at $\hbar=1$, $m^{2}=\pm 1$ and $N=2$. The presented results are the first non-trivial orders of the collective LE and OPT with the three different tested optimization procedures. See also the caption of Fig.~\ref{fig:O1PTcoll} for the meaning of the indication ``$\mathcal{O}\big(\hbar^{n}\big)$'' for the collective LE results.}
    \label{fig:O2OPTvsLE_1}
  \end{center}
\end{figure}
%%%%%%%%%%%%%%%%%%%
%%%%%%%%%%%%%%%%%%%
\begin{figure}[!t]
  \begin{center}
      \includegraphics[width=0.95\linewidth]{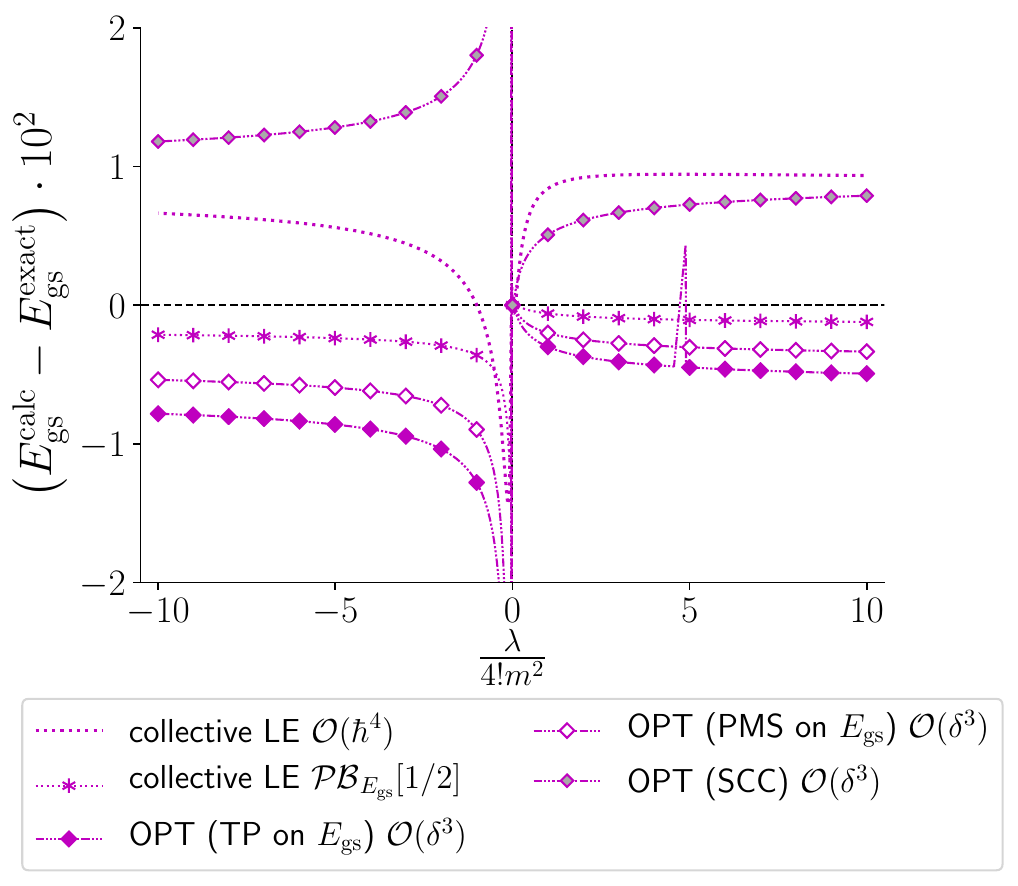}
      \includegraphics[width=0.95\linewidth]{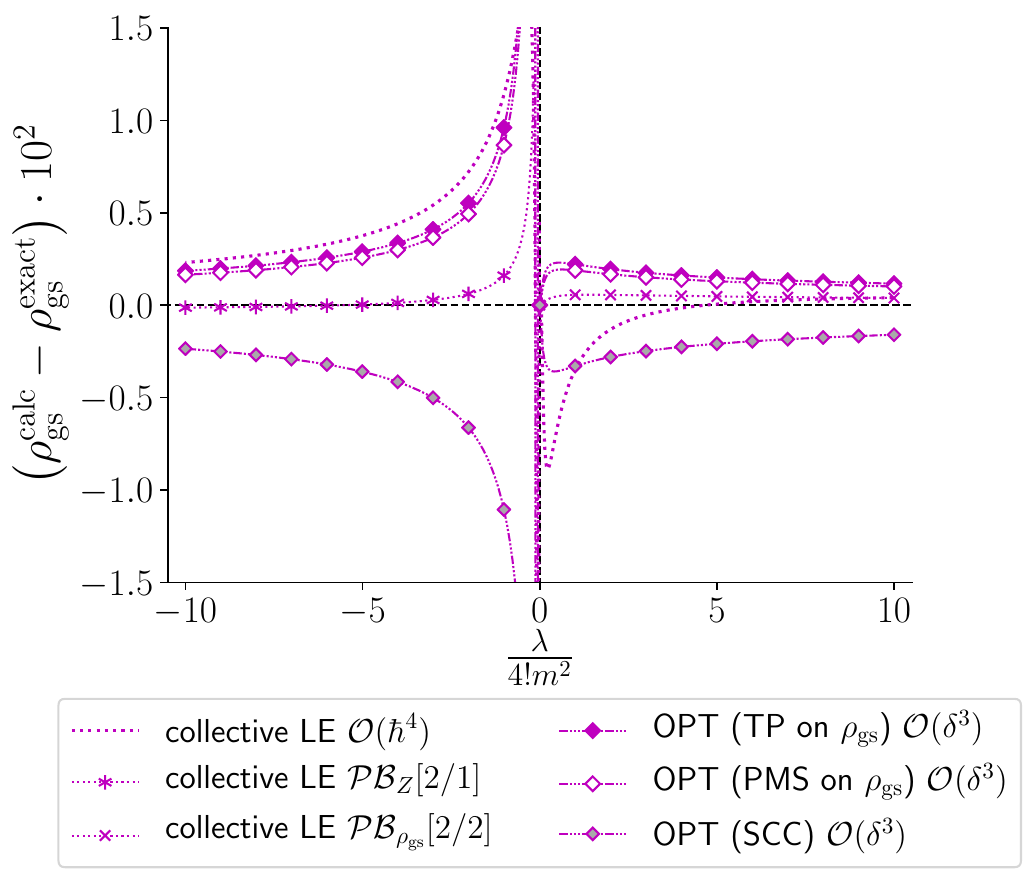}
    \caption{Same as Fig.~\ref{fig:O2OPTvsLE_1} but for the third non-trivial orders of the collective LE (including the best tested resummation procedures among the Pad\'e-Borel and Borel-hypergeometric schemes) and of OPT.}
    \label{fig:O2OPTvsLE_3}
  \end{center}
\end{figure}
%%%%%%%%%%%%%%%%%%%

Our OPT results are compared at $N=2$ with those of the collective LE (combined with the best tested resummation procedure for the third non-trivial order) in Figs.~\ref{fig:O2OPTvsLE_1} and~\ref{fig:O2OPTvsLE_3}. An excellent reproduction of the gs energy and density is achieved with OPT based on the PMS, which notably results in an accuracy around $0.5\%$ at the third non-trivial order in both the unbroken- and broken-symmetry regimes at $N=2$, as shown notably by Fig.~\ref{fig:O2OPTvsLE_3}. At the first non-trivial order, the PMS and the SCC lead to identical results, as justified below Eq.~\eqref{eq:SolutionSCC0DON} and illustrated by Fig.~\ref{fig:O2OPTvsLE_1}. However, Fig.~\ref{fig:O2OPTvsLE_3} shows (still at $N=2$) that the PMS slightly outperforms the SCC: for example, the corresponding estimates for $E_{\mathrm{gs}}$ are respectively around $0.6\%$ and $1.1\%$ throughout most of the tested range of values for the coupling constant. This loss of accuracy of the SCC as compared to the PMS is rather small considering the simplicity of the underlying equations to solve: at the third non-trivial order (and regardless of the values of $N$ and $m^2$), the SCC still amounts to find the roots of a quadratic polynomial whereas the PMS criterion is now a polynomial equation of order $6$. Note also that, according to Figs.~\ref{fig:O2OPTvsLE_1} and~\ref{fig:O2OPTvsLE_3}, the TP method is clearly less performing than the SCC and the PMS at the first non-trivial order of the OPT expansion but becomes comparable to SCC at the third one, whereas the underpinning equations to solve are closer to those of the PMS in terms of complexity. The TP method is thus disappointing (as compared to both the PMS and the SCC) in that respect.

Comparing the two best approaches investigated so far, i.e. OPT and the collective LE, Figs.~\ref{fig:O2OPTvsLE_1} and~\ref{fig:O2OPTvsLE_3} show that, after combination with resummation, the collective LE outperforms OPT at both the first and third non-trivial orders for both $E_{\mathrm{gs}}$ and $\rho_{\mathrm{gs}}$, at $N=2$. Furthermore, as discussed earlier from Figs.~\ref{fig:O1PTcoll} and~\ref{fig:O2PTcoll}, the performances of the collective LE are expected to improve with $N$ due to its connection with the $1/N$-expansion whereas such an argument does not hold for OPT (for instance, the performances of OPT in Fig.~\ref{fig:OPTPMSonZorE} do not differ significantly at $N=1$ and $2$). However, regarding the formalisms underpinning these two techniques, the diagrammatic representations of their respective expansions is much more demanding to determine on the side of the collective LE: whereas OPT diagrams are directly obtained by adding the square vertex~\eqref{eq:FeynRulesOPTvertexSquare} to the diagrams of the original LE at $\vec{\varphi}_{\mathrm{cl}}=\vec{0}$ (i.e. in the unbroken-symmetry regime) in all possible ways, the collective LE requires to construct the diagrammatic expressions of all vertex functions $S^{(n)}_{\mathrm{col},\mathcal{J}}$ for $n=2,\cdots,2k$ to determine the corresponding Schwinger functional up to order $\mathcal{O}\big(\hbar^{k}\big)$. As can be inferred from Eqs.~\eqref{eq:SbosonicKLoopExpansionS3} and~\eqref{eq:SbosonicKLoopExpansionS4} expressing respectively $S^{(3)}_{\mathrm{col},\mathcal{J}}$ and $S^{(4)}_{\mathrm{col},\mathcal{J}}$, the determination of the diagrammatic expressions of $S^{(n)}_{\mathrm{col},\mathcal{J}}$ becomes quickly lengthy as $n$ increases. This cumbersomeness directly results from the logarithm structure of the collective classical action. Hence, the simplicity of the construction of diagrammatic series is an important advantage of OPT, over the collective LE in particular.

In conclusion, OPT offers an interesting framework to describe strongly-coupled many-body systems at low cost. As opposed to LEs, OPT results are directly systematically improvable in the sense that they do not rely on resummation procedures: they take the form of (tremendously) fast convergent series (see notably Refs.~\cite{buc93,dun93,ben94,arv95} for detailed studies on the convergence behavior of OPT series). However, the energy and the density are not tied in a functional as in the EDF approach. We now investigate the third and last family of approaches considered in the present study of diagrammatic PI techniques. These approaches, coined as SCPT, are based on EAs from which one can work with functionals of the correlation functions of the theory, or their local versions that can coincide with e.g. the density of the system. We will also see that SCPT relies, as its name suggests, on self-consistent equations (in finite dimensions), another important feature shared with the EDF approach.

%==============================
\subsection{\label{sec:SCPT}Self-consistent perturbation theory}
%~~~
\subsubsection{\label{sec:Orig2PIEA}Original 2PI effective action}

\begin{figure}[!t]
  \begin{center}
      \includegraphics[width=0.80\linewidth]{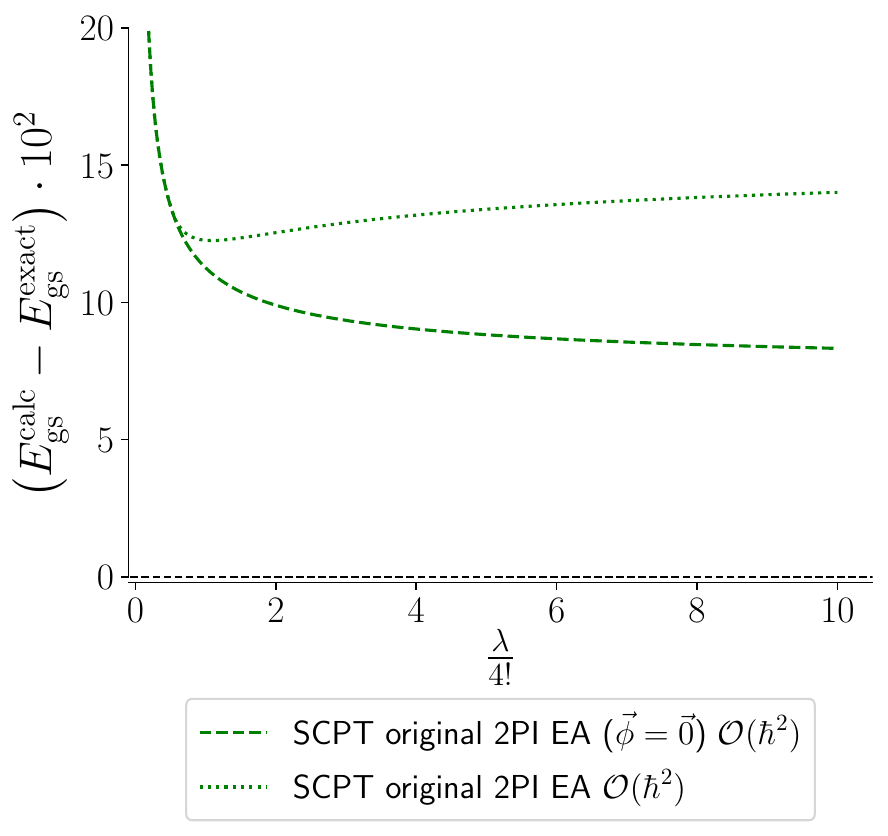}
      \includegraphics[width=0.80\linewidth]{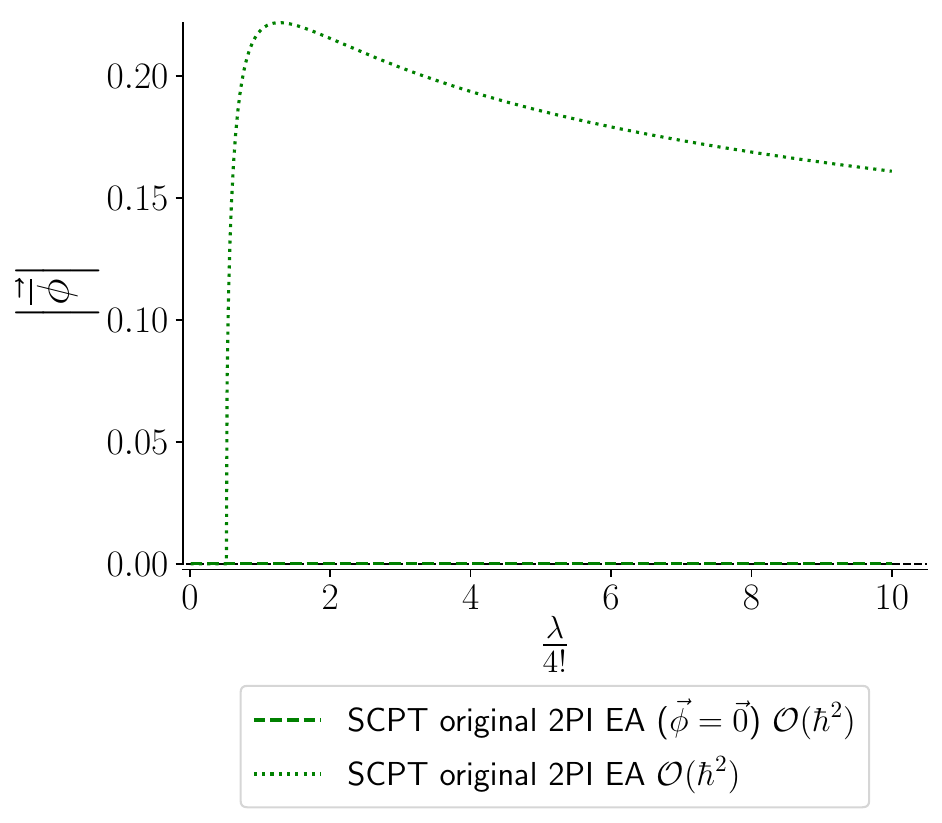}
    \caption{Two different solutions of the gap equations of the 2PI EA $\Gamma^{(\mathrm{2PI})}\big(\vec{\phi},\boldsymbol{G}\big)$ at its first non-trivial order for the gs energy $E_{\mathrm{gs}}$ or the 1-point correlation function $\vec{\overline{\phi}}$ at $\hbar=1$, $m^{2}=-1$ and $N=2$. More precisely, the upper panel shows the difference between the gs energy $E_{\mathrm{gs}}^{\mathrm{calc}}$ calculated from each of these solutions on the one hand and the corresponding exact solution $E_{\mathrm{gs}}^{\mathrm{exact}}$ on the other hand. In general, the indication ``$\mathcal{O}\big(\hbar^{n}\big)$'' for the SCPT results specifies that the series representing the corresponding EA has been exploited up to order $\mathcal{O}\big(\hbar^{n}\big)$ (which implies notably that the corresponding series for the gs energy is calculated up to order $\mathcal{O}(\hbar^{n-1})$ according e.g. to Eq.~\eqref{eq:2PIorigE} for the original 2PI EA).}
    \label{fig:2PIEAzeroVsNonzerovev}
  \end{center}
\end{figure}

The simplest and numerically less demanding implementations of SCPT are based on 1PI EAs but the latter are actually not relevant to grasp non-perturbative features for the studied model, especially because of the constraint of the $O(N)$ symmetry (see appendix~\ref{app:SCPT1PIEA} for further explanations, illustrated with numerical applications for $E_{\mathrm{gs}}$ and $\rho_{\mathrm{gs}}$ in the original and collective representations). This is not the case of 2PI EAs, which are closely related to DFT, and therefore to EDFs, for reasons discussed below. We thus first consider the 2PI EA in the original representation, i.e. the original 2PI EA, which has already been investigated for the $O(N)$-symmetric $\varphi^{4}$-theory in finite dimensions \cite{baa03bis2,baa04,baa05}, as well as for the (0+0)\nobreakdash-D case at $N=1$ \cite{bro15}. The original 2PI EA is defined as the Legendre transform:
\begin{equation}
\begin{split}
\Gamma^{(\mathrm{2PI})}\Big[\vec{\phi},\boldsymbol{G}\Big] \equiv & -W\Big[\vec{J},\boldsymbol{K}\Big] + \int_{x}J^{a}(x) \frac{\delta W\big[\vec{J},\boldsymbol{K}\big]}{\delta J^{a}(x)} \\
& + \int_{x,y}\boldsymbol{K}^{ab}(x,y) \frac{\delta W\big[\vec{J},\boldsymbol{K}\big]}{\delta \boldsymbol{K}^{ba}(x,y)} \\
= & -W\Big[\vec{J},\boldsymbol{K}\Big] + \int_{x}J^{a}(x) \phi_{a}(x) \\
& + \frac{1}{2} \int_{x,y} \phi_{a}(x) \boldsymbol{K}^{ab}(x,y) \phi_{b}(y) \\
& + \frac{\hbar}{2} \int_{x,y} \boldsymbol{K}^{ab}(x,y) \boldsymbol{G}_{ba}(y,x) \;,
\end{split}
\label{eq:pure2PIEAdefinition0DONmain}
\end{equation}
with
\begin{equation}
\phi_{a}(x) = \frac{\delta W\big[\vec{J},\boldsymbol{K}\big]}{\delta J^{a}(x)} \;,
\label{eq:pure2PIEAdefinitionbis0DONmain}
\end{equation}
\begin{equation}
\begin{split}
\boldsymbol{G}_{ab}(x,y) = & \ \frac{\delta^{2} W\big[\vec{J},\boldsymbol{K}\big]}{\delta J^{a}(x)\delta J^{b}(y)} \\
= & \ \frac{2}{\hbar} \frac{\delta W\big[\vec{J},\boldsymbol{K}\big]}{\delta\boldsymbol{K}^{ab}(x,y)} - \frac{1}{\hbar} \phi_{a}(x) \phi_{b}(y) \;,
\end{split}
\label{eq:pure2PIEAdefinitionbis20DONmain}
\end{equation}
and $W\big[\vec{J},\boldsymbol{K}\big]$ is expressed by Eq.~\eqref{eq:ZJKfiniteD}. The original 2PI EA can be expressed in terms of 2PI diagrams only, which translates up to second order in $\hbar$ into:
\begin{equation}
\begin{split}
\Gamma^{(\mathrm{2PI})}\Big[\vec{\phi},\boldsymbol{G}\Big] = & \ S\Big[\vec{\phi}\Big] -\frac{\hbar}{2}\mathrm{STr}\left[\ln\big(\boldsymbol{G}\big)\right] \\
& + \frac{\hbar}{2}\mathrm{STr}\left[\boldsymbol{G}^{-1}_\phi\boldsymbol{G}-\mathbb{I}\right] \\
& + \hbar^{2} \left(\rule{0cm}{1.2cm}\right. \frac{1}{24} \hspace{0.08cm} \scalebox{0.85}{${\displaystyle \begin{gathered}
\begin{fmffile}{Diagrams/1PIEA_Hartree}
\begin{fmfgraph}(30,20)
\fmfleft{i}
\fmfright{o}
\fmf{phantom,tension=10}{i,i1}
\fmf{phantom,tension=10}{o,o1}
\fmf{plain,left,tension=0.5,foreground=(1,,0,,0)}{i1,v1,i1}
\fmf{plain,right,tension=0.5,foreground=(1,,0,,0)}{o1,v2,o1}
\fmf{zigzag,foreground=(0,,0,,1)}{v1,v2}
\end{fmfgraph}
\end{fmffile}
\end{gathered}}$}
+\frac{1}{12}\scalebox{0.85}{${\displaystyle\begin{gathered}
\begin{fmffile}{Diagrams/1PIEA_Fock}
\begin{fmfgraph}(15,15)
\fmfleft{i}
\fmfright{o}
\fmf{phantom,tension=11}{i,v1}
\fmf{phantom,tension=11}{v2,o}
\fmf{plain,left,tension=0.4,foreground=(1,,0,,0)}{v1,v2,v1}
\fmf{zigzag,foreground=(0,,0,,1)}{v1,v2}
\end{fmfgraph}
\end{fmffile}
\end{gathered} }$} \\
& - \frac{1}{18} \scalebox{0.85}{${\displaystyle \begin{gathered}
\begin{fmffile}{Diagrams/1PIEA_Diag1}
\begin{fmfgraph}(27,15)
\fmfleft{i}
\fmfright{o}
\fmftop{vUp}
\fmfbottom{vDown}
\fmfv{decor.shape=cross,decor.size=3.5thick,foreground=(1,,0,,0)}{v1}
\fmfv{decor.shape=cross,decor.size=3.5thick,foreground=(1,,0,,0)}{v2}
\fmf{phantom,tension=10}{i,i1}
\fmf{phantom,tension=10}{o,o1}
\fmf{phantom,tension=2.2}{vUp,v5}
\fmf{phantom,tension=2.2}{vDown,v6}
\fmf{phantom,tension=0.5}{v3,v4}
\fmf{phantom,tension=10.0}{i1,v1}
\fmf{phantom,tension=10.0}{o1,v2}
\fmf{dashes,tension=2.0,foreground=(0,,0,,1),foreground=(1,,0,,0)}{v1,v3}
\fmf{dots,left=0.4,tension=0.5,foreground=(0,,0,,1)}{v3,v5}
\fmf{plain,left=0.4,tension=0.5,foreground=(1,,0,,0)}{v5,v4}
\fmf{plain,right=0.4,tension=0.5,foreground=(1,,0,,0)}{v3,v6}
\fmf{dots,right=0.4,tension=0.5,foreground=(0,,0,,1)}{v6,v4}
\fmf{dashes,tension=2.0,foreground=(0,,0,,1),foreground=(1,,0,,0)}{v4,v2}
\fmf{plain,tension=0,foreground=(1,,0,,0)}{v5,v6}
\end{fmfgraph}
\end{fmffile}
\end{gathered} }$} - \frac{1}{36} \hspace{-0.15cm} \scalebox{0.85}{${\displaystyle\begin{gathered}
\begin{fmffile}{Diagrams/1PIEA_Diag2}
\begin{fmfgraph}(25,20)
\fmfleft{i}
\fmfright{o}
\fmftop{vUp}
\fmfbottom{vDown}
\fmfv{decor.shape=cross,decor.angle=45,decor.size=3.5thick,foreground=(1,,0,,0)}{vUpbis}
\fmfv{decor.shape=cross,decor.angle=45,decor.size=3.5thick,foreground=(1,,0,,0)}{vDownbis}
\fmf{phantom,tension=0.8}{vUp,vUpbis}
\fmf{phantom,tension=0.8}{vDown,vDownbis}
\fmf{dashes,tension=0.5,foreground=(0,,0,,1),foreground=(1,,0,,0)}{v3,vUpbis}
\fmf{phantom,tension=0.5}{v4,vUpbis}
\fmf{phantom,tension=0.5}{v3,vDownbis}
\fmf{dashes,tension=0.5,foreground=(0,,0,,1),foreground=(1,,0,,0)}{v4,vDownbis}
\fmf{phantom,tension=11}{i,v1}
\fmf{phantom,tension=11}{v2,o}
\fmf{plain,left,tension=0.5,foreground=(1,,0,,0)}{v1,v2,v1}
\fmf{dots,tension=1.7,foreground=(0,,0,,1)}{v1,v3}
\fmf{plain,foreground=(1,,0,,0)}{v3,v4}
\fmf{dots,tension=1.7,foreground=(0,,0,,1)}{v4,v2}
\end{fmfgraph}
\end{fmffile}
\end{gathered} }$} \hspace{-0.22cm} \left.\rule{0cm}{1.2cm}\right) \\
& + \mathcal{O}\big(\hbar^{3}\big)\;,
\end{split}
\label{eq:2PIEAfinalexpression}
\end{equation}
with
\begin{equation}
\begin{split}
\boldsymbol{G}^{-1}_{\phi;ab}(x,y) \equiv & \ \left.\frac{\delta^{2} S[\vec{\widetilde{\varphi}}]}{\delta \widetilde{\varphi}^{a}(x)\delta \widetilde{\varphi}^{b}(y)}\right|_{\vec{\widetilde{\varphi}}=\vec{\phi}} \\
= & \left(-\nabla_x^2 + m^2 + \frac{\lambda}{6} \vec{\phi}^2(x)\right)\delta_{ab}\delta(x-y) \\
& + \frac{\lambda}{3}\phi_{a}(x)\phi_{b}(x)\delta(x-y) \;,
\end{split}
\label{eq:DefinitionG1PIEAhbarExpansionbis}
\end{equation}
$\mathbb{I}$ denotes the identity with respect to both spacetime and color indices (i.e. $\mathbb{I}_{ab}(x,y)=\delta_{ab}\delta(x-y)$) and Eq.~\eqref{eq:2PIEAfinalexpression} is based on the Feynman rules:
\begin{subequations}
\begin{align}
%\begin{equation}
\begin{gathered}
\begin{fmffile}{Diagrams/1PIEA_G}
\begin{fmfgraph*}(20,20)
\fmfleft{i0,i1,i2,i3}
\fmfright{o0,o1,o2,o3}
\fmflabel{$x, a$}{v1}
\fmflabel{$y, b$}{v2}
\fmf{phantom}{i1,v1}
\fmf{phantom}{i2,v1}
\fmf{plain,tension=0.6,foreground=(1,,0,,0)}{v1,v2}
\fmf{phantom}{v2,o1}
\fmf{phantom}{v2,o2}
\end{fmfgraph*}
\end{fmffile}
\end{gathered} \quad &\rightarrow \boldsymbol{G}_{ab}(x,y)\;,
\label{eq:FeynRuleorig2PIEAG} \\
%\end{equation} 
%\begin{equation}
\begin{gathered}
\begin{fmffile}{Diagrams/1PIEA_V3}
\begin{fmfgraph*}(20,20)
\fmfleft{i0,i1,i2,i3}
\fmfright{o0,o1,o2,o3}
\fmfv{decor.shape=cross,decor.angle=45,decor.size=3.5thick,foreground=(1,,0,,0)}{o2}
\fmf{phantom,tension=2.0}{i1,i1bis}
\fmf{plain,tension=2.0,foreground=(1,,0,,0)}{i1bis,v1}
\fmf{phantom,tension=2.0}{i2,i2bis}
\fmf{plain,tension=2.0,foreground=(1,,0,,0)}{i2bis,v1}
\fmf{dots,label=$x$,tension=0.6,foreground=(0,,0,,1)}{v1,v2}
\fmf{phantom,tension=2.0}{o1bis,o1}
\fmf{plain,tension=2.0,foreground=(1,,0,,0)}{v2,o1bis}
\fmf{phantom,tension=2.0}{o2bis,o2}
\fmf{phantom,tension=2.0,foreground=(1,,0,,0)}{v2,o2bis}
\fmf{dashes,tension=0.0,foreground=(1,,0,,0)}{v2,o2}
\fmflabel{$a$}{i1bis}
\fmflabel{$b$}{i2bis}
\fmflabel{$c$}{o1bis}
\fmflabel{$N$}{o2bis}
\end{fmfgraph*}
\end{fmffile}
\end{gathered} \quad &\rightarrow \lambda\left|\vec{\phi}(x)\right|\delta_{a b}\delta_{c N}\;,
\label{eq:FeynRules2PIEA3legVertexSourceJ0main} \\
%\end{equation}
%\begin{equation}
\begin{gathered}
\begin{fmffile}{Diagrams/1PIEA_V4}
\begin{fmfgraph*}(20,20)
\fmfleft{i0,i1,i2,i3}
\fmfright{o0,o1,o2,o3}
\fmf{phantom,tension=2.0}{i1,i1bis}
\fmf{plain,tension=2.0,foreground=(1,,0,,0)}{i1bis,v1}
\fmf{phantom,tension=2.0}{i2,i2bis}
\fmf{plain,tension=2.0,foreground=(1,,0,,0)}{i2bis,v1}
\fmf{zigzag,label=$x$,tension=0.6,foreground=(0,,0,,1)}{v1,v2}
\fmf{phantom,tension=2.0}{o1bis,o1}
\fmf{plain,tension=2.0,foreground=(1,,0,,0)}{v2,o1bis}
\fmf{phantom,tension=2.0}{o2bis,o2}
\fmf{plain,tension=2.0,foreground=(1,,0,,0)}{v2,o2bis}
\fmflabel{$a$}{i1bis}
\fmflabel{$b$}{i2bis}
\fmflabel{$c$}{o1bis}
\fmflabel{$d$}{o2bis}
\end{fmfgraph*}
\end{fmffile}
\end{gathered} \quad &\rightarrow \lambda\delta_{a b}\delta_{c d}\;,
\label{eq:FeynRuleorig2PIEAV4}
%\end{equation}
\end{align}
\end{subequations}
and we fix our coordinates in color space once again such that a spontaneous breakdown of the $O(N)$ symmetry can only occur in the direction set by $a=N$ (still without any loss of generality), which implies that $\big|\vec{\phi}(x)\big|=\phi_{N}(x)$. The self-consistent equations to solve in the present implementation of SCPT are obtained by extremizing the original 2PI EA:
\begin{equation}
\frac{\delta\Gamma^{(\mathrm{2PI})}\big[\vec{\phi},\boldsymbol{G}\big]}{\delta\vec{\phi}(x)} = \vec{0} \quad \forall x \;,
\label{eq:GapEq2PIEAdphiFiniteDim}
\end{equation}
\begin{equation}
\frac{\delta\Gamma^{(\mathrm{2PI})}\big[\vec{\phi},\boldsymbol{G}\big]}{\delta\boldsymbol{G}(x,y)} = \boldsymbol{0} \quad \forall x,y \;.
\label{eq:GapEq2PIEAdGFiniteDim}
\end{equation}
We have carefully checked that the optimal solutions found from such gap equations in (0+0)\nobreakdash-D always conserve the $O(N)$ symmetry of our model in the tested range $\lambda/4! \in [0,10]$. This is illustrated by Fig.~\ref{fig:2PIEAzeroVsNonzerovev} which shows two different solutions of the system made of Eqs.~\eqref{eq:GapEq2PIEAdphiFiniteDim} and~\eqref{eq:GapEq2PIEAdGFiniteDim} with $\Gamma^{(\mathrm{2PI})}$ considered up to second order in $\hbar$: only one of these two solutions exhibits a spontaneous breakdown of the $O(N)$ symmetry, with a finite 1-point correlation function $\vec{\overline{\phi}}$ in the non-perturbative regime. It is however the solution without SSB (at least over the whole range of values for $\lambda/4!$ shown in Fig.~\ref{fig:2PIEAzeroVsNonzerovev}) that minimizes the gs energy and can thus be coined as physical solution. This illustrates that, for the $O(N)$ model under consideration whose exact solution does not break the $O(N)$ symmetry, we can safely consider the 2PI EA at $\vec{\phi}=\vec{0}$ defined as $\Gamma^{(\mathrm{2PI})}[\boldsymbol{G}]\equiv\Gamma^{(\mathrm{2PI})}\big[\vec{\phi}=\vec{0},\boldsymbol{G}\big]$, without any loss of accuracy for a given truncation order. Such a restriction enables us to reach more readily higher truncation orders of the 2PI EA since it imposes that all diagrams containing vertex~\eqref{eq:FeynRules2PIEA3legVertexSourceJ0main} vanish. We thus obtain in this way:
\begin{equation}
\begin{split}
\Gamma^{(\mathrm{2PI})}[\boldsymbol{G}] = & -\frac{\hbar}{2}\mathrm{STr}\left[\ln\big(\boldsymbol{G}\big)\right] + \frac{\hbar}{2}\mathrm{STr}\left[\boldsymbol{G}^{-1}_{0}\boldsymbol{G}-\mathbb{I}\right] \\
& + \hbar^{2} \left(\rule{0cm}{1.2cm}\right. \hspace{-0.1cm} \frac{1}{24} \hspace{0.08cm} \scalebox{0.95}{${\displaystyle \begin{gathered}
\begin{fmffile}{Diagrams/1PIEA_Hartree}
\begin{fmfgraph}(30,20)
\fmfleft{i}
\fmfright{o}
\fmf{phantom,tension=10}{i,i1}
\fmf{phantom,tension=10}{o,o1}
\fmf{plain,left,tension=0.5,foreground=(1,,0,,0)}{i1,v1,i1}
\fmf{plain,right,tension=0.5,foreground=(1,,0,,0)}{o1,v2,o1}
\fmf{zigzag,foreground=(0,,0,,1)}{v1,v2}
\end{fmfgraph}
\end{fmffile}
\end{gathered}}$}
+\frac{1}{12}\scalebox{0.95}{${\displaystyle\begin{gathered}
\begin{fmffile}{Diagrams/1PIEA_Fock}
\begin{fmfgraph}(15,15)
\fmfleft{i}
\fmfright{o}
\fmf{phantom,tension=11}{i,v1}
\fmf{phantom,tension=11}{v2,o}
\fmf{plain,left,tension=0.4,foreground=(1,,0,,0)}{v1,v2,v1}
\fmf{zigzag,foreground=(0,,0,,1)}{v1,v2}
\end{fmfgraph}
\end{fmffile}
\end{gathered}}$} \hspace{-0.15cm} \left.\rule{0cm}{1.2cm}\right) \\
& - \hbar^{3} \left(\rule{0cm}{1.2cm}\right. \hspace{-0.1cm} \frac{1}{72} \hspace{0.38cm} \begin{gathered}
\begin{fmffile}{Diagrams/2PIEAzerovev_Diag1}
\begin{fmfgraph}(12,12)
\fmfleft{i0,i1}
\fmfright{o0,o1}
\fmftop{v1,vUp,v2}
\fmfbottom{v3,vDown,v4}
\fmf{phantom,tension=20}{i0,v1}
\fmf{phantom,tension=20}{i1,v3}
\fmf{phantom,tension=20}{o0,v2}
\fmf{phantom,tension=20}{o1,v4}
\fmf{plain,left=0.4,tension=0.5,foreground=(1,,0,,0)}{v3,v1}
\fmf{phantom,left=0.1,tension=0.5}{v1,vUp}
\fmf{phantom,left=0.1,tension=0.5}{vUp,v2}
\fmf{plain,left=0.4,tension=0.0,foreground=(1,,0,,0)}{v1,v2}
\fmf{plain,left=0.4,tension=0.5,foreground=(1,,0,,0)}{v2,v4}
\fmf{phantom,left=0.1,tension=0.5}{v4,vDown}
\fmf{phantom,left=0.1,tension=0.5}{vDown,v3}
\fmf{plain,left=0.4,tension=0.0,foreground=(1,,0,,0)}{v4,v3}
\fmf{zigzag,tension=0.5,foreground=(0,,0,,1)}{v1,v4}
\fmf{zigzag,tension=0.5,foreground=(0,,0,,1)}{v2,v3}
\end{fmfgraph}
\end{fmffile}
\end{gathered} \hspace{0.3cm} + \frac{1}{144} \hspace{0.38cm} \begin{gathered}
\begin{fmffile}{Diagrams/2PIEAzerovev_Diag2}
\begin{fmfgraph}(12,12)
\fmfleft{i0,i1}
\fmfright{o0,o1}
\fmftop{v1,vUp,v2}
\fmfbottom{v3,vDown,v4}
\fmf{phantom,tension=20}{i0,v1}
\fmf{phantom,tension=20}{i1,v3}
\fmf{phantom,tension=20}{o0,v2}
\fmf{phantom,tension=20}{o1,v4}
\fmf{plain,left=0.4,tension=0.5,foreground=(1,,0,,0)}{v3,v1}
\fmf{phantom,left=0.1,tension=0.5}{v1,vUp}
\fmf{phantom,left=0.1,tension=0.5}{vUp,v2}
\fmf{zigzag,left=0.4,tension=0.0,foreground=(0,,0,,1)}{v1,v2}
\fmf{plain,left=0.4,tension=0.5,foreground=(1,,0,,0)}{v2,v4}
\fmf{phantom,left=0.1,tension=0.5}{v4,vDown}
\fmf{phantom,left=0.1,tension=0.5}{vDown,v3}
\fmf{zigzag,left=0.4,tension=0.0,foreground=(0,,0,,1)}{v4,v3}
\fmf{plain,left=0.4,tension=0.5,foreground=(1,,0,,0)}{v1,v3}
\fmf{plain,right=0.4,tension=0.5,foreground=(1,,0,,0)}{v2,v4}
\end{fmfgraph}
\end{fmffile}
\end{gathered} \hspace{0.2cm} \left.\rule{0cm}{1.2cm}\right) \\
& + \hbar^{4} \left(\rule{0cm}{1.2cm}\right. \hspace{-0.1cm} \frac{1}{324} \hspace{0.2cm} \begin{gathered}\begin{fmffile}{Diagrams/2PIEAzerovev_Diag3}
\begin{fmfgraph}(16,16)
\fmfleft{i}
\fmfright{o}
\fmftop{vUpLeft,vUp,vUpRight}
\fmfbottom{vDownLeft,vDown,vDownRight}
\fmf{phantom,tension=1}{i,v1}
\fmf{phantom,tension=1}{v2,o}
\fmf{phantom,tension=14.0}{v3,vUpLeft}
\fmf{phantom,tension=2.0}{v3,vUpRight}
\fmf{phantom,tension=4.0}{v3,i}
\fmf{phantom,tension=2.0}{v4,vUpLeft}
\fmf{phantom,tension=14.0}{v4,vUpRight}
\fmf{phantom,tension=4.0}{v4,o}
\fmf{phantom,tension=14.0}{v5,vDownLeft}
\fmf{phantom,tension=2.0}{v5,vDownRight}
\fmf{phantom,tension=4.0}{v5,i}
\fmf{phantom,tension=2.0}{v6,vDownLeft}
\fmf{phantom,tension=14.0}{v6,vDownRight}
\fmf{phantom,tension=4.0}{v6,o}
\fmf{zigzag,tension=0,foreground=(0,,0,,1)}{v1,v2}
\fmf{zigzag,tension=0.6,foreground=(0,,0,,1)}{v3,v6}
\fmf{zigzag,tension=0.6,foreground=(0,,0,,1)}{v5,v4}
\fmf{plain,left=0.18,tension=0,foreground=(1,,0,,0)}{v1,v3}
\fmf{plain,left=0.42,tension=0,foreground=(1,,0,,0)}{v3,v4}
\fmf{plain,left=0.18,tension=0,foreground=(1,,0,,0)}{v4,v2}
\fmf{plain,left=0.18,tension=0,foreground=(1,,0,,0)}{v2,v6}
\fmf{plain,left=0.42,tension=0,foreground=(1,,0,,0)}{v6,v5}
\fmf{plain,left=0.18,tension=0,foreground=(1,,0,,0)}{v5,v1}
\end{fmfgraph}
\end{fmffile}
\end{gathered} \hspace{0.15cm} + \frac{1}{108} \hspace{0.5cm} \begin{gathered}
\begin{fmffile}{Diagrams/2PIEAzerovev_Diag4}
\begin{fmfgraph}(12.5,12.5)
\fmfleft{i0,i1}
\fmfright{o0,o1}
\fmftop{v1,vUp,v2}
\fmfbottom{v3,vDown,v4}
\fmf{phantom,tension=20}{i0,v1}
\fmf{phantom,tension=20}{i1,v3}
\fmf{phantom,tension=20}{o0,v2}
\fmf{phantom,tension=20}{o1,v4}
\fmf{phantom,tension=0.005}{v5,v6}
\fmf{zigzag,left=0.4,tension=0,foreground=(0,,0,,1)}{v3,v1}
\fmf{phantom,left=0.1,tension=0}{v1,vUp}
\fmf{phantom,left=0.1,tension=0}{vUp,v2}
\fmf{plain,left=0.25,tension=0,foreground=(1,,0,,0)}{v1,v2}
\fmf{zigzag,left=0.4,tension=0,foreground=(0,,0,,1)}{v2,v4}
\fmf{phantom,left=0.1,tension=0}{v4,vDown}
\fmf{phantom,left=0.1,tension=0}{vDown,v3}
\fmf{plain,right=0.25,tension=0,foreground=(1,,0,,0)}{v3,v4}
\fmf{plain,left=0.2,tension=0.01,foreground=(1,,0,,0)}{v1,v5}
\fmf{plain,left=0.2,tension=0.01,foreground=(1,,0,,0)}{v5,v3}
\fmf{plain,right=0.2,tension=0.01,foreground=(1,,0,,0)}{v2,v6}
\fmf{plain,right=0.2,tension=0.01,foreground=(1,,0,,0)}{v6,v4}
\fmf{zigzag,tension=0,foreground=(0,,0,,1)}{v5,v6}
\end{fmfgraph}
\end{fmffile}
\end{gathered} \\
& + \frac{1}{324} \hspace{0.4cm} \begin{gathered}
\begin{fmffile}{Diagrams/2PIEAzerovev_Diag5}
\begin{fmfgraph}(12.5,12.5)
\fmfleft{i0,i1}
\fmfright{o0,o1}
\fmftop{v1,vUp,v2}
\fmfbottom{v3,vDown,v4}
\fmf{phantom,tension=20}{i0,v1}
\fmf{phantom,tension=20}{i1,v3}
\fmf{phantom,tension=20}{o0,v2}
\fmf{phantom,tension=20}{o1,v4}
\fmf{phantom,tension=0.005}{v5,v6}
\fmf{plain,left=0.4,tension=0,foreground=(1,,0,,0)}{v3,v1}
\fmf{phantom,left=0.1,tension=0}{v1,vUp}
\fmf{phantom,left=0.1,tension=0}{vUp,v2}
\fmf{zigzag,left=0.25,tension=0,foreground=(0,,0,,1)}{v1,v2}
\fmf{plain,left=0.4,tension=0,foreground=(1,,0,,0)}{v2,v4}
\fmf{phantom,left=0.1,tension=0}{v4,vDown}
\fmf{phantom,left=0.1,tension=0}{vDown,v3}
\fmf{zigzag,right=0.25,tension=0,foreground=(0,,0,,1)}{v3,v4}
\fmf{plain,left=0.2,tension=0.01,foreground=(1,,0,,0)}{v1,v5}
\fmf{plain,left=0.2,tension=0.01,foreground=(1,,0,,0)}{v5,v3}
\fmf{plain,right=0.2,tension=0.01,foreground=(1,,0,,0)}{v2,v6}
\fmf{plain,right=0.2,tension=0.01,foreground=(1,,0,,0)}{v6,v4}
\fmf{zigzag,tension=0,foreground=(0,,0,,1)}{v5,v6}
\end{fmfgraph}
\end{fmffile}
\end{gathered} \hspace{0.35cm} + \frac{1}{216} \hspace{-0.35cm} \begin{gathered}
\begin{fmffile}{Diagrams/2PIEAzerovev_Diag6}
\begin{fmfgraph}(30,15)
\fmfleft{i0,i,i1}
\fmfright{o0,o,o1}
\fmftop{v1b,vUp,v2b}
\fmfbottom{v3b,vDown,v4b}
\fmf{phantom,tension=20}{i0,v1b}
\fmf{phantom,tension=20}{i1,v3b}
\fmf{phantom,tension=20}{o0,v2b}
\fmf{phantom,tension=20}{o1,v4b}
\fmf{phantom,tension=0.511}{i,v7}
\fmf{phantom,tension=0.11}{o,v7}
\fmf{phantom,tension=0.1}{v1,v1b}
\fmf{phantom,tension=0.1}{v2,v2b}
\fmf{phantom,tension=0.1}{v3,v3b}
\fmf{phantom,tension=0.1}{v4,v4b}
\fmf{phantom,tension=0.005}{v5,v6}
\fmf{phantom,left=0.1,tension=0.1}{v1,vUp}
\fmf{phantom,left=0.1,tension=0.1}{vUp,v2}
\fmf{zigzag,left=0.15,tension=0,foreground=(0,,0,,1)}{v1,v2}
\fmf{plain,left=0.4,tension=0,foreground=(1,,0,,0)}{v2,v4}
\fmf{plain,right=0.4,tension=0,foreground=(1,,0,,0)}{v2,v4}
\fmf{phantom,left=0.1,tension=0.1}{v4,vDown}
\fmf{phantom,left=0.1,tension=0.1}{vDown,v3}
\fmf{zigzag,left=0.15,tension=0,foreground=(0,,0,,1)}{v4,v3}
\fmf{plain,left=0.2,tension=0.01,foreground=(1,,0,,0)}{v1,v5}
\fmf{plain,left=0.2,tension=0.01,foreground=(1,,0,,0)}{v5,v3}
\fmf{plain,right=0.2,tension=0,foreground=(1,,0,,0)}{v1,v7}
\fmf{plain,right=0.2,tension=0,foreground=(1,,0,,0)}{v7,v3}
\fmf{phantom,right=0.2,tension=0.01}{v2,v6}
\fmf{phantom,right=0.2,tension=0.01}{v6,v4}
\fmf{zigzag,tension=0,foreground=(0,,0,,1)}{v5,v7}
\end{fmfgraph}
\end{fmffile}
\end{gathered} \\
& + \frac{1}{1296} \hspace{-0.32cm} \begin{gathered}
\begin{fmffile}{Diagrams/2PIEAzerovev_Diag7}
\begin{fmfgraph}(30,14)
\fmfleft{i0,i,i1}
\fmfright{o0,o,o1}
\fmftop{v1b,vUp,v2b}
\fmfbottom{v3b,vDown,v4b}
\fmf{phantom,tension=5}{vUp,v5}
\fmf{phantom,tension=1}{v1b,v5}
\fmf{phantom,tension=5}{vUp,v6}
\fmf{phantom,tension=1}{v2b,v6}
\fmf{phantom,tension=20}{i0,v1b}
\fmf{phantom,tension=20}{i1,v3b}
\fmf{phantom,tension=20}{o0,v2b}
\fmf{phantom,tension=20}{o1,v4b}
\fmf{phantom,tension=0.1}{v1,v1b}
\fmf{phantom,tension=0.1}{v2,v2b}
\fmf{phantom,tension=0.1}{v3,v3b}
\fmf{phantom,tension=0.1}{v4,v4b}
\fmf{phantom,tension=0.005}{v5,v6}
\fmf{phantom,left=0.1,tension=0.1}{v1,vUp}
\fmf{phantom,left=0.1,tension=0.1}{vUp,v2}
\fmf{plain,left=0.4,tension=0.005,foreground=(1,,0,,0)}{v2,v4}
\fmf{plain,right=0.4,tension=0.005,foreground=(1,,0,,0)}{v2,v4}
\fmf{plain,left=0.4,tension=0.005,foreground=(1,,0,,0)}{v1,v3}
\fmf{plain,right=0.4,tension=0.005,foreground=(1,,0,,0)}{v1,v3}
\fmf{phantom,left=0.1,tension=0.1}{v4,vDown}
\fmf{phantom,left=0.1,tension=0.1}{vDown,v3}
\fmf{zigzag,left=0.05,tension=0,foreground=(0,,0,,1)}{v1,v5}
\fmf{plain,left,tension=0,foreground=(1,,0,,0)}{v5,v6}
\fmf{plain,right,tension=0,foreground=(1,,0,,0)}{v5,v6}
\fmf{zigzag,left=0.05,tension=0,foreground=(0,,0,,1)}{v6,v2}
\fmf{zigzag,left=0.15,tension=0,foreground=(0,,0,,1)}{v4,v3}
\end{fmfgraph}
\end{fmffile}
\end{gathered} \hspace{-0.5cm} \left.\rule{0cm}{1.2cm}\right) \\
& + \mathcal{O}\big(\hbar^{5}\big)\;,
\end{split}
\label{eq:2PIEAzerovevfinalexpression}
\end{equation}
with
\begin{equation}
\boldsymbol{G}^{-1}_{0;ab}(x,y) = \left(-\nabla_x^2 + m^2 \right)\delta_{ab}\delta(x-y) \;.
\label{eq:1PIEAlambdaExppDefG0}
\end{equation}
Requiring that $\vec{\phi}=\vec{0}$ also imposes that the structure of the propagator vanishes in color space, i.e. $\boldsymbol{G}_{a b}= G \ \delta_{a b}$ $\forall a,b$. Hence, Eq.~\eqref{eq:2PIEAzerovevfinalexpression} reduces in (0+0)\nobreakdash-D to the simple expression:
\begin{equation}
\begin{split}
\Gamma^{(\mathrm{2PI})}(\boldsymbol{G}) = & \ \hbar\left(-\frac{N}{2}\ln(2\pi G) + \frac{N}{2}\left(m^{2} G - 1\right)\right) \\
& + \hbar^{2} \left(\frac{N^{2}+2N}{24}\lambda G^{2}\right) \\
& -\hbar^{3}\left(\frac{N^{2}+2N}{144}\lambda^{2}G^{4}\right) \\
& + \hbar^{4} \left(\frac{N^{3}+10N^{2}+16N}{1296}\lambda^{3}G^{6}\right) \\
& + \mathcal{O}\big(\hbar^{5}\big)\;,
\end{split}
\label{eq:2PIEAzerovevfinalexpression0DON}
\end{equation}
and the corresponding gap equation reads:
\begin{equation}
\begin{split}
0 = & \left.\frac{\partial\Gamma^{(\mathrm{2PI})}(\boldsymbol{G})}{\partial G}\right|_{\boldsymbol{G}=\overline{\boldsymbol{G}}} \\
= & \ \hbar\left(-\frac{N}{2}\overline{G}^{-1} + \frac{N}{2}m^{2}\right) + \hbar^{2} \left(\frac{N^{2}+2N}{12}\lambda \overline{G}\right) \\
& -\hbar^{3}\left(\frac{N^{2}+2N}{36}\lambda^{2}\overline{G}^{3}\right) \\
& + \hbar^{4} \left(\frac{N^{3}+10N^{2}+16N}{216}\lambda^{3}\overline{G}^{5}\right) \\
& + \mathcal{O}\big(\hbar^{5}\big)\;,
\end{split}
\label{eq:2PIEAzerovevgapequation0DON}
\end{equation}
with $\overline{\boldsymbol{G}}_{a b}=\overline{G} \ \delta_{a b}$ $\forall a,b$. Finally, the gs energy and density are obtained from the solution $\overline{G}$ via the relations:
\begin{equation}
E^\text{2PI EA;orig;$\vec{\phi}=\vec{0}$}_\text{gs} = \frac{1}{\hbar}\Gamma^{(\mathrm{2PI})}\big(\boldsymbol{G}=\overline{\boldsymbol{G}}\big) \;,
\label{eq:2PIorigE}
\end{equation}
\begin{equation}
\rho^\text{2PI EA;orig;$\vec{\phi}=\vec{0}$}_\text{gs} = \frac{\hbar}{N} \mathrm{Tr}_{a}\big(\overline{\boldsymbol{G}}\big) = \hbar\overline{G} \;.
\label{eq:2PIorigRho}
\end{equation}

%%%%%%%%%%%%%%%%%%%
%%%%%%%%%%%%%%%%%%%

%%%%%%%%%%%%%%%%%%%
\begin{figure}[!t]
  \begin{center}
      \includegraphics[width=0.95\linewidth]{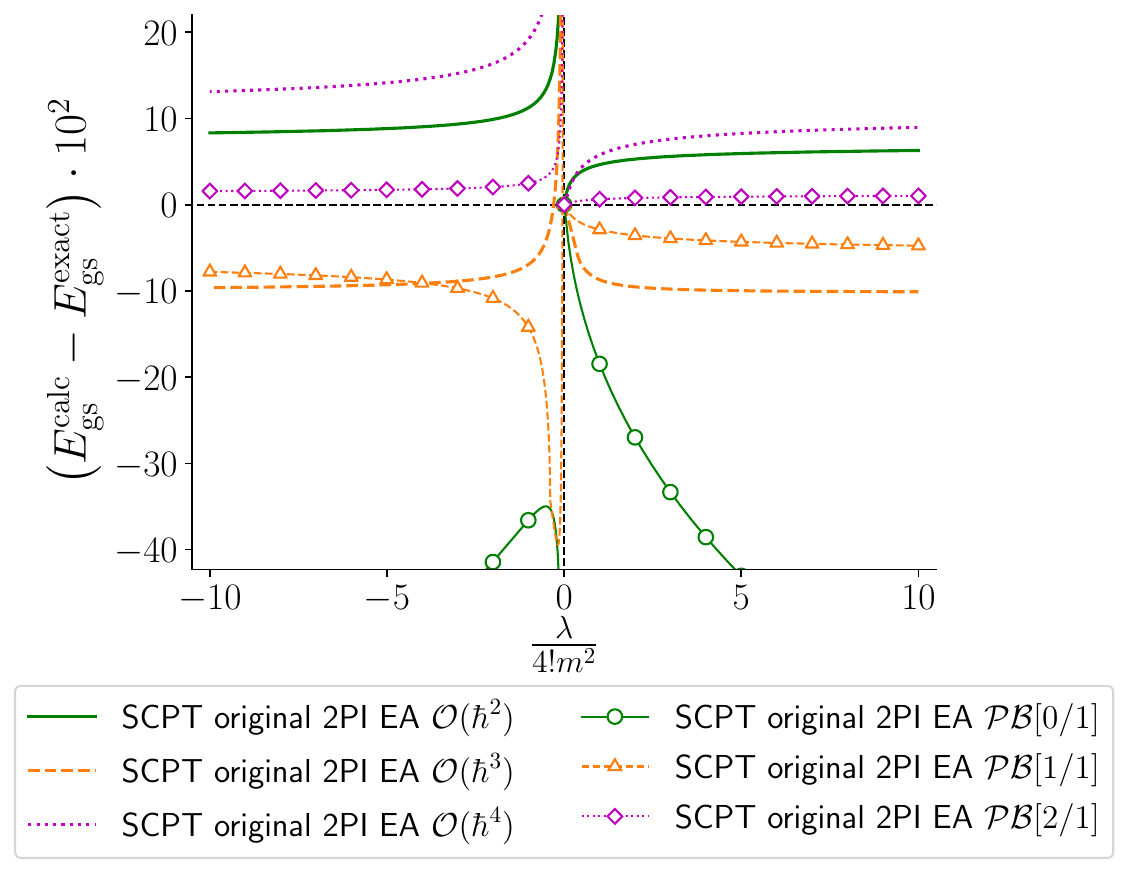}
      \includegraphics[width=0.95\linewidth]{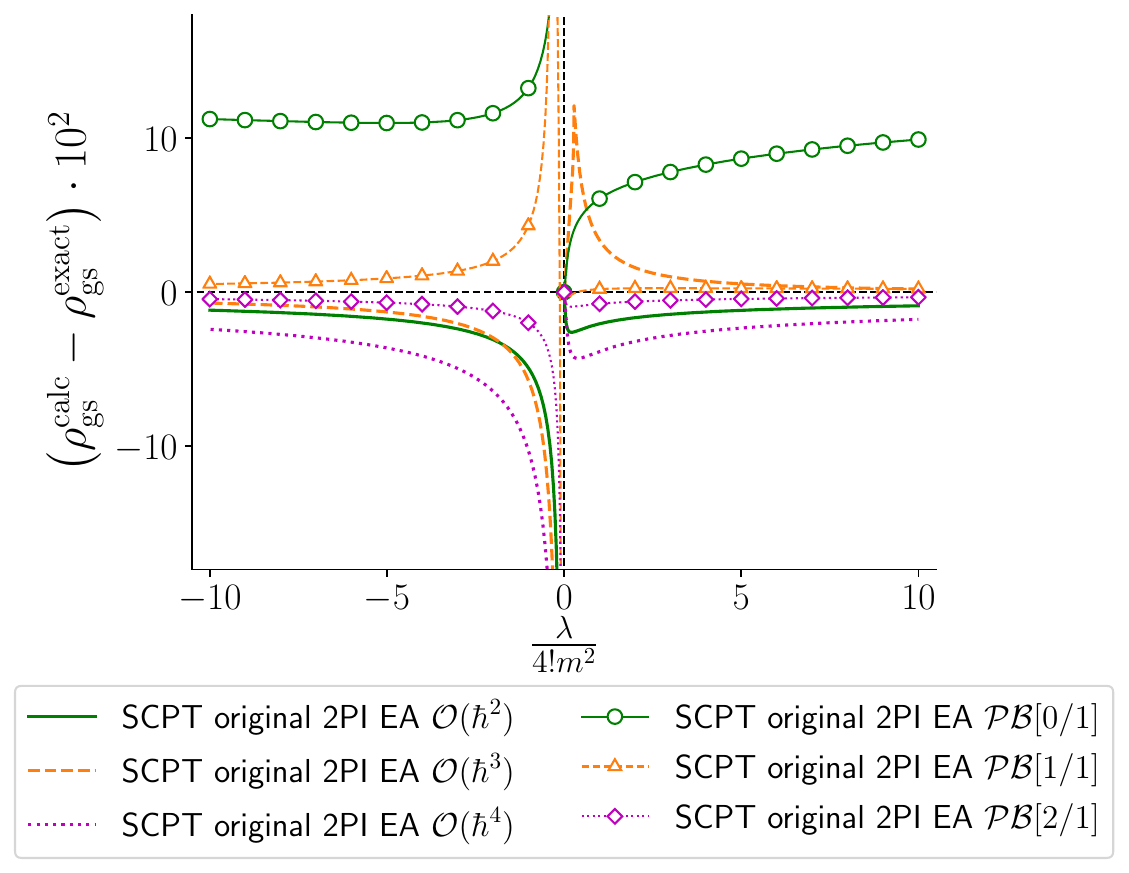}
    \caption{Difference between the calculated gs energy $E_{\mathrm{gs}}^{\mathrm{calc}}$ or density $\rho_{\mathrm{gs}}^{\mathrm{calc}}$ and the corresponding exact solution $E_{\mathrm{gs}}^{\mathrm{exact}}$ or $\rho_{\mathrm{gs}}^{\mathrm{exact}}$ at $\hbar=1$, $m^{2}=\pm 1$ and $N=2$. See also the caption of Fig.~\ref{fig:2PIEAzeroVsNonzerovev} for the meaning of the indication ``$\mathcal{O}\big(\hbar^{n}\big)$'' for SCPT results.}
    \label{fig:2PIEAorigN2}
  \end{center}
\end{figure}
%%%%%%%%%%%%%%%%%%%

The results thus obtained from $\Gamma^{(\mathrm{2PI})}(\boldsymbol{G})$ are displayed in Fig.~\ref{fig:2PIEAorigN2}. The first non-trivial order, which is implemented by truncating $\Gamma^{(\mathrm{2PI})}(\boldsymbol{G})$ right beyond order $\mathcal{O}\big(\hbar^{2}\big)$, coincides with the standard Hartree-Fock result as it can be shown that the gap equation~\eqref{eq:2PIEAzerovevgapequation0DON} is equivalent to a Dyson equation with Hartree-Fock self-energy if all terms of order $\mathcal{O}\big(\hbar^{3}\big)$ or higher are ignored. According to Fig.~\ref{fig:2PIEAorigN2}, this truncation is barely affected as the coupling constant $\lambda$ increases in the regime set by $\lambda/4! \gtrsim 1$ (for both signs of $m^{2}$) and achieves an accuracy of about $10\%$ for both $E_{\mathrm{gs}}$ and $\rho_{\mathrm{gs}}$ at $N=2$ in this situation. We have illustrated in this way the non-perturbative character of the Hartree-Fock theory and we can thus see that the present 2PI EA approach is designed to improve this Hartree-Fock result in a systematic fashion. However, if we consider the solutions of the gap equation~\eqref{eq:2PIEAzerovevgapequation0DON} at the next two orders in $\hbar$, we actually observe the reverse: in almost the entire interval $\lambda/4! \in [0,10]$ for both signs of $m^{2}$, the resulting estimates of $E_{\mathrm{gs}}$ and $\rho_{\mathrm{gs}}$ worsen as the truncation order with respect to $\hbar$ increases. Note that the series~\eqref{eq:2PIEAzerovevfinalexpression0DON} representing $\Gamma^{(\mathrm{2PI})}(\boldsymbol{G})$ is also asymptotic and divergent, even after setting $\boldsymbol{G}=\overline{\boldsymbol{G}}$.

We therefore exploit a resummation procedure, and more specifically the Pad\'{e}-Borel resummation scheme, to illustrate the expected improvement with respect to the Hartree-Fock result. It amounts to modifying the underlying procedure as follows: the expression of $\Gamma^{(\mathrm{2PI})}(\boldsymbol{G})$ (given by Eq.~\eqref{eq:2PIEAzerovevfinalexpression0DON}) truncated at the chosen order with respect to $\hbar$ is replaced by a given Pad\'{e} approximant to subsequently derive the gap equations. In this way, the solution $\overline{\boldsymbol{G}}$ is systematically improved via resummation. As a next step, the gs energy and density are still inferred from Eqs.~\eqref{eq:2PIorigE} and~\eqref{eq:2PIorigRho}, with one additional peculiarity for $E_{\mathrm{gs}}$: $\Gamma^{(\mathrm{2PI})}\big(\boldsymbol{G}=\overline{\boldsymbol{G}}\big)$ is rewritten in Eq.~\eqref{eq:2PIorigE} with the Pad\'{e}-Borel resummation procedure. We refer to this entire procedure as Pad\'{e}-Borel resummation of the EA, even though there is no Borel transform involved in the determination of $\rho_{\mathrm{gs}}$. The implementation of the Borel-hypergeometric resummation is not that straightforward for SCPT since we do not have analytical formulae to rewrite derivatives of Meijer G-functions with respect to each of their entries \cite{fic80}. Considering the good performances of this resummation procedure at the level of the LE, we can definitely expect it to be relevant in the framework of SCPT as well, but we postpone such an investigation to future works.

Regarding the numerical results thus obtained with the Pad\'{e}-Borel resummation, we can indeed see in Fig.~\ref{fig:2PIEAorigN2}, which shows the results obtained from the best Pad\'{e} approximants at each of the three first non-trivial orders of $\Gamma^{(\mathrm{2PI})}(\boldsymbol{G})$, that a $[2/1]$ Pad\'{e} approximant reaches an accuracy of $1\%$ for both $E_{\mathrm{gs}}$ and $\rho_{\mathrm{gs}}$ for $\lambda/4!\in[0,10]$, which is to be compared with the $10\%$ of the Hartree-Fock result. Note however that the best Pad\'{e} approximants at the first two non-trivial orders, i.e. the $[0/1]$ and $[1/1]$ approximants, do not manage to clearly improve the corresponding bare results (labeled respectively ``SCPT original 2PI EA $\mathcal{O}(\hbar^{2})$'' and ``SCPT original 2PI EA $\mathcal{O}(\hbar^{3})$'' in Fig.~\ref{fig:2PIEAorigN2}) for all values of the coupling constant $\lambda$ in both the unbroken- and broken-symmetry phases. There are actually very few studies \cite{bro15} investigating SCPT in combination with resummation theory. However, the latter remains a key aspect of the present approach as it is the resummation that enables us to turn this EA technique into a systematically improvable approach. We will therefore not content ourselves with the present resummation analysis and perform similar applications to what will turn out to be the most performing SCPT method of this study, which relies on the 2PI EA in the mixed representation, i.e. the mixed 2PI EA.

%~~~
\subsubsection{\label{sec:Mixed2PIEA}Mixed 2PI effective action}

Although mixed EAs of $O(N)$ models were pioneered by the work of Coleman, Jackiw and Politzer \cite{col74}, the developments of their 2PI versions via SCPT were carried out by Cooper and collaborators \cite{ben77,cho99,mih01,bla01,coo03,coo05} and later by Aarts \textit{et al.} \cite{aar02} for instance. Note also some applications of this formalism to the study of chiral symmetry restoration \cite{see12}. As for the collective LE discussed previously, the present study is to our knowledge the first pushing SCPT based on the mixed 2PI EA up to its third non-trivial order (i.e. up to order $\mathcal{O}(\hbar^{4})$ for the expansion of the mixed 2PI EA\footnote{To clarify, the present mixed 2PI EA approach has to our knowledge never been pushed up to its third non-trivial order \emph{regardless of the chosen expansion parameter}, even in the framework of the $1/N$-expansion which is often considered for 2PI EA studies notably \cite{mih01,aar02,bla01,coo03}.}) and to combine it with a resummation procedure. Furthermore, a well-known implementation of SCPT based on the mixed 2PI EA is the bare vertex approximation (BVA) \cite{mih01,bla01,coo03}, which is equivalent to the first non-trivial order for our expansion in terms of $\hbar$. The BVA was notably shown to be successful in the framework of QFTs at finite temperature. For an $O(N)$-symmetric $\varphi^4$-theory in (1+1)\nobreakdash-D for instance, the problematic absence of thermalization found in the framework of the Hartree approximation is cured by the BVA \cite{coo03}.

\pagebreak

The definition of the mixed 2PI EA can be inferred from that of the original 2PI EA by replacing correlation functions and sources by their supercounterparts, thus leading to:
\begin{equation}
\begin{split}
\Gamma_{\mathrm{mix}}^{(\mathrm{2PI})}\big[\Phi,\mathcal{G}\big] \equiv & -W_{\mathrm{mix}}\big[\mathcal{J},\mathcal{K}\big] + \int_{x}\mathcal{J}^{\alpha}(x) \frac{\delta W_{\mathrm{mix}}\big[\mathcal{J},\mathcal{K}\big]}{\delta \mathcal{J}^{\alpha}(x)} \\
& + \int_{x,y}\mathcal{K}^{\alpha\beta}(x,y) \frac{\delta W_{\mathrm{mix}}\big[\mathcal{J},\mathcal{K}\big]}{\delta \mathcal{K}^{\beta\alpha}(x,y)} \\
= & -W_{\mathrm{mix}}\big[\mathcal{J},\mathcal{K}\big] + \int_{x}\mathcal{J}^{\alpha}(x) \Phi_{\alpha}(x) \\
& + \frac{1}{2} \int_{x,y} \Phi_{\alpha}(x) \mathcal{K}^{\alpha\beta}(x,y) \Phi_{\beta}(y) \\
& + \frac{\hbar}{2} \int_{x,y} \mathcal{K}^{\alpha\beta}(x,y) \mathcal{G}_{\beta\alpha}(y,x) \;,
\end{split}
\label{eq:mixed2PIEAdefinition0DONmain}
\end{equation}
with
\begin{equation}
\Phi_{\alpha}(x) = \frac{\delta W_{\mathrm{mix}}\big[\mathcal{J},\mathcal{K}\big]}{\delta \mathcal{J}^{\alpha}(x)} \;,
\label{eq:mixed2PIEAdefinitionbis0DONmain}
\end{equation}
\begin{equation}
\begin{split}
\mathcal{G}_{\alpha\beta}(x,y) = & \ \frac{\delta^{2} W_{\mathrm{mix}}\big[\mathcal{J},\mathcal{K}\big]}{\delta \mathcal{J}^{\alpha}(x)\delta \mathcal{J}^{\beta}(y)} \\
= & \ \frac{2}{\hbar} \frac{\delta W_{\mathrm{mix}}\big[\mathcal{J},\mathcal{K}\big]}{\delta\mathcal{K}^{\alpha\beta}(x,y)} - \frac{1}{\hbar} \Phi_{\alpha}(x) \Phi_{\beta}(y) \;,
\end{split}
\label{eq:mixed2PIEAdefinitionbis20DONmain}
\end{equation}
or, to further specify our supernotations (still involving the 1-point correlation functions $\vec{\phi}(x) = \left\langle\vec{\widetilde{\varphi}}(x)\right\rangle$ and $\eta(x)=\left\langle\widetilde{\sigma}(x)\right\rangle$),
\begin{equation}
\Phi = \begin{pmatrix}
\vec{\phi} \\
\eta
\end{pmatrix} \;,
\end{equation}

\pagebreak

\begin{equation}
\mathcal{G} = \begin{pmatrix}
\boldsymbol{G} & \vec{F} \\
\vec{F}^{\mathrm{T}} & D
\end{pmatrix} \;,
\label{eq:mixed2PIEADefSuperpropagator}
\end{equation}
and the Schwinger functional $W_{\mathrm{mix}}\big[\mathcal{J},\mathcal{K}\big]$ is notably expressed by Eq.~\eqref{eq:Zmix}. For the $O(N)$ model under consideration, the mixed 2PI EA can also be written as:
\begin{equation}
\begin{split}
& \Gamma^{(\mathrm{2PI})}_{\mathrm{mix}}\big[\Phi,\mathcal{G}\big] \\
& = S_{\mathrm{mix}}[\Phi] -\frac{\hbar}{2}\mathcal{ST}r\left[\ln\big(\mathcal{G}\big)\right] + \frac{\hbar}{2}\mathcal{ST}r\left[\mathcal{G}^{-1}_{\Phi}\mathcal{G}-\mathfrak{I}\right] \\
& \hspace{0.3cm} + \hbar^{2} \left(\rule{0cm}{1.2cm}\right. \frac{1}{12} \hspace{0.1cm} \begin{gathered}
\begin{fmffile}{Diagrams/Mixed2PIEA_Fock}
\begin{fmfgraph}(15,15)
\fmfleft{i}
\fmfright{o}
\fmfv{decor.shape=circle,decor.size=2.0thick,foreground=(0,,0,,1)}{v1}
\fmfv{decor.shape=circle,decor.size=2.0thick,foreground=(0,,0,,1)}{v2}
\fmf{phantom,tension=11}{i,v1}
\fmf{phantom,tension=11}{v2,o}
\fmf{plain,left,tension=0.4,foreground=(1,,0,,0)}{v1,v2,v1}
\fmf{wiggly,foreground=(1,,0,,0)}{v1,v2}
\end{fmfgraph}
\end{fmffile}
\end{gathered} + \frac{1}{6} \hspace{0.1cm} \begin{gathered}
\begin{fmffile}{Diagrams/Mixed2PIEA_Diag1}
\begin{fmfgraph}(15,15)
\fmfleft{i}
\fmfright{o}
\fmfv{decor.shape=circle,decor.size=2.0thick,foreground=(0,,0,,1)}{v1}
\fmfv{decor.shape=circle,decor.size=2.0thick,foreground=(0,,0,,1)}{v2}
\fmf{phantom,tension=11}{i,v1}
\fmf{phantom,tension=11}{v2,o}
\fmf{dashes,left,tension=0.4,foreground=(1,,0,,0)}{v1,v2,v1}
\fmf{plain,foreground=(1,,0,,0)}{v1,v2}
\end{fmfgraph}
\end{fmffile}
\end{gathered} \left.\rule{0cm}{1.2cm}\right) \\
& \hspace{0.3cm} - \hbar^{3} \left(\rule{0cm}{1.2cm}\right. \frac{1}{72} \hspace{0.35cm} \begin{gathered}
\begin{fmffile}{Diagrams/Mixed2PIEA_Diag2}
\begin{fmfgraph}(10,10)
\fmfleft{i0,i1}
\fmfright{o0,o1}
\fmftop{v1,vUp,v2}
\fmfbottom{v3,vDown,v4}
\fmfv{decor.shape=circle,decor.size=2.0thick,foreground=(0,,0,,1)}{v1}
\fmfv{decor.shape=circle,decor.size=2.0thick,foreground=(0,,0,,1)}{v2}
\fmfv{decor.shape=circle,decor.size=2.0thick,foreground=(0,,0,,1)}{v3}
\fmfv{decor.shape=circle,decor.size=2.0thick,foreground=(0,,0,,1)}{v4}
\fmf{phantom,tension=20}{i0,v1}
\fmf{phantom,tension=20}{i1,v3}
\fmf{phantom,tension=20}{o0,v2}
\fmf{phantom,tension=20}{o1,v4}
\fmf{plain,left=0.4,tension=0.5,foreground=(1,,0,,0)}{v3,v1}
\fmf{phantom,left=0.1,tension=0.5}{v1,vUp}
\fmf{phantom,left=0.1,tension=0.5}{vUp,v2}
\fmf{plain,left=0.4,tension=0.0,foreground=(1,,0,,0)}{v1,v2}
\fmf{plain,left=0.4,tension=0.5,foreground=(1,,0,,0)}{v2,v4}
\fmf{phantom,left=0.1,tension=0.5}{v4,vDown}
\fmf{phantom,left=0.1,tension=0.5}{vDown,v3}
\fmf{plain,left=0.4,tension=0.0,foreground=(1,,0,,0)}{v4,v3}
\fmf{wiggly,tension=0.5,foreground=(1,,0,,0)}{v1,v4}
\fmf{wiggly,tension=0.5,foreground=(1,,0,,0)}{v2,v3}
\end{fmfgraph}
\end{fmffile}
\end{gathered} \hspace{0.35cm} + \frac{1}{36} \hspace{0.35cm} \begin{gathered}
\begin{fmffile}{Diagrams/Mixed2PIEA_Diag3}
\begin{fmfgraph}(10,10)
\fmfleft{i0,i1}
\fmfright{o0,o1}
\fmftop{v1,vUp,v2}
\fmfbottom{v3,vDown,v4}
\fmfv{decor.shape=circle,decor.size=2.0thick,foreground=(0,,0,,1)}{v1}
\fmfv{decor.shape=circle,decor.size=2.0thick,foreground=(0,,0,,1)}{v2}
\fmfv{decor.shape=circle,decor.size=2.0thick,foreground=(0,,0,,1)}{v3}
\fmfv{decor.shape=circle,decor.size=2.0thick,foreground=(0,,0,,1)}{v4}
\fmf{phantom,tension=20}{i0,v1}
\fmf{phantom,tension=20}{i1,v3}
\fmf{phantom,tension=20}{o0,v2}
\fmf{phantom,tension=20}{o1,v4}
\fmf{dashes,left=0.4,tension=0.5,foreground=(1,,0,,0)}{v3,v1}
\fmf{phantom,left=0.1,tension=0.5}{v1,vUp}
\fmf{phantom,left=0.1,tension=0.5}{vUp,v2}
\fmf{dashes,left=0.4,tension=0.0,foreground=(1,,0,,0)}{v1,v2}
\fmf{dashes,left=0.4,tension=0.5,foreground=(1,,0,,0)}{v2,v4}
\fmf{phantom,left=0.1,tension=0.5}{v4,vDown}
\fmf{phantom,left=0.1,tension=0.5}{vDown,v3}
\fmf{dashes,left=0.4,tension=0.0,foreground=(1,,0,,0)}{v4,v3}
\fmf{plain,tension=0.5,foreground=(1,,0,,0)}{v1,v4}
\fmf{plain,tension=0.5,foreground=(1,,0,,0)}{v2,v3}
\end{fmfgraph}
\end{fmffile}
\end{gathered} \\
& \hspace{0.3cm} + \frac{1}{18} \hspace{0.35cm} \begin{gathered}
\begin{fmffile}{Diagrams/Mixed2PIEA_Diag4}
\begin{fmfgraph}(10,10)
\fmfleft{i0,i1}
\fmfright{o0,o1}
\fmftop{v1,vUp,v2}
\fmfbottom{v3,vDown,v4}
\fmfv{decor.shape=circle,decor.size=2.0thick,foreground=(0,,0,,1)}{v1}
\fmfv{decor.shape=circle,decor.size=2.0thick,foreground=(0,,0,,1)}{v2}
\fmfv{decor.shape=circle,decor.size=2.0thick,foreground=(0,,0,,1)}{v3}
\fmfv{decor.shape=circle,decor.size=2.0thick,foreground=(0,,0,,1)}{v4}
\fmf{phantom,tension=20}{i0,v1}
\fmf{phantom,tension=20}{i1,v3}
\fmf{phantom,tension=20}{o0,v2}
\fmf{phantom,tension=20}{o1,v4}
\fmf{plain,left=0.4,tension=0.5,foreground=(1,,0,,0)}{v3,v1}
\fmf{phantom,left=0.1,tension=0.5}{v1,vUp}
\fmf{phantom,left=0.1,tension=0.5}{vUp,v2}
\fmf{plain,left=0.4,tension=0.0,foreground=(1,,0,,0)}{v1,v2}
\fmf{dashes,left=0.4,tension=0.5,foreground=(1,,0,,0)}{v2,v4}
\fmf{phantom,left=0.1,tension=0.5}{v4,vDown}
\fmf{phantom,left=0.1,tension=0.5}{vDown,v3}
\fmf{dashes,left=0.4,tension=0.0,foreground=(1,,0,,0)}{v4,v3}
\fmf{wiggly,tension=0.5,foreground=(1,,0,,0)}{v1,v4}
\fmf{plain,tension=0.5,foreground=(1,,0,,0)}{v2,v3}
\end{fmfgraph}
\end{fmffile}
\end{gathered} \hspace{0.35cm} + \frac{1}{9} \hspace{0.35cm} \begin{gathered}
\begin{fmffile}{Diagrams/Mixed2PIEA_Diag5}
\begin{fmfgraph}(10,10)
\fmfleft{i0,i1}
\fmfright{o0,o1}
\fmftop{v1,vUp,v2}
\fmfbottom{v3,vDown,v4}
\fmfv{decor.shape=circle,decor.size=2.0thick,foreground=(0,,0,,1)}{v1}
\fmfv{decor.shape=circle,decor.size=2.0thick,foreground=(0,,0,,1)}{v2}
\fmfv{decor.shape=circle,decor.size=2.0thick,foreground=(0,,0,,1)}{v3}
\fmfv{decor.shape=circle,decor.size=2.0thick,foreground=(0,,0,,1)}{v4}
\fmf{phantom,tension=20}{i0,v1}
\fmf{phantom,tension=20}{i1,v3}
\fmf{phantom,tension=20}{o0,v2}
\fmf{phantom,tension=20}{o1,v4}
\fmf{dashes,left=0.4,tension=0.5,foreground=(1,,0,,0)}{v3,v1}
\fmf{phantom,left=0.1,tension=0.5}{v1,vUp}
\fmf{phantom,left=0.1,tension=0.5}{vUp,v2}
\fmf{plain,left=0.4,tension=0.0,foreground=(1,,0,,0)}{v1,v2}
\fmf{dashes,left=0.4,tension=0.5,foreground=(1,,0,,0)}{v2,v4}
\fmf{phantom,left=0.1,tension=0.5}{v4,vDown}
\fmf{phantom,left=0.1,tension=0.5}{vDown,v3}
\fmf{plain,left=0.4,tension=0.0,foreground=(1,,0,,0)}{v4,v3}
\fmf{wiggly,tension=0.5,foreground=(1,,0,,0)}{v1,v4}
\fmf{plain,tension=0.5,foreground=(1,,0,,0)}{v2,v3}
\end{fmfgraph}
\end{fmffile}
\end{gathered} \hspace{0.35cm} + \frac{1}{9} \hspace{0.35cm} \begin{gathered}
\begin{fmffile}{Diagrams/Mixed2PIEA_Diag6}
\begin{fmfgraph}(10,10)
\fmfleft{i0,i1}
\fmfright{o0,o1}
\fmftop{v1,vUp,v2}
\fmfbottom{v3,vDown,v4}
\fmfv{decor.shape=circle,decor.size=2.0thick,foreground=(0,,0,,1)}{v1}
\fmfv{decor.shape=circle,decor.size=2.0thick,foreground=(0,,0,,1)}{v2}
\fmfv{decor.shape=circle,decor.size=2.0thick,foreground=(0,,0,,1)}{v3}
\fmfv{decor.shape=circle,decor.size=2.0thick,foreground=(0,,0,,1)}{v4}
\fmf{phantom,tension=20}{i0,v1}
\fmf{phantom,tension=20}{i1,v3}
\fmf{phantom,tension=20}{o0,v2}
\fmf{phantom,tension=20}{o1,v4}
\fmf{plain,left=0.4,tension=0.5,foreground=(1,,0,,0)}{v3,v1}
\fmf{phantom,left=0.1,tension=0.5}{v1,vUp}
\fmf{phantom,left=0.1,tension=0.5}{vUp,v2}
\fmf{plain,left=0.4,tension=0.0,foreground=(1,,0,,0)}{v1,v2}
\fmf{dashes,left=0.4,tension=0.5,foreground=(1,,0,,0)}{v2,v4}
\fmf{phantom,left=0.1,tension=0.5}{v4,vDown}
\fmf{phantom,left=0.1,tension=0.5}{vDown,v3}
\fmf{dashes,left=0.4,tension=0.0,foreground=(1,,0,,0)}{v4,v3}
\fmf{dashes,tension=0.5,foreground=(1,,0,,0)}{v1,v4}
\fmf{dashes,tension=0.5,foreground=(1,,0,,0)}{v2,v3}
\end{fmfgraph}
\end{fmffile}
\end{gathered} \hspace{0.35cm} \left.\rule{0cm}{1.2cm}\right) \\
& \hspace{0.3cm} + \hbar^{4} \left(\rule{0cm}{1.2cm}\right. \frac{1}{324} \hspace{0.3cm} \begin{gathered}
\begin{fmffile}{Diagrams/Mixed2PIEA_Diag7}
\begin{fmfgraph}(15,15)
\fmfleft{i}
\fmfright{o}
\fmftop{vUpLeft,vUp,vUpRight}
\fmfbottom{vDownLeft,vDown,vDownRight}
\fmfv{decor.shape=circle,decor.size=2.0thick,foreground=(0,,0,,1)}{v1}
\fmfv{decor.shape=circle,decor.size=2.0thick,foreground=(0,,0,,1)}{v2}
\fmfv{decor.shape=circle,decor.size=2.0thick,foreground=(0,,0,,1)}{v3}
\fmfv{decor.shape=circle,decor.size=2.0thick,foreground=(0,,0,,1)}{v4}
\fmfv{decor.shape=circle,decor.size=2.0thick,foreground=(0,,0,,1)}{v5}
\fmfv{decor.shape=circle,decor.size=2.0thick,foreground=(0,,0,,1)}{v6}
\fmf{phantom,tension=1}{i,v1}
\fmf{phantom,tension=1}{v2,o}
\fmf{phantom,tension=14.0}{v3,vUpLeft}
\fmf{phantom,tension=2.0}{v3,vUpRight}
\fmf{phantom,tension=4.0}{v3,i}
\fmf{phantom,tension=2.0}{v4,vUpLeft}
\fmf{phantom,tension=14.0}{v4,vUpRight}
\fmf{phantom,tension=4.0}{v4,o}
\fmf{phantom,tension=14.0}{v5,vDownLeft}
\fmf{phantom,tension=2.0}{v5,vDownRight}
\fmf{phantom,tension=4.0}{v5,i}
\fmf{phantom,tension=2.0}{v6,vDownLeft}
\fmf{phantom,tension=14.0}{v6,vDownRight}
\fmf{phantom,tension=4.0}{v6,o}
\fmf{wiggly,tension=0,foreground=(1,,0,,0)}{v1,v2}
\fmf{wiggly,tension=0.6,foreground=(1,,0,,0)}{v3,v6}
\fmf{wiggly,tension=0.6,foreground=(1,,0,,0)}{v5,v4}
\fmf{plain,left=0.18,tension=0,foreground=(1,,0,,0)}{v1,v3}
\fmf{plain,left=0.42,tension=0,foreground=(1,,0,,0)}{v3,v4}
\fmf{plain,left=0.18,tension=0,foreground=(1,,0,,0)}{v4,v2}
\fmf{plain,left=0.18,tension=0,foreground=(1,,0,,0)}{v2,v6}
\fmf{plain,left=0.42,tension=0,foreground=(1,,0,,0)}{v6,v5}
\fmf{plain,left=0.18,tension=0,foreground=(1,,0,,0)}{v5,v1}
\end{fmfgraph}
\end{fmffile}
\end{gathered} \hspace{0.3cm} + \frac{1}{108} \hspace{0.5cm} \begin{gathered}
\begin{fmffile}{Diagrams/Mixed2PIEA_Diag8}
\begin{fmfgraph}(12.5,12.5)
\fmfleft{i0,i1}
\fmfright{o0,o1}
\fmftop{v1,vUp,v2}
\fmfbottom{v3,vDown,v4}
\fmfv{decor.shape=circle,decor.size=2.0thick,foreground=(0,,0,,1)}{v1}
\fmfv{decor.shape=circle,decor.size=2.0thick,foreground=(0,,0,,1)}{v2}
\fmfv{decor.shape=circle,decor.size=2.0thick,foreground=(0,,0,,1)}{v3}
\fmfv{decor.shape=circle,decor.size=2.0thick,foreground=(0,,0,,1)}{v4}
\fmfv{decor.shape=circle,decor.size=2.0thick,foreground=(0,,0,,1)}{v5}
\fmfv{decor.shape=circle,decor.size=2.0thick,foreground=(0,,0,,1)}{v6}
\fmf{phantom,tension=20}{i0,v1}
\fmf{phantom,tension=20}{i1,v3}
\fmf{phantom,tension=20}{o0,v2}
\fmf{phantom,tension=20}{o1,v4}
\fmf{phantom,tension=0.005}{v5,v6}
\fmf{wiggly,left=0.4,tension=0,foreground=(1,,0,,0)}{v3,v1}
\fmf{phantom,left=0.1,tension=0}{v1,vUp}
\fmf{phantom,left=0.1,tension=0}{vUp,v2}
\fmf{plain,left=0.25,tension=0,foreground=(1,,0,,0)}{v1,v2}
\fmf{wiggly,left=0.4,tension=0,foreground=(1,,0,,0)}{v2,v4}
\fmf{phantom,left=0.1,tension=0}{v4,vDown}
\fmf{phantom,left=0.1,tension=0}{vDown,v3}
\fmf{plain,right=0.25,tension=0,foreground=(1,,0,,0)}{v3,v4}
\fmf{plain,left=0.2,tension=0.01,foreground=(1,,0,,0)}{v1,v5}
\fmf{plain,left=0.2,tension=0.01,foreground=(1,,0,,0)}{v5,v3}
\fmf{plain,right=0.2,tension=0.01,foreground=(1,,0,,0)}{v2,v6}
\fmf{plain,right=0.2,tension=0.01,foreground=(1,,0,,0)}{v6,v4}
\fmf{wiggly,tension=0,foreground=(1,,0,,0)}{v5,v6}
\end{fmfgraph}
\end{fmffile}
\end{gathered} \\
& \hspace{0.3cm} + \frac{1}{324} \hspace{0.4cm} \begin{gathered}
\begin{fmffile}{Diagrams/Mixed2PIEA_Diag9}
\begin{fmfgraph}(12.5,12.5)
\fmfleft{i0,i1}
\fmfright{o0,o1}
\fmftop{v1,vUp,v2}
\fmfbottom{v3,vDown,v4}
\fmfv{decor.shape=circle,decor.size=2.0thick,foreground=(0,,0,,1)}{v1}
\fmfv{decor.shape=circle,decor.size=2.0thick,foreground=(0,,0,,1)}{v2}
\fmfv{decor.shape=circle,decor.size=2.0thick,foreground=(0,,0,,1)}{v3}
\fmfv{decor.shape=circle,decor.size=2.0thick,foreground=(0,,0,,1)}{v4}
\fmfv{decor.shape=circle,decor.size=2.0thick,foreground=(0,,0,,1)}{v5}
\fmfv{decor.shape=circle,decor.size=2.0thick,foreground=(0,,0,,1)}{v6}
\fmf{phantom,tension=20}{i0,v1}
\fmf{phantom,tension=20}{i1,v3}
\fmf{phantom,tension=20}{o0,v2}
\fmf{phantom,tension=20}{o1,v4}
\fmf{phantom,tension=0.005}{v5,v6}
\fmf{plain,left=0.4,tension=0,foreground=(1,,0,,0)}{v3,v1}
\fmf{phantom,left=0.1,tension=0}{v1,vUp}
\fmf{phantom,left=0.1,tension=0}{vUp,v2}
\fmf{wiggly,left=0.25,tension=0,foreground=(1,,0,,0)}{v1,v2}
\fmf{plain,left=0.4,tension=0,foreground=(1,,0,,0)}{v2,v4}
\fmf{phantom,left=0.1,tension=0}{v4,vDown}
\fmf{phantom,left=0.1,tension=0}{vDown,v3}
\fmf{wiggly,right=0.25,tension=0,foreground=(1,,0,,0)}{v3,v4}
\fmf{plain,left=0.2,tension=0.01,foreground=(1,,0,,0)}{v1,v5}
\fmf{plain,left=0.2,tension=0.01,foreground=(1,,0,,0)}{v5,v3}
\fmf{plain,right=0.2,tension=0.01,foreground=(1,,0,,0)}{v2,v6}
\fmf{plain,right=0.2,tension=0.01,foreground=(1,,0,,0)}{v6,v4}
\fmf{wiggly,tension=0,foreground=(1,,0,,0)}{v5,v6}
\end{fmfgraph}
\end{fmffile}
\end{gathered} \hspace{0.4cm} + \mathcal{O}\Big(\vec{F}^{2}\Big) \left.\rule{0cm}{1.2cm}\right) \\
& \hspace{0.3cm} + \mathcal{O}\big(\hbar^{5}\big)\;,
\end{split}
\label{eq:mixed2PIEAfinalexpression}
\end{equation}
with the supertrace $\mathcal{ST}r=\mathrm{Tr}_{\alpha}\mathrm{Tr}_{x}$, the superidentity $\mathfrak{I}_{\alpha\beta}(x,y)=\delta_{\alpha\beta}\delta(x-y)$ and the Feynman rules:
\begin{subequations}
\begin{align}
%\begin{equation}
\left.
\begin{array}{ll}
\begin{gathered}
\begin{fmffile}{Diagrams/mixed2PIEA_FeynRuleVertexbis1}
\begin{fmfgraph*}(4,4)
\fmfleft{i0,i1,i2,i3}
\fmfright{o0,o1,o2,o3}
\fmfv{label=$x$,label.angle=90,label.dist=4}{v1}
\fmfbottom{v2}
\fmf{phantom}{i1,v1}
\fmf{plain,foreground=(1,,0,,0)}{i2,v1}
\fmf{phantom}{v1,o1}
\fmf{plain,foreground=(1,,0,,0)}{v1,o2}
\fmf{wiggly,tension=0.6,foreground=(1,,0,,0)}{v1,v2}
\fmfv{decor.shape=circle,decor.size=2.0thick,foreground=(0,,0,,1)}{v1}
\fmflabel{$a$}{i2}
\fmflabel{$b$}{o2}
\end{fmfgraph*}
\end{fmffile}
\end{gathered} \\
\\
\begin{gathered}
\begin{fmffile}{Diagrams/mixed2PIEA_FeynRuleVertexbis2}
\begin{fmfgraph*}(4,4)
\fmfleft{i0,i1,i2,i3}
\fmfright{o0,o1,o2,o3}
\fmfv{label=$x$,label.angle=90,label.dist=4}{v1}
\fmfbottom{v2}
\fmf{phantom}{i1,v1}
\fmf{plain,foreground=(1,,0,,0)}{i2,v1}
\fmf{phantom}{v1,o1}
\fmf{plain,foreground=(1,,0,,0)}{v1,o2}
\fmf{dots,tension=0.6,foreground=(1,,0,,0)}{v1,v2}
\fmfv{decor.shape=circle,decor.size=2.0thick,foreground=(0,,0,,1)}{v1}
\fmflabel{$a$}{i2}
\fmflabel{$b$}{o2}
\end{fmfgraph*}
\end{fmffile}
\end{gathered} \\
\\
\begin{gathered}
\begin{fmffile}{Diagrams/mixed2PIEA_FeynRuleVertexbis3}
\begin{fmfgraph*}(4,4)
\fmfleft{i0,i1,i2,i3}
\fmfright{o0,o1,o2,o3}
\fmfv{label=$x$,label.angle=90,label.dist=4}{v1}
\fmfbottom{v2}
\fmf{phantom}{i1,v1}
\fmf{plain,foreground=(1,,0,,0)}{i2,v1}
\fmf{phantom}{v1,o1}
\fmf{dots,foreground=(1,,0,,0)}{v1,o2}
\fmf{dots,tension=0.6,foreground=(1,,0,,0)}{v1,v2}
\fmfv{decor.shape=circle,decor.size=2.0thick,foreground=(0,,0,,1)}{v1}
\fmflabel{$a$}{i2}
\fmflabel{$b$}{o2}
\end{fmfgraph*}
\end{fmffile}
\end{gathered}
\end{array}
\quad \right\rbrace &\rightarrow \sqrt{\lambda} \ \delta_{ab} \;, 
\label{eq:mixed2PIEAvertex} \\
%\end{equation}
%\begin{equation}
\begin{gathered}
\begin{fmffile}{Diagrams/mixed2PIEA_FeynRuleGbis}
\begin{fmfgraph*}(20,20)
\fmfleft{i0,i1,i2,i3}
\fmfright{o0,o1,o2,o3}
\fmflabel{$x, a$}{v1}
\fmflabel{$y, b$}{v2}
\fmf{phantom}{i1,v1}
\fmf{phantom}{i2,v1}
\fmf{plain,tension=0.6,foreground=(1,,0,,0)}{v1,v2}
\fmf{phantom}{v2,o1}
\fmf{phantom}{v2,o2}
\end{fmfgraph*}
\end{fmffile}
\end{gathered} \quad &\rightarrow \boldsymbol{G}_{ab}(x,y) \;,
\label{eq:mixed2PIEAFeynRuleG} \\
%\end{equation}
%\begin{equation}
\begin{gathered}
\begin{fmffile}{Diagrams/mixed2PIEA_FeynRuleDbis}
\begin{fmfgraph*}(20,20)
\fmfleft{i0,i1,i2,i3}
\fmfright{o0,o1,o2,o3}
\fmfv{label=$x$}{v1}
\fmfv{label=$y$}{v2}
\fmf{phantom}{i1,v1}
\fmf{phantom}{i2,v1}
\fmf{wiggly,tension=0.6,foreground=(1,,0,,0)}{v1,v2}
\fmf{phantom}{v2,o1}
\fmf{phantom}{v2,o2}
\end{fmfgraph*}
\end{fmffile}
\end{gathered} \quad &\rightarrow D(x,y) \;,
\label{eq:mixed2PIEAFeynRuleD} \\
%\end{equation}
%\begin{equation}
\begin{gathered}
\begin{fmffile}{Diagrams/mixed2PIEA_FeynRuleFbis}
\begin{fmfgraph*}(20,20)
\fmfleft{i0,i1,i2,i3}
\fmfright{o0,o1,o2,o3}
\fmflabel{$x, a$}{v1}
\fmfv{label=$y$}{v2}
\fmf{phantom}{i1,v1}
\fmf{phantom}{i2,v1}
\fmf{dashes,tension=0.6,foreground=(1,,0,,0)}{v1,v2}
\fmf{phantom}{v2,o1}
\fmf{phantom}{v2,o2}
\end{fmfgraph*}
\end{fmffile}
\end{gathered} \quad &\rightarrow F_{a}(x,y) \;,
\label{eq:mixed2PIEAFeynRuleF}
%\end{equation}
\end{align}
\end{subequations}
whereas the propagator $\mathcal{G}_{\Phi}$ satisfies:
\begin{equation}
\begin{split}
& \mathcal{G}^{-1}_{\Phi}(x,y) \equiv \left.\frac{\delta^2 S_{\text{mix}}\big[\widetilde{\Psi}\big]}{\delta \widetilde{\Psi}(x)\delta \widetilde{\Psi}(y)}\right|_{\widetilde{\Psi} = \Phi} \\
& = \begin{pmatrix}
\left(-\nabla_x^2 + m^2 + i\sqrt{\frac{\lambda}{3}}\eta(x)\right)\mathbb{I}_{N} & i\sqrt{\frac{\lambda}{3}}\vec{\phi}(x) \\
i\sqrt{\frac{\lambda}{3}}\vec{\phi}^\mathrm{T}(x) & 1 \end{pmatrix}\delta(x-y) \;,
\end{split}
\label{eq:mixed2PIEAMathcalGPhi}
\end{equation}
with $\mathbb{I}_{D}$ being the $D$-dimensional identity matrix. One can check that all diagrams involved in Eq.~\eqref{eq:mixed2PIEAfinalexpression} are 2PI with respect to each of the three propagator lines~\eqref{eq:mixed2PIEAFeynRuleG},~\eqref{eq:mixed2PIEAFeynRuleD} and~\eqref{eq:mixed2PIEAFeynRuleF}. Furthermore, we have ignored all $\vec{F}$-dependent diagrams contributing at order $\mathcal{O}\big(\hbar^{4}\big)$ in the right-hand side of Eq.~\eqref{eq:mixed2PIEAfinalexpression} as we assume no spontaneous breakdown of the $O(N)$ symmetry for the present implementation of SCPT at third non-trivial order with respect to $\hbar$. This will be justified thereafter.

For the sake of comparison with the equations obtained from the original 2PI EA, we will just consider the (0+0)\nobreakdash-D limit of Eq.~\eqref{eq:mixed2PIEAfinalexpression} at $N=1$, although we will present numerical applications for $N=2$ as well. The mixed 2PI EA of the studied (0+0)\nobreakdash-D $O(N)$ model thus satisfies at $N=1$:
\begin{equation}
\begin{split}
& \Gamma_{\mathrm{mix}}^{(\mathrm{2PI})}\big(\Phi,\mathcal{G}\big) \\
& = S_{\mathrm{mix}}(\Phi) + \hbar \Bigg[ -\frac{1}{2}\ln\big(2\pi\boldsymbol{G}_{11}\big) -\frac{1}{2}\ln(D) \\
& \hspace{0.3cm} + \frac{1}{2} \Bigg(\left(m^{2}+i\sqrt{\frac{\lambda}{3}}\eta\right)\boldsymbol{G}_{11} + 2 i \sqrt{\frac{\lambda}{3}} \phi_{1} F_{1} + D \Bigg) - 1 \Bigg] \\
& \hspace{0.3cm} + \hbar^{2}\Bigg[ \frac{1}{12} \lambda \boldsymbol{G}_{11} \left(2 F_{1}^{2} + D \boldsymbol{G}_{11}\right) \Bigg] \\
& \hspace{0.3cm} + \hbar^{3}\Bigg[ \frac{1}{72} \lambda^{2} \boldsymbol{G}_{11}^2 \left(10 F_{1}^{4} + 12 D F_{1}^{2} \boldsymbol{G}_{11} + D^2 \boldsymbol{G}_{11}^2\right) \Bigg] \\
& \hspace{0.3cm} + \hbar^{4}\Bigg[ \frac{5}{324} \lambda^{3} D^3 \boldsymbol{G}_{11}^6 + \mathcal{O}\Big(\vec{F}^{2}\Big) \Bigg] \\
& \hspace{0.3cm} + \mathcal{O}\big(\hbar^{5}\big)\;,
\end{split}
\label{eq:mixed2PIEAfinalexpressionN10DON}
\end{equation}
and the corresponding gap equations are:
\begin{equation}
0 = \left.\frac{\partial \Gamma_{\mathrm{mix}}^{(\mathrm{2PI})}\big(\Phi,\mathcal{G}\big)}{\partial \phi_{1}}\right|_{\Phi=\overline{\Phi} \atop \mathcal{G}=\overline{\mathcal{G}}} \;,
\end{equation}
\begin{equation}
0 = \left.\frac{\partial \Gamma_{\mathrm{mix}}^{(\mathrm{2PI})}\big(\Phi,\mathcal{G}\big)}{\partial \eta}\right|_{\Phi=\overline{\Phi} \atop \mathcal{G}=\overline{\mathcal{G}}} \;,
\end{equation}
\begin{equation}
0 = \left.\frac{\partial \Gamma_{\mathrm{mix}}^{(\mathrm{2PI})}\big(\Phi,\mathcal{G}\big)}{\partial \boldsymbol{G}_{11}}\right|_{\Phi=\overline{\Phi} \atop \mathcal{G}=\overline{\mathcal{G}}} \;,
\end{equation}
\begin{equation}
0 = \left.\frac{\partial \Gamma_{\mathrm{mix}}^{(\mathrm{2PI})}\big(\Phi,\mathcal{G}\big)}{\partial D}\right|_{\Phi=\overline{\Phi} \atop \mathcal{G}=\overline{\mathcal{G}}} \;,
\end{equation}
\begin{equation}
0 = \left.\frac{\partial \Gamma_{\mathrm{mix}}^{(\mathrm{2PI})}\big(\Phi,\mathcal{G}\big)}{\partial F_{1}}\right|_{\Phi=\overline{\Phi} \atop \mathcal{G}=\overline{\mathcal{G}}} \;.
\end{equation}
The gs energy and density are then inferred from the solutions of the gap equations $\overline{\Phi}$ and $\overline{\mathcal{G}}$ alongside with the equalities:
\begin{equation}
E^\text{2PI EA;mix}_\text{gs} = \frac{1}{\hbar}\Gamma_{\mathrm{mix}}^{(\mathrm{2PI})}\big(\Phi=\overline{\Phi},\mathcal{G}=\overline{\mathcal{G}}\big) \;,
\label{eq:2PImixE}
\end{equation}
\begin{equation}
\rho^\text{2PI EA;mix}_\text{gs} = \frac{1}{N} \left(\hbar\mathrm{Tr}_{a}\big(\overline{\boldsymbol{G}}\big) + \vec{\overline{\phi}}^2\right) \;,
\label{eq:2PImixRho}
\end{equation}
which are both valid for all $N$.

%%%%%%%%%%%%%%%%%%%
\begin{figure}[!t]
  \begin{center}
      \includegraphics[width=0.95\linewidth]{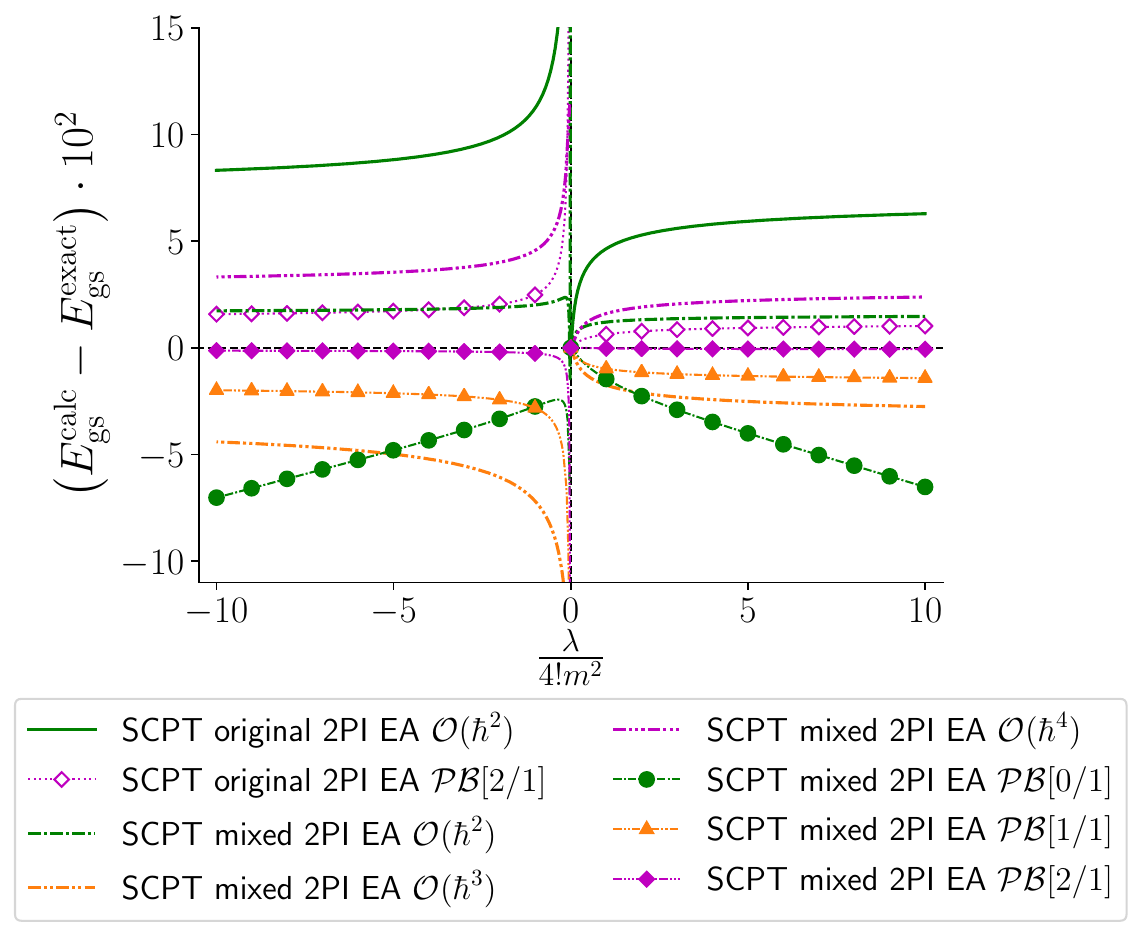}
      \includegraphics[width=0.95\linewidth]{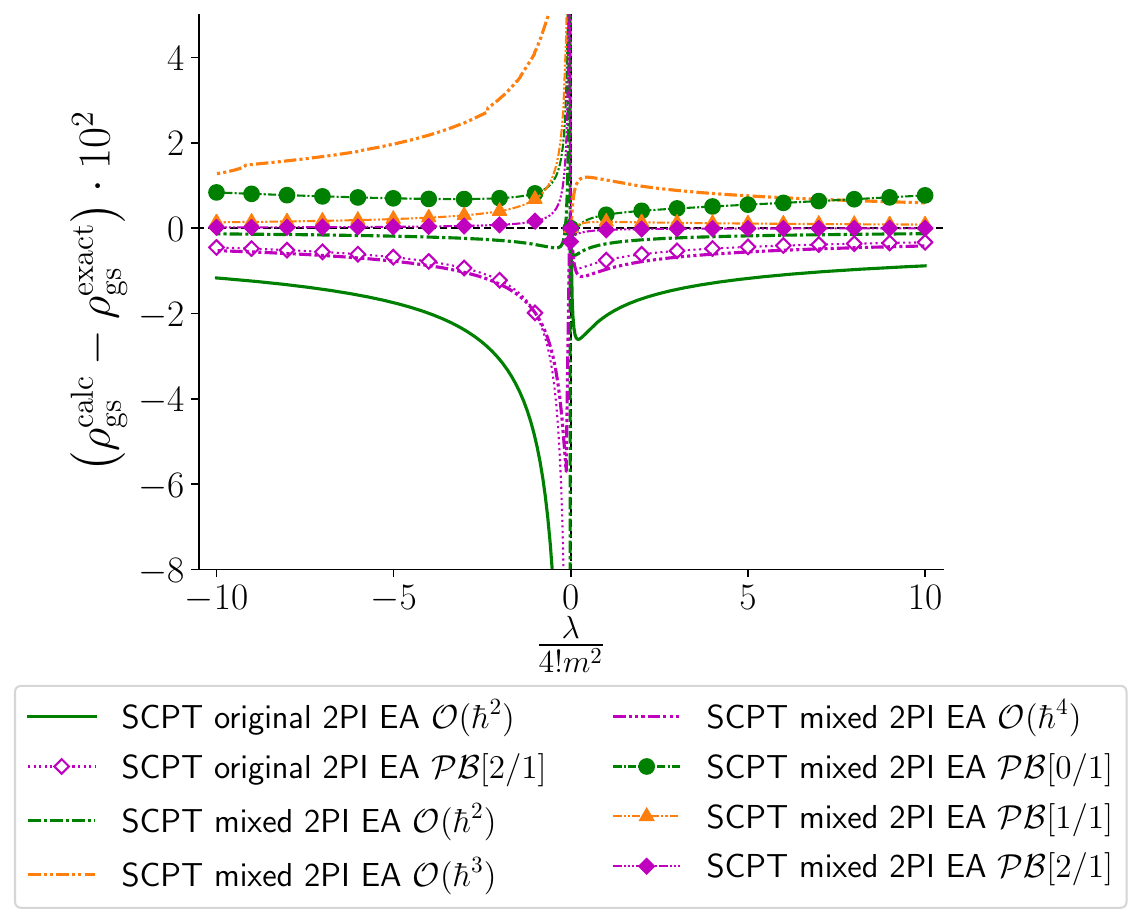}
    \caption{Difference between the calculated gs energy $E_{\mathrm{gs}}^{\mathrm{calc}}$ or density $\rho_{\mathrm{gs}}^{\mathrm{calc}}$ and the corresponding exact solution $E_{\mathrm{gs}}^{\mathrm{exact}}$ or $\rho_{\mathrm{gs}}^{\mathrm{exact}}$ at $\hbar=1$, $m^{2}=\pm 1$ and $N=2$. See also the caption of Fig.~\ref{fig:2PIEAzeroVsNonzerovev} for the meaning of the indication ``$\mathcal{O}\big(\hbar^{n}\big)$'' for SCPT results.}
    \label{fig:2PIEAorigVsmixN2}
  \end{center}
\end{figure}
%%%%%%%%%%%%%%%%%%%

Our estimations of the gs energy and density determined from SCPT based on the mixed 2PI EA at $N=2$ are presented in Fig.~\ref{fig:2PIEAorigVsmixN2}. With or without resummation, the mixed 2PI EA results outperform in general those of the original 2PI EA $\Gamma^{(\mathrm{2PI})}(\boldsymbol{G})$ for a given truncation with respect to $\hbar$. This is illustrated for both $E_{\mathrm{gs}}$ and $\rho_{\mathrm{gs}}$ in Fig.~\ref{fig:2PIEAorigVsmixN2} for the first non-trivial orders of these approaches and, after resummation, for their third non-trivial orders (with the $[2/1]$ Pad\'{e} approximants). In particular, this figure shows that, in the non-perturbative regime of the studied model at $N=2$, the mixed 2PI EA achieves an accuracy of about $2\%$ for $E_{\mathrm{gs}}$ (to be compared with about $5\%$ to $8\%$ for the corresponding results of the original 2PI EA), and even less for $\rho_{\mathrm{gs}}$, already at its first non-trivial order, which corresponds to the BVA result mentioned previously. Furthermore, as the solutions of the gap equations leading to our mixed 2PI EA results never break the $O(N)$ symmetry (and notably always satisfy $\vec{\overline{F}}=\vec{0}$) at the first two non-trivial orders (i.e. when the mixed 2PI EA is considered up to orders $\mathcal{O}(\hbar^2)$ and $\mathcal{O}(\hbar^3)$), we can reasonably expect no SSB at the next non-trivial order, which motivates our previous assumption ignoring the contribution of $\vec{F}$-dependent diagrams at order $\mathcal{O}\big(\hbar^{4}\big)$ in Eq.~\eqref{eq:mixed2PIEAfinalexpression}. Interestingly, the absence of SSB at first non-trivial order with respect to $\hbar$ for the mixed 2PI EA also implies that only the Fock diagram with a wiggly line (i.e. with a $D$ propagator) contributes to the BVA result, which implies that the latter is equivalent to the result obtained from the same mixed 2PI EA at order $\mathcal{O}\big(1/N\big)$. Although we do not treat $1/N$-expansions of EAs in this study, it is interesting to know that its first non-trivial order in the expansion of the mixed 2PI EA coincides with the excellent BVA approximation for the studied model, which illustrates that $1/N$ can be a viable alternative to $\hbar$ as expansion parameter.

Although the BVA has already been applied to many QFTs, this is certainly not the case of higher truncation orders of SCPT based on the mixed 2PI EA (i.e. for truncations beyond order $\mathcal{O}(\hbar^2)$). In addition, we are performing to our knowledge the first applications of resummation theory to SCPT based on the mixed 2PI EA. Choosing once again the Pad\'{e}-Borel scheme as resummation procedure, we find that the best Pad\'{e} approximant obtained from the first non-trivial order of the mixed 2PI EA does not manage to clearly improve the corresponding bare results (i.e. the BVA results) in general, as illustrated by Fig.~\ref{fig:2PIEAorigVsmixN2}. Pad\'{e}-Borel resummation thus starts being efficient at order $\mathcal{O}\big(\hbar^{3}\big)$ and, at the third non-trivial order (and more specifically from $[2/1]$ Pad\'{e} approximants), yields excellent results which are barely distinguishable from the exact solution in Fig.~\ref{fig:2PIEAorigVsmixN2}.

In conclusion, the significant improvement as compared to the original 2PI EA results can be attributed to both the 1-point correlation function $\eta$ and the propagator $D$ of the Hubbard-Stratonovich field since we always find $\vec{\overline{\phi}}=\vec{\overline{F}}=\vec{0}$ in the solutions of the mixed 2PI EA's gap equations leading to the results of Fig.~\ref{fig:2PIEAorigVsmixN2} (obtained at $N=2$)\footnote{Note that we have thoroughly checked that the same holds at $N=1$~\cite{fra21}.}. To conclude on the technical side, SCPT based on 2PI EAs is clearly the closest in spirit to the EDF framework among all PI techniques tested in this study since they both rely on variational equations derived from similar functionals: for the EDF, this functional is the energy of the system expressed in terms of its density whereas 2PI EAs (from which energies can also be directly extracted) are actually functionals of Green's functions whose diagonal parts in both space and time coordinates also coincide with the density of the system. This connection is even accentuated in the local versions of 2PI EAs, called 2PPI EAs, which are themselves functionals of the density of the system. As mentioned in the introduction, 2PPI EAs have already been exploited within the nuclear physics community, but mostly at the level of toy models \cite{pug03,bha05,fur07,dru10,fur12,kem13,kem16,yok19,yok19bis}. However, SCPT based on a 2PI EA and that based on its homologous 2PPI EA lead to identical results for the (0+0)\nobreakdash-D model considered here owing to the conservation of the $O(N)$ symmetry by the optimal solutions of the corresponding gap equations. Note also that the connection between SCPT and the EDF method is even more apparent within the single-reference scheme of the latter, where the density functional is typically treated within the Hartree-Fock-Bogoliubov theory~\cite{bog58,val58,deg66,rin80}. Indeed, in the case of SCPT based on the original 2PI EA, we have seen that the Hartree-Fock contribution corresponds to the first non-trivial order and, as opposed to the EDF approach, the Hartree-Fock result thus obtained can be systematically improved reliably by expanding the EA beyond second order in $\hbar$ (as was done in Eq.~\eqref{eq:2PIEAzerovevfinalexpression}) and then using resummation theory.

We will then turn to our conclusions for the present comparative study in order to compare the techniques that stand out from it and to summarize the main take-away message, with respect to our aim of reformulating the EDF formalism but also to prepare the ground for future applications to quantum many-body systems in general.

%%%%%%%%%%%%%%%%%%%%%%%%%%%%%%%
%%%%%%%%%%%%%%%%%%%%%%%%%%%%%%%
\section{\label{sec:Conclusion}Conclusion}

%%%%%%%%%%%%%%%%%%%
\begin{figure}[!t]
  \begin{center}
      \includegraphics[width=0.95\linewidth]{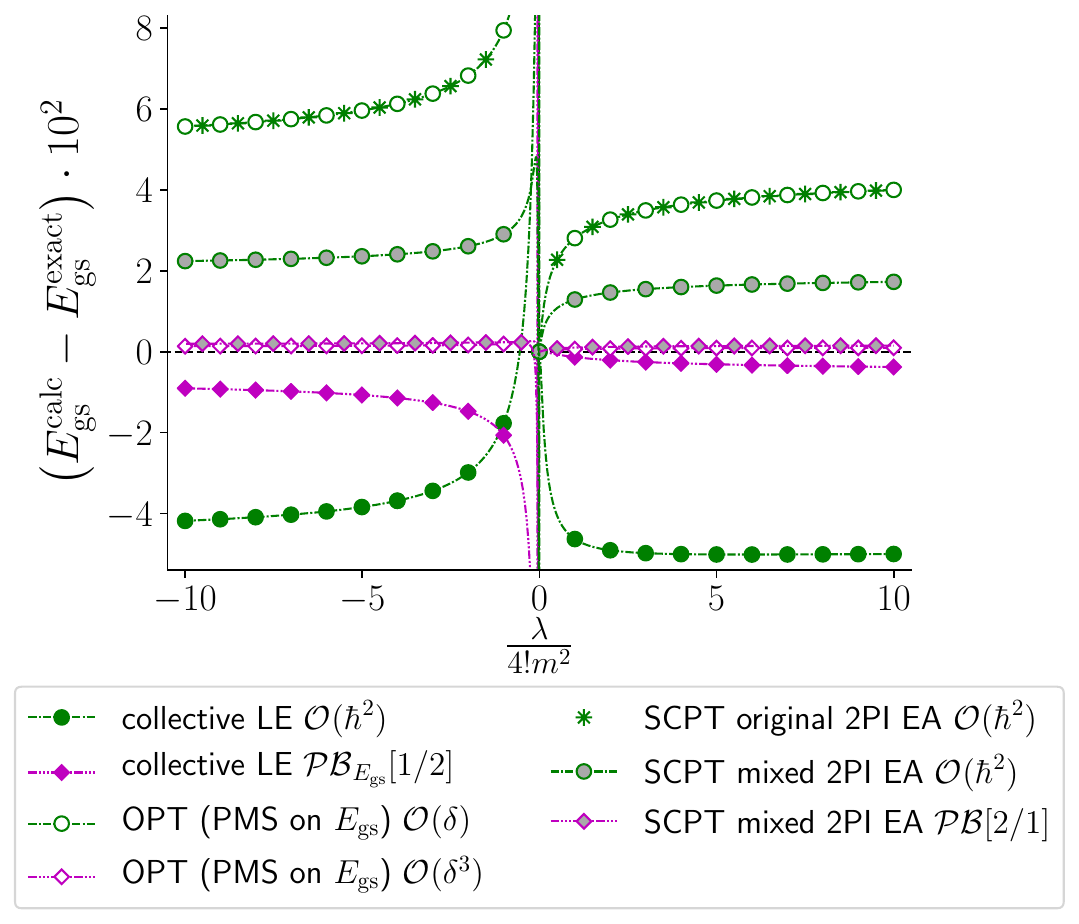}
      \includegraphics[width=0.95\linewidth]{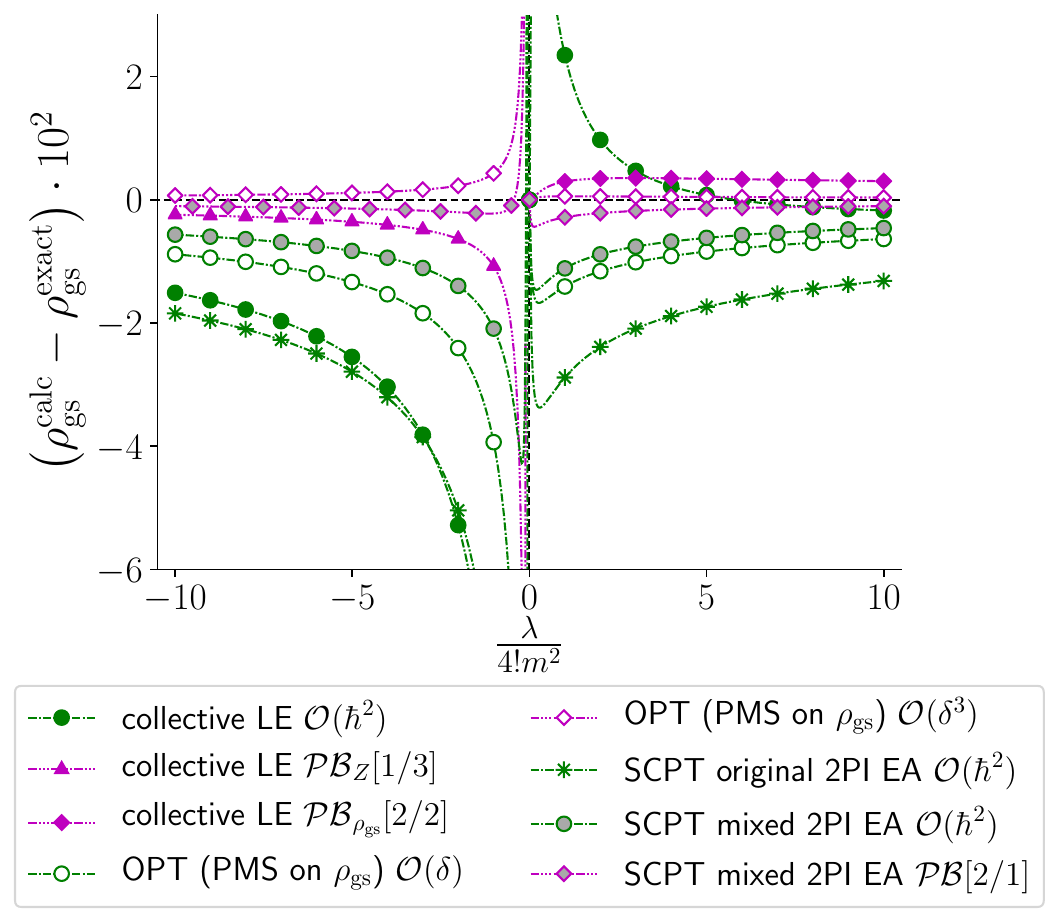}
    \caption{Difference between the calculated gs energy $E_{\mathrm{gs}}^{\mathrm{calc}}$ or density $\rho_{\mathrm{gs}}^{\mathrm{calc}}$ and the corresponding exact solution $E_{\mathrm{gs}}^{\mathrm{exact}}$ or $\rho_{\mathrm{gs}}^{\mathrm{exact}}$ at $\hbar=1$, $m^{2}=\pm 1$ and $N=1$. See also the captions of Figs.~\ref{fig:O1PTcoll} and~\ref{fig:2PIEAzeroVsNonzerovev} for the meaning of the indication ``$\mathcal{O}\big(\hbar^{n}\big)$'' for the results obtained from the collective LE and SCPT.}
    \label{fig:ConcluN1}
  \end{center}
\end{figure}
%%%%%%%%%%%%%%%%%%%

%%%%%%%%%%%%%%%%%%%
\begin{figure}[!t]
  \begin{center}
      \includegraphics[width=0.95\linewidth]{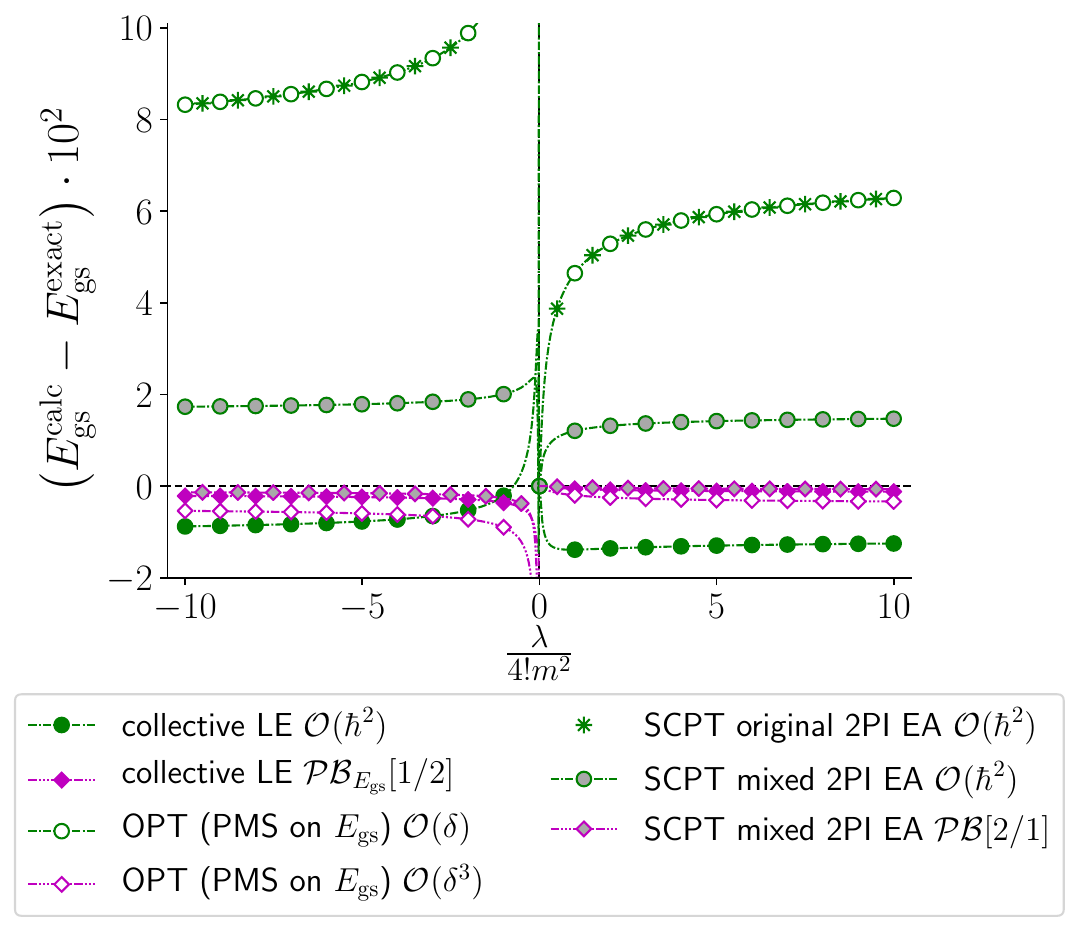}
      \includegraphics[width=0.95\linewidth]{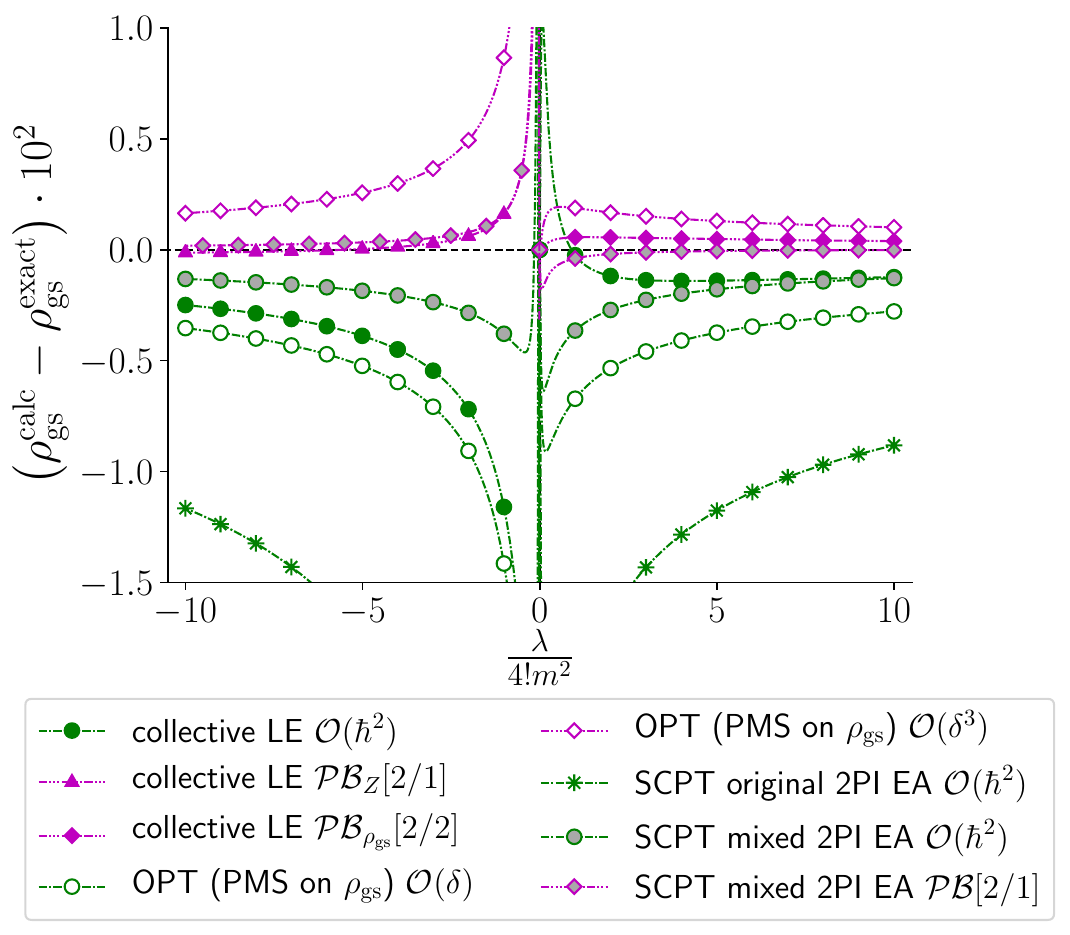}
    \caption{Same as Fig.~\ref{fig:ConcluN1} with $N=2$ instead.}
    \label{fig:ConcluN2}
  \end{center}
\end{figure}
%%%%%%%%%%%%%%%%%%%

In this comparative study, we have investigated three families of diagrammatic PI techniques, i.e. the LE, OPT and SCPT, by calculating the gs energy and density of the (0+0)\nobreakdash-D $O(N)$-symmetric $\varphi^{4}$-theory. We have also examined the situation where a HST has been applied to this model. This was done by considering three representations of the latter: its original representation (where no HST has been performed), its mixed representation (where a collective field has been introduced via a HST in addition to the original one) and the collective representation (where the original field of the mixed representation has been integrated out). Furthermore, most of the tested PI techniques have been pushed up to their third non-trivial orders, some of them for the first time like the collective LE and SCPT based on the mixed 2PI EA. Considering that both LE and SCPT approaches rely on asymptotic series, another important feature of the presented study is the use of resummation techniques. Three have been tested: the Pad\'e-Borel resummation, the conformal mapping and the Borel-hypergeometric resummation, where the latter was developed rather recently \cite{mer15,mer16,ped16a,ped16b,san17,mer18,ant19}. These approaches were applied to both LE and SCPT techniques with great success at their third non-trivial orders, especially in the description of the non-perturbative regime of the model under consideration. We can thus clearly see at this point the advantage of choosing a simple zero-dimensional model as a playground since it allows to readily (from a numerical point of view at least) reach rather high truncation orders and thus explore the behavior of the computed series under resummation, whereas such implementations might be much lenghtier and harder to achieve in more realistic frameworks. Besides its zero-dimensional character, the $O(N)$ symmetry of our toy model was also at the heart of our discussions. Besides the original LE, no other PI technique tested here was found to manifest a spurious breakdown of the $O(N)$ symmetry, as expected from the exact solution of the studied model which can thus be seen as a trivial counterpart of mesoscopic systems which can not exhibit SSB either.

Let us finally turn to a final comparison of the obtained results. As the mixed 2PI EA method clearly stands out among the tested SCPT approaches, we compare it in Figs.~\ref{fig:ConcluN1} and~\ref{fig:ConcluN2} with the most performing implementations of LE and OPT methods treated previously, i.e. the collective LE and OPT based on the PMS for $N=1$ and $2$, at first and third non-trivial orders. We can see on these figures that, after resummation if necessary, the excellent performances of the collective LE, OPT with PMS and SCPT based on the mixed 2PI EA are very close both at first and third non-trivial orders, with the exception of the first non-trivial order for OPT with PMS which is known to coincide with the Hartree-Fock result of SCPT for the original 2PI EA $\Gamma^{(\mathrm{2PI})}(\boldsymbol{G})$ for $E_{\mathrm{gs}}$ (note however that the original 2PI EA as treated here relies on resummation procedures, whereas OPT does not), as can be seen from Figs.~\ref{fig:ConcluN1} and~\ref{fig:ConcluN2} as well. These performances should of course be put in contrast with the ability to treat different channels in more realistic systems. This is indeed of crucial importance for the description of many realistic many-body systems, and of most atomic nuclei in particular. Although the PI techniques tested here were applied to a toy model involving a single channel, their implementation can in principle be extended to treat several channels in an equitable fashion. For example, for methods based on the mixed or the collective representation, one might use multi-channel HSTs, as was done e.g. in Refs.~\cite{lan15,kug18,bon22}, and OPT can be implemented by introducing a classical field coupled to each relevant bilinear in the original field(s) of the theory under consideration~\cite{kle11}\footnote{OPT is sometimes put forward (notably in Ref.~\cite{kle11}) as a technique designed to solve an old standing problem of HSTs leading to their inability to describe efficiently competing channels. One should stress however that the more recently introduced multi-channel HSTs also allow for a democratic treatment of several competing channels.}. That said, we postpone applications to more realistic systems with competing channels to future works but we conclude from the present study that the LE, OPT and SCPT all show promising predispositions to treat such systems.

Furthermore, as explained earlier, SCPT is, among all PI techniques considered here, the closest approach to the nuclear EDF method, and especially to its single-reference implementation, since it also relies on variational (self-consistent) equations and, depending on the EA under consideration, on density functionals. However, it should be stressed that, as opposed to the EDF method, SCPT and all other PI techniques exploited in this study offer the possibility to take a classical action describing the bare nuclear interaction (derived e.g. from an EFT of QCD) as an input and also to improve the obtained results in a systematic way. It should also be stressed that, besides SCPT, the tested PI techniques are radically different from the EDF approach in the sense that they rely on completely different treatments of collective (bosonic) dofs: in particular, the optimization of the classical field(s) introduced within OPT or even the introduction of collective field(s) via HST have no counterpart in the EDF framework. Note also that, as compared to the EDF method, LE and SCPT approaches possess an additional freedom through the use of resummation procedures that can be chosen according to the functions we are trying to describe. The present study has thus illustrated the richness of the PI framework and in what ways the latter can be used to reformulate the nuclear EDF formalism in a more reliable manner. Another important family of PI techniques that is also a relevant candidate to achieve the latter purpose is the FRG, which can exploit e.g. SCPT results as input and thus potentially offer a direct connection with the multi-reference scheme of the EDF method. These approaches will be discussed in a subsequent comparative study \cite{fra22}.

\section*{Acknowledgments}
The authors thank Jean-Paul Blaizot, Thomas Duguet, Vittorio Som\`a and Nicolas Dupuis for fruitful discussions.

\subsection*{Data Availability}
The datasets generated during and/or analyzed during the current study are available from the corresponding author on reasonable request.

%%%%%%%%%%%%%%%%%%%%%%%%%%%%%%%
%%%%%%%%%%%%%%%%%%%%%%%%%%%%%%%
%Appendix

\appendix

%%%%%%%%%%%%%%%%%%%%%%%%%%%%%%%
%%%%%%%%%%%%%%%%%%%%%%%%%%%%%%%
\section{\label{app:GeneratingFuncPIformalism}Generating functionals of interest in the path-integral formalism}

We present in this appendix a brief introduction to the PI formalism. In the canonical formulation of quantum mechanics, the constituents of the theory are represented by \emph{operator-valued} distributions $\hat{\varphi}_{\alpha}$, with a generic index $\alpha \equiv (a,r,\tau,m_s,c) \equiv (a,x)$ collecting spacetime coordinates $(r,\tau)$ (with $r$ and $\tau$ being respectively the $(D-1)$-dimensional space position vector and the imaginary time) and, if relevant, spin projection $m_s$, charge $c$, and internal $a$ labels, such that:
\begin{equation}
\hat{\varphi}_{\alpha}=\hat{\varphi}_{a,x}=\left\{
\begin{array}{lll}
        \displaystyle{\hat{\varphi}_{a,m_s}(r,\tau) \quad \mathrm{for}~c=-\;.} \\
        \\
        \displaystyle{\hat{\varphi}^{\dagger}_{a,m_s}(r,\tau) \quad \mathrm{for}~c=+\;.}
    \end{array}
\right.
\end{equation}
The $n$-point correlation function $G^{(n)}$ then stems from the expectation value, in the (interacting) many-body system gs $\ket{\mathrm{vac}}$, of a (time-ordered) product of $n$ field operators:
\begin{equation}
G^{(n)}_{\alpha_{1}\alpha_{2}\cdots\alpha_{n}} \equiv \braket{\mathrm{vac}|\mathrm{T}\hat{\varphi}_{\alpha_{1}}\hat{\varphi}_{\alpha_{2}}\cdots\hat{\varphi}_{\alpha_{n}}|\mathrm{vac}} \;,
\label{eq:CanonicalFormalismCorrelationFunction}
\end{equation}
where $\mathrm{T}$ stands for the time-ordered product.

Alternatively, correlation functions can be computed from standard generating functionals \cite{sch51}, conveniently expressed as sum-over-histories in configuration space within Feynman's PI formulation of quantum mechanics \cite{fey10,fri20}, where the dofs of the theory are now realized via \emph{number-valued} fields, such as $\widetilde{\varphi}_\alpha$\footnote{It is assumed in the entire appendix~\ref{app:GeneratingFuncPIformalism} that $\widetilde{\varphi}_{\alpha}$ is a bosonic (i.e. non-Grassmann) field, as all fields involved in the presented toy model study are bosonic. More exhaustive introductions on the PI formalism for both bosonic and fermionic field theories can be found e.g. in Refs.~\cite{neg98,kop10bis3}.}. Coupling the field $\widetilde{\varphi}_\alpha$ to a test (external) source $J_{\alpha}$ yields the action:
\begin{equation}
S_{J}[\widetilde{\varphi}] = S[\widetilde{\varphi}] - J_{\alpha} \widetilde{\varphi}_{\alpha} \;,
\label{eq:GeneralSJ}
\end{equation}
where $S[\widetilde{\varphi}]$ stands for the (Euclidean) classical action of the system and summation over repeated indices is assumed, i.e.:
\begin{equation}
\begin{split}
& J_{\alpha} \widetilde{\varphi}_{\alpha} \equiv J_{a,x} \widetilde{\varphi}_{a,x} \\
& \equiv \sum_{m_s,c,a}\int_0^{\hbar/T} d\tau \int_{\mathbb{R}^{D-1}}d^{D-1}r \ J_{a,m_s,c}(r,\tau) \widetilde{\varphi}_{a,m_s,-c}(r,\tau) \;, \\
\end{split}
\end{equation}
in a $D$-dimensional spacetime and at temperature $T$. The (Euclidean) PI representation of the system's partition function in presence of the source derives from these ingredients and is given by the PI:
\begin{equation}
Z[J]=\mathcal{N}\int_\mathcal{C}\mathcal{D}\widetilde{\varphi} \ e^{-\frac{1}{\hbar}S_{J}[\widetilde{\varphi}]} \;,
\label{eq:GeneralGeneratingFunctional}
\end{equation}
where $\mathcal{N}$ is a normalization factor, $\mathcal{C}$ the space of configurations\footnote{The integration domain defining each PI will be left implicit in most cases.} for the fluctuating field $\widetilde{\varphi}_{\alpha}$ and $\mathcal{D}\widetilde{\varphi}$ the PI measure. The functional $Z[J]$ is the generating functional of correlation functions, namely:
\begin{equation}
\begin{split}
G^{(n)}_{\alpha_{1} \alpha_{2} \cdots \alpha_{n}} \equiv & \ \braket{\widetilde{\varphi}_{\alpha_{1}}\widetilde{\varphi}_{\alpha_{2}}\cdots\widetilde{\varphi}_{\alpha_{n}}}_{\mathrm{vac}} \\
= & \ \frac{\int \mathcal{D}\widetilde{\varphi} \ \widetilde{\varphi}_{\alpha_{1}}\widetilde{\varphi}_{\alpha_{2}}\cdots\widetilde{\varphi}_{\alpha_{n}} e^{-\frac{1}{\hbar} S\left[\widetilde{\varphi}\right]}}{\int \mathcal{D}\widetilde{\varphi} \ e^{-\frac{1}{\hbar} S\left[\widetilde{\varphi}\right]}} \\
= & \ \frac{\hbar^{n}}{Z[J=0]}\left.\frac{\delta^{n} Z[J]}{\delta J_{\alpha_{1}}\delta J_{\alpha_{2}}\cdots \delta J_{\alpha_{n}}}\right|_{J=0} \;.
\end{split}
\label{eq:Gn}
\end{equation}
A diagrammatic representation of $Z[J]$ consists of the sum of \emph{all} vacuum diagrams, implying that the correlation functions~\eqref{eq:Gn} contain both connected and disconnected contributions. On the other hand, physically relevant observables often only involve the fully connected part of $Z[J]$, which can be summarized in terms of another generating functional called the Schwinger functional\footnote{In analogy with thermodynamics, $Z[J=0]$ is the partition function while $W[J=0]$ corresponds to (minus) the free energy (up to a constant proportional to the temperature).} $W[J]$ defined via:
\begin{equation}
Z[J] \equiv e^{\frac{1}{\hbar}W[J]} \;.
\end{equation}
The cumulants or \emph{connected} correlation functions $G^{(n),c}$ then follow from the functional derivatives of $W[J]$:
\begin{equation}
G^{(n),c}_{\alpha_{1} \alpha_{2} \cdots \alpha_{n}} = \hbar^{n-1}\left.\frac{\delta^{n} W[J]}{\delta J_{\alpha_{1}}\delta J_{\alpha_{2}}\cdots \delta J_{\alpha_{n}}}\right|_{J=0} \;.
\end{equation}

An exact and compact representation of the generating functional $Z[J]$ can be achieved through the $n$PI EA \cite{jon64,ded64,cor74}. While the diagrammatic representation of $Z[J]$ consists of vacuum diagrams involving the \emph{bare} propagator and vertices of the theory, the $n$PI EA provides a systematic method to perform non-perturbative resummations on the $m$-point correlation functions ($m\leq n$) of the theory, yielding a diagrammatic series in terms of the \emph{dressed} propagator and $m$-point vertex functions with $m\leq n$. The diagrammatic series expressing a $n$PI EA involve $n$PI diagrams only, hence the name ``$n$PI'' EA. In that regard, we point out the works of Vasil'ev and collaborators \cite{vas72,vas73,vas73bis,vas74,pis74,vas74bis}, which prove notably:
\begin{itemize}
\item The 1PI and 2PI natures of the diagrams expressing the 1PI and 2PI EAs, respectively \cite{vas72}.
\item The close relation between the Schwinger-Dyson equations \cite{dys49,sch51} and the gap equations of the $n$PI EA formalism \cite{vas73}.
\item The convexity of $n$PI EAs with respect to each of their arguments \cite{vas73bis}.
\item Expressions of $n$PI vertices in terms of 1PI vertices \cite{vas74}.
\item The 3PI nature of the diagrams expressing the 3PI EA \cite{pis74}.
\item The 4PI nature of the diagrams expressing the 4PI EA \cite{vas74bis}.
\end{itemize}
The 2PI EA framework\footnote{See Ref.~\cite{ber04} for a pedagogical introduction on the 2PI EA.} is a direct reformulation of the Green's function formalism based on Dyson equation and the Luttinger-Ward functional \cite{dys49,lut60}. It was pioneered by the work of Lee, Yang, De Dominicis and others in statistical physics \cite{lee60,lut60,ded62,ded64,ded64bis,bay62}, and subsequently extended by Cornwall, Jackiw and Tomboulis \cite{cor74} to the framework of field theory discussed here, which is why the 2PI EA approach is also coined as CJT formalism.

Regarding the mathematical definitions of $n$PI EAs, one first introduces (external) sources $J_{\alpha}$, $K_{\alpha_1 \alpha_2}$, $L^{(3)}_{\alpha_1 \alpha_2 \alpha_3}$, ... , $L^{(n)}_{\alpha_1 \cdots \alpha_n}$ coupled to the local field $\widetilde{\varphi}_{\alpha}$ and the composite bilocal field $\widetilde{\varphi}_{\alpha_1}\widetilde{\varphi}_{\alpha_2}$, trilocal field $\widetilde{\varphi}_{\alpha_1}\widetilde{\varphi}_{\alpha_2}\widetilde{\varphi}_{\alpha_3}$, ..., $n$-local field $\widetilde{\varphi}_{\alpha_1}\cdots\widetilde{\varphi}_{\alpha_n}$, respectively:
\begin{equation}
\begin{split}
S_{JKL^{(3)}\cdots L^{(n)}}[\widetilde{\varphi}] \equiv & \ S\left[\widetilde{\varphi}\right] - J_{\alpha} \widetilde{\varphi}_{\alpha} - \frac{1}{2}  K_{\alpha_{1} \alpha_{2}} \widetilde{\varphi}_{\alpha_{1}} \widetilde{\varphi}_{\alpha_{2}} \\
& - \frac{1}{3!}  L^{(3)}_{\alpha_{1} \alpha_{2} \alpha_{3}} \widetilde{\varphi}_{\alpha_{1}} \widetilde{\varphi}_{\alpha_{2}}\widetilde{\varphi}_{\alpha_{3}}-\cdots \\
& - \frac{1}{n!}  L^{(n)}_{\alpha_1\cdots \alpha_n} \widetilde{\varphi}_{\alpha_1}\cdots \widetilde{\varphi}_{\alpha_n} \;,
\end{split}
\label{eq:GeneralSJKL}
\end{equation}
and
\begin{equation}
\begin{split}
Z\big[J,K,L^{(3)},\cdots\big] \equiv & \ e^{\frac{1}{\hbar}W\left[J,K,L^{(3)},\cdots\right]} \\
= & \ \mathcal{N}\int_\mathcal{C}\mathcal{D}\widetilde{\varphi} \ e^{-\frac{1}{\hbar}S_{JKL^{(3)}\cdots}[\widetilde{\varphi}]} \;.
\end{split}
\end{equation}
The $n$PI EA $\Gamma^{(n\mathrm{PI})}$ is then obtained after Legendre transforming the Schwinger functional with respect to the sources:
\begin{equation}
\begin{split}
\Gamma^{(n\mathrm{PI})} [\phi,G,V,\cdots] = & -W\big[J,K,L^{(3)},\cdots\big] \\
& + J_{\alpha} \frac{\delta W\big[J,K,L^{(3)},\cdots\big]}{\delta J_{\alpha}} \\
& + K_{\alpha_{1}\alpha_{2}} \frac{\delta W\big[J,K,L^{(3)},\cdots\big]}{\delta K_{\alpha_{1}\alpha_{2}}} \\
& + L^{(3)}_{\alpha_{1}\alpha_{2}\alpha_{3}} \frac{\delta W\big[J,K,L^{(3)},\cdots\big]}{\delta L^{(3)}_{\alpha_{1}\alpha_{2}\alpha_{3}}} + \cdots \\
= & -W\big[J,K,L^{(3)},\cdots\big] + J_{\alpha} \phi_{\alpha} \\
& + \frac{1}{2} K_{\alpha_{1}\alpha_{2}} \left(\phi_{\alpha_{1}}\phi_{\alpha_{2}} + \hbar G_{\alpha_{1}\alpha_{2}}\right) \\
& +\frac{1}{6}L^{(3)}_{\alpha_{1}\alpha_{2}\alpha_{3}}\big(\phi_{\alpha_{1}}\phi_{\alpha_{2}}\phi_{\alpha_{3}} \\
& + \hbar G_{\alpha_{1}\alpha_{2}}\phi_{\alpha_{3}} + \hbar G_{\alpha_{1}\alpha_{3}}\phi_{\alpha_{2}} \\
& + \hbar G_{\alpha_{2}\alpha_{3}}\phi_{\alpha_{1}} + \hbar^2 V_{\alpha_{1}\alpha_{2}\alpha_{3}}\big) + \cdots \;,
\end{split}
\end{equation}
where the 1-point correlation function $\phi_{\alpha}$, the propagator $G_{\alpha_{1}\alpha_{2}}$, the 3-point vertex $V_{\alpha_{1}\alpha_{2}\alpha_{3}}$, ... satisfy:
\begin{equation}
\frac{\delta W\big[J,K,L^{(3)},\cdots\big]}{\delta J_{\alpha}} = \phi_{\alpha} \;,
\end{equation}
\begin{equation}
\frac{\delta W\big[J,K,L^{(3)},\cdots\big]}{\delta K_{\alpha_{1}\alpha_{2}}} = \frac{1}{2}\left[\phi_{\alpha_{1}} \phi_{\alpha_{2}}+\hbar G_{\alpha_{1}\alpha_{2}}\right] \;,
\end{equation}
\begin{equation}
\begin{split}
\frac{\delta W\big[J,K,L^{(3)},\cdots\big]}{\delta L^{(3)}_{\alpha_{1}\alpha_{2}\alpha_{3}}} = & \ \frac{1}{6}\Big[\phi_{\alpha_{1}}\phi_{\alpha_{2}}\phi_{\alpha_{3}} + \hbar G_{\alpha_{1}\alpha_{2}}\phi_{\alpha_{3}} \\
& + \hbar G_{\alpha_{1}\alpha_{3}}\phi_{\alpha_{2}} + \hbar G_{\alpha_{2}\alpha_{3}}\phi_{\alpha_{1}} \\
& +\hbar^2 V_{\alpha_{1}\alpha_{2}\alpha_{3}}\Big] \;,
\end{split}
\end{equation}
\begin{equation*}
\vdots
\end{equation*}
Within the $n$PI EA framework, the physical $m$-point functions (i.e. the $m$-point functions at vanishing sources) of the theory with $m\leq n$ are self-consistently dressed through a variational principle, i.e. by solving the gap equations:
\begin{equation}
\left.\frac{\delta\Gamma^{(n\mathrm{PI})}[\phi,G,V,\cdots]}{\delta\phi_{\alpha}}\right|_{\phi=\overline{\phi},G=\overline{G},V=\overline{V},\cdots} = 0 \quad \forall \alpha \;,
\label{eq:GapEquationphi}
\end{equation}
\begin{equation}
\left.\frac{\delta\Gamma^{(n\mathrm{PI})}[\phi,G,V,\cdots]}{\delta G_{\alpha_{1}\alpha_{2}}}\right|_{\phi=\overline{\phi},G=\overline{G},V=\overline{V},\cdots} = 0 \quad \forall \alpha_{1},\alpha_{2} \;,
\label{eq:GapEquationG}
\end{equation}
\begin{equation}
\left.\frac{\delta\Gamma^{(n\mathrm{PI})}[\phi,G,V,\cdots]}{\delta V_{\alpha_{1}\alpha_{2}\alpha_{3}}}\right|_{\phi=\overline{\phi},G=\overline{G},V=\overline{V},\cdots} = 0 \quad \forall \alpha_{1},\alpha_{2},\alpha_{3} \;,
\label{eq:GapEquationV}
\end{equation}
\begin{equation*}
\vdots
\end{equation*}
while the higher $m$-point functions (with $m>n$) coincide with the bare ones.

Basic properties about the system of interest can be obtained from the above generating functionals. Among these, we have notably:
\begin{itemize}
\item The gs energy $E_{\mathrm{gs}}$ of the interacting system:
\begin{equation}
\begin{split}
E_{\mathrm{gs}} = & \ \underset{T\rightarrow 0}{\lim} \left(-T \ln( Z[J=0,\cdots])\right) \\
= & \ \underset{T\rightarrow 0}{\lim} \left(-\frac{T}{\hbar} W\left[J=0,\cdots\right]\right) \\
= & \ \underset{T\rightarrow 0}{\lim} \left(\frac{T}{\hbar} \Gamma^{(n\mathrm{PI})}\big[\phi=\overline{\phi},\cdots\big]\right) \;.
\end{split}
\label{eq:Ener}
\end{equation}

\item The gs density $\rho_{\mathrm{gs}}$ of the interacting system:
\begin{equation}
\begin{split}
\rho_{\mathrm{gs}}(r,\tau) = & \ \braket{\widetilde{\varphi}_{\alpha}\widetilde{\varphi}_{\alpha}}_{\mathrm{vac}} \\
= & \ \frac{\hbar^2}{Z[J=0,K=0,\cdots]} \\
& \times \left.\frac{\delta^2 Z[J,K,\cdots]}{\delta J_{\alpha} \delta J_{\alpha}}\right|_{J=0,K=0,\cdots} \\
= & \ \hbar\left.\frac{\delta^2 W\left[J,K,\cdots\right]}{\delta J_{\alpha} \delta J_{\alpha}}\right|_{J=0,K=0,\cdots}+\overline{\phi}_{\alpha} \overline{\phi}_{\alpha} \\
= & \ 2\left.\frac{\delta W[J,K,\cdots]}{\delta K_{\alpha\alpha}}\right|_{J=0,K=0,\cdots} \;.
\end{split}
\label{eq:Dens}
\end{equation}

\item The effective potential $V_\text{eff}(\phi)$, which is determined from the 1PI EA evaluated at a uniform (i.e. spacetime-independent) field configuration $\phi_{\mathrm{u}}$ after factorizing the volume of Euclidean spacetime \cite{col73}:
\begin{equation}
\Gamma^{(\mathrm{1PI})}[\phi_{\mathrm{u}}] = \int d\tau d^{D-1}r \ V_{\mathrm{eff}}(\phi_{\mathrm{u}}) \;.
\label{eq:effpot}
\end{equation}
\end{itemize}

%%%%%%%%%%%%%%%%%%%%%%%%%%%%%%%
%%%%%%%%%%%%%%%%%%%%%%%%%%%%%%%
\section{\label{app:1overNexpansion0DON}$1/N$-expansion for the (0+0)-D $O(N)$-symmetric $\varphi^{4}$-theory}

We summarize below the derivation of the $1/N$-expansion for the (0+0)\nobreakdash-D $O(N)$-symmetric $\varphi^{4}$-theory given by Keitel and Bartosch in Ref.~\cite{kei12} which is also exploited by Rosa \textit{et al.} in Ref.~\cite{ros16}. We will pursue the derivations up to higher orders for the gs density and also explain how to extend their results in the broken-symmetry regime. Let us start by considering the partition function of the studied toy model:
\begin{equation}
Z = \int_{\mathbb{R}^{N}} d^{N}\vec{\widetilde{\varphi}} \ e^{-S\big(\vec{\widetilde{\varphi}}\big)} = \int_{\mathbb{R}^{N}} d^{N}\vec{\widetilde{\varphi}} \ e^{-\frac{m^{2}}{2}\vec{\widetilde{\varphi}}^{2}-\frac{\lambda}{4!}\left(\vec{\widetilde{\varphi}}^{2}\right)^{2}} \;,
\label{eq:1overNpartitionfunction}
\end{equation}
where we have set $\hbar=1$, which is a convention followed throughout the entire appendix~\ref{app:1overNexpansion0DON}. In the limit $N\rightarrow\infty$, our zero-dimensional toy model possesses an infinite number of dofs and fluctuations are therefore suppressed in this case. This translates into the constraint that the integrand vanishes in Eq.~\eqref{eq:1overNpartitionfunction} in this limit, which can be achieved by imposing that the classical action $S\big(\vec{\widetilde{\varphi}}\big)$ is of order $\mathcal{O}(N)$. In this way, we deduce that $\vec{\widetilde{\varphi}}^{2}=\mathcal{O}(N)$ since the mass $m$ does not depend on $N$. In order to ensure dimensional consistency, we infer from this that $\lambda=\mathcal{O}\big(N^{-1}\big)$. This dimensional analysis suggests to introduce the $N$-independent quantities $\widetilde{y}=N^{-1}\vec{\widetilde{\varphi}}^{2}$ and $\breve{\lambda}\equiv N\lambda$. After defining the norm $\widetilde{u}\equiv\Big|\vec{\widetilde{\varphi}}\Big|=\sqrt{N\widetilde{y}}$, we rewrite Eq.~\eqref{eq:1overNpartitionfunction} in hyperspherical coordinates as follows:
\begin{equation}
\begin{split}
Z^{\text{$1/N$-exp}} = & \ \Omega_{N} \int_{0}^{\infty} d\widetilde{u} \ \widetilde{u}^{N-1} e^{-\frac{m^{2}}{2}\widetilde{u}^{2}-\frac{\lambda}{4!}\widetilde{u}^{4}} \\
= & \ \Omega_{N} N^{\frac{N-1}{2}} \\
& \times \int_{0}^{\infty} \left(\frac{1}{2}\sqrt{\frac{N}{\widetilde{y}}} \ d\widetilde{y}\right) \widetilde{y}^{\frac{N-1}{2}} e^{-N\left(\frac{m^{2}}{2}\widetilde{y}+\frac{\breve{\lambda}}{4!}\widetilde{y}^{2}\right)} \\
= & \ \frac{1}{2} \Omega_{N} N^{\frac{N}{2}} \int_{0}^{\infty} \frac{d\widetilde{y}}{\widetilde{y}} \ \widetilde{y}^{\frac{N}{2}} e^{-N f(\widetilde{y})} \;,
\end{split}
\label{eq:1overNintroducefy}
\end{equation}
where $\Omega_{N}$ is expressed by Eq.~\eqref{eq:SN} and we have notably used the following definition in the last line:
\begin{equation}
f\big(\widetilde{y}\big) \equiv \frac{m^{2}}{2}\widetilde{y} + \frac{\breve{\lambda}}{4!}\widetilde{y}^{2} -\frac{1}{2}\ln(\widetilde{y})\;.
\label{eq:1overNdefinitiony}
\end{equation}
At this stage, we are in a situation very similar to that of the starting point of the original LE in Eq.~\eqref{eq:ZJKfiniteD}, except that the role of the expansion parameter is now played by $1/N$ instead of $\hbar$. Hence, we will carry out a saddle point approximation by solving:
\begin{equation}
\left.\frac{\partial f\big(\widetilde{y}\big)}{\partial\widetilde{y}}\right|_{\widetilde{y}=\overline{y}} = 0\;,
\end{equation}
which has two solutions, whose physical relevance depends on the value of $m^{2}$ (and that of $\lambda$), as in Eq.~\eqref{eq:SolutionsmodulusrhoLE} (combined with Eq.~\eqref{eq:DefPhiclModulusRho}). We find for $\lambda\neq 0$:
\begin{equation}
\overline{y} = \left\{
\begin{array}{lll}
		\displaystyle{\frac{3 m^{2}}{\breve{\lambda}}\left(\sqrt{1+\frac{2\breve{\lambda}}{3 m^{4}}} - 1\right) \quad \forall m^2 > 0\;.} \\
		\\
		\displaystyle{\frac{3 m^{2}}{\breve{\lambda}}\left(-\sqrt{1+\frac{2\breve{\lambda}}{3 m^{4}}} - 1\right) \quad \forall m^2 < 0\;.}
    \end{array}
\right.
\label{eq:1overNsaddlepointapprox0DON}
\end{equation}
Only the upper solution was exploited in Refs.~\cite{kei12} and \cite{ros16} as these works focus on the unbroken-symmetry phase in their applications of the $1/N$-expansion. Then, the rest of the procedure is also quite similar to that of a LE. We Taylor expand $f\big(\widetilde{y}\big)$ as well as $1/\widetilde{y}$ around $\widetilde{y}=\overline{y}$ up to a chosen order in $\widetilde{y}-\overline{y}$ and finally expand the partition function thus obtained. If $f\big(\widetilde{y}\big)$ and $1/\widetilde{y}$ are respectively expanded up to fourth and second orders in $\widetilde{y}-\overline{y}$, this leads to:
\begin{equation}
\begin{split}
Z^{\text{$1/N$-exp}} = & \ \Omega_{N}N^{\frac{N}{2}}\sqrt{\frac{2\pi}{4\overline{y}^{2}\overline{f}^{(2)}}} \ e^{-N \overline{f}} \\
& \times \left[1+\frac{12 m^{4} \overline{y}^{2}-27m^{2}\overline{y}+16}{6N\left(2-m^{2}\overline{y}\right)^{3}}\right] \\
& \times \left[1+\mathcal{O}\hspace{-0.07cm}\left(\frac{1}{N}\right)\right]\;,
\end{split}
\label{eq:1overNZseries0DON}
\end{equation}
where the function $f$ and its second-order derivative can be rewritten as follows when evaluated at $\widetilde{y}=\overline{y}$:
\begin{equation}
\overline{f} \equiv f\big(\widetilde{y}=\overline{y}\big) = \frac{m^{2}}{4}\overline{y}+\frac{1}{4}-\frac{1}{2}\ln(\overline{y})\;,
\end{equation}
\begin{equation}
\overline{f}^{(2)} \equiv \left.\frac{\partial^{2} f\big(\widetilde{y}\big)}{\partial\widetilde{y}^{2}}\right|_{\widetilde{y}=\overline{y}} = \frac{1}{\overline{y}^{2}}-\frac{m^{2}}{2\overline{y}}\;.
\end{equation}
The gs energy $E^{\text{$1/N$-exp}}_{\mathrm{gs}}=-\ln\big(Z^{\text{$1/N$-exp}}\big)$ can be expressed by expanding $\ln\big(Z^{\text{$1/N$-exp}}\big)$, with $Z^{\text{$1/N$-exp}}$ given by Eq.~\eqref{eq:1overNZseries0DON}:
\begin{equation}
\begin{split}
E^{\text{$1/N$-exp}}_{\mathrm{gs}} = & \ N\Bigg[\frac{m^{2}}{4}\overline{y}-\frac{1}{4}-\frac{1}{2}\ln(2\pi\overline{y})\Bigg]+\frac{1}{2}\ln\hspace{-0.07cm}\Big(2-m^{2}\overline{y}\Big) \\
& -\frac{1}{N}\Bigg[\frac{\left(8+m^{2}\overline{y}\right)\left(m^{2}\overline{y}-1\right)^{2}}{6\left(2-m^{2}\overline{y}\right)^{3}}\Bigg] \\
& +\mathcal{O}\hspace{-0.07cm}\left(\frac{1}{N^{2}}\right)\;,
\end{split}
\label{eq:1overNEgsseries0DON}
\end{equation}
and the gs density is determined by differentiating Eq.~\eqref{eq:1overNEgsseries0DON} with respect to $m^{2}$ according to the relation\footnote{The dependence of $\overline{y}$ with respect to $m^{2}$, which is set by Eq.~\eqref{eq:1overNsaddlepointapprox0DON}, must be taken into account when differentiating $E^{\text{$1/N$-exp}}_{\mathrm{gs}}$ to deduce $\rho^{\text{$1/N$-exp}}_{\mathrm{gs}}$.} $\rho^{\text{$1/N$-exp}}_{\mathrm{gs}}=\frac{2}{N}\frac{\partial E^{\text{$1/N$-exp}}_{\mathrm{gs}}}{\partial m^{2}}$ (which follows from Eqs.~\eqref{eq:DefEgsExactZexact0DON} to~\eqref{eq:vacuumExpectationValue0DON}):
\begin{equation}
\begin{split}
& \rho^{\text{$1/N$-exp}}_{\mathrm{gs}} \\
& = \overline{y} +\frac{1}{N}\left[\frac{\frac{2}{\frac{\breve{\lambda}}{3}\overline{y}+m^{2}}-2\overline{y}}{2-m^{2}\overline{y}}\right] \\
& \hspace{0.3cm} +\frac{1}{N^{2}}\left[\frac{4\left(1-m^{2}\overline{y}\right)\left(1+2m^{2}\overline{y}\right)\left(-\sqrt{3}+m^{2}\overline{y}+\frac{\breve{\lambda}}{3}\overline{y}^{2}\right)}{\left(\frac{\breve{\lambda}}{3}\overline{y}+m^{2}\right)\left(-2+m^{2}\overline{y}\right)^{4}}\right] \\
& \hspace{0.3cm} + \mathcal{O}\hspace{-0.07cm}\left(\frac{1}{N^{3}}\right)\;.
\end{split}
\label{eq:1overNrhogsseries0DON}
\end{equation}
For $m^{2}>0$, the first-order coefficient in Eq.~\eqref{eq:1overNrhogsseries0DON} reduces to the free gs density (i.e. the exact gs density at $\lambda=0$) in the limit of vanishing coupling constant, i.e.:
\begin{equation}
\textcolor{white}{\quad \forall m^{2}>0} \underset{\lambda \rightarrow 0}{\lim} ~ \overline{y} = \frac{1}{m^{2}} \quad \forall m^{2}>0\;,
\end{equation}
as expected. Note that an alternative derivation of result~\eqref{eq:1overNEgsseries0DON} was developed by Schelstraete and Verschelde in Ref.~\cite{sch94}.

%%%%%%%%%%%%%%%%%%%%%%%%%%%%%%%
%%%%%%%%%%%%%%%%%%%%%%%%%%%%%%%
\section{\label{app:Resum}Resummation procedures}
%~~~
\subsection{\label{app:PBResum}Pad\'e-Borel resummation}

We first briefly present the Pad\'e-Borel-Le Roy resummation~\cite{pad1892,ler1900,bor28,ell96,ben99,kle01}. To that end, let us first consider a generic physical quantity of interest $P(g)$, represented by a factorially divergent asymptotic series:
\begin{equation}
P(g)\sim\sum_{n=0}^\infty p_n g^n \;,
\label{eq:P}
\end{equation}
and its Borel transform:
\begin{equation}
\mathcal{B}[P](\zeta) = \sum_{n=0}^\infty \frac{p_n}{\Gamma(n+1)}\zeta^n \quad \forall\zeta\in\mathbb{C} \;,
\label{eq:BtransFirst}
\end{equation}
which is a specific case of the Borel-Le Roy transform:
\begin{equation}
\mathcal{B}_s[P](\zeta) = \sum_{n=0}^\infty \frac{p_n}{\Gamma(n+s+1)}\zeta^n \quad \forall s\in\mathbb{R}, \forall\zeta\in\mathbb{C} \;.
\label{eq:Btrans}
\end{equation}
From this, one can compute the inverse Borel-Le Roy transform:
\begin{equation}
P_{\mathcal{B}_s}(g) = \int_0^{\infty} d\zeta\ \zeta^s e^{-\zeta}\mathcal{B}_s[P](g\zeta) \;,
\label{eq:Bsum}
\end{equation}
whose definition derives from the identity:
\begin{equation}
1=\frac{\int_0^\infty d\zeta \ \zeta^{n+s} e^{-\zeta}}{\Gamma(n+s+1)} \;.
\end{equation}
In practice, we only have access to a few $p_{n}$ coefficients and therefore to approximated versions of $P(g)$. The idea behind Pad\'e-Borel-Le Roy resummation is to approximate the Borel-Le Roy transform $\mathcal{B}_s[P]$ with a Pad\'e approximant $\mathcal{P}^{U/V}\mathcal{B}_s[P]$, which is a rational function and can therefore develop a richer analytic behavior (with singularities in particular) as compared to the polynomial representing $\mathcal{B}_s[P]$ initially. The Pad\'e approximant $\mathcal{P}^{U/V}\mathcal{B}_s[P]$ is constructed from the knowledge of the partial sum representing $\mathcal{B}_s[P]$ up to order $M$ as:
\begin{equation}
\mathcal{P}^{U/V}\mathcal{B}_s[P](\zeta) = \frac{\sum_{n=0}^U a_n \zeta^n}{1+\sum_{n=0}^V b_n \zeta^n} \;,
\label{eq:PBL}
\end{equation}
with $U+V=M$. The coefficients $\{a_{n}\}$ and $\{b_{n}\}$ are fixed by equating order by order the Taylor series of the right-hand side of Eq.~\eqref{eq:PBL} with the expansion~\eqref{eq:Btrans}, up to the desired order. The original function $P(g)$ is estimated after substituting the Borel-Le Roy sum $\mathcal{B}_s[P]$ by its Pad\'e approximant $\mathcal{P}^{U/V}\mathcal{B}_s[P]$ in the integral of Eq.~\eqref{eq:Bsum}. For our numerical applications, we have focused on the Pad\'e-Borel resummation, i.e. on the Pad\'e-Borel-Le Roy resummation with $s=0$, for which we define $\mathcal{P}\mathcal{B}_{P}[U/V]\equiv\mathcal{P}^{U/V}\mathcal{B}_{s=0}[P]$. We point out however that there exists recent studies, such as that of Ref.~\cite{adz19}, discussing the determination of optimal values for the $s$ parameter within the framework of Pad\'e-Borel-Le Roy resummation, which is a task that we defer to future works for the $O(N)$ model considered in the presented study.

%~~~
\subsection{\label{app:ConfMappResum}Conformal mapping}

While the Pad\'e-Borel(-Le Roy) resummation procedure only involves the knowledge about the first terms of the initial perturbative series~\eqref{eq:P}, the method of Borel(-Le Roy) transform with conformal mapping \cite{leg80} aims at more reliable results by incorporating in addition the knowledge on the large-order behavior of $p_n$, which is of the form:
\begin{equation}
p_n\underset{n\rightarrow\infty}{\sim}(-1)^n n! a^n n^b c \left(1+\mathcal{O}\left(\frac{1}{n}\right)\right) \;.
\label{eq:LOB}
\end{equation}
The parameters $a$, $b$ and $c$ in Eq.~\eqref{eq:LOB} can be computed via, e.g., the Lipatov method where the coefficients $p_n$ are represented by the contour integral $p_n=\frac{1}{2\pi i}\oint_\mathcal{C} d\zeta\frac{P(\zeta)}{\zeta^{n+1}}$ calculated for large $n$ through steepest descent. In particular, the coefficient $a$ determines the position of the singularity of the Borel-Le Roy transform~\eqref{eq:Btrans} which is the closest to the origin, i.e. $\mathcal{B}_s[P](\zeta)$ is analytical in a circle of radius $1/a$, with $a=2/3$ ($a=4/3$) at $N=1$ ($N=2$) for the studied (0+0)\nobreakdash-D $O(N)$ model. One of the methods for continuing the Borel-Le Roy transform beyond its circle of convergence, and (to a certain extent) accelerating the convergence of $P_{\mathcal{B}_s}(g)$, relies on the conformal mapping of the Borel plane:
\begin{equation}
\begin{cases} \displaystyle{w(\zeta) = \frac{\sqrt{1+a\zeta}-1}{\sqrt{1+a\zeta}+1} \;.} \\
\\
\displaystyle{\zeta = \frac{4}{a}\frac{w}{(1-w)^2} \;.} \end{cases}
\end{equation}
Under this transformation, a point in the $\zeta$-complex plane is mapped within a disk of unit radius $|w|=1$. In particular, the origin is left invariant and the branch-cut singularity $\zeta\in]-\infty,-1/a]$ is mapped to the boundary of the $w$-unit disk, thus turning the Taylor expansion of the function $\widetilde{\mathcal{B}}_s[P](w)\equiv \mathcal{B}_s[P](\zeta(w))$ into a convergent one for $|w|<1$. The original quantity $P(g)$ is then determined after re-expanding $\mathcal{B}_s[P](g\zeta)$ in the new variable $w(g\zeta)$ in Eq.~\eqref{eq:Bsum}, i.e. after writing (assuming that only the first $M$ terms of the original series are known):
\begin{equation}
\mathcal{B}_s[P](\zeta) = \sum_{n=0}^M W_n \left(w(\zeta)\right)^n \;,
\end{equation}
with
\begin{equation}
W_n = \sum_{k=0}^n \frac{p_k}{\Gamma(k+s+1)} \left(\frac{4}{a}\right)^k \frac{(k+n-1)!}{(n-k)!(2k-1)!} \;,
\end{equation}
and we have also considered only the situation with $s=0$ for our calculations.

%~~~
\subsection{\label{app:BorelHyperGeoResum}Borel-hypergeometric resummation}

As was stressed in section~\ref{sec:Resummation}, the Borel-hypergeometric resummation \cite{mer15,mer16,ped16a,ped16b,san17,mer18,ant19} is based on approximants that can exhibit branch cuts via hypergeometric functions, and therefore constitutes an alternative to all resummation methods based on Pad\'e approximants, as the latter can only display poles. Regardless of the model under consideration, the recipe underlying the Borel-hypergeometric resummation procedure can be presented as follows \cite{mer18,ant19}:
\begin{enumerate}
\item[1.] As for the other resummation techniques, starting from the asymptotic series representing $P(g)$ truncated at an odd order $M$ (the case of even truncation orders will be discussed subsequently), one first computes the coefficients $b_n\equiv p_n/n! = p_n/\Gamma(n+1)$ of the Borel transform $\mathcal{B}_{P}\equiv\mathcal{B}_{s=0}[P]$.

\item[2.] One then computes the $M$ ratios $b_1/b_0,\cdots,b_M/b_{M-1}$ of two consecutive coefficients of the Borel series and makes the ansatz that such ratios $b_{n+1}/b_n$ are rational functions of $n$, coined as $r_M(n)$ and defined as:
\begin{equation}
r_M(n)\equiv\frac{\sum_{k=0}^{l}u_k n^k}{1+\sum_{k=1}^{l}v_k n^k} \;,
\end{equation}
with $l=(M-1)/2$. The $\frac{M+1}{2}+\frac{M-1}{2}=M$ unknowns $u_k$ and $v_k$ are determined from the $M$ equations:
\begin{equation}
\frac{b_{n+1}}{b_n} = r_M(n) \;,
\end{equation} 
where $n$ runs from $0$ to $M-1$. 

\item[3.] Hypergeometric vectors $x=(1,-x_1,\cdots,-x_l)$ and $y=(-y_1,\cdots,-y_l)$ are then constructed via the equations:
\begin{equation}
\sum_{k=0}^l u_k x^k = 0 \;,
\end{equation}
\begin{equation}
1+\sum_{k=1}^l v_k y^k = 0 \;,
\end{equation}
and used to define the hypergeometric approximant of the Borel transform $\mathcal{B}_{P}$ in terms of the generalized hypergeometric function:
\begin{equation}
\mathcal{H}^M\mathcal{B}_{P}(\zeta) \equiv \prescript{}{l+1}{F}_l\bigg(x,y,\frac{u_l}{v_l}\zeta\bigg) \;.
\end{equation}

\item[4.] One finally recovers the original function $P(g)$ through an inverse Borel transform, which can be represented in terms of a Meijer G-function $G^{m,n}_{p,q}
\Big(\begin{smallmatrix}
a_1,\cdots,a_p \\
b_1,\cdots,b_q
\end{smallmatrix}\Big| z \Big)$, i.e.:
\begin{equation}
\begin{split}
P_\mathcal{HB}(g) = & \ \int_0^{\infty} d\zeta \ e^{-\zeta}\mathcal{H}^M\mathcal{B}_{P}(g\zeta) \\
= & \ \frac{\prod_{k=1}^l\Gamma(-y_k)}{\prod_{k=1}^l\Gamma(-x_k)} \\
& \times G^{l+2,1}_{l+1,l+2}
\bigg(\begin{smallmatrix}
1,-y_1,\cdots,-y_l \\
1,1,-x_1,\cdots,-x_l
\end{smallmatrix}\bigg|
-\frac{v_l}{u_l g}\bigg) \;.
\end{split}
\end{equation}
\end{enumerate}
For an even truncation order $M$, one first subtracts the constant zeroth-order term from the original series, then factors out the first-order term and finally follows the above recipe on the resulting series with an odd ($M-1$) truncation order. The final answer is obtained after re-multiplying the resummed series by the first-order term and re-adding the constant.

We then examine in more detail the results presented in Fig.~\ref{fig:MeijerON} for the studied toy model, and more specifically those obtained at $N=2$ and $4$ that were found to be exact. In other words, we found that, at $N=2$ or $4$ and in the unbroken-symmetry regime, the partition function of the (0+0)\nobreakdash-D $O(N)$ model is exactly reproduced at third order of the original (or mixed) LE treated via Borel-hypergeometric resummation. In the unbroken-symmetry regime (i.e. in the phase where $m^{2}>0$), the original LE series representing the partition function of this $O(N)$ model at vanishing sources reads (see appendix~\ref{app:DiagramsOrigLEOPT} and notably Eq.~\eqref{eq:ResultZLoopExpansion0DONAppendix} for a more general version of Eq.~\eqref{eq:ResultZLE0DONm2posAppendix}):
\begin{strip}
\begin{equation}
\begin{split}
Z^\text{LE;orig}\Big(\vec{J}=\vec{0},\boldsymbol{K}=\boldsymbol{0}\Big) = & \ \left(\frac{2\pi\hbar}{m^{2}}\right)^{\frac{N}{2}} \Bigg[ 1 - \frac{g}{24}\big(2N+N^2\big) + \frac{g^2}{1152} \big(48N+44N^2+12N^3+N^4\big) \\
& - \frac{g^3}{82944} \big(3840N+4384N^2+1800N^3+340N^4 + 30N^5 + N^6\big) + \mathcal{O}\Big(g^3\Big) \Bigg] \;,
\end{split}
\label{eq:ResultZLE0DONm2posAppendix}
\end{equation}
with $g\equiv \hbar\lambda/m^4$. Following the recipe outlined in the present appendix, treating the series~\eqref{eq:ResultZLE0DONm2posAppendix} up to order $\mathcal{O}\big(g^{3}\big)$ with Borel-hypergeometric resummation leads to:
\begin{equation}
\begin{split}
Z^{\text{LE;orig}}_\mathcal{HB}(g) = & \ \left(\frac{2\pi\hbar}{m^{2}}\right)^{\frac{N}{2}} \frac{\Gamma\Big(\frac{104-18N-5N^2}{56+18N-11N^2}\Big)}{\Gamma\Big(\frac{N(2+N)(104-18N-5N^2)}{960+192N-124N^2-12N^3+N^4}\Big)} \\
& \times G^{3,1}_{2,3}
\Bigg(\begin{smallmatrix}
1,\frac{104-18N-5N^2}{56+18N-11N^2} \\
1,1,\frac{N(2+N)(104-18N-5N^2)}{960+192N-124N^2-12N^3+N^4}
\end{smallmatrix}\Bigg|
\frac{24\big(56+18N-11N^2\big)}{g\big(960+192N-124N^2-12N^3+N^4\big)}\Bigg) \;.
\end{split}
\label{eq:ResultZLE0DONm2posBorHypResum}
\end{equation}
\end{strip}
We then define the difference between result~\eqref{eq:ResultZLE0DONm2posBorHypResum} and the corresponding exact result $Z^\text{exact}$ for the partition function (expressed by Eqs.~\eqref{eq:Z0DONexactsolution} and~\eqref{eq:RN}) as:
\begin{equation}
\Delta Z^{\text{LE;orig}}_{\mathcal{HB}}(N) \equiv Z^{\text{LE;orig}}_\mathcal{HB}(g) - Z^\text{exact}\Big(\vec{J}=\vec{0},\boldsymbol{K}=\boldsymbol{0}\Big)\;.
\end{equation}
In the specific situations where $N=2$ or $4$, the Meijer G-function and the confluent hypergeometric functions, brought by the resummed result~\eqref{eq:ResultZLE0DONm2posBorHypResum} and by the exact one~\eqref{eq:Z0DONexactsolution}, can be replaced in $\Delta Z^{\text{LE;orig}}_{\mathcal{HB}}(N)$ (still for $m^{2}>0$) as follows:
\begin{equation}
\begin{split}
\Delta Z^{\text{LE;orig}}_{\mathcal{HB}}(N=2) = & \ \frac{\pi\hbar}{m^{2}} e^{\frac{3}{2g}} \sqrt{\frac{6}{g}} \Bigg[\sqrt{\pi}\Bigg(\mathrm{erf}\Bigg(\sqrt{\frac{3}{2g}}\Bigg) -1\Bigg) \\
& + \Gamma\bigg(\frac{1}{2},\frac{3}{2g}\bigg) \Bigg] \;,
\end{split}
\label{eq:DeltaZLEorigN2}
\end{equation}
\begin{equation}
\begin{split}
\Delta Z^{\text{LE;orig}}_{\mathcal{HB}}(N=4) = & \ \frac{3\pi^2 \hbar^2}{g^2 m^4} \Bigg\lbrace -4g \\
& + e^{\frac{3}{2g}} \sqrt{6g} \Bigg[2\sqrt{\pi}\Bigg(1 - \mathrm{erf}\Bigg(\sqrt{\frac{3}{2g}}\Bigg)\Bigg) \\
& + \Gamma\bigg(-\frac{1}{2},\frac{3}{2g}\bigg) \Bigg] \Bigg\rbrace \;,
\end{split}
\label{eq:DeltaZLEorigN4}
\end{equation}
where we have just introduced error functions and upper incomplete gamma functions~\cite{abr65}, denoted respectively as $\mathrm{erf}(z)$ and $\Gamma(z,x)$. By making use of the recurrence relation $\Gamma\big(z+1,x\big)=z\Gamma\big(z,x\big)+x^z e^{-x}$ and the property $\mathrm{erf}(x)=1-\frac{1}{\sqrt{\pi}}\Gamma\big(\frac{1}{2},x^{2}\big)$ (for $x$ real and positive), one can then directly prove from Eqs.~\eqref{eq:DeltaZLEorigN2} and~\eqref{eq:DeltaZLEorigN4} that:
\begin{equation}
\Delta Z^{\text{LE;orig}}_{\mathcal{HB}}(N=2) = \Delta Z^{\text{LE;orig}}_{\mathcal{HB}}(N=4) = 0 \;,
\end{equation}
which explains the exactness of the results shown in Fig.~\ref{fig:MeijerON} at $N=2$ and $4$.

%%%%%%%%%%%%%%%%%%%%%%%%%%%%%%%
%%%%%%%%%%%%%%%%%%%%%%%%%%%%%%%
\section{\label{app:Transseries}Transseries derived from Picard-Lefschetz theory}

Picard-Lefschetz theory provides an elegant framework for generating an ambiguous-free representation of a perturbative series \cite{wit10,tan15}. For the sake of simplicity, we illustrate its application to the studied zero-dimensional model at $N=1$. Extensions to higher-dimensional PIs and systems invariant under continuous symmetries are detailed, e.g., in Refs.~\cite{wit10,tan15}. Adding a multiplicative constant $1/\sqrt{\hbar}$ for later convenience, the partition function of the studied (0+0)\nobreakdash-D $O(N)$ model at $N=1$ reads:
\begin{equation}
Z(m^2,\lambda,\hbar) = \frac{1}{\sqrt{\hbar}}\int_\mathbb{R} d\widetilde{\varphi} \ e^{-\frac{1}{\hbar} S(\widetilde{\varphi})} \;,
\end{equation}         
with classical action:
\begin{equation}
S(\widetilde{\varphi}) = \frac{m^2}{2}\widetilde{\varphi}^2 + \frac{\lambda}{4!}\widetilde{\varphi}^4 \;.
\end{equation}
The analysis can be straightforwardly extended to more general integrals of the form:
\begin{equation}
\int_\mathbb{R} d\widetilde{\varphi} \ e^{-\frac{1}{\hbar} S(\widetilde{\varphi})-J\widetilde{\varphi}-\frac{K}{2}\widetilde{\varphi}^2} \;,
\end{equation}
or
\begin{equation}
\int_\mathbb{R} d\widetilde{\varphi}\ p(\widetilde{\varphi})  e^{-\frac{1}{\hbar} S(\widetilde{\varphi})} \;,
\end{equation}
with $p(\widetilde{\varphi})$ a polynomial in the field $\widetilde{\varphi}$. Redefining the field via $\widetilde{\varphi} \rightarrow \widetilde{\varphi}/\sqrt{\lambda}$ yields:
\begin{equation}
Z(m^2,g) = \frac{1}{\sqrt{g}}\int_\mathbb{R} d\widetilde{\varphi}\ e^{-\frac{1}{g} V(\widetilde{\varphi})} \;,
\label{eq:oriint}
\end{equation}         
with $g\equiv\hbar\lambda$ and
\begin{equation}
V(\widetilde{\varphi})\equiv \left.S(\widetilde{\varphi})\right|_{\lambda=1} =  \frac{m^2}{2}\widetilde{\varphi}^2 + \frac{1}{4!}\widetilde{\varphi}^4 \;.
\end{equation}
Even if the integral $Z(m^2,g)\equiv Z(m^2,\lambda,\hbar)$ is defined over real variables, the natural space in which the saddle (or critical) points of the action $V(\widetilde{\varphi})$ and their corresponding integration cycles live is the complexification of the original space. Understanding the behavior of $Z(m^2,g)$ for $g\in\mathbb{R}^+$ therefore passes by the study of its analytic continuation where $g=|g|e^{i\theta} \in\mathbb{C}$. The argument of the exponential in Eq.~\eqref{eq:oriint} becomes complex-valued, thus turning the partition function into a violently oscillating integral whose evaluation is difficult. One can significantly improve the properties of the integral by:
\begin{itemize}
\item Continuing the integrand into the complex plane, i.e. viewing the action $V(z)$ as a holomorphic function of the complex variable $z$, such that $Z$ now reads as an open contour integral:
\begin{equation}
Z(m^2,g) = \frac{1}{\sqrt{g}}\int_\mathcal{C} dz\ e^{-\frac{1}{g} V(z)} \;,
\label{eq:ACint}
\end{equation}
where $\mathcal{C}$ is a cycle with real dimension 1, coinciding with the real line when $g\in\mathbb{R}^+$.

\item Continuously deforming the integration domain as $g$ varies such that the integral~\eqref{eq:ACint} is convergent.

\end{itemize}

%%%%%%%%%%%%%%%%%%%
\begin{figure*}
\begin{center}
      \includegraphics[width=0.75\linewidth]{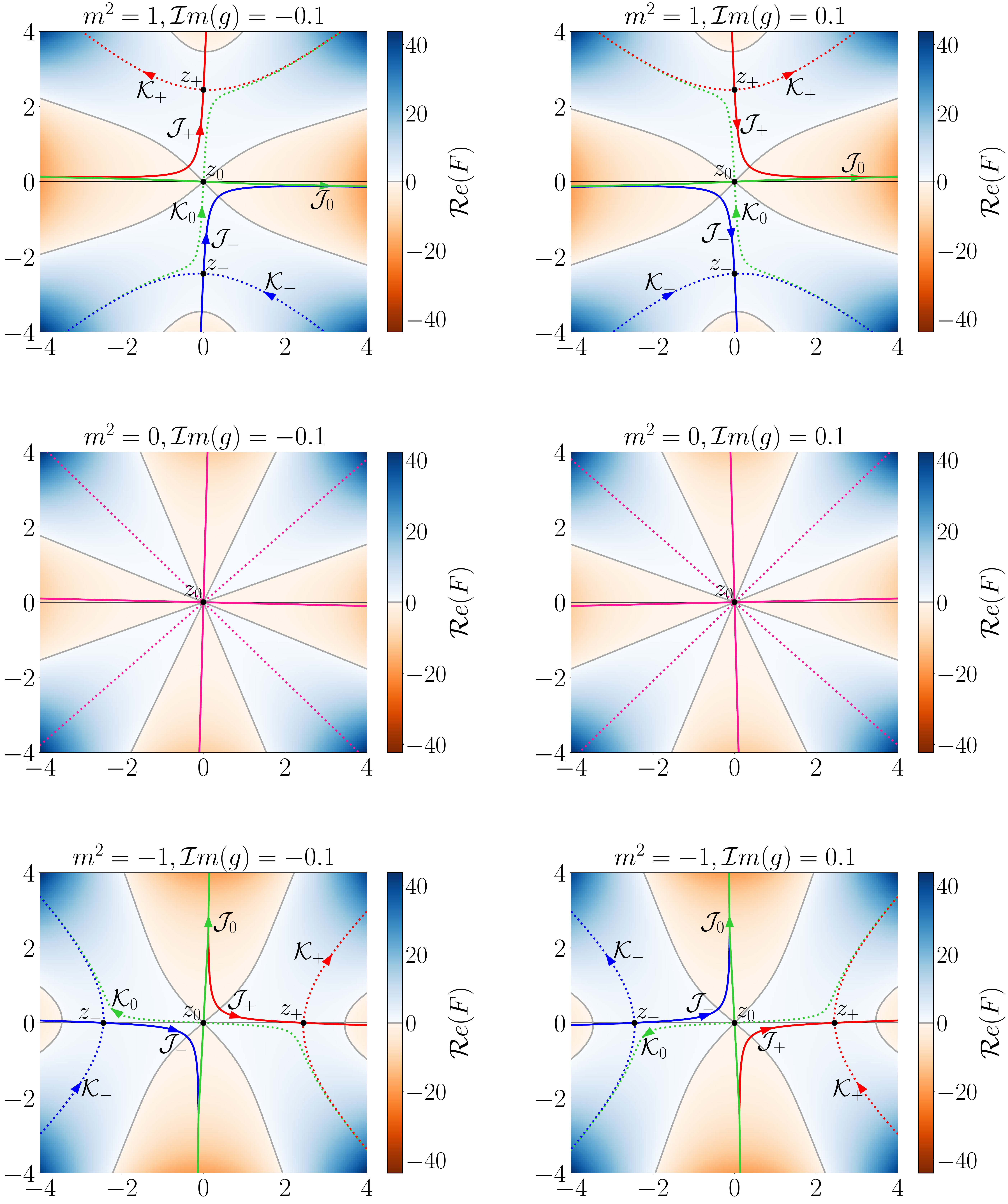}
      \end{center}
      \caption{Critical points (black dots) of $F(z)$ and their downward (solid lines) and  upward (dotted lines) flows  in  the $z$-complex plane, for $g= 1 - 0.1i$ (left column) and $g= 1+ 0.1i$ (right column),  and for  $m^2 = +1$ (upper panels), the degenerate case $m^2=0$ (middle panels) and  $m^2 = -1$ (lower panels). The value of $\mathcal{R}e(F)$ is given by the colormap. Recall that $g\equiv\hbar\lambda$ here.}
      \label{fig:PL}
\end{figure*}
%%%%%%%%%%%%%%%%%%%

Picard-Lefschetz integration method provides a decomposition of the integration cycle $\mathcal{C}$ into a linear combination $\mathcal{C} = \sum_i n_i \mathcal{J}_i$ (with $n_i\in \mathbb{Z}$) of nicer cycles (over which the integral is convergent) $\mathcal{J}_i$ attached to the saddle points $z^\star_i$ of $V(z)$, and obtained after solving the gradient flow (or steepest descent) equations:
\begin{equation}
\begin{cases} \displaystyle{\frac{\partial z}{\partial\tau} = - \frac{\partial \bar{F}}{\partial \bar{z}} \;,} \\
\\
\displaystyle{\frac{\partial \bar{z}}{\partial\tau} = - \frac{\partial F}{\partial z} \;,} \end{cases}
\label{eq:PLd}
\end{equation}
where $F(z)\equiv -V(z)/g$, $\tau$ is the flow parameter and the upper bars denote the complex conjugation. The cycles $\mathcal{J}_i$ are called Lefschetz thimbles or simply thimbles. Along the flow, $\mathcal{R}e(F)$ is strictly decreasing (except for the trivial solution that sits at a saddle point $z^\star_i$ for all $\tau$) and $\mathcal{I}m(F)$ is constant and equal to $\mathcal{I}m(F(z^\star_i))$. In the studied case, the saddle points of $V(z)$ are $z^\star_0 = 0$ and $z^\star_\pm = \pm\sqrt{-6m^2}$, and sit on the imaginary (real) axis when $m^2\geq 0$ ($m^2\leq 0$). Saddle points and solutions of Eqs.~\eqref{eq:PLd} are displayed in Fig.~\ref{fig:PL} for different values of $m^2$ and $g = 1 \pm 0.1i$.

The integer coefficients $n_i$ are found after considering the upward flows $\mathcal{K}_i$ (which are called anti-thimbles), solutions of the converse (steepest ascent) equations:
\begin{equation}
\begin{cases} \displaystyle{\frac{\partial z}{\partial\tau} =+\frac{\partial \bar{F}}{\partial \bar{z}} \;,} \\
\\
\displaystyle{\frac{\partial \bar{z}}{\partial\tau} = +\frac{\partial F}{\partial z} \;,} \end{cases}
\label{eq:PLu}
\end{equation}
shown as dotted lines in Fig.~\ref{fig:PL}. Along the anti-thimbles $\mathcal{K}_i$, $\mathcal{R}e(F)$ is monotonically increasing (making the integral divergent) and $\mathcal{I}m(F)$ is constant and equals to $\mathcal{I}m(F(z^\star_i))$. According to Picard-Lefschetz theory, $n_i$ corresponds to the intersection pairing of the original contour $\mathcal{C}$ and the upward flow $\mathcal{K}_i$. The partition function~\eqref{eq:ACint} can then be written as:
\begin{equation}
Z(m^2,g) = \sum_i n_i Z_i(m^2,g) \;,
\end{equation}
where 
\begin{equation}
Z_i(m^2,g) \equiv \frac{1}{\sqrt{g}}\int_{\mathcal{J}_i(\theta)} dz \ e^{-\frac{1}{g} V(z)} \;,
\end{equation}
admits an asymptotic power series expansion around the saddle point $z^\star_i$, which is Borel-summable to the exact result \cite{ber91} (recall that $\theta \equiv \mathrm{arg}(g)$).

As can be seen in Fig.~\ref{fig:PL}, the intersection numbers $(n_0=+1, n_\pm=0)$ of the upward flows $\mathcal{K}_i$ with the original integration cycle (i.e. the real axis) do not depend on the sign of $\mathcal{I}m(g)$ in the phase with $m^2>0$. In this case, the integration cycle coincides with a single thimble, which translates into:
\begin{equation}
Z(m^2>0,g) = Z_0(m^2>0,g)= \frac{1}{\sqrt{g}}\int_{\mathcal{J}_0(\theta)} dz\ e^{-\frac{1}{g} V(z)} \;.
\end{equation}
By expanding around the saddle point $z^\star_0=0$, the partition function of the theory can therefore be unambiguously represented by an asymptotic power series $ Z(m^2>0,g)= Z^{(0),m^2>0}(g)= \sum_n Z^{(0),m^2>0}_n g^n$  which is Borel-summable to the exact result, i.e. $Z(m^2>0,g) = Z^{(0),m^2>0}_{\mathcal{B}_s}(g)$ with:
\begin{equation}
\begin{split}
Z^{(0),m^2>0}_{\mathcal{B}_s}(g) = & \ \frac{1}{\sqrt{g}} e^{-\frac{1}{g}V(z^\star_0)} \\
& \times \int_0^{\infty}d\zeta\ \zeta^s e^{-\zeta}\mathcal{B}_s\Big[Z^{(0),m^2>0}\Big](g\zeta) \;,
\end{split}
\end{equation}
and
\begin{equation}
\mathcal{B}_s\Big[Z^{(0),m^2>0}\Big](\zeta) = \sum_{n=0}^{\infty} \frac{Z^{(0),m^2>0}_n}{\Gamma(n+s+1)}\zeta^n \;.
\label{eq:BL}
\end{equation}

In the phase with $m^2<0$, the integral is on the Stokes line, which is reflected by the jump of the intersection numbers from $(n_0=+1, n_\pm=+1)$ for $\mathcal{I}m(g)<0$ to $(n_0=-1, n_\pm=+1)$ for $\mathcal{I}m(g)>0$. Since the integrals over the thimbles $\mathcal{J}_+$ and $\mathcal{J}_-$ yield the same result, the partition function of the theory can be written as follows:
\begin{equation}
\begin{split}
Z(m^2<0,g) = & \pm Z_0(m^2<0,g) \\
& + 2 Z_+(m^2<0,g) \quad \forall \hspace{0.1cm} \mathcal{I}m(g)\lessgtr 0 \;,
\end{split}
\end{equation}
where each $Z_i$ can again be represented by an asymptotic series $Z^{(i),m^2<0}(g)= \sum_n Z^{(i),m^2<0}_n g^n$ after being expanded around the corresponding saddle point $z_i^\star$ and being Borel resummed, thus yielding the resurgent transseries:
\begin{equation}
\begin{split}
& Z(m^2<0,g) \\
& = \frac{1}{\sqrt{g}}\Bigg\lbrace \pm e^{-\frac{1}{g}V(z^\star_0)}\int_0^{\infty}d\zeta\ \zeta^s e^{-\zeta}\mathcal{B}_s\Big[Z^{(0),m^2<0}\Big](g\zeta) \\
& \hspace{0.3cm} + 2e^{-\frac{1}{g}V(z^\star_+)}\int_0^{\infty}d\zeta\ \zeta^s e^{-\zeta}\mathcal{B}_s\Big[Z^{(+),m^2<0}\Big](g\zeta)\Bigg\rbrace \\
& \hspace{0.3cm} \forall \hspace{0.1cm} \mathcal{I}m(g)\lessgtr 0 \;.
\end{split}
\end{equation}

In practice, only the first terms of the asymptotic series $Z^{(i)}(g)$ are known, such that calculating the integral $\int_0^\infty d\zeta\ \zeta^s e^{-\zeta}\mathcal{B}_s\big[Z^{(i)}\big](g\zeta)$ only amounts to reinserting the $\Gamma(n+s+1)$ factors and leads back to the initial diverging series. Getting non-trivial results thus requires to make some assumptions about the unknown coefficients of the series, e.g. by re-expressing $\mathcal{B}_s\big[Z^{(i)}\big](\zeta)$ in terms of non-polynomial functions that play the role of approximants in the resummation procedures treated e.g. in appendix~\ref{app:Resum}.

%%%%%%%%%%%%%%%%%%%%%%%%%%%%%%%
%%%%%%%%%%%%%%%%%%%%%%%%%%%%%%%
\section{\label{app:SCPT1PIEA}Self-consistent perturbation theory based on 1PI effective actions for the (0+0)-D $O(N)$-symmetric $\varphi^{4}$-theory}
\subsection{\label{sec:Orig1PIEA}Original 1PI effective action}

As usual, we will start our discussion in arbitrary dimensions before focusing on the (0+0)\nobreakdash-D limit for our numerical applications. We first consider the 1PI EA in the original representation of our $O(N)$ model, i.e. the original 1PI EA. The latter is defined by the Legendre transform of the corresponding Schwinger functional, i.e.:
\begin{equation}
\begin{split}
\Gamma^{(\mathrm{1PI})}\Big[\vec{\phi}\Big] \equiv & -W\Big[\vec{J}\Big]+\int_{x}J^{a}(x) \frac{\delta W\big[\vec{J}\big]}{\delta J^{a}(x)} \\
= & -W\Big[\vec{J}\Big]+\int_{x}J^{a}(x) \phi_{a}(x) \;,
\end{split}
\label{eq:pure1PIEAdefinition0DONmain}
\end{equation}
with
\begin{equation}
\phi_{a}(x)=\frac{\delta W\big[\vec{J}\big]}{\delta J^{a}(x)} \;,
\label{eq:pure1PIEAdefinitionbis0DONmain}
\end{equation}
and $W\big[\vec{J}\big]$ coincides with the original Schwinger functional introduced in Eq.~\eqref{eq:ZJKfiniteD} at $\boldsymbol{K}=\boldsymbol{0}$, i.e. $W\big[\vec{J}\big]=W\big[\vec{J},\boldsymbol{K}=\boldsymbol{0}\big]$. This EA can also be written exclusively in terms of 1PI diagrams. Up to the first non-trivial order (i.e. up to order $\mathcal{O}(\hbar^{2})$), this leads to:
\begin{equation}
\begin{split}
\Gamma^{(\mathrm{1PI})}\Big[\vec{\phi}\Big] = & \ S\Big[\vec{\phi}\Big] -\frac{\hbar}{2}\mathrm{STr}\left[\ln\big(\boldsymbol{G}_{\phi}\big)\right] \\
& + \hbar^{2} \left(\rule{0cm}{1.2cm}\right. \frac{1}{24} \hspace{0.08cm} \scalebox{0.85}{${\displaystyle \begin{gathered}
\begin{fmffile}{Diagrams/1PIEA_Hartree}
\begin{fmfgraph}(30,20)
\fmfleft{i}
\fmfright{o}
\fmf{phantom,tension=10}{i,i1}
\fmf{phantom,tension=10}{o,o1}
\fmf{plain,left,tension=0.5,foreground=(1,,0,,0)}{i1,v1,i1}
\fmf{plain,right,tension=0.5,foreground=(1,,0,,0)}{o1,v2,o1}
\fmf{zigzag,foreground=(0,,0,,1)}{v1,v2}
\end{fmfgraph}
\end{fmffile}
\end{gathered}}$}
+\frac{1}{12}\scalebox{0.85}{${\displaystyle\begin{gathered}
\begin{fmffile}{Diagrams/1PIEA_Fock}
\begin{fmfgraph}(15,15)
\fmfleft{i}
\fmfright{o}
\fmf{phantom,tension=11}{i,v1}
\fmf{phantom,tension=11}{v2,o}
\fmf{plain,left,tension=0.4,foreground=(1,,0,,0)}{v1,v2,v1}
\fmf{zigzag,foreground=(0,,0,,1)}{v1,v2}
\end{fmfgraph}
\end{fmffile}
\end{gathered} }$} \\
& - \frac{1}{18} \scalebox{0.85}{${\displaystyle \begin{gathered}
\begin{fmffile}{Diagrams/1PIEA_Diag1}
\begin{fmfgraph}(27,15)
\fmfleft{i}
\fmfright{o}
\fmftop{vUp}
\fmfbottom{vDown}
\fmfv{decor.shape=cross,decor.size=3.5thick,foreground=(1,,0,,0)}{v1}
\fmfv{decor.shape=cross,decor.size=3.5thick,foreground=(1,,0,,0)}{v2}
\fmf{phantom,tension=10}{i,i1}
\fmf{phantom,tension=10}{o,o1}
\fmf{phantom,tension=2.2}{vUp,v5}
\fmf{phantom,tension=2.2}{vDown,v6}
\fmf{phantom,tension=0.5}{v3,v4}
\fmf{phantom,tension=10.0}{i1,v1}
\fmf{phantom,tension=10.0}{o1,v2}
\fmf{dashes,tension=2.0,foreground=(0,,0,,1),foreground=(1,,0,,0)}{v1,v3}
\fmf{dots,left=0.4,tension=0.5,foreground=(0,,0,,1)}{v3,v5}
\fmf{plain,left=0.4,tension=0.5,foreground=(1,,0,,0)}{v5,v4}
\fmf{plain,right=0.4,tension=0.5,foreground=(1,,0,,0)}{v3,v6}
\fmf{dots,right=0.4,tension=0.5,foreground=(0,,0,,1)}{v6,v4}
\fmf{dashes,tension=2.0,foreground=(0,,0,,1),foreground=(1,,0,,0)}{v4,v2}
\fmf{plain,tension=0,foreground=(1,,0,,0)}{v5,v6}
\end{fmfgraph}
\end{fmffile}
\end{gathered} }$} - \frac{1}{36} \hspace{-0.15cm} \scalebox{0.85}{${\displaystyle\begin{gathered}
\begin{fmffile}{Diagrams/1PIEA_Diag2}
\begin{fmfgraph}(25,20)
\fmfleft{i}
\fmfright{o}
\fmftop{vUp}
\fmfbottom{vDown}
\fmfv{decor.shape=cross,decor.angle=45,decor.size=3.5thick,foreground=(1,,0,,0)}{vUpbis}
\fmfv{decor.shape=cross,decor.angle=45,decor.size=3.5thick,foreground=(1,,0,,0)}{vDownbis}
\fmf{phantom,tension=0.8}{vUp,vUpbis}
\fmf{phantom,tension=0.8}{vDown,vDownbis}
\fmf{dashes,tension=0.5,foreground=(0,,0,,1),foreground=(1,,0,,0)}{v3,vUpbis}
\fmf{phantom,tension=0.5}{v4,vUpbis}
\fmf{phantom,tension=0.5}{v3,vDownbis}
\fmf{dashes,tension=0.5,foreground=(0,,0,,1),foreground=(1,,0,,0)}{v4,vDownbis}
\fmf{phantom,tension=11}{i,v1}
\fmf{phantom,tension=11}{v2,o}
\fmf{plain,left,tension=0.5,foreground=(1,,0,,0)}{v1,v2,v1}
\fmf{dots,tension=1.7,foreground=(0,,0,,1)}{v1,v3}
\fmf{plain,foreground=(1,,0,,0)}{v3,v4}
\fmf{dots,tension=1.7,foreground=(0,,0,,1)}{v4,v2}
\end{fmfgraph}
\end{fmffile}
\end{gathered} }$} \hspace{-0.22cm} \left.\rule{0cm}{1.2cm}\right) \\
& + \mathcal{O}\big(\hbar^{3}\big)\;,
\end{split}
\label{eq:1PIEAfinalexpression}
\end{equation}
with the Feynman rules:
\begin{subequations}
\begin{align}
%\begin{equation}
\begin{gathered}
\begin{fmffile}{Diagrams/1PIEA_G}
\begin{fmfgraph*}(20,20)
\fmfleft{i0,i1,i2,i3}
\fmfright{o0,o1,o2,o3}
\fmflabel{$x, a$}{v1}
\fmflabel{$y, b$}{v2}
\fmf{phantom}{i1,v1}
\fmf{phantom}{i2,v1}
\fmf{plain,tension=0.6,foreground=(1,,0,,0)}{v1,v2}
\fmf{phantom}{v2,o1}
\fmf{phantom}{v2,o2}
\end{fmfgraph*}
\end{fmffile}
\end{gathered} \quad &\rightarrow \boldsymbol{G}_{\phi;ab}(x,y)\;,
\label{eq:DefinitionG1PIEAhbarExpansion} \\
%\end{equation} 
%\begin{equation}
\begin{gathered}
\begin{fmffile}{Diagrams/1PIEA_V3}
\begin{fmfgraph*}(20,20)
\fmfleft{i0,i1,i2,i3}
\fmfright{o0,o1,o2,o3}
\fmfv{decor.shape=cross,decor.angle=45,decor.size=3.5thick,foreground=(1,,0,,0)}{o2}
\fmf{phantom,tension=2.0}{i1,i1bis}
\fmf{plain,tension=2.0,foreground=(1,,0,,0)}{i1bis,v1}
\fmf{phantom,tension=2.0}{i2,i2bis}
\fmf{plain,tension=2.0,foreground=(1,,0,,0)}{i2bis,v1}
\fmf{dots,label=$x$,tension=0.6,foreground=(0,,0,,1)}{v1,v2}
\fmf{phantom,tension=2.0}{o1bis,o1}
\fmf{plain,tension=2.0,foreground=(1,,0,,0)}{v2,o1bis}
\fmf{phantom,tension=2.0}{o2bis,o2}
\fmf{phantom,tension=2.0,foreground=(1,,0,,0)}{v2,o2bis}
\fmf{dashes,tension=0.0,foreground=(1,,0,,0)}{v2,o2}
\fmflabel{$a$}{i1bis}
\fmflabel{$b$}{i2bis}
\fmflabel{$c$}{o1bis}
\fmflabel{$N$}{o2bis}
\end{fmfgraph*}
\end{fmffile}
\end{gathered} \quad &\rightarrow \lambda\left|\vec{\phi}(x)\right|\delta_{a b}\delta_{c N}\;,
\label{eq:FeynRules1PIEA3legVertexSourceJ0main} \\
%\end{equation}
%\begin{equation}
\begin{gathered}
\begin{fmffile}{Diagrams/1PIEA_V4}
\begin{fmfgraph*}(20,20)
\fmfleft{i0,i1,i2,i3}
\fmfright{o0,o1,o2,o3}
\fmf{phantom,tension=2.0}{i1,i1bis}
\fmf{plain,tension=2.0,foreground=(1,,0,,0)}{i1bis,v1}
\fmf{phantom,tension=2.0}{i2,i2bis}
\fmf{plain,tension=2.0,foreground=(1,,0,,0)}{i2bis,v1}
\fmf{zigzag,label=$x$,tension=0.6,foreground=(0,,0,,1)}{v1,v2}
\fmf{phantom,tension=2.0}{o1bis,o1}
\fmf{plain,tension=2.0,foreground=(1,,0,,0)}{v2,o1bis}
\fmf{phantom,tension=2.0}{o2bis,o2}
\fmf{plain,tension=2.0,foreground=(1,,0,,0)}{v2,o2bis}
\fmflabel{$a$}{i1bis}
\fmflabel{$b$}{i2bis}
\fmflabel{$c$}{o1bis}
\fmflabel{$d$}{o2bis}
\end{fmfgraph*}
\end{fmffile}
\end{gathered} \quad &\rightarrow \lambda\delta_{a b}\delta_{c d}\;,
\label{eq:FeynRules1PIEA4legVertexSourceJ0main}
%\end{equation}
\end{align}
\end{subequations}
where $\big|\vec{\phi}(x)\big|=\phi_{N}(x)$ as in section~\ref{sec:SCPT}. We also point out that the propagator $\boldsymbol{G}_{\phi}$, defined by Eq.~\eqref{eq:DefinitionG1PIEAhbarExpansionbis}, is not dressed by the classical configuration $\vec{\varphi}_\text{cl}$ of the original field as in Eq.~\eqref{eq:LEG} for the original LE, but by its 1-point correlation function $\vec{\phi}$ (satisfying Eq.~\eqref{eq:pure1PIEAdefinitionbis0DONmain}), which contains quantal or radiative corrections.

Let us then evaluate the original 1PI EA in the (0+0)\nobreakdash-D situation. As for the original LE (with Eqs.~\eqref{eq:origLEGoldstoneProp} to~\eqref{eq:origLEmassiveProp} more specifically), we separate the (inverse) propagator $\boldsymbol{G}^{-1}_{\phi}$ into the Goldstone modes one $G^{-1}_{\phi;\mathfrak{g}} = \mathfrak{G}^{-1}_{\phi;\mathfrak{g}} \mathbb{I}_{N-1} = \left(m^{2} + \lambda \vec{\phi}^{2}/6\right) \mathbb{I}_{N-1}$ and that of the Higgs mode $\boldsymbol{G}^{-1}_{\phi;NN}= m^{2} + \lambda \vec{\phi}^{2}/2$. According to this, Eq.~\eqref{eq:1PIEAfinalexpression} reduces in the (0+0)\nobreakdash-D limit to:
\begin{equation}
\begin{split}
\Gamma^{(\mathrm{1PI})}\Big(\vec{\phi}\Big) = & \ S\Big(\vec{\phi}\Big) - \frac{\hbar}{2}\Big[\left(N-1\right)\ln\big(2\pi \mathfrak{G}_{\phi;\mathfrak{g}}\big) \\
& +\ln\big(2\pi \boldsymbol{G}_{\phi;NN}\big)\Big] \\
& + \hbar^{2}\Bigg[\frac{\lambda}{72}\Big(9\left(\boldsymbol{G}_{\phi;NN}\right)^{2}+3\left(\mathfrak{G}_{\phi;\mathfrak{g}}\right)^{2}\left(N^{2}-1\right) \\
& -6\left(\boldsymbol{G}_{\phi;NN}\right)^{3}\lambda\phi_{N}^{2} -2\boldsymbol{G}_{\phi;NN}\mathfrak{G}_{\phi;\mathfrak{g}}\left(N-1\right) \\
& \times \left(-3+\mathfrak{G}_{\phi;\mathfrak{g}}\lambda\phi_{N}^{2}\right)\Big)\Bigg] \\
& + \mathcal{O}\big(\hbar^{3}\big) \;,
\end{split}
\label{eq:1PIEAfinalexpression0DON}
\end{equation}
with $S\big(\vec{\phi}\big)=m^{2}\phi^{2}_{N}/2+\lambda\phi^{4}_{N}/4!$ here. This expression of the original 1PI EA is then exploited by fixing the configuration of the 1-point correlation function $\vec{\phi}$ and more specifically of its component $\phi_{N}$. This is achieved by extremizing this EA through the gap equation:
\begin{equation}
0 = \left.\frac{\partial \Gamma^{(\mathrm{1PI})}\big(\vec{\phi}\big)}{\partial \phi_{N}}\right|_{\vec{\phi}=\vec{\overline{\phi}}} \;,
\end{equation}
with $\vec{\overline{\phi}}=\begin{pmatrix}
\overline{\phi}_{1} & \cdots & \overline{\phi}_{N-1} & \overline{\phi}_{N}
\end{pmatrix}^{\mathrm{T}}=\begin{pmatrix}
0 & \cdots & 0 & \overline{\phi}_{N}
\end{pmatrix}^{\mathrm{T}}$. The gs energy and density are subsequently inferred from the solution $\vec{\overline{\phi}}$ together with Eq.~\eqref{eq:1PIEAfinalexpression0DON} according to:
\begin{equation}
E^\text{1PI EA;orig}_{\mathrm{gs}} = \frac{1}{\hbar} \Gamma^{\mathrm{(1PI)}}\Big(\vec{\phi}=\vec{\overline{\phi}}\Big) \;,
\label{eq:DeduceEgs1PIEAdiag}
\end{equation}
\begin{equation}
\rho^\text{1PI EA;orig}_{\mathrm{gs}} = \frac{1}{N} \left(\hbar\mathrm{Tr}_{a}\big(\boldsymbol{G}_{\overline{\phi}}\big) + \vec{\overline{\phi}}^{2}\right) \;,
\label{eq:DeduceRhogs1PIEAdiag}
\end{equation}
where $\boldsymbol{G}_{\overline{\phi}}$ corresponds to the propagator $\boldsymbol{G}_{\phi}$ evaluated at $\vec{\phi}=\vec{\overline{\phi}}$.

%%%%%%%%%%%%%%%%%%%
\begin{figure}[!t]
  \begin{center}
      \includegraphics[width=0.95\linewidth]{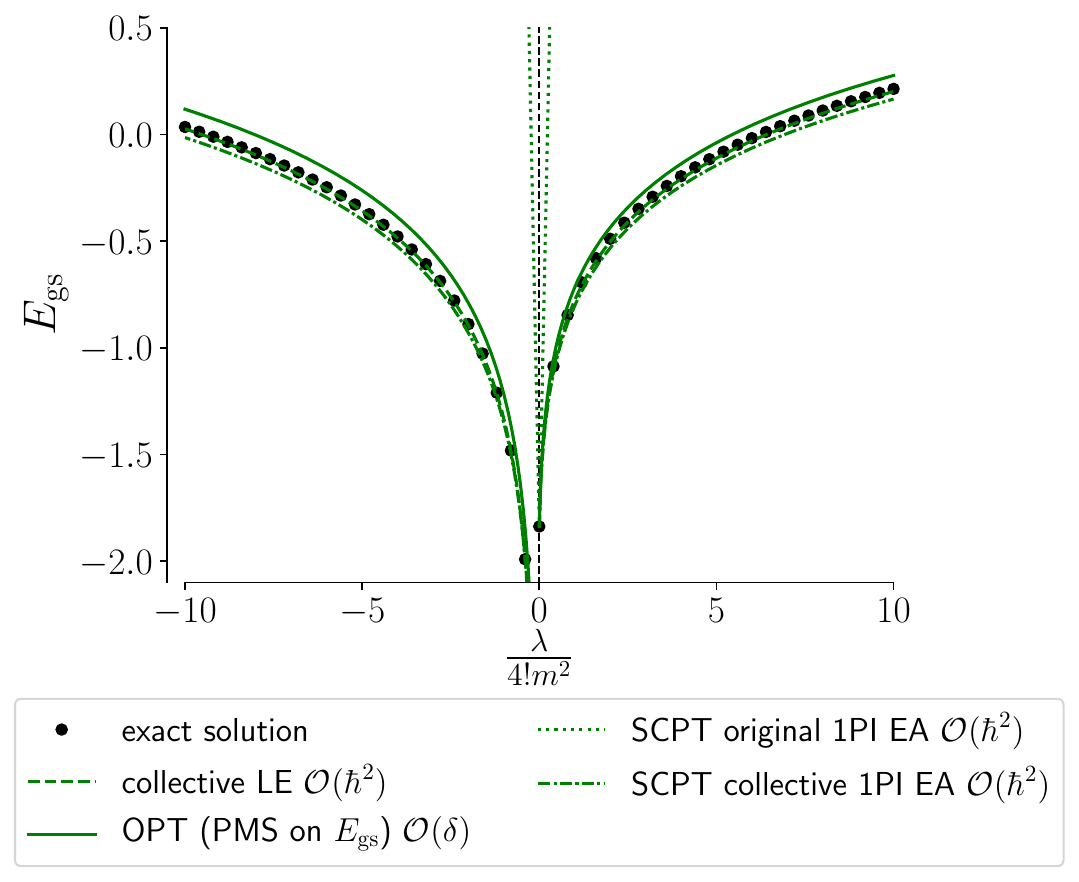}
      \includegraphics[width=0.95\linewidth]{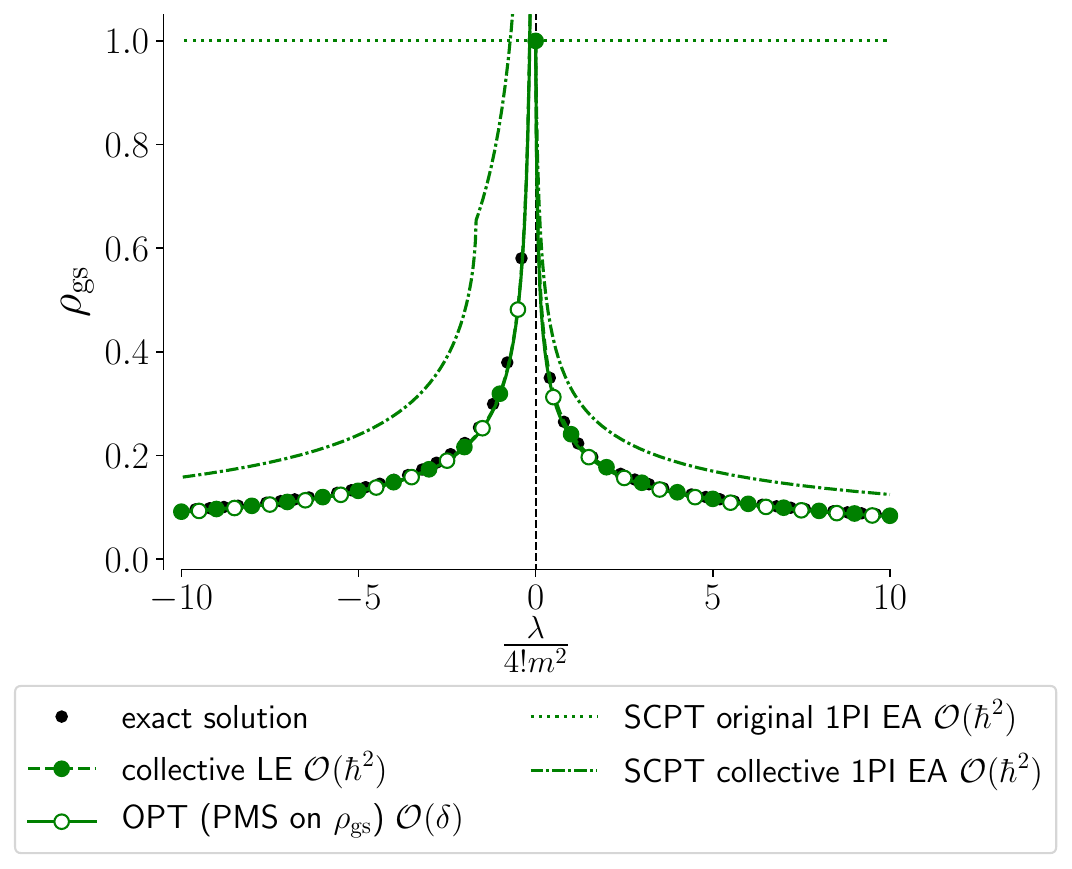}
    \caption{Gs energy $E_{\mathrm{gs}}$ or density $\rho_{\mathrm{gs}}$ calculated at $\hbar=1$, $m^{2}=\pm 1$ and $N=2$. See also the caption of Fig.~\ref{fig:2PIEAzeroVsNonzerovev} for the meaning of the indication ``$\mathcal{O}\big(\hbar^{n}\big)$'' for SCPT results.}
    \label{fig:1PIEA}
  \end{center}
\end{figure}
%%%%%%%%%%%%%%%%%%%

For both signs of $m^{2}$, the optimal (i.e. physical) solutions found from the extremization of the original 1PI EA satisfy $\vec{\overline{\phi}}=\vec{0}$. Most importantly, this means that the present implementation of SCPT at its first non-trivial order does not yield any spurious spontaneous breakdown of the $O(N)$ symmetry, in the sense that it is in accordance with the absence of SSB in the exact solution of the studied model. Even though this is a reassuring feature, it is also fatal for SCPT based on the original 1PI EA since $\vec{\overline{\phi}}$ is the only adjustable variable that can be used to grasp correlations in this framework. This is illustrated by Fig.~\ref{fig:1PIEA} where the gs energy and density estimated from SCPT based on the original 1PI EA clearly do not manage to reproduce, even qualitatively, the corresponding exact solution. Hence, this implementation of SCPT is notably completely irrelevant to tackle our $O(N)$ model, regardless of the calculated quantity.

%~~~
\subsection{\label{sec:Coll1PIEA}Collective 1PI effective action}

Similarly to the analysis made in the section dedicated to the LE and for the same motivations (namely investigating whether the introduction of an auxiliary collective field helps in grasping more efficiently non-trivial correlations at low orders), we now focus on SCPT based on the collective 1PI EA, i.e. the 1PI EA implemented in the collective representation where the original field $\vec{\widetilde{\varphi}}$ has been integrated out in favor of the collective Hubbard-Stratonovich field $\widetilde{\sigma}$. There are numerous works exploiting the collective 1PI EA. These applications take multiple forms for the following reasons:
\begin{itemize}
\item One can use the Schwinger-Dyson equations formalism, equivalent to the implementation of SCPT treated here \cite{ben77,coo78,tam78,cam79,hay79,coo79,gol81,mun82,sar84,ben85,kul87,coo99,coo04,coo16}. The collective 1PI EA formalism in this form is referred to as mean-field perturbation theory, mean-field theory or self-consistent field approximation.
\item The EA can be expanded using different expansion parameters like $1/N$ \cite{can79,gol81,coo94,coo95,fei95,coo97,coo99,coo04,chi12} or $\hbar$ \cite{tam78bis,gur79,fur83,sac88,coo10,coo11,mih11bis,mih11,coo12,daw12,chi14,coo14,coo14bis,coo16}.
\end{itemize}
We stress that all of these approaches are equivalent as long as the truncation of the EA is organized with respect to the same parameter, e.g. $1/N$ or $\hbar$ typically\footnote{See Ref.~\cite{coo04} for an exhaustive discussion on truncation schemes of Schwinger-Dyson equations.}. If the parameter in question is $\hbar$, the resulting approach is sometimes called mean-field expansion \cite{tam78bis,gur79} or, more recently, auxiliary field LE (LOAF) \cite{coo10,coo11,mih11bis,mih11,coo12,daw12,chi14,coo14,coo14bis,coo16}\footnote{See Ref.~\cite{coo16} for a detailed discussion on the LOAF.}.

The collective 1PI EA is defined by Legendre transforming the Schwinger functional $W_{\mathrm{col}}\big[\mathcal{J}\big]$. This translates into:
\begin{equation}
\begin{split}
\Gamma_{\mathrm{col}}^{(\mathrm{1PI})}[\Phi] \equiv & -W_{\mathrm{col}}\big[\mathcal{J}\big] + \int_{x} \mathcal{J}^{\alpha}(x) \frac{\delta W_{\mathrm{col}}\big[\mathcal{J}\big]}{\delta\mathcal{J}^{\alpha}(x)} \\
= & -W_{\mathrm{col}}\big[\mathcal{J}\big]+\int_{\alpha}\mathcal{J}^{\alpha}(x)\Phi_{\alpha}(x)\;,
\end{split}
\label{eq:bosonic1PIEAdefinition0DONmain}
\end{equation}
with
\begin{equation}
\Phi_{\alpha}(x)=\frac{\delta W_{\mathrm{col}}\big[\mathcal{J}\big]}{\delta \mathcal{J}^{\alpha}(x)}\;,
\label{eq:bosonic1PIEAdefinitionbis0DONmain}
\end{equation}
or, in terms of the 1-point correlation functions of the original and Hubbard-Stratonovich fields (i.e. $\vec{\phi}(x) = \left\langle\vec{\widetilde{\varphi}}(x)\right\rangle$ and $\eta(x)=\left\langle\widetilde{\sigma}(x)\right\rangle$, respectively),
\begin{equation}
\Phi(x) = \begin{pmatrix}
\vec{\phi}(x) \\
\eta(x)
\end{pmatrix}\;.
\end{equation}
The collective 1PI EA can be represented diagrammatically according to:
\begin{equation}
\begin{split}
& \Gamma_{\mathrm{col}}^{(\mathrm{1PI})}[\Phi] \\
& = S_{\mathrm{col}}[\eta] + \frac{1}{2} \int_{x,y}\phi^a(x) \boldsymbol{G}^{-1}_{\Phi;ab}(x,y) \phi^b(y) - \frac{\hbar}{2}\mathrm{Tr}\left[\ln\big(D_{\Phi}\big)\right] \\
& \hspace{0.3cm} - \hbar^{2}  \left(\rule{0cm}{1.1cm}\right. \frac{1}{2} \hspace{-0.1cm} \begin{gathered}
\begin{fmffile}{Diagrams/bosonic1PIEA_Gamma2_Diag1bis}
\begin{fmfgraph*}(25,25)
\fmfleft{i1,i2}
\fmfright{o1,o2}
\fmfbottom{i0,o0}
\fmftop{i3,o3}
\fmfv{decor.shape=circle,decor.size=2.0thick,foreground=(0,,0,,1)}{v1}
\fmfv{decor.shape=circle,decor.size=2.0thick,foreground=(0,,0,,1)}{v2}
\fmfv{decor.shape=circle,decor.size=2.0thick,foreground=(0,,0,,1)}{v3}
\fmfv{decor.shape=circle,decor.size=2.0thick,foreground=(0,,0,,1)}{v4}
\fmfv{decor.shape=cross,decor.size=0.25cm,decor.angle=35,foreground=(1,,0,,0)}{v1b}
\fmfv{decor.shape=cross,decor.size=0.25cm,decor.angle=-35,foreground=(1,,0,,0)}{v4b}
\fmf{phantom}{i1,v1}
\fmf{phantom}{i2,v4}
\fmf{phantom}{o1,v2}
\fmf{phantom}{o2,v3}
\fmf{phantom}{i3,v4b}
\fmf{phantom}{o3,v3b}
\fmf{phantom}{i0,v1b}
\fmf{phantom}{o0,v2b}
\fmf{plain,tension=1.6,foreground=(1,,0,,0)}{v1,v2}
\fmf{plain,tension=1.6,foreground=(1,,0,,0)}{v3,v4}
\fmf{wiggly,tension=2.0,foreground=(1,,0,,0)}{v1,v4}
\fmf{wiggly,tension=2.0,foreground=(1,,0,,0)}{v2,v3}
\fmf{phantom,tension=0}{v1,v3}
\fmf{phantom,tension=0}{v2,v4}
\fmf{plain,right=0.8,tension=0,foreground=(1,,0,,0)}{v2,v3}
\fmf{phantom,left=0.8,tension=0}{v1,v4}
\fmf{dashes,foreground=(1,,0,,0)}{v1,v1b}
\fmf{phantom}{v2,v2b}
\fmf{phantom}{v3,v3b}
\fmf{dashes,foreground=(1,,0,,0)}{v4,v4b}
\end{fmfgraph*}
\end{fmffile}
\end{gathered} \hspace{-0.4cm} + \frac{1}{2} \hspace{-0.1cm} \begin{gathered}
\begin{fmffile}{Diagrams/bosonic1PIEA_Gamma2_Diag2bis}
\begin{fmfgraph*}(25,25)
\fmfleft{i1,i2}
\fmfright{o1,o2}
\fmfbottom{i0,o0}
\fmftop{i3,o3}
\fmfv{decor.shape=circle,decor.size=2.0thick,foreground=(0,,0,,1)}{v1}
\fmfv{decor.shape=circle,decor.size=2.0thick,foreground=(0,,0,,1)}{v2}
\fmfv{decor.shape=circle,decor.size=2.0thick,foreground=(0,,0,,1)}{v3}
\fmfv{decor.shape=circle,decor.size=2.0thick,foreground=(0,,0,,1)}{v4}
\fmfv{decor.shape=cross,decor.size=0.25cm,decor.angle=35,foreground=(1,,0,,0)}{v3b}
\fmfv{decor.shape=cross,decor.size=0.25cm,decor.angle=-35,foreground=(1,,0,,0)}{v4b}
\fmf{phantom}{i1,v1}
\fmf{phantom}{i2,v4}
\fmf{phantom}{o1,v2}
\fmf{phantom}{o2,v3}
\fmf{phantom}{i3,v4b}
\fmf{phantom}{o3,v3b}
\fmf{phantom}{i0,v1b}
\fmf{phantom}{o0,v2b}
\fmf{plain,tension=1.6,foreground=(1,,0,,0)}{v1,v2}
\fmf{phantom,tension=1.6}{v3,v4}
\fmf{wiggly,tension=2.0,foreground=(1,,0,,0)}{v1,v4}
\fmf{wiggly,tension=2.0,foreground=(1,,0,,0)}{v2,v3}
\fmf{plain,tension=0,foreground=(1,,0,,0)}{v1,v3}
\fmf{plain,tension=0,foreground=(1,,0,,0)}{v2,v4}
\fmf{phantom,right=0.8,tension=0}{v2,v3}
\fmf{phantom,left=0.8,tension=0}{v1,v4}
\fmf{phantom}{v1,v1b}
\fmf{phantom}{v2,v2b}
\fmf{dashes,foreground=(1,,0,,0)}{v3,v3b}
\fmf{dashes,foreground=(1,,0,,0)}{v4,v4b}
\end{fmfgraph*}
\end{fmffile}
\end{gathered} \\
& \hspace{0.3cm} + \frac{1}{2} \hspace{-0.1cm} \begin{gathered}
\begin{fmffile}{Diagrams/bosonic1PIEA_Gamma2_Diag5bis}
\begin{fmfgraph*}(35,20)
\fmfleft{i1,i2}
\fmfright{o1,o2}
\fmfbottom{i0,o0}
\fmfbottom{b0}
\fmfbottom{b1}
\fmfbottom{b2}
\fmftop{i3,o3}
\fmfv{decor.shape=circle,decor.size=2.0thick,foreground=(0,,0,,1)}{v1}
\fmfv{decor.shape=circle,decor.size=2.0thick,foreground=(0,,0,,1)}{v2}
\fmfv{decor.shape=circle,decor.size=2.0thick,foreground=(0,,0,,1)}{v3}
\fmfv{decor.shape=circle,decor.size=2.0thick,foreground=(0,,0,,1)}{v4}
\fmfv{decor.shape=circle,decor.size=2.0thick,foreground=(0,,0,,1)}{v5}
\fmfv{decor.shape=circle,decor.size=2.0thick,foreground=(0,,0,,1)}{v6}
\fmfv{decor.shape=cross,decor.size=0.25cm,decor.angle=16,foreground=(1,,0,,0)}{v1b}
\fmfv{decor.shape=cross,decor.size=0.25cm,foreground=(1,,0,,0)}{v3b}
\fmfv{decor.shape=cross,decor.size=0.25cm,decor.angle=71,foreground=(1,,0,,0)}{v4b}
\fmfv{decor.shape=cross,decor.size=0.25cm,foreground=(1,,0,,0)}{v6b}
\fmf{phantom,tension=1.4}{i1,v1}
\fmf{phantom}{i2,v3b}
\fmf{phantom}{i0,v1b}
\fmf{phantom}{o2,v6b}
\fmf{phantom,tension=1.4}{o1,v5}
\fmf{phantom}{o0,v5b}
\fmf{phantom,tension=1.11}{v3b,v6b}
\fmf{phantom,tension=1.38}{i0,v2}
\fmf{phantom,tension=1.38}{o0,v2}
\fmf{phantom,tension=1.8}{i0,v2b}
\fmf{phantom,tension=1.2}{o0,v2b}
\fmf{phantom,tension=1.2}{b0,v2b}
\fmf{phantom,tension=1.2}{b1,v2b}
\fmf{phantom,tension=1.2}{b2,v2b}
\fmf{phantom,tension=1.38}{i0,v4}
\fmf{phantom,tension=1.38}{o0,v4}
\fmf{phantom,tension=1.2}{i0,v4b}
\fmf{phantom,tension=1.8}{o0,v4b}
\fmf{phantom,tension=1.2}{b0,v4b}
\fmf{phantom,tension=1.2}{b1,v4b}
\fmf{phantom,tension=1.2}{b2,v4b}
\fmf{phantom,tension=2}{i3,v3}
\fmf{phantom,tension=2}{o3,v6}
\fmf{phantom,tension=2}{i3,v3b}
\fmf{phantom,tension=0.8}{o3,v3b}
\fmf{phantom,tension=0.8}{i3,v6b}
\fmf{phantom,tension=2}{o3,v6b}
\fmf{plain,tension=1.4,foreground=(1,,0,,0)}{v1,v2}
\fmf{plain,tension=1.4,foreground=(1,,0,,0)}{v4,v5}
\fmf{phantom}{v1,v3}
\fmf{phantom,left=0.8,tension=0}{v1,v3}
\fmf{plain,foreground=(1,,0,,0)}{v5,v6}
\fmf{phantom,right=0.8,tension=0}{v5,v6}
\fmf{plain,foreground=(1,,0,,0)}{v2,v3}
\fmf{phantom}{v4,v6}
\fmf{wiggly,tension=0.5,foreground=(1,,0,,0)}{v2,v4}
\fmf{wiggly,tension=2,foreground=(1,,0,,0)}{v3,v6}
\fmf{wiggly,right=0.8,tension=0,foreground=(1,,0,,0)}{v1,v5}
\fmf{dashes,tension=1,foreground=(1,,0,,0)}{v1,v1b}
\fmf{phantom,tension=0.2}{v2,v2b}
\fmf{dashes,tension=1.5,foreground=(1,,0,,0)}{v3,v3b}
\fmf{dashes,tension=0.2,foreground=(1,,0,,0)}{v4,v4b}
\fmf{phantom,tension=1}{v5,v5b}
\fmf{dashes,tension=1.5,foreground=(1,,0,,0)}{v6,v6b}
\end{fmfgraph*}
\end{fmffile}
\end{gathered} \hspace{-0.5cm} +\frac{1}{4} \begin{gathered}
\begin{fmffile}{Diagrams/bosonic1PIEA_Gamma2_Diag6bis}
\begin{fmfgraph*}(35,20)
\fmfleft{i1,i2}
\fmfright{o1,o2}
\fmfbottom{i0,o0}
\fmfbottom{b0}
\fmfbottom{b1}
\fmfbottom{b2}
\fmftop{i3,o3}
\fmfv{decor.shape=circle,decor.size=2.0thick,foreground=(0,,0,,1)}{v1}
\fmfv{decor.shape=circle,decor.size=2.0thick,foreground=(0,,0,,1)}{v2}
\fmfv{decor.shape=circle,decor.size=2.0thick,foreground=(0,,0,,1)}{v3}
\fmfv{decor.shape=circle,decor.size=2.0thick,foreground=(0,,0,,1)}{v4}
\fmfv{decor.shape=circle,decor.size=2.0thick,foreground=(0,,0,,1)}{v5}
\fmfv{decor.shape=circle,decor.size=2.0thick,foreground=(0,,0,,1)}{v6}
\fmfv{decor.shape=cross,decor.size=0.25cm,decor.angle=16,foreground=(1,,0,,0)}{v1b}
\fmfv{decor.shape=cross,decor.size=0.25cm,foreground=(1,,0,,0)}{v3b}
\fmfv{decor.shape=cross,decor.size=0.25cm,decor.angle=-16,foreground=(1,,0,,0)}{v5b}
\fmfv{decor.shape=cross,decor.size=0.25cm,foreground=(1,,0,,0)}{v6b}
\fmf{phantom,tension=1.4}{i1,v1}
\fmf{phantom}{i2,v3b}
\fmf{phantom}{i0,v1b}
\fmf{phantom}{o2,v6b}
\fmf{phantom,tension=1.4}{o1,v5}
\fmf{phantom}{o0,v5b}
\fmf{phantom,tension=1.11}{v3b,v6b}
\fmf{phantom,tension=1.38}{i0,v2}
\fmf{phantom,tension=1.38}{o0,v2}
\fmf{phantom,tension=1.8}{i0,v2b}
\fmf{phantom,tension=1.2}{o0,v2b}
\fmf{phantom,tension=1.2}{b0,v2b}
\fmf{phantom,tension=1.2}{b1,v2b}
\fmf{phantom,tension=1.2}{b2,v2b}
\fmf{phantom,tension=1.38}{i0,v4}
\fmf{phantom,tension=1.38}{o0,v4}
\fmf{phantom,tension=1.2}{i0,v4b}
\fmf{phantom,tension=1.8}{o0,v4b}
\fmf{phantom,tension=1.2}{b0,v4b}
\fmf{phantom,tension=1.2}{b1,v4b}
\fmf{phantom,tension=1.2}{b2,v4b}
\fmf{phantom,tension=2}{i3,v3}
\fmf{phantom,tension=2}{o3,v6}
\fmf{phantom,tension=2}{i3,v3b}
\fmf{phantom,tension=0.8}{o3,v3b}
\fmf{phantom,tension=0.8}{i3,v6b}
\fmf{phantom,tension=2}{o3,v6b}
\fmf{plain,tension=1.4,foreground=(1,,0,,0)}{v1,v2}
\fmf{plain,tension=1.4,foreground=(1,,0,,0)}{v4,v5}
\fmf{phantom}{v1,v3}
\fmf{phantom,left=0.8,tension=0}{v1,v3}
\fmf{phantom}{v5,v6}
\fmf{phantom,right=0.8,tension=0}{v5,v6}
\fmf{plain,foreground=(1,,0,,0)}{v2,v3}
\fmf{plain,foreground=(1,,0,,0)}{v4,v6}
\fmf{wiggly,tension=0.5,foreground=(1,,0,,0)}{v2,v4}
\fmf{wiggly,tension=2,foreground=(1,,0,,0)}{v3,v6}
\fmf{wiggly,right=0.8,tension=0,foreground=(1,,0,,0)}{v1,v5}
\fmf{dashes,tension=1,foreground=(1,,0,,0)}{v1,v1b}
\fmf{phantom,tension=0.2}{v2,v2b}
\fmf{dashes,tension=1.5,foreground=(1,,0,,0)}{v3,v3b}
\fmf{phantom,tension=0.2}{v4,v4b}
\fmf{dashes,tension=1,foreground=(1,,0,,0)}{v5,v5b}
\fmf{dashes,tension=1.5,foreground=(1,,0,,0)}{v6,v6b}
\end{fmfgraph*}
\end{fmffile}
\end{gathered} \\
\\
& \hspace{0.3cm} + \frac{1}{2} \begin{gathered}
\begin{fmffile}{Diagrams/bosonic1PIEA_Gamma2_Diag7bis}
\begin{fmfgraph*}(35,20)
\fmfleft{i1,i2}
\fmfright{o1,o2}
\fmfbottom{i0,o0}
\fmfbottom{b0}
\fmfbottom{b1}
\fmfbottom{b2}
\fmftop{i3,o3}
\fmfv{decor.shape=circle,decor.size=2.0thick,foreground=(0,,0,,1)}{v1}
\fmfv{decor.shape=circle,decor.size=2.0thick,foreground=(0,,0,,1)}{v2}
\fmfv{decor.shape=circle,decor.size=2.0thick,foreground=(0,,0,,1)}{v3}
\fmfv{decor.shape=circle,decor.size=2.0thick,foreground=(0,,0,,1)}{v4}
\fmfv{decor.shape=circle,decor.size=2.0thick,foreground=(0,,0,,1)}{v5}
\fmfv{decor.shape=circle,decor.size=2.0thick,foreground=(0,,0,,1)}{v6}
\fmfv{decor.shape=cross,decor.size=0.25cm,decor.angle=16,foreground=(1,,0,,0)}{v1b}
\fmfv{decor.shape=cross,decor.size=0.25cm,foreground=(1,,0,,0)}{v3b}
\fmf{phantom,tension=1.4}{i1,v1}
\fmf{phantom}{i2,v3b}
\fmf{phantom}{i0,v1b}
\fmf{phantom}{o2,v6b}
\fmf{phantom,tension=1.4}{o1,v5}
\fmf{phantom}{o0,v5b}
\fmf{phantom,tension=1.11}{v3b,v6b}
\fmf{phantom,tension=1.38}{i0,v2}
\fmf{phantom,tension=1.38}{o0,v2}
\fmf{phantom,tension=1.8}{i0,v2b}
\fmf{phantom,tension=1.2}{o0,v2b}
\fmf{phantom,tension=1.2}{b0,v2b}
\fmf{phantom,tension=1.2}{b1,v2b}
\fmf{phantom,tension=1.2}{b2,v2b}
\fmf{phantom,tension=1.38}{i0,v4}
\fmf{phantom,tension=1.38}{o0,v4}
\fmf{phantom,tension=1.2}{i0,v4b}
\fmf{phantom,tension=1.8}{o0,v4b}
\fmf{phantom,tension=1.2}{b0,v4b}
\fmf{phantom,tension=1.2}{b1,v4b}
\fmf{phantom,tension=1.2}{b2,v4b}
\fmf{phantom,tension=2}{i3,v3}
\fmf{phantom,tension=2}{o3,v6}
\fmf{phantom,tension=2}{i3,v3b}
\fmf{phantom,tension=0.8}{o3,v3b}
\fmf{phantom,tension=0.8}{i3,v6b}
\fmf{phantom,tension=2}{o3,v6b}
\fmf{plain,tension=1.4,foreground=(1,,0,,0)}{v1,v2}
\fmf{plain,tension=1.4,foreground=(1,,0,,0)}{v4,v5}
\fmf{phantom}{v1,v3}
\fmf{phantom,left=0.8,tension=0}{v1,v3}
\fmf{plain,foreground=(1,,0,,0)}{v5,v6}
\fmf{phantom,right=0.8,tension=0}{v5,v6}
\fmf{plain,foreground=(1,,0,,0)}{v2,v3}
\fmf{plain,foreground=(1,,0,,0)}{v4,v6}
\fmf{wiggly,tension=0.5,foreground=(1,,0,,0)}{v2,v4}
\fmf{wiggly,tension=2,foreground=(1,,0,,0)}{v3,v6}
\fmf{wiggly,right=0.8,tension=0,foreground=(1,,0,,0)}{v1,v5}
\fmf{dashes,tension=1,foreground=(1,,0,,0)}{v1,v1b}
\fmf{phantom,tension=0.2}{v2,v2b}
\fmf{dashes,tension=1.5,foreground=(1,,0,,0)}{v3,v3b}
\fmf{phantom,tension=0.2}{v4,v4b}
\fmf{phantom,tension=1}{v5,v5b}
\fmf{phantom,tension=1.5}{v6,v6b}
\end{fmfgraph*}
\end{fmffile}
\end{gathered} \hspace{-0.5cm} +\frac{1}{12} \hspace{-0.5cm} \begin{gathered}
\begin{fmffile}{Diagrams/bosonic1PIEA_Gamma2_Diag8bis}
\begin{fmfgraph*}(35,20)
\fmfleft{i1,i2}
\fmfright{o1,o2}
\fmfbottom{i0,o0}
\fmfbottom{b0}
\fmfbottom{b1}
\fmfbottom{b2}
\fmftop{i3,o3}
\fmfv{decor.shape=circle,decor.size=2.0thick,foreground=(0,,0,,1)}{v1}
\fmfv{decor.shape=circle,decor.size=2.0thick,foreground=(0,,0,,1)}{v2}
\fmfv{decor.shape=circle,decor.size=2.0thick,foreground=(0,,0,,1)}{v3}
\fmfv{decor.shape=circle,decor.size=2.0thick,foreground=(0,,0,,1)}{v4}
\fmfv{decor.shape=circle,decor.size=2.0thick,foreground=(0,,0,,1)}{v5}
\fmfv{decor.shape=circle,decor.size=2.0thick,foreground=(0,,0,,1)}{v6}
\fmf{phantom,tension=1.4}{i1,v1}
\fmf{phantom}{i2,v3b}
\fmf{phantom}{i0,v1b}
\fmf{phantom}{o2,v6b}
\fmf{phantom,tension=1.4}{o1,v5}
\fmf{phantom}{o0,v5b}
\fmf{phantom,tension=1.11}{v3b,v6b}
\fmf{phantom,tension=1.38}{i0,v2}
\fmf{phantom,tension=1.38}{o0,v2}
\fmf{phantom,tension=1.8}{i0,v2b}
\fmf{phantom,tension=1.2}{o0,v2b}
\fmf{phantom,tension=1.2}{b0,v2b}
\fmf{phantom,tension=1.2}{b1,v2b}
\fmf{phantom,tension=1.2}{b2,v2b}
\fmf{phantom,tension=1.38}{i0,v4}
\fmf{phantom,tension=1.38}{o0,v4}
\fmf{phantom,tension=1.2}{i0,v4b}
\fmf{phantom,tension=1.8}{o0,v4b}
\fmf{phantom,tension=1.2}{b0,v4b}
\fmf{phantom,tension=1.2}{b1,v4b}
\fmf{phantom,tension=1.2}{b2,v4b}
\fmf{phantom,tension=2}{i3,v3}
\fmf{phantom,tension=2}{o3,v6}
\fmf{phantom,tension=2}{i3,v3b}
\fmf{phantom,tension=0.8}{o3,v3b}
\fmf{phantom,tension=0.8}{i3,v6b}
\fmf{phantom,tension=2}{o3,v6b}
\fmf{plain,tension=1.4,foreground=(1,,0,,0)}{v1,v2}
\fmf{plain,tension=1.4,foreground=(1,,0,,0)}{v4,v5}
\fmf{plain,foreground=(1,,0,,0)}{v1,v3}
\fmf{phantom,left=0.8,tension=0}{v1,v3}
\fmf{plain,foreground=(1,,0,,0)}{v5,v6}
\fmf{phantom,right=0.8,tension=0}{v5,v6}
\fmf{plain,foreground=(1,,0,,0)}{v2,v3}
\fmf{plain,foreground=(1,,0,,0)}{v4,v6}
\fmf{wiggly,tension=0.5,foreground=(1,,0,,0)}{v2,v4}
\fmf{wiggly,tension=2,foreground=(1,,0,,0)}{v3,v6}
\fmf{wiggly,right=0.8,tension=0,foreground=(1,,0,,0)}{v1,v5}
\fmf{phantom,tension=1}{v1,v1b}
\fmf{phantom,tension=0.2}{v2,v2b}
\fmf{phantom,tension=1.5}{v3,v3b}
\fmf{phantom,tension=0.2}{v4,v4b}
\fmf{phantom,tension=1}{v5,v5b}
\fmf{phantom,tension=1.5}{v6,v6b}
\end{fmfgraph*}
\end{fmffile}
\end{gathered} \\
& \hspace{0.3cm} +\frac{1}{4} \hspace{-0.4cm} \begin{gathered}
\begin{fmffile}{Diagrams/bosonic1PIEA_Gamma2_Diag3bis}
\begin{fmfgraph*}(25,25)
\fmfleft{i1,i2}
\fmfright{o1,o2}
\fmfbottom{i0,o0}
\fmftop{i3,o3}
\fmfv{decor.shape=circle,decor.size=2.0thick,foreground=(0,,0,,1)}{v1}
\fmfv{decor.shape=circle,decor.size=2.0thick,foreground=(0,,0,,1)}{v2}
\fmfv{decor.shape=circle,decor.size=2.0thick,foreground=(0,,0,,1)}{v3}
\fmfv{decor.shape=circle,decor.size=2.0thick,foreground=(0,,0,,1)}{v4}
\fmf{phantom}{i1,v1}
\fmf{phantom}{i2,v4}
\fmf{phantom}{o1,v2}
\fmf{phantom}{o2,v3}
\fmf{phantom}{i3,v4b}
\fmf{phantom}{o3,v3b}
\fmf{phantom}{i0,v1b}
\fmf{phantom}{o0,v2b}
\fmf{plain,tension=1.6,foreground=(1,,0,,0)}{v1,v2}
\fmf{plain,tension=1.6,foreground=(1,,0,,0)}{v3,v4}
\fmf{wiggly,tension=2.0,foreground=(1,,0,,0)}{v1,v4}
\fmf{wiggly,tension=2.0,foreground=(1,,0,,0)}{v2,v3}
\fmf{phantom,tension=0}{v1,v3}
\fmf{phantom,tension=0}{v2,v4}
\fmf{plain,right=0.8,tension=0,foreground=(1,,0,,0)}{v2,v3}
\fmf{plain,left=0.8,tension=0,foreground=(1,,0,,0)}{v1,v4}
\fmf{phantom}{v1,v1b}
\fmf{phantom}{v2,v2b}
\fmf{phantom}{v3,v3b}
\fmf{phantom}{v4,v4b}
\end{fmfgraph*}
\end{fmffile}
\end{gathered} \hspace{-0.4cm} +\frac{1}{8} \hspace{-0.5cm} \begin{gathered}
\begin{fmffile}{Diagrams/bosonic1PIEA_Gamma2_Diag4bis}
\begin{fmfgraph*}(25,25)
\fmfleft{i1,i2}
\fmfright{o1,o2}
\fmfbottom{i0,o0}
\fmftop{i3,o3}
\fmfv{decor.shape=circle,decor.size=2.0thick,foreground=(0,,0,,1)}{v1}
\fmfv{decor.shape=circle,decor.size=2.0thick,foreground=(0,,0,,1)}{v2}
\fmfv{decor.shape=circle,decor.size=2.0thick,foreground=(0,,0,,1)}{v3}
\fmfv{decor.shape=circle,decor.size=2.0thick,foreground=(0,,0,,1)}{v4}
\fmf{phantom}{i1,v1}
\fmf{phantom}{i2,v4}
\fmf{phantom}{o1,v2}
\fmf{phantom}{o2,v3}
\fmf{phantom}{i3,v4b}
\fmf{phantom}{o3,v3b}
\fmf{phantom}{i0,v1b}
\fmf{phantom}{o0,v2b}
\fmf{plain,tension=1.6,foreground=(1,,0,,0)}{v1,v2}
\fmf{plain,tension=1.6,foreground=(1,,0,,0)}{v3,v4}
\fmf{wiggly,tension=2.0,foreground=(1,,0,,0)}{v1,v4}
\fmf{wiggly,tension=2.0,foreground=(1,,0,,0)}{v2,v3}
\fmf{plain,tension=0,foreground=(1,,0,,0)}{v1,v3}
\fmf{plain,tension=0,foreground=(1,,0,,0)}{v2,v4}
\fmf{phantom,right=0.8,tension=0}{v2,v3}
\fmf{phantom,left=0.8,tension=0}{v1,v4}
\fmf{phantom}{v1,v1b}
\fmf{phantom}{v2,v2b}
\fmf{phantom}{v3,v3b}
\fmf{phantom}{v4,v4b}
\end{fmfgraph*}
\end{fmffile}
\end{gathered} \hspace{-0.5cm} \left.\rule{0cm}{1.1cm}\right) \\
& \hspace{0.3cm} + \mathcal{O}\big(\hbar^{3}\big) \;,
\end{split}
\label{eq:bosonic1PIEAIMGamma0DON}
\end{equation}
where all diagrams are 1PI with respect to each of the two types of propagator lines. The latter correspond to $\boldsymbol{G}_{\Phi}$ and $D_{\Phi}$ which are the original and collective field propagators respectively, conveniently collected in the superpropagator $\mathcal{G}_{\Phi}$ as follows:
\begin{equation}
\mathcal{G}_{\Phi} = \begin{pmatrix}
\boldsymbol{G}_{\Phi} & \vec{0} \\
\vec{0}^{\mathrm{T}} & D_{\Phi}
\end{pmatrix}\;,
\end{equation}
\begin{equation}
\boldsymbol{G}^{-1}_{\Phi;ab}(x,y) = \left(-\nabla^2_x + m^{2} + i\sqrt{\frac{\lambda}{3}}\eta(x)\right) \delta_{ab}\delta(x-y) \;,
\label{eq:bosonic1PIEApropagatorGJ0main}
\end{equation}
\begin{equation}
D^{-1}_{\Phi}(x,y) = \left.\frac{\delta^{2} S_{\mathrm{col},\mathcal{J}}[\widetilde{\sigma}]}{\delta\widetilde{\sigma}(x) \delta\widetilde{\sigma}(y)}\right|_{\vec{\widetilde{\varphi}}=\vec{\phi} \atop \widetilde{\sigma}=\eta} \;.
\label{eq:bosonic1PIEApropagatorHJ0main}
\end{equation}
Result~\eqref{eq:bosonic1PIEAIMGamma0DON} relies on the Feynman rules:
\begin{subequations}
\begin{align}
\begin{gathered}
\begin{fmffile}{Diagrams/1PIEAcol-G}
\begin{fmfgraph*}(20,16)
\fmfleft{i0,i1,i2,i3}
\fmfright{o0,o1,o2,o3}
\fmflabel{$x, a$}{v1}
\fmflabel{$y, b$}{v2}
\fmf{phantom}{i1,v1}
\fmf{phantom}{i2,v1}
\fmf{plain,tension=0.6,foreground=(1,,0,,0)}{v1,v2}
\fmf{phantom}{v2,o1}
\fmf{phantom}{v2,o2}
\end{fmfgraph*}
\end{fmffile}
\end{gathered} \quad &\rightarrow \boldsymbol{G}_{\Phi;ab}(x,y)\;, 
\label{eq:FeynRulesBosonic1PIEAJ0Gmain}\\
\begin{gathered}
\begin{fmffile}{Diagrams/1PIEAcol-D}
\begin{fmfgraph*}(20,20)
\fmfleft{i0,i1,i2,i3}
\fmfright{o0,o1,o2,o3}
\fmfv{label=$x$}{v1}
\fmfv{label=$y$}{v2}
\fmf{phantom}{i1,v1}
\fmf{phantom}{i2,v1}
\fmf{wiggly,tension=0.6,foreground=(1,,0,,0)}{v1,v2}
\fmf{phantom}{v2,o1}
\fmf{phantom}{v2,o2}
\end{fmfgraph*}
\end{fmffile}
\end{gathered} \quad &\rightarrow D_{\Phi}(x,y)\;,
\label{eq:FeynRulesBosonic1PIEAJ0Hmain} \\
\begin{gathered}
\begin{fmffile}{Diagrams/IPIEAcol-V}
\begin{fmfgraph*}(5,5)
\fmfleft{i1}
\fmfright{o1}
\fmfv{label=$x$,label.angle=-90,label.dist=4,foreground=(0,,0,,1)}{v1}
\fmf{plain,foreground=(1,,0,,0)}{i1,v1}
\fmf{plain,foreground=(1,,0,,0)}{v1,o1}
\fmflabel{$a$}{i1}
\fmflabel{$b$}{o1}
\fmfdot{v1}
\end{fmfgraph*}
\end{fmffile}
\end{gathered}\qquad &\rightarrow i\sqrt{\frac{\lambda}{3}}\delta_{ab}\;.
\label{eq:FeynRulesBosonic1PIEAJ0vertexDotmain}
\end{align}
\end{subequations}

As a next step, we study the collective 1PI EA in (0+0)\nobreakdash-D. In this limit, the propagators~\eqref{eq:bosonic1PIEApropagatorGJ0main} and~\eqref{eq:bosonic1PIEApropagatorHJ0main} respectively satisfy:
\begin{equation}
\boldsymbol{G}_{\Phi;ab}^{-1} = G_{\Phi}^{-1}\delta_{ab}=\left(m^{2}+i\sqrt{\frac{\lambda}{3}} \eta\right)\delta_{ab} \;,
\label{eq:1PIEAcolGpropagator}
\end{equation}
\begin{equation}
D_{\Phi}^{-1} = \frac{\lambda}{3} G_{\Phi} \phi_{N}^{2}+\frac{\lambda}{6} N G_{\Phi}^{2} + 1 \;,
\label{eq:1PIEAcolDpropagator}
\end{equation}
and Eq.~\eqref{eq:bosonic1PIEAIMGamma0DON} reduces to:
\begin{equation}
\begin{split}
\Gamma_{\mathrm{col}}^{(\mathrm{1PI})}\big(\Phi\big) = & \ S_{\mathrm{col}}(\eta) + \frac{1}{2}G_{\Phi}^{-1} \phi_{N}^{2} - \frac{\hbar}{2}\ln\big(D_{\Phi}\big) \\
& + \hbar^{2}\Bigg[-\frac{\lambda^{2}}{9}G_{\Phi}^{3}D_{\Phi}^{2}\phi_{N}^{2} -\frac{\lambda^{2}}{24} N G_{\Phi}^{4} D_{\Phi}^{2} \\
& + \frac{\lambda^{3}}{36}G_{\Phi}^{4} D_{\Phi}^{3} \phi_{N}^{4} + \frac{\lambda^{3}}{54} N G_{\Phi}^{5}D_\Phi^{3}\phi_{N}^{2} \\
& + \frac{\lambda^{3}}{324} N^{2} G_{\Phi}^{6} D_{\Phi}^{3}\Bigg] \\
& + \mathcal{O}\big(\hbar^{3}\big)\;,
\end{split}
\label{eq:bosonic1PIEAfinalexpression0DON}
\end{equation}
with $S_{\mathrm{col}}(\eta)=\frac{1}{2}\eta^{2}-\frac{N}{2}\ln\big(2\pi G_{\Phi}\big)$. As before, the expression for the 1PI EA becomes exploitable after fixing the relevant 1-point correlation function(s) (i.e. the components of $\Phi$ here), which is now done via the gap equations:
\begin{equation}
0 = \left.\frac{\partial \Gamma_{\mathrm{col}}^{(\mathrm{1PI})}\big(\Phi\big)}{\partial \phi_{N}}\right|_{\Phi=\overline{\Phi}} \;,
\label{eq:bosonic1PIEAGapEquationphiN0DON}
\end{equation}
and
\begin{equation}
0 = \left.\frac{\partial \Gamma_{\mathrm{col}}^{(\mathrm{1PI})}\big(\Phi\big)}{\partial \eta}\right|_{\Phi=\overline{\Phi}} \;,
\label{eq:bosonic1PIEAGapEquationeta0DON}
\end{equation}
with $\overline{\Phi}=\begin{pmatrix}
\vec{\overline{\phi}} & \overline{\eta} \end{pmatrix}^{\mathrm{T}} = \begin{pmatrix}
\overline{\phi}_{1} & \cdots & \overline{\phi}_{N-1} & \overline{\phi}_{N} & \overline{\eta}
\end{pmatrix}^{\mathrm{T}} = \begin{pmatrix}
0 & \cdots & 0 & \overline{\phi}_{N} & \overline{\eta}
\end{pmatrix}^{\mathrm{T}}$. Finally, the gs energy and density are respectively deduced after plugging the solution $\overline{\Phi}$ into:
\begin{equation}
E^\text{1PI EA;col}_{\mathrm{gs}} = \frac{1}{\hbar} \Gamma_{\mathrm{col}}^{(\mathrm{1PI})}\big(\Phi=\overline{\Phi}\big) \;,
\end{equation}
\begin{equation}
\rho^\text{1PI EA;col}_{\mathrm{gs}} = \frac{i}{N} \sqrt{\frac{12}{\lambda}} \overline{\eta} \;.
\label{eq:DeducerhogsbosonicGeneral1PIdiagEA0DON}
\end{equation}
The latter relation follows by taking the expectation value of the following classical equation of motion in the mixed representation:
\begin{equation}
\frac{\partial S_{\mathrm{mix}}\big(\vec{\widetilde{\varphi}},\widetilde{\sigma}\big)}{\partial\widetilde{\sigma}} = \widetilde{\sigma} + i\sqrt{\frac{\lambda}{12}} \ \vec{\widetilde{\varphi}}^{2} = 0 \;.
\label{eq:DerivSmixforrhogsinCollTheory}
\end{equation}

As for the original 1PI EA, the physical solutions found from the gap equations~\eqref{eq:bosonic1PIEAGapEquationphiN0DON} and~\eqref{eq:bosonic1PIEAGapEquationeta0DON} always satisfy $\vec{\overline{\phi}}=\vec{0}$. However, while the constraint of the $O(N)$ symmetry is too strong for the original 1PI EA, the collective one manages to capture non-perturbative physics thanks to the 1-point correlation function $\eta$ of the Hubbard-Stratonovich field. This illustrates a key advantage of HSTs by which one introduces a new field in the arena that is not constrained by the symmetries of the model under consideration. The Hubbard-Stratonovich field being a scalar with respect to the $O(N)$ transformations in the present case, its expectation value can be finite without spoiling the $O(N)$ symmetry and can therefore dress the propagator $\boldsymbol{G}_{\Phi}$ with non-perturbative physics. However, we can question the efficiency of the collective 1PI EA from another angle as Fig.~\ref{fig:1PIEA} shows for both $E_{\mathrm{gs}}$ and $\rho_{\mathrm{gs}}$ at $N=2$ that this implementation of SCPT is outperformed by the collective LE over the whole range of tested values for the coupling constant. Even though the determination of the diagrammatic representation of the 1PI EA is less demanding than that of the Schwinger functional (as the latter includes connected 1-particle-reducible (1PR) diagrams as opposed to the 1PI EA), the EA framework requires to solve also gap equations which are self-consistent for realistic models. This is a significant reason to favor the LE as compared to SCPT in this situation, especially considering the good performances of the collective LE beyond its first non-trivial order illustrated notably by Figs.~\ref{fig:O2OPTvsLE_1} and~\ref{fig:O2OPTvsLE_3}.

%%%%%%%%%%%%%%%%%%%%%%%%%%%%%%%
%%%%%%%%%%%%%%%%%%%%%%%%%%%%%%%
\section{\label{app:Diagrams}Determination of diagrams}
\subsection{\label{app:DiagramsOrigLEOPT}Original loop expansion and optimized perturbation theory}

In the framework of the original LE, each diagram contributing to the Schwinger functional $W^\text{LE;orig}\big[\vec{J},\boldsymbol{K}\big]$ (expressed by Eq.~\eqref{eq:WKjLoopExpansionStep3}) comes with a multiplicity (i.e. the number of Wick's contractions) given by:
\begin{equation}
\mathcal{M}_{\mathrm{LE},\mathrm{orig}}=\frac{(2p)!!4^{p}(2q)!!}{(2!)^{S+D} N_{\mathrm{V}}} \;,
\label{eq:MultiplicityDiagLoopExpansion}
\end{equation}
which can be deduced from the work of Ref.~\cite{kle00}. The integers $S$ and $D$ denote respectively the number of self and double connections with the propagator lines~\eqref{eq:FeynRulesLoopExpansionPropagator} (representing $\boldsymbol{G}_{\varphi_{\mathrm{cl}};JK;ab}(x,y)$) whereas $q$ and $p$ are the number of vertices~\eqref{eq:FeynRulesLoopExpansion3legVertex} and~\eqref{eq:FeynRulesLoopExpansion4legVertex} involved in the diagram under consideration. Finally, $N_{\mathrm{V}}$ corresponds to the number of vertex permutations that leave the diagram unchanged. These numbers are given in Tab.~\ref{tab:MultiplicityOPTdiagramsON} for the diagrams of Eq.~\eqref{eq:WKjLoopExpansionStep3}. Formula~\eqref{eq:MultiplicityDiagLoopExpansion} can be generalized for the OPT expansion by taking into account that the underlying diagrams might possess 2-leg vertices as well:
\begin{equation}
\mathcal{M}_{\mathrm{OPT}}=\frac{(2!)^{r}(r!)(2p)!!4^{p}(2q)!!}{(2!)^{S+D} N_{\mathrm{V}}}\;,
\label{eq:MultiplicityDiagOPT}
\end{equation}
where $r$ denotes the number of 2-leg vertices~\eqref{eq:FeynRulesOPTvertexSquare}, whereas $S$ and $D$ are now respectively the number of self and double connections with the propagator lines~\eqref{eq:FeynRulesOPTPropagator} (representing $\boldsymbol{G}_{\sigma;ab}(x,y)$). Every diagram contributing to $W^\text{OPT}$ (expressed by Eq.~\eqref{eq:WKjLoopExpansionStep3OPT}) up to order $\mathcal{O}\big(\delta^{3}\big)$ is given in Tab.~\ref{tab:MultiplicityOPTdiagramsON} together with the corresponding $S$, $D$ and $N_{\mathrm{V}}$ factors. Note that we always have $q=0$ in the framework of OPT as treated in section~\ref{sec:OPT} since the OPT expansion was performed around a trivial saddle point (i.e. for $\vec{\varphi}_{\mathrm{cl}}=\vec{0}$) in this section. It would be possible to design an OPT expansion around a non-trivial (i.e. finite) saddle point, just like the original LE for the studied toy model in the broken-symmetry regime (for which $\vec{\varphi}^{2}_{\mathrm{cl}}=-6m^{2}/\lambda$). Such an expansion is of little relevance for the zero-dimensional $O(N)$ model under consideration since the minimum of its exact effective potential $V^{\text{exact}}_{\text{eff}}\big(\vec{\phi}\big)$ always lies at $\vec{\phi}=\vec{0}$, as was discussed in section~\ref{sec:0DONmodel}. However, the additional OPT diagrams appearing when $\vec{\varphi}_{\mathrm{cl}}\neq\vec{0}$ are given in Tab.~\ref{tab:MultiplicityOPTdiagramsON} as well.

Finally, we illustrate an advantage of the studied toy model resulting from its (0+0)\nobreakdash-D nature: in (0+0)\nobreakdash-D and as opposed to finite-dimensional problems, partition functions or Schwinger functionals can be expressed in terms of a Lebesgue integral which can be directly expanded with respect to $\hbar$ (or another chosen parameter) after a relevant change of coordinates, thus bypassing the increasing complexity of the diagrammatic with the truncation order. For the original LE, the change of coordinates we are referring to consists in rewriting in hyperspherical coordinates the zero-dimensional counterpart of the generating functional~\eqref{eq:ZJKfiniteD} after the saddle point approximation introducing $\vec{\widetilde{\chi}}$ via $\vec{\widetilde{\varphi}}=\vec{\varphi}_{\mathrm{cl}}+\sqrt{\hbar} \ \vec{\widetilde{\chi}}$. This generating functional can be put in the form:
\begin{strip}
\begin{equation}
\begin{split}
Z\Big(\vec{J},\boldsymbol{K}\Big) = & \ e^{-\frac{1}{\hbar}S_{JK}(\vec{\varphi}_{\mathrm{cl}})} \int_{\mathbb{R}^N} d^{N}\vec{\widetilde{\chi}} \ e^{-\frac{1}{2} \widetilde{\chi}^a \boldsymbol{G}^{-1}_{\varphi_{\text{cl}};JK;ab} \widetilde{\chi}^b -\frac{\hbar^{1/2}\lambda}{3!} \vec{\widetilde{\chi}}^{2} \left(\vec{\varphi}_{\mathrm{cl}}\cdot\vec{\widetilde{\chi}}\right) - \frac{\hbar\lambda}{4!}\left(\vec{\widetilde{\chi}}^{2}\right)^{2}} \\
= & \ e^{-\frac{1}{\hbar}S_{JK}(\vec{\varphi}_{\mathrm{cl}})} \int_{\mathbb{R}^N} d^{N}\vec{\widetilde{\chi}} \ e^{-\frac{1}{2} \mathfrak{G}^{-1}_{\varphi_\text{cl};JK;\mathfrak{g}} \vec{\widetilde{A}}^{2} - \frac{1}{2} \boldsymbol{G}^{-1}_{\varphi_\text{cl};JK;NN} \widetilde{\chi}^{2}_{N} -\frac{\hbar^{1/2}\lambda}{3!}\varrho\widetilde{\chi}^{}_{N}\left(\vec{\widetilde{A}}^{2}+\widetilde{\chi}^{2}_{N}\right)-\frac{\hbar\lambda}{4!} \left(\left(\vec{\widetilde{A}}^{2}\right)^{2}+2\vec{\widetilde{A}}^{2}\widetilde{\chi}^{2}_{N}+\widetilde{\chi}^{4}_{N}\right)}\;,
\end{split}
\label{eq:ZKjHypersphericalCoordinatesStep1}
\end{equation}
\end{strip}
where $\vec{\widetilde{A}}$ satisfies:
\begin{equation}
\vec{\widetilde{\chi}} \equiv \begin{pmatrix}
\widetilde{\chi}^{}_{1} \\
\vdots \\
\widetilde{\chi}^{}_{N-1} \\
\widetilde{\chi}^{}_{N}
\end{pmatrix} \equiv \begin{pmatrix}
\vec{\widetilde{A}} \\
\widetilde{\chi}^{}_{N}
\end{pmatrix}\;,
\label{eq:AvectorLoopExpansion0DON}
\end{equation}
and $\mathfrak{G}^{-1}_{\varphi_\text{cl};JK;\mathfrak{g}} = m^2 + \frac{\lambda}{6} \varrho^2 - K$ is the diagonal part of the inverse Goldstone propagator defined by Eqs.~\eqref{eq:origLEGoldstoneProp} and~\eqref{eq:origLEGoldstonePropBis} whereas $\boldsymbol{G}^{-1}_{\varphi_\text{cl};JK;NN}=m^2 + \frac{\lambda}{2} \varrho^2 - K$ is associated to the Higgs mode (and expressed by Eq.~\eqref{eq:LEGN} in arbitrary dimensions), still assuming that the source $\boldsymbol{K}$ is a scalar in color space (i.e. $\boldsymbol{K}_{ab}=K\delta_{ab}$). Note also that $\varrho$ is the modulus introduced via the choice of coordinates~\eqref{eq:DefPhiclModulusRho} and expressed by Eq.~\eqref{eq:SolutionsmodulusrhoLE} at vanishing sources. Taking into account that the $O(N)$ symmetry of our toy model can be spontaneously broken (in the direction set by $a=N$ in color space according to Eq.~\eqref{eq:DefPhiclModulusRho}), isotropy in color space is only exhibited in the subspace of dimension $N-1$ in which $\vec{\widetilde{A}}$ lives. Performing the aforementioned change of coordinates within this subspace and carrying out integration over angular variables lead to:
\begin{equation}
\begin{split}
Z\Big(\vec{J},\boldsymbol{K}\Big) = & \ \Omega_{N-1} \ e^{-\frac{1}{\hbar}S_{JK}(\vec{\varphi}_{\mathrm{cl}})} \\
& \times \int_{-\infty}^{\infty} d\widetilde{\chi}^{}_{N} \ \mathcal{P}_{N-2}(\widetilde{\chi}^{}_{N},\varrho) \ e^{-\frac{1}{2} \boldsymbol{G}^{-1}_{\varphi_\text{cl};JK;NN} \widetilde{\chi}^{2}_{N}} \\
& \times e^{-\frac{\hbar^{1/2}\lambda}{3!}\varrho\widetilde{\chi}^{3}_{N}-\frac{\hbar\lambda}{4!}\widetilde{\chi}^{4}_{N}}\;,
\end{split}
\label{eq:ZKjHypersphericalCoordinatesStep2}
\end{equation}
where $\Omega_{N}$ is defined by Eq.~\eqref{eq:SN} and we have also:
\begin{equation}
\begin{split}
\mathcal{P}_{N}(\widetilde{\chi}^{}_{N},\varrho) = & \ \int_{0}^{\infty} d\widetilde{a} \ \widetilde{a}^{N} e^{-\frac{1}{2} \mathfrak{G}^{-1}_{\varphi_\text{cl};JK;\mathfrak{g}} \widetilde{a}^{2}-\frac{\hbar^{1/2}\lambda}{3!}\varrho\widetilde{\chi}^{}_{N} \widetilde{a}^{2}} \\
& \times e^{-\frac{\hbar\lambda}{12}\widetilde{\chi}^{2}_{N} \widetilde{a}^{2}-\frac{\hbar\lambda}{4!} \widetilde{a}^{4}}\;,
\end{split}
\label{eq:ZKjHypersphericalCoordinatesPN}
\end{equation}
with $\widetilde{a}$ being the norm of $\vec{\widetilde{A}}$, i.e. $\widetilde{a}\equiv\Big|\vec{\widetilde{A}}\Big|$. Expanding the right-hand side of Eq.~\eqref{eq:ZKjHypersphericalCoordinatesStep2} with respect to $\hbar$ yields:
\begin{strip}
\begin{equation*}
\begin{split}
\scalebox{0.85}{${\displaystyle Z^\text{LE;orig}\Big(\vec{J},\boldsymbol{K}\Big) = }$} & \scalebox{0.85}{${\displaystyle \ e^{-\frac{1}{\hbar}S_{JK}(\vec{\varphi}_{\mathrm{cl}})} (2\pi)^{\frac{N}{2}} \mathfrak{G}_{\varphi_\text{cl};JK;\mathfrak{g}}^{\frac{N-1}{2}} \sqrt{\boldsymbol{G}_{\varphi_\text{cl};JK;NN}} \ \Bigg\{ 1 }$} \\
& \scalebox{0.85}{${\displaystyle + \frac{\hbar\lambda}{72} \Big[-3 \mathfrak{G}_{\varphi_\text{cl};JK;\mathfrak{g}}^2 \left(-1 + N^2\right) + 15 \boldsymbol{G}_{\varphi_\text{cl};JK;NN}^3 \lambda \varrho^{2} + \boldsymbol{G}_{\varphi_\text{cl};JK;NN}^2 \left(-9 + 6 \mathfrak{G}_{\varphi_\text{cl};JK;\mathfrak{g}} \left(-1 + N\right) \lambda \varrho^{2}\right) }$} \\
& \hspace{0.85cm} \scalebox{0.85}{${\displaystyle + \boldsymbol{G}_{\varphi_\text{cl};JK;NN} \mathfrak{G}_{\varphi_\text{cl};JK;\mathfrak{g}} \left(-1 + N\right) \left(-6 + \mathfrak{G}_{\varphi_\text{cl};JK;\mathfrak{g}} \left(1 + N\right) \lambda \varrho^{2}\right) \Big] }$} \\
& \scalebox{0.85}{${\displaystyle + \frac{\hbar^{2}\lambda^{2}}{10368} \Big[ 9 \mathfrak{G}_{\varphi_\text{cl};JK;\mathfrak{g}}^4 \left(-15 - 8 N + 14 N^2 + 8 N^3 + N^4\right) + 3465 \boldsymbol{G}_{\varphi_\text{cl};JK;NN}^6 \lambda^2 \varrho^{4} }$} \\
& \hspace{1.12cm} \scalebox{0.85}{${\displaystyle + 630 \boldsymbol{G}_{\varphi_\text{cl};JK;NN}^5 \lambda \varrho^{2} \left(-9 + 2 \mathfrak{G}_{\varphi_\text{cl};JK;\mathfrak{g}} \left(-1 + N\right) \lambda \varrho^{2}\right) }$} \\
& \hspace{1.12cm} \scalebox{0.85}{${\displaystyle - 6 \boldsymbol{G}_{\varphi_\text{cl};JK;NN} \mathfrak{G}_{\varphi_\text{cl};JK;\mathfrak{g}}^3 \left(-3 - N + 3 N^2 + N^3\right) \left(-6 + \mathfrak{G}_{\varphi_\text{cl};JK;\mathfrak{g}} \left(5 + N\right) \lambda \varrho^{2}\right) }$} \\
& \hspace{1.12cm} \scalebox{0.85}{${\displaystyle + 105 \boldsymbol{G}_{\varphi_\text{cl};JK;NN}^4 \left(9 - 24 \mathfrak{G}_{\varphi_\text{cl};JK;\mathfrak{g}} \left(-1 + N\right) \lambda \varrho^{2} + 2 \mathfrak{G}_{\varphi_\text{cl};JK;\mathfrak{g}}^2 \left(-1 + N^2\right) \lambda^2 \varrho^{4} \right) }$} \\
& \hspace{1.12cm} \scalebox{0.85}{${\displaystyle + 20 \boldsymbol{G}_{\varphi_\text{cl};JK;NN}^3 \mathfrak{G}_{\varphi_\text{cl};JK;\mathfrak{g}} \left(-1 + N\right) \left(27 - 27 \mathfrak{G}_{\varphi_\text{cl};JK;\mathfrak{g}} \left(1 + N\right) \lambda \varrho^{2} + \mathfrak{G}_{\varphi_\text{cl};JK;\mathfrak{g}}^2 \left(3 + 4 N + N^2\right) \lambda^2 \varrho^{4} \right) }$} \\
& \hspace{1.12cm} \scalebox{0.85}{${\displaystyle + \boldsymbol{G}_{\varphi_\text{cl};JK;NN}^2 \mathfrak{G}_{\varphi_\text{cl};JK;\mathfrak{g}}^2 \left(-1 + N^2\right) \left(162 - 72 \mathfrak{G}_{\varphi_\text{cl};JK;\mathfrak{g}} \left(3 + N\right) \lambda \varrho^{2} + \mathfrak{G}_{\varphi_\text{cl};JK;\mathfrak{g}}^2 \left(15 + 8 N + N^2\right) \lambda^2 \varrho^{4} \right) \Big] }$} \\
& \scalebox{0.85}{${\displaystyle + \frac{\hbar^{3}\lambda^{3}}{2239488} \Big[ -27 \mathfrak{G}_{\varphi_\text{cl};JK;\mathfrak{g}}^6 \left(-945 - 744 N + 739 N^2 + 720 N^3 + 205 N^4 + 24 N^5 + N^6\right) }$} \\
& \hspace{1.56cm} \scalebox{0.85}{${\displaystyle + 2297295 \boldsymbol{G}_{\varphi_\text{cl};JK;NN}^9 \lambda^3 \varrho^{6} + 405405 \boldsymbol{G}_{\varphi_\text{cl};JK;NN}^8 \lambda^2 \varrho^{4} \left(-15 + 2 \mathfrak{G}_{\varphi_\text{cl};JK;\mathfrak{g}} \left(-1 + N\right) \lambda \varrho^{2}\right) }$} \\
& \hspace{1.56cm} \scalebox{0.85}{${\displaystyle + 27 \boldsymbol{G}_{\varphi_\text{cl};JK;NN} \mathfrak{G}_{\varphi_\text{cl};JK;\mathfrak{g}}^5 \left(-1 + N\right) \left(105 + 176 N + 86 N^2 + 16 N^3 + N^4\right) \left(-6 + \mathfrak{G}_{\varphi_\text{cl};JK;\mathfrak{g}} \left(9 + N\right) \lambda \varrho^{2}\right) }$} \\
& \hspace{1.56cm} \scalebox{0.85}{${\displaystyle + 135135 \boldsymbol{G}_{\varphi_\text{cl};JK;NN}^7 \lambda \varrho^{2} \left(27 - 18 \mathfrak{G}_{\varphi_\text{cl};JK;\mathfrak{g}} \left(-1 + N\right) \lambda \varrho^{2} + \mathfrak{G}_{\varphi_\text{cl};JK;\mathfrak{g}}^2 \left(-1 + N^2\right) \lambda^2 \varrho^{4} \right) }$} \\
& \hspace{1.56cm} \scalebox{0.85}{${\displaystyle - 9 \boldsymbol{G}_{\varphi_\text{cl};JK;NN}^2 \mathfrak{G}_{\varphi_\text{cl};JK;\mathfrak{g}}^4 \left(-15 - 8 N + 14 N^2 + 8 N^3 + N^4\right) \Big(135 - 54 \mathfrak{G}_{\varphi_\text{cl};JK;\mathfrak{g}} \left(7 + N\right) \lambda \varrho^{2} }$} \\
& \hspace{1.56cm} \scalebox{0.85}{${\displaystyle + \mathfrak{G}_{\varphi_\text{cl};JK;\mathfrak{g}}^2 \left(63 + 16 N + N^2\right) \lambda^2 \varrho^{4} \Big) + 3465 \boldsymbol{G}_{\varphi_\text{cl};JK;NN}^6 \Big(-81 + 486 \mathfrak{G}_{\varphi_\text{cl};JK;\mathfrak{g}} \left(-1 + N\right) \lambda \varrho^{2} }$} \\
& \hspace{1.56cm} \scalebox{0.85}{${\displaystyle - 135 \mathfrak{G}_{\varphi_\text{cl};JK;\mathfrak{g}}^2 \left(-1 + N^2\right) \lambda^2 \varrho^{4} + 4 \mathfrak{G}_{\varphi_\text{cl};JK;\mathfrak{g}}^3 \left(-3 - N + 3 N^2 + N^3\right) \lambda^3 \varrho^{6}\Big) }$} \\
& \hspace{1.56cm} \scalebox{0.85}{${\displaystyle + \boldsymbol{G}_{\varphi_\text{cl};JK;NN}^3 \mathfrak{G}_{\varphi_\text{cl};JK;\mathfrak{g}}^3 \left(-3 - N + 3 N^2 + N^3\right) \Big(-8100 + 6075 \mathfrak{G}_{\varphi_\text{cl};JK;\mathfrak{g}} \left(5 + N\right) \lambda \varrho^{2} }$} \\
& \hspace{1.56cm} \scalebox{0.85}{${\displaystyle - 270 \mathfrak{G}_{\varphi_\text{cl};JK;\mathfrak{g}}^2 \left(35 + 12 N + N^2\right) \lambda^2 \varrho^{4} + \mathfrak{G}_{\varphi_\text{cl};JK;\mathfrak{g}}^3 \left(315 + 143 N + 21 N^2 + N^3\right) \lambda^3 \varrho^{6} \Big) }$} \\
& \hspace{1.56cm} \scalebox{0.85}{${\displaystyle + 945 \boldsymbol{G}_{\varphi_\text{cl};JK;NN}^5 \mathfrak{G}_{\varphi_\text{cl};JK;\mathfrak{g}} \Big(-162 \left(-1 + N\right) + 405 \mathfrak{G}_{\varphi_\text{cl};JK;\mathfrak{g}} \left(-1 + N^2\right) \lambda \varrho^{2} }$} \\
& \hspace{1.56cm} \scalebox{0.85}{${\displaystyle - 60 \mathfrak{G}_{\varphi_\text{cl};JK;\mathfrak{g}}^2 \left(-3 - N + 3 N^2 + N^3\right) \lambda^2 \varrho^{4} + \mathfrak{G}_{\varphi_\text{cl};JK;\mathfrak{g}}^3 \left(-15 - 8 N + 14 N^2 + 8 N^3 + N^4\right) \lambda^3 \varrho^{6} \Big) }$}
\end{split}
\end{equation*}
\pagebreak
\begin{equation}
\begin{split}
& \hspace{1.56cm} \scalebox{0.85}{${\displaystyle + 21 \boldsymbol{G}_{\varphi_\text{cl};JK;NN}^4 \mathfrak{G}_{\varphi_\text{cl};JK;\mathfrak{g}}^2 \Big(-2025 \left(-1 + N^2\right) + 2700 \mathfrak{G}_{\varphi_\text{cl};JK;\mathfrak{g}} \left(-3 - N + 3 N^2 + N^3\right) \lambda \varrho^{2} }$} \\
& \hspace{1.56cm} \scalebox{0.85}{${\displaystyle - 225 \mathfrak{G}_{\varphi_\text{cl};JK;\mathfrak{g}}^2 \left(-15 - 8 N + 14 N^2 + 8 N^3 + N^4\right) \lambda^2 \varrho^{4} }$} \\
& \hspace{1.56cm} \scalebox{0.85}{${\displaystyle + 2 \mathfrak{G}_{\varphi_\text{cl};JK;\mathfrak{g}}^3 \left(-105 - 71 N + 90 N^2 + 70 N^3 + 15 N^4 + N^5\right) \lambda^3 \varrho^{6} \Big) \Big] }$} \\
& \scalebox{0.85}{${\displaystyle + \mathcal{O}\big(\hbar^{4}\big) \Bigg\}\;,}$}
\end{split}
\label{eq:ResultZLoopExpansion0DONAppendix}
\end{equation}
or, considering $W^\text{LE;orig}\big(\vec{J},\boldsymbol{K}\big)\equiv\hbar\ln\Big(Z^\text{LE;orig}\big(\vec{J},\boldsymbol{K}\big)\Big)$ instead of $Z^\text{LE;orig}\big(\vec{J},\boldsymbol{K}\big)$:
\begin{equation}
\begin{split}
\scalebox{0.85}{${\displaystyle W^\text{LE;orig}\Big(\vec{J},\boldsymbol{K}\Big) = }$} & \scalebox{0.85}{${\displaystyle -S_{JK}(\vec{\varphi}_{\mathrm{cl}}) + \frac{\hbar}{2} \left[(N-1)\ln\big(2\pi \mathfrak{G}_{\varphi_\text{cl};JK;\mathfrak{g}}\big)+\ln\big(2\pi \boldsymbol{G}_{\varphi_\text{cl};JK;NN}\big)\right] }$} \\
& \scalebox{0.85}{${\displaystyle + \frac{\hbar^{2}\lambda}{72} \Big[ -3 \mathfrak{G}_{\varphi_\text{cl};JK;\mathfrak{g}}^2 \left(-1 + N^2\right) + 15 \boldsymbol{G}_{\varphi_\text{cl};JK;NN}^3 \lambda \varrho^2 + \boldsymbol{G}_{\varphi_\text{cl};JK;NN}^2 \left(-9 + 6 \mathfrak{G}_{\varphi_\text{cl};JK;\mathfrak{g}} \left(-1 + N\right) \lambda \varrho^2\right) }$} \\
& \hspace{0.98cm} \scalebox{0.85}{${\displaystyle + \boldsymbol{G}_{\varphi_\text{cl};JK;NN} \mathfrak{G}_{\varphi_\text{cl};JK;\mathfrak{g}} \left(-1 + N\right) \left(-6 + \mathfrak{G}_{\varphi_\text{cl};JK;\mathfrak{g}} \left(1 + N\right) \lambda \varrho^2\right) \Big] }$} \\
& \scalebox{0.85}{${\displaystyle + \frac{\hbar^{3}\lambda^{2}}{1296} \Big[ 9 \mathfrak{G}_{\varphi_\text{cl};JK;\mathfrak{g}}^4 \left(-2 - N + 2 N^2 + N^3\right) + 405 \boldsymbol{G}_{\varphi_\text{cl};JK;NN}^6 \lambda^2 \varrho^{4} + 135 \boldsymbol{G}_{\varphi_\text{cl};JK;NN}^5 \lambda \varrho^2 \left(-5 + \mathfrak{G}_{\varphi_\text{cl};JK;\mathfrak{g}} \left(-1 + N\right) \lambda \varrho^2\right) }$} \\
& \hspace{0.98cm} \scalebox{0.85}{${\displaystyle - 6 \boldsymbol{G}_{\varphi_\text{cl};JK;NN} \mathfrak{G}_{\varphi_\text{cl};JK;\mathfrak{g}}^3 \left(-1 + N^2\right) \left(-3 + \mathfrak{G}_{\varphi_\text{cl};JK;\mathfrak{g}} \left(2 + N\right) \lambda \varrho^2\right) }$} \\
& \hspace{0.98cm} \scalebox{0.85}{${\displaystyle + \boldsymbol{G}_{\varphi_\text{cl};JK;NN}^3 \mathfrak{G}_{\varphi_\text{cl};JK;\mathfrak{g}} \left(-1 + N\right) \left(54 - 9 \mathfrak{G}_{\varphi_\text{cl};JK;\mathfrak{g}} \left(7 + 5 N\right) \lambda \varrho^2 + \mathfrak{G}_{\varphi_\text{cl};JK;\mathfrak{g}}^2 \left(9 + 10 N + N^2\right) \lambda^2 \varrho^{4}\right) }$} \\
& \hspace{0.98cm} \scalebox{0.85}{${\displaystyle + 9 \boldsymbol{G}_{\varphi_\text{cl};JK;NN}^4 \left(12 - 31 \mathfrak{G}_{\varphi_\text{cl};JK;\mathfrak{g}} \left(-1 + N \right) \lambda \varrho^2 + \mathfrak{G}_{\varphi_\text{cl};JK;\mathfrak{g}}^2 \left(-3 + N + 2 N^2\right) \lambda^2 \varrho^{4}\right) }$} \\
& \hspace{0.98cm} \scalebox{0.85}{${\displaystyle + \boldsymbol{G}_{\varphi_\text{cl};JK;NN}^2 \mathfrak{G}_{\varphi_\text{cl};JK;\mathfrak{g}}^2 \left(-1 + N\right) \Big( 18 - 33 \mathfrak{G}_{\varphi_\text{cl};JK;\mathfrak{g}} \lambda \varrho^2 + 2 \mathfrak{G}_{\varphi_\text{cl};JK;\mathfrak{g}}^2 \lambda^2 \varrho^{4} + \mathfrak{G}_{\varphi_\text{cl};JK;\mathfrak{g}} N^2 \lambda \varrho^2 \left(-3 + \mathfrak{G}_{\varphi_\text{cl};JK;\mathfrak{g}} \lambda \varrho^2\right) }$} \\
& \hspace{0.98cm} \scalebox{0.85}{${\displaystyle + 3 N \left(3 - 12 \mathfrak{G}_{\varphi_\text{cl};JK;\mathfrak{g}} \lambda \varrho^2 + \mathfrak{G}_{\varphi_\text{cl};JK;\mathfrak{g}}^2 \lambda^2 \varrho^{4}\right) \Big) \Big] }$} \\
\scalebox{0.85}{${\displaystyle \textcolor{white}{W^\text{LE;orig}\Big(\vec{J},\boldsymbol{K}\Big) =} }$} & \scalebox{0.85}{${\displaystyle + \frac{\hbar^{4}\lambda^{3}}{279936} \Big[ -108 \mathfrak{G}_{\varphi_\text{cl};JK;\mathfrak{g}}^6 \left(-31 - 24 N + 26 N^2 + 24 N^3 + 5 N^4\right) + 268515 \boldsymbol{G}_{\varphi_\text{cl};JK;NN}^9 \lambda^3 \varrho^{6} }$} \\
& \hspace{1.408cm} \scalebox{0.85}{${\displaystyle + 3645 \boldsymbol{G}_{\varphi_\text{cl};JK;NN}^8 \lambda^2 \varrho^{4} \left(-197 + 24 \mathfrak{G}_{\varphi_\text{cl};JK;\mathfrak{g}} \left(-1 + N\right) \lambda \varrho^2\right) }$} \\
& \hspace{1.408cm} \scalebox{0.85}{${\displaystyle + 2430 \boldsymbol{G}_{\varphi_\text{cl};JK;NN}^7 \lambda \varrho^2 \left(178 - 110 \mathfrak{G}_{\varphi_\text{cl};JK;\mathfrak{g}} \left(-1 + N\right) \lambda \varrho^2 + \mathfrak{G}_{\varphi_\text{cl};JK;\mathfrak{g}}^2 \left(-7 + 2 N + 5 N^2\right) \lambda^2 \varrho^{4}\right) }$} \\
& \hspace{1.408cm} \scalebox{0.85}{${\displaystyle + 27 \boldsymbol{G}_{\varphi_\text{cl};JK;NN}^5 \mathfrak{G}_{\varphi_\text{cl};JK;\mathfrak{g}} \left(-1 + N\right) \Big(-576 + 12 \mathfrak{G}_{\varphi_\text{cl};JK;\mathfrak{g}} \left(139 + 102 N\right) \lambda \varrho^2 }$} \\
& \hspace{1.408cm} \scalebox{0.85}{${\displaystyle - 4 \mathfrak{G}_{\varphi_\text{cl};JK;\mathfrak{g}}^2 \left(217 + 242 N + 31 N^2\right) \lambda^2 \varrho^{4} + \mathfrak{G}_{\varphi_\text{cl};JK;\mathfrak{g}}^3 \left(71 + 99 N + 29 N^2 + N^3\right) \lambda^3 \varrho^{6}\Big) }$} \\
& \hspace{1.408cm} \scalebox{0.85}{${\displaystyle + 54 \boldsymbol{G}_{\varphi_\text{cl};JK;NN}^6 \Big(-594 + 3438 \mathfrak{G}_{\varphi_\text{cl};JK;\mathfrak{g}} \left(-1 + N\right) \lambda \varrho^2 - 3 \mathfrak{G}_{\varphi_\text{cl};JK;\mathfrak{g}}^2 \left(-351 + 92 N + 259 N^2\right) \lambda^2 \varrho^{4} }$} \\
& \hspace{1.408cm} \scalebox{0.85}{${\displaystyle + 2 \mathfrak{G}_{\varphi_\text{cl};JK;\mathfrak{g}}^3 \left(-53 - 6 N + 51 N^2 + 8 N^3\right) \lambda^3 \varrho^{6}\Big) }$} \\
& \hspace{1.408cm} \scalebox{0.85}{${\displaystyle + 108 \boldsymbol{G}_{\varphi_\text{cl};JK;NN} \mathfrak{G}_{\varphi_\text{cl};JK;\mathfrak{g}}^5 \left(-1 + N^2\right) \Big(-24 + 31 \mathfrak{G}_{\varphi_\text{cl};JK;\mathfrak{g}} \lambda \varrho^2 + 5 \mathfrak{G}_{\varphi_\text{cl};JK;\mathfrak{g}} N^2 \lambda \varrho^2 }$} \\
& \hspace{1.408cm} \scalebox{0.85}{${\displaystyle + 12 N \left(-1 + 2 \mathfrak{G}_{\varphi_\text{cl};JK;\mathfrak{g}} \lambda \varrho^2\right)\Big) }$} \\
& \hspace{1.408cm} \scalebox{0.85}{${\displaystyle -36 \boldsymbol{G}_{\varphi_\text{cl};JK;NN}^2 \mathfrak{G}_{\varphi_\text{cl};JK;\mathfrak{g}}^4 \left(-1 + N^2\right) \Big( 63 - 201 \mathfrak{G}_{\varphi_\text{cl};JK;\mathfrak{g}} \lambda \varrho^2 + 31 \mathfrak{G}_{\varphi_\text{cl};JK;\mathfrak{g}}^2 \lambda^2 \varrho^{4} }$} \\
& \hspace{1.408cm} \scalebox{0.85}{${\displaystyle + \mathfrak{G}_{\varphi_\text{cl};JK;\mathfrak{g}} N^2 \lambda \varrho^2 \left(-9 + 5 \mathfrak{G}_{\varphi_\text{cl};JK;\mathfrak{g}} \lambda \varrho^2\right) + 6 N \left(3 - 19 \mathfrak{G}_{\varphi_\text{cl};JK;\mathfrak{g}} \lambda \varrho^2 + 4 \mathfrak{G}_{\varphi_\text{cl};JK;\mathfrak{g}}^2 \lambda^2 \varrho^{4}\right)\Big) }$} \\
& \hspace{1.408cm} \scalebox{0.85}{${\displaystyle + 4 \boldsymbol{G}_{\varphi_\text{cl};JK;NN}^3 \mathfrak{G}_{\varphi_\text{cl};JK;\mathfrak{g}}^3 \left(-1 + N\right) \Big( -864 + 2943 \mathfrak{G}_{\varphi_\text{cl};JK;\mathfrak{g}} \lambda \varrho^2 - 990 \mathfrak{G}_{\varphi_\text{cl};JK;\mathfrak{g}}^2 \lambda^2 \varrho^{4} + 31 \mathfrak{G}_{\varphi_\text{cl};JK;\mathfrak{g}}^3 \lambda^3 \varrho^{6} }$} \\
& \hspace{1.408cm} \scalebox{0.85}{${\displaystyle + \mathfrak{G}_{\varphi_\text{cl};JK;\mathfrak{g}} N^3 \lambda \varrho^2 \left(27 - 54 \mathfrak{G}_{\varphi_\text{cl};JK;\mathfrak{g}} \lambda \varrho^2 + 5 \mathfrak{G}_{\varphi_\text{cl};JK;\mathfrak{g}}^2 \lambda^2 \varrho^{4}\right) }$} \\
& \hspace{1.408cm} \scalebox{0.85}{${\displaystyle + N^2 \left(-54 + 1188 \mathfrak{G}_{\varphi_\text{cl};JK;\mathfrak{g}} \lambda \varrho^2 - 630 \mathfrak{G}_{\varphi_\text{cl};JK;\mathfrak{g}}^2 \lambda^2 \varrho^{4} + 29 \mathfrak{G}_{\varphi_\text{cl};JK;\mathfrak{g}}^3 \lambda^3 \varrho^{6}\right) }$} \\
& \hspace{1.408cm} \scalebox{0.85}{${\displaystyle + N \left(-864 + 4104 \mathfrak{G}_{\varphi_\text{cl};JK;\mathfrak{g}} \lambda \varrho^2 - 1566 \mathfrak{G}_{\varphi_\text{cl};JK;\mathfrak{g}}^2 \lambda^2 \varrho^{4} + 55 \mathfrak{G}_{\varphi_\text{cl};JK;\mathfrak{g}}^3 \lambda^3 \varrho^{6}\right) \Big) }$} \\
& \hspace{1.408cm} \scalebox{0.85}{${\displaystyle + 9 \boldsymbol{G}_{\varphi_\text{cl};JK;NN}^4 \mathfrak{G}_{\varphi_\text{cl};JK;\mathfrak{g}}^2 \left(-1 + N\right) \Big( -540 + 2664 \mathfrak{G}_{\varphi_\text{cl};JK;\mathfrak{g}} \lambda \varrho^2 - 1047 \mathfrak{G}_{\varphi_\text{cl};JK;\mathfrak{g}}^2 \lambda^2 \varrho^{4} + 68 \mathfrak{G}_{\varphi_\text{cl};JK;\mathfrak{g}}^3 \lambda^3 \varrho^{6} }$} \\
& \hspace{1.408cm} \scalebox{0.85}{${\displaystyle + \mathfrak{G}_{\varphi_\text{cl};JK;\mathfrak{g}}^2 N^3 \lambda^2 \varrho^{4} \left(-13 + 4 \mathfrak{G}_{\varphi_\text{cl};JK;\mathfrak{g}} \lambda \varrho^2\right) + \mathfrak{G}_{\varphi_\text{cl};JK;\mathfrak{g}} N^2 \lambda \varrho^2 \left(324 - 429 \mathfrak{G}_{\varphi_\text{cl};JK;\mathfrak{g}} \lambda \varrho^2 + 44 \mathfrak{G}_{\varphi_\text{cl};JK;\mathfrak{g}}^2 \lambda^2 \varrho^{4}\right) }$} \\
& \hspace{1.408cm} \scalebox{0.85}{${\displaystyle + N \left(-324 + 2916 \mathfrak{G}_{\varphi_\text{cl};JK;\mathfrak{g}} \lambda \varrho^2 - 1463 \mathfrak{G}_{\varphi_\text{cl};JK;\mathfrak{g}}^2 \lambda^2 \varrho^{4} + 108 \mathfrak{G}_{\varphi_\text{cl};JK;\mathfrak{g}}^3 \lambda^3 \varrho^{6}\right) \Big) \Big] }$} \\
& \scalebox{0.85}{${\displaystyle + \mathcal{O}\big(\hbar^{5}\big)\;.}$}
\end{split}
\label{eq:ResultWLoopExpansion0DONAppendix}
\end{equation}
\end{strip}

\pagebreak
\noindent
It is then straightforward to deduce an expression for the gs energy and density from Eq.~\eqref{eq:ResultZLoopExpansion0DONAppendix} or~\eqref{eq:ResultWLoopExpansion0DONAppendix} combined with the relations:
\begin{equation}
\begin{split}
E^\text{LE;orig}_{\mathrm{gs}} = & - \ln\left( Z^\text{LE;orig}\Big(\vec{J}=\vec{0},\boldsymbol{K}=\boldsymbol{0}\Big)\right) \\
= & - \frac{1}{\hbar} W^\text{LE;orig}\Big(\vec{J}=\vec{0},\boldsymbol{K}=\boldsymbol{0}\Big) \;,
\end{split}
\label{eq:DefEgsExactZexact0DONAppendix}
\end{equation}
\begin{equation}
\rho^\text{LE;orig}_{\mathrm{gs}} = \frac{2\hbar}{N} \frac{\partial E^\text{LE;orig}_{\mathrm{gs}}}{\partial m^{2}} \;,
\label{eq:DefrhogsExactwithExpectationValue0DONAppendix}
\end{equation}
which follow respectively from Eqs.~\eqref{eq:DefEgsExactZexact0DON} and~\eqref{eq:DefrhogsExactwithExpectationValue0DON}. A similar procedure can also be followed to derive the corresponding expansions within the collective LE or OPT.

%==============================
\subsection{Mixed loop expansion}
\label{sec:DiagLEM}

After following the recipe outlined notably in section~\ref{sec:OrigLE} for the original LE, one can show that the mixed LE yields the following diagrammatic expression of the corresponding Schwinger functional up to second order in $\hbar$:
\begin{equation}
\begin{split}
W^\text{LE;mix}\big[\mathcal{J},\mathcal{K}\big] = & -S_{\mathrm{mix},\mathcal{J}\mathcal{K}}[\Psi_{\mathrm{cl}}] + \frac{\hbar}{2} \mathcal{ST}r\left[\ln\big(\mathcal{G}_{\Psi_\text{cl};\mathcal{J}\mathcal{K}}\big)\right] \\
& - \hbar^{2} \left(\rule{0cm}{1.0cm}\right. \hspace{-0.05cm} \frac{1}{24}\begin{gathered}
\begin{fmffile}{Diagrams/LoopExpansionMixedHS_Hartree}
\begin{fmfgraph}(30,20)
\fmfleft{i}
\fmfright{o}
\fmfv{decor.shape=circle,decor.size=2.0thick,foreground=(0,,0,,1)}{v1}
\fmfv{decor.shape=circle,decor.size=2.0thick,foreground=(0,,0,,1)}{v2}
\fmf{phantom,tension=10}{i,i1}
\fmf{phantom,tension=10}{o,o1}
\fmf{plain,left,tension=0.5}{i1,v1,i1}
\fmf{plain,right,tension=0.5}{o1,v2,o1}
\fmf{wiggly}{v1,v2}
\end{fmfgraph}
\end{fmffile}
\end{gathered} \\
& +\frac{1}{12}\begin{gathered}
\begin{fmffile}{Diagrams/LoopExpansionMixedHS_Fock}
\begin{fmfgraph}(15,15)
\fmfleft{i}
\fmfright{o}
\fmfv{decor.shape=circle,decor.size=2.0thick,foreground=(0,,0,,1)}{v1}
\fmfv{decor.shape=circle,decor.size=2.0thick,foreground=(0,,0,,1)}{v2}
\fmf{phantom,tension=11}{i,v1}
\fmf{phantom,tension=11}{v2,o}
\fmf{plain,left,tension=0.4}{v1,v2,v1}
\fmf{wiggly}{v1,v2}
\end{fmfgraph}
\end{fmffile}
\end{gathered} + \frac{1}{6}\begin{gathered}
\begin{fmffile}{Diagrams/LoopExpansionMixedHS_Diag1}
\begin{fmfgraph}(30,20)
\fmfleft{i}
\fmfright{o}
\fmfv{decor.shape=circle,decor.size=2.0thick,foreground=(0,,0,,1)}{v1}
\fmfv{decor.shape=circle,decor.size=2.0thick,foreground=(0,,0,,1)}{v2}
\fmf{phantom,tension=10}{i,i1}
\fmf{phantom,tension=10}{o,o1}
\fmf{dashes,left,tension=0.5}{i1,v1,i1}
\fmf{dashes,right,tension=0.5}{o1,v2,o1}
\fmf{plain}{v1,v2}
\end{fmfgraph}
\end{fmffile}
\end{gathered} \\
& + \frac{1}{6}\begin{gathered}
\begin{fmffile}{Diagrams/LoopExpansionMixedHS_Diag2}
\begin{fmfgraph}(30,20)
\fmfleft{i}
\fmfright{o}
\fmfv{decor.shape=circle,decor.size=2.0thick,foreground=(0,,0,,1)}{v1}
\fmfv{decor.shape=circle,decor.size=2.0thick,foreground=(0,,0,,1)}{v2}
\fmf{phantom,tension=10}{i,i1}
\fmf{phantom,tension=10}{o,o1}
\fmf{plain,left,tension=0.5}{i1,v1,i1}
\fmf{dashes,right,tension=0.5}{o1,v2,o1}
\fmf{dashes}{v1,v2}
\end{fmfgraph}
\end{fmffile}
\end{gathered} + \frac{1}{6}\begin{gathered}
\begin{fmffile}{Diagrams/LoopExpansionMixedHS_Diag3}
\begin{fmfgraph}(15,15)
\fmfleft{i}
\fmfright{o}
\fmfv{decor.shape=circle,decor.size=2.0thick,foreground=(0,,0,,1)}{v1}
\fmfv{decor.shape=circle,decor.size=2.0thick,foreground=(0,,0,,1)}{v2}
\fmf{phantom,tension=11}{i,v1}
\fmf{phantom,tension=11}{v2,o}
\fmf{dashes,left,tension=0.4}{v1,v2,v1}
\fmf{plain}{v1,v2}
\end{fmfgraph}
\end{fmffile}
\end{gathered} \hspace{-0.2cm} \left.\rule{0cm}{1.0cm}\right) \\
& + \mathcal{O}\big(\hbar^{3}\big)\;,
\end{split}
\label{eq:WmixedKjLoopExpansionStep2}
\end{equation}
with the superpropagator:
\begin{equation}
\mathcal{G}_{\Psi_\text{cl};\mathcal{JK}} = \begin{pmatrix}
\boldsymbol{G}_{\sigma_\text{cl};\mathcal{JK}} & \vec{F}_{\varphi_\text{cl};\mathcal{JK}} \\
\vec{F}^{\mathrm{T}}_{\varphi_\text{cl};\mathcal{JK}} & D_{\sigma_\text{cl};\mathcal{JK}}
\end{pmatrix}\;,
\end{equation}
and the Feynman rules:
\begin{subequations}
\begin{align}
%\begin{equation}
\left.
\begin{array}{ll}
\begin{gathered}
\begin{fmffile}{Diagrams/LoopExpansionMixedHS_FeynRuleVertexbis}
\begin{fmfgraph*}(4,4)
\fmfleft{i0,i1,i2,i3}
\fmfright{o0,o1,o2,o3}
\fmfv{label=$x$,label.angle=90,label.dist=4}{v1}
\fmfbottom{v2}
\fmf{phantom}{i1,v1}
\fmf{plain}{i2,v1}
\fmf{phantom}{v1,o1}
\fmf{plain}{v1,o2}
\fmf{wiggly,tension=0.6}{v1,v2}
\fmfv{decor.shape=circle,decor.size=2.0thick,foreground=(0,,0,,1)}{v1}
\fmflabel{$a$}{i2}
\fmflabel{$b$}{o2}
\end{fmfgraph*}
\end{fmffile}
\end{gathered} \\
\\
\begin{gathered}
\begin{fmffile}{Diagrams/LoopExpansionMixedHS_FeynRuleVertex2bis}
\begin{fmfgraph*}(4,4)
\fmfleft{i0,i1,i2,i3}
\fmfright{o0,o1,o2,o3}
\fmfv{label=$x$,label.angle=90,label.dist=4}{v1}
\fmfbottom{v2}
\fmf{phantom}{i1,v1}
\fmf{plain}{i2,v1}
\fmf{phantom}{v1,o1}
\fmf{plain}{v1,o2}
\fmf{dots,tension=0.6}{v1,v2}
\fmfv{decor.shape=circle,decor.size=2.0thick,foreground=(0,,0,,1)}{v1}
\fmflabel{$a$}{i2}
\fmflabel{$b$}{o2}
\end{fmfgraph*}
\end{fmffile}
\end{gathered} \\
\\
\begin{gathered}
\begin{fmffile}{Diagrams/LoopExpansionMixedHS_FeynRuleVertex3bis}
\begin{fmfgraph*}(4,4)
\fmfleft{i0,i1,i2,i3}
\fmfright{o0,o1,o2,o3}
\fmfv{label=$x$,label.angle=90,label.dist=4}{v1}
\fmfbottom{v2}
\fmf{phantom}{i1,v1}
\fmf{plain}{i2,v1}
\fmf{phantom}{v1,o1}
\fmf{dots}{v1,o2}
\fmf{dots,tension=0.6}{v1,v2}
\fmfv{decor.shape=circle,decor.size=2.0thick,foreground=(0,,0,,1)}{v1}
\fmflabel{$a$}{i2}
\fmflabel{$b$}{o2}
\end{fmfgraph*}
\end{fmffile}
\end{gathered}
\end{array}
\quad \right\rbrace &\rightarrow \sqrt{\lambda} \ \delta_{ab} \;, 
\label{eq:FeynRulesLoopExpansionMixedHSvertex} \\
%\end{equation}
%\begin{equation}
\begin{gathered}
\begin{fmffile}{Diagrams/LoopExpansionMixedHS_FeynRuleGbis}
\begin{fmfgraph*}(20,20)
\fmfleft{i0,i1,i2,i3}
\fmfright{o0,o1,o2,o3}
\fmflabel{$x, a$}{v1}
\fmflabel{$y, b$}{v2}
\fmf{phantom}{i1,v1}
\fmf{phantom}{i2,v1}
\fmf{plain,tension=0.6}{v1,v2}
\fmf{phantom}{v2,o1}
\fmf{phantom}{v2,o2}
\end{fmfgraph*}
\end{fmffile}
\end{gathered} \quad &\rightarrow \boldsymbol{G}_{\sigma_\text{cl};\mathcal{JK};ab}(x,y)\;,
\label{eq:FeynRulesLoopExpansionMixedHSG} \\
%\end{equation}
%\begin{equation}
\begin{gathered}
\begin{fmffile}{Diagrams/LoopExpansionMixedHS_FeynRuleDbis}
\begin{fmfgraph*}(20,20)
\fmfleft{i0,i1,i2,i3}
\fmfright{o0,o1,o2,o3}
\fmfv{label=$x$}{v1}
\fmfv{label=$y$}{v2}
\fmf{phantom}{i1,v1}
\fmf{phantom}{i2,v1}
\fmf{wiggly,tension=0.6}{v1,v2}
\fmf{phantom}{v2,o1}
\fmf{phantom}{v2,o2}
\end{fmfgraph*}
\end{fmffile}
\end{gathered} \quad &\rightarrow D_{\sigma_\text{cl};\mathcal{JK}}(x,y)\;,
\label{eq:FeynRulesLoopExpansionMixedHSD} \\
%\end{equation}
%\begin{equation}
\begin{gathered}
\begin{fmffile}{Diagrams/LoopExpansionMixedHS_FeynRuleFbis}
\begin{fmfgraph*}(20,20)
\fmfleft{i0,i1,i2,i3}
\fmfright{o0,o1,o2,o3}
\fmflabel{$x, a$}{v1}
\fmfv{label=$y$}{v2}
\fmf{phantom}{i1,v1}
\fmf{phantom}{i2,v1}
\fmf{dashes,tension=0.6}{v1,v2}
\fmf{phantom}{v2,o1}
\fmf{phantom}{v2,o2}
\end{fmfgraph*}
\end{fmffile}
\end{gathered} \quad &\rightarrow F_{\varphi_\text{cl};\mathcal{JK};a}(x,y)\;.
\label{eq:FeynRulesLoopExpansionMixedHSF1}
%\end{equation}
\end{align}
\end{subequations}

We also work out a formula expressing the multiplicities of all diagrams resulting from the mixed LE. In this way, we obtain:
\begin{equation}
\mathcal{M}_{\mathrm{LE},\mathrm{mix}}=\frac{(2p)!4^{p} N_{\mathrm{F}}}{(2!)^{S+D} N_{\mathrm{V}}}\;,
\label{eq:MultiplicityDiagLoopExpansionMixed}
\end{equation}
where $S$ and $D$ are now respectively the number of self and double connections with the propagator lines~\eqref{eq:FeynRulesLoopExpansionMixedHSG} and~\eqref{eq:FeynRulesLoopExpansionMixedHSF1} representing respectively $\boldsymbol{G}_{\sigma_{\mathrm{cl}};\mathcal{J}\mathcal{K}}$ and $\vec{F}_{\varphi_{\mathrm{cl}};\mathcal{J}\mathcal{K}}$ (self and double connections made of $D_{\sigma_{\mathrm{cl}};\mathcal{J}\mathcal{K}}$ propagators, i.e. made of wiggly lines according to rule~\eqref{eq:FeynRulesLoopExpansionMixedHSD}, are not possible). Note that $N_{\mathrm{V}}$ still denotes the number of vertex permutations that leave the diagram unchanged, $N_{\mathrm{F}}$ is to be specified below and $p$ equals half the number of vertices~\eqref{eq:FeynRulesLoopExpansionMixedHSvertex} involved in the diagram under consideration. Result~\eqref{eq:MultiplicityDiagLoopExpansionMixed} can actually be determined through slight modifications of Eq.~\eqref{eq:MultiplicityDiagLoopExpansion}. The $\boldsymbol{G}_{\sigma_{\mathrm{cl}};\mathcal{J}\mathcal{K}}$ and $D_{\sigma_{\mathrm{cl}};\mathcal{J}\mathcal{K}}$ propagators (i.e.~\eqref{eq:FeynRulesLoopExpansionMixedHSG} and~\eqref{eq:FeynRulesLoopExpansionMixedHSD}) contribute to $\mathcal{M}_{\mathrm{LE},\mathrm{mix}}$ respectively in the same way as the $\boldsymbol{G}_{\varphi_{\mathrm{cl}};JK}$ propagator (i.e.~\eqref{eq:FeynRulesLoopExpansionPropagator}) and the zigzag vertex (i.e.~\eqref{eq:FeynRulesLoopExpansion4legVertex}) to $\mathcal{M}_{\mathrm{LE},\mathrm{orig}}$. Due to the HST performed in the mixed representation, we have swapped the two interaction terms (with associated vertices~\eqref{eq:FeynRulesLoopExpansion3legVertex} and~\eqref{eq:FeynRulesLoopExpansion4legVertex}) of the original $\varphi^{4}$-theory for a Yukawa interaction (corresponding to rule~\eqref{eq:FeynRulesLoopExpansionMixedHSvertex}), thus inducing that the factor $(2p)!!(2q)!!$ in Eq.~\eqref{eq:MultiplicityDiagLoopExpansion} is replaced by $(2p)!$ in Eq.~\eqref{eq:MultiplicityDiagLoopExpansionMixed}. The other and last difference between the diagrammatic of the original and mixed LEs is the presence in the mixed case of the $\vec{F}_{\varphi_{\mathrm{cl}};\mathcal{J}\mathcal{K}}$ propagator which has no counterpart in the original LE. As opposed to all other propagators introduced so far, we must account for the possibility to exchange the extremities of the $\vec{F}_{\varphi_{\mathrm{cl}};\mathcal{J}\mathcal{K}}$ propagator (even though $\vec{F}^{\mathrm{T}}_{\varphi_{\mathrm{cl}};\mathcal{J}\mathcal{K};a}(x,y) \equiv \vec{F}_{\varphi_{\mathrm{cl}};\mathcal{J}\mathcal{K};a}(y,x) = \vec{F}_{\varphi_{\mathrm{cl}};\mathcal{J}\mathcal{K};a}(x,y)$ to ensure that $\mathcal{G}_{\Phi_{\mathrm{cl}};\mathcal{J}\mathcal{K}}^{\mathrm{T}}=\mathcal{G}_{\Phi_{\mathrm{cl}};\mathcal{J}\mathcal{K}}$) since only one of its two extremities is associated to a color index, as can be seen from rule~\eqref{eq:FeynRulesLoopExpansionMixedHSF1}. More specifically, for every loop exclusively made of $\vec{F}_{\varphi_{\mathrm{cl}};\mathcal{J}\mathcal{K}}$ propagators, we can switch the extremities of all these propagators at once without affecting the nature of the studied diagram. Such a property contributes a factor 2 to the multiplicity of the latter. This explains the dependence of $\mathcal{M}_{\mathrm{LE},\mathrm{mix}}$ with respect to the number $N_{\mathrm{F}}$ of loops exclusively made of $\vec{F}_{\varphi_{\mathrm{cl}};\mathcal{J}\mathcal{K}}$ propagators. The factors $S$, $D$, $N_{\mathrm{V}}$ and $N_{\mathrm{F}}$ associated to 2PI diagrams expressing the mixed 2PI EA $\Gamma_{\mathrm{mix}}^{(\mathrm{2PI})}\big[\Phi,\mathcal{G}\big]$ are given by Tab.~\ref{tab:MultiplicityMixedEAdiagramsON} and their multiplicities can also be inferred from Eq.~\eqref{eq:MultiplicityDiagLoopExpansionMixed} (2PI diagrams contributing to $W^\text{LE;mix}[\mathcal{J},\mathcal{K}]$ determined from the mixed LE are also essentially given by Tab.~\ref{tab:MultiplicityMixedEAdiagramsON}, except that all of their propagator lines are black and not red).

\onecolumn

\begin{longtable}{|M{0.15\textwidth}|M{0.14\textwidth}M{0.14\textwidth}M{0.14\textwidth}M{0.14\textwidth}M{0.14\textwidth}|}
\caption{\label{tab:MultiplicityOPTdiagramsON} Diagrams involved in the expansion of $W^\text{OPT}$ (expressed by Eq.~\eqref{eq:WKjLoopExpansionStep3OPT}) up to order $\mathcal{O}\big(\delta^{3}\big)$. The diagrams are sorted according to the connected correlation functions that generate them in the OPT expansion (the corresponding expectation value $\big\langle\cdots\big\rangle_{0,\sigma}$ is defined by Eqs.~\eqref{eq:OPTsigmaDepExpValueNumber1} and~\eqref{eq:OPTsigmaDepExpValueNumber2}), as indicated by the left-hand column. The multiplicity $\mathcal{M}_{\mathrm{OPT}}$ of a given diagram can be deduced from Eq.~\eqref{eq:MultiplicityDiagOPT} together with the corresponding vector $(S,D;N_{\mathrm{V}})$ given below (see text below Eqs.~\eqref{eq:MultiplicityDiagLoopExpansion} and~\eqref{eq:MultiplicityDiagOPT} for the definitions of $S$, $D$ and $N_{\mathrm{V}}$).} \\ \hline
 & & & & & \\
 \scalebox{1.2}{Contribution} & & & \scalebox{1.2}{Diagrams} & & \\ \hline
  & & & & & \\
 \scalebox{0.8}{$\sigma\left\langle\vec{\widetilde{\chi}}^{2}\right\rangle_{0,\sigma}^{\text{c}}$} & & & \begin{fmffile}{Diagrams/OPTtable_Diag1}
\begin{fmfgraph}(10,10)
\fmfleft{i}
\fmfright{o}
\fmfv{decor.shape=square,decor.size=2.5thick,foreground=(0,,0,,1)}{v1}
\fmftop{v3}
\fmfbottom{v4}
\fmf{phantom,tension=30}{i,v1}
\fmf{phantom,tension=30}{v2,o}
\fmf{plain,left=0.43,tension=0.5}{v1,v3}
\fmf{plain,left=0.43,tension=0.5}{v3,v2}
\fmf{plain,left=0.43,tension=0.5}{v2,v4}
\fmf{plain,left=0.43,tension=0.5}{v4,v1}
\end{fmfgraph}
\end{fmffile} & & \\
  & & & $(1,0;1)$ & & \\
  & & & & & \\ \hline
  & & & & & \\
\hline
\end{longtable}

\begin{longtable}{|M{0.15\textwidth}|M{0.14\textwidth}M{0.14\textwidth}M{0.14\textwidth}M{0.14\textwidth}M{0.14\textwidth}|}
\caption{\label{tab:MultiplicityMixedEAdiagramsON} Diagrams contributing to the 2PI EA $\Gamma_{\mathrm{mix}}^{(\mathrm{2PI})}\big[\Phi,\mathcal{G}\big]$ up to order $\mathcal{O}\big(\hbar^{4}\big)$ (with $\vec{F}=\vec{0}$ at order $\mathcal{O}(\hbar^{4})$). The left-hand column indicates the order at which the diagrams contribute to the expansion of $\Gamma_{\mathrm{mix}}^{(\mathrm{2PI})}\big[\Phi,\mathcal{G}\big]$. The multiplicity $\mathcal{M}_{\mathrm{LE},\mathrm{mix}}$ of a given diagram can be deduced from Eq.~\eqref{eq:MultiplicityDiagLoopExpansionMixed} together with the corresponding vector $(S,D;N_{\mathrm{V}},N_{\mathrm{F}})$ given below (see text below Eq.~\eqref{eq:MultiplicityDiagLoopExpansionMixed} for the definitions of $S$, $D$, $N_{\mathrm{V}}$ and $N_{\mathrm{F}}$).} \\ \hline
 & & & & & \\
 \scalebox{1.2}{Order} & & & \scalebox{1.2}{Diagrams} & & \\ \hline
  & & & & & \\
 $\mathcal{O}\big(\hbar^{2}\big)$ & & \begin{fmffile}{Diagrams/MixedEAtable_Fock}
\begin{fmfgraph}(12.5,12.5)
\fmfleft{i}
\fmfright{o}
\fmfv{decor.shape=circle,decor.size=2.0thick,foreground=(0,,0,,1)}{v1}
\fmfv{decor.shape=circle,decor.size=2.0thick,foreground=(0,,0,,1)}{v2}
\fmf{phantom,tension=11}{i,v1}
\fmf{phantom,tension=11}{v2,o}
\fmf{plain,left,tension=0.4,foreground=(1,,0,,0)}{v1,v2,v1}
\fmf{wiggly,foreground=(1,,0,,0)}{v1,v2}
\end{fmfgraph}
\end{fmffile} & & \begin{fmffile}{Diagrams/MixedEAtable_Diag1}
\begin{fmfgraph}(12.5,12.5)
\fmfleft{i}
\fmfright{o}
\fmfv{decor.shape=circle,decor.size=2.0thick,foreground=(0,,0,,1)}{v1}
\fmfv{decor.shape=circle,decor.size=2.0thick,foreground=(0,,0,,1)}{v2}
\fmf{phantom,tension=11}{i,v1}
\fmf{phantom,tension=11}{v2,o}
\fmf{dashes,left,tension=0.4,foreground=(1,,0,,0)}{v1,v2,v1}
\fmf{plain,foreground=(1,,0,,0)}{v1,v2}
\end{fmfgraph}
\end{fmffile} & \\
  & & \hspace{0.08cm}$(0,1;2,1)$ & & \hspace{0.08cm}$(0,1;2,2)$ & \\
\hline
\end{longtable}

\twocolumn

%%%%%%%%%%%%%%%%%%%%%%%%%%%%%%%
%%%%%%%%%%%%%%%%%%%%%%%%%%%%%%%
%%%%%%%%%%%%%%%%%%%%%%%%%%%%%%%
\clearpage

\bibliography{paperDiag_0DON}

\end{document}